\documentclass[12pt]{cit_thesis}
\usepackage{multirow}
\usepackage{rotating}
\usepackage{epsfig}
\usepackage{graphicx}
\usepackage{amsmath}

\def\gtorder{\mathrel{\raise.3ex\hbox{$>$}\mkern-14mu
             \lower0.6ex\hbox{$\sim$}}}
\def\ltorder{\mathrel{\raise.3ex\hbox{$<$}\mkern-14mu
             \lower0.6ex\hbox{$\sim$}}}
\def\micro{\mu}

\title{\bfseries Precision Rosenbluth Measurement of the Proton Elastic Electromagnetic 
Form Factors and Their Ratio at $Q^2$ = 2.64, 3.20, and 4.10 GeV$^2$}

\author{Issam A. Qattan}

\date{December 2005}

\begin{document}

\begin{frontmatter}

\maketitle

\makecopyright

\vfill
\pagebreak

\begin{abstract}
\begin{center}
\vspace{-.6cm}
\large \bf{ 
Precision Rosenbluth Measurement of the Proton Elastic Electromagnetic
Form Factors and Their Ratio at $Q^2$ = 2.64, 3.20, and 4.10 GeV$^2$}\\
{\large Issam A. Qattan}\\
\end{center}
\vspace{.05cm}
Due to the inconsistency in the results of the $\frac{\mu_{p} G_{Ep}}{G_{Mp}}$ ratio
of the proton, as extracted from the Rosenbluth and recoil polarization techniques, high precision 
measurements of the e-p elastic scattering cross sections were made at $Q^2$ = 2.64, 3.20, and 4.10 GeV$^2$.
Protons were detected, in contrast to previous measurements where the scattered electrons were 
detected, which dramatically decreased $\varepsilon$ dependent systematic uncertainties and 
corrections. A single spectrometer measured the scattered protons of interest while simultaneous measurements at 
$Q^2$ = 0.5 GeV$^2$ were carried out using another spectrometer which served as a luminosity monitor in order to remove 
any uncertainties due to beam charge and target density fluctuations. The absolute uncertainty in the measured cross 
sections is $\approx$ 3\% for both spectrometers and with relative uncertainties, random and slope, below 1\% 
for the higher $Q^2$ protons, and below 1\% random and 6\% slope for the monitor spectrometer. 
The extracted electric and magnetic form factors were determined to 4\%-7\% for $G_{Ep}$ and 1.5\% for $G_{Mp}$. 
The ratio $\frac{\mu_{p}G_{Ep}}{G_{Mp}}$ was determined to 4\%-7\% and showed $\frac{\mu_{p}G_{Ep}}{G_{Mp}} \approx$ 1.0. 
The results of this work are in agreement with the previous Rosenbluth data and inconsistent with high-$Q^2$ recoil polarization 
results, implying a systematic difference between the two techniques.

\end{abstract}

\begin{acknowledgements}
My most important acknowledgment is to my caring and loving family and in particular my mother 
and my father, for without their love and support none of this would have been possible. 
Their tremendous support all the time motivated, inspired, and kept me alive and helped keep my sane.
No words can express my thanks and gratitude to my parents for all that they have gone through and
done for me. {\bf So, father and mother: this work is dedicated to you with great love and admiration.}

This thesis is a direct result of the dedication of a great number of people. In particular, I am greatly indebted to my Ph.D.
and research advisors Professor Ralph E. Segel from Northwestern University and Dr. John R. Arrington from Argonne National
Laboratory, for without their help, steady support, and encouragement 
this work would not have been possible as well. Their dedication to this work, patience, and ``way'' too many heated 
and rousing discussions guided me through too many dark days and tough times. {\bf Ralph and John to you I say: 
I am extremely grateful for all you have done for me and in particular giving me the opportunity to work on 
a such high-profile experiment.} Also I like to thank Professors Heidi Schellman and David Buchholz for their enthusiasm for this work
and being a committee members.

I gratefully acknowledge the staff of the Accelerator Division, the Hall A technical staff, the members of the
survey and cryotarget groups at the Thomas Jefferson National Laboratory for their efforts in making this experiment possible. 
I like to acknowledge all of those ``without going through the full list of names'' who took shifts on the E01-001 experiment and kept 
the data flowing. In particular I would like to acknowledge the hard work and long hours on shifts put in by my fellow graduate 
students at JLAB. Needless to say that I am extremely grateful to Dr. Mark K. Jones for his guidance and patience during the 
initial stages of the data analysis while I resided at JLAB for almost year and a half. 
I am grateful to the physics division at Argonne National Laboratory and in particular the medium energy group for providing
me with the best working environment to carry out the data analysis for this work. In particular I like to thank
Dr. Xiaochao Zheng for performing the simulations for the E01-001 experiment. 

Finally, how can I forget the constant support of my sister ``Rima'', brother ``Diaa'', and brother-in-law ``Rammy'' 
who kept me alive to date. Rima has been the world to me and filled my life with love and joy. Of course my niece ''Lana'', 
nephew ``Ryan'', and niece ``Tala'' have been everything I ever wanted in life. Thank you for all the therapeutic and useful 
distractions during the course of this work. Thank you for opening your California home to me to visit and seek refuge.

\begin{figure}[htb]
\begin{center}
\epsfig{file=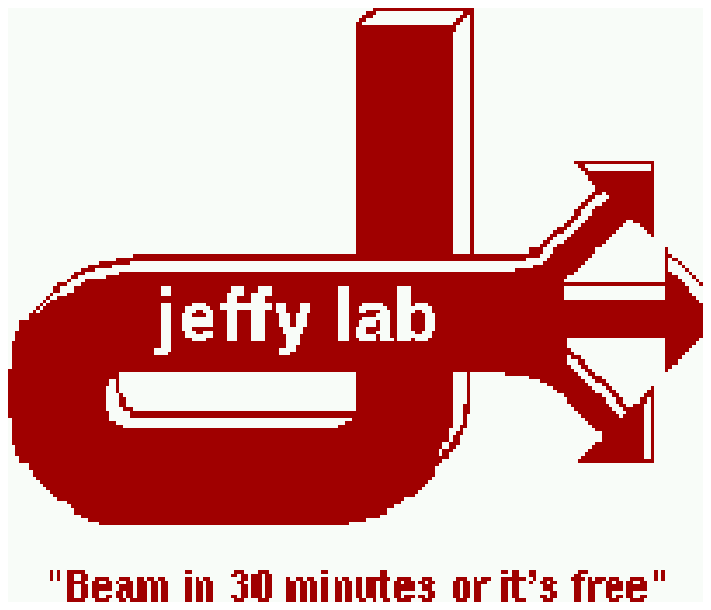}
\end{center}
\end{figure}

\end{acknowledgements}
\tableofcontents
\newpage
\addcontentsline{toc}{chapter}{List of Figures}
\listoffigures
\newpage
\addcontentsline{toc}{chapter}{List of Tables}
\listoftables
\end{frontmatter}

\chapter{Introduction} \label{chap_intro}
\pagenumbering{arabic}
\section{Overview}

The modern concept of the atom as an object composed of a dense nucleus 
surrounded by an electron cloud came to life as a result of Rutherford's experiments 
in 1911. As time went on, it became clear that the nucleus itself is composed of 
even smaller objects or particles which we call nucleons. The fact that the proton's magnetic moment
is 2.79$\mu_{N}$ ($\mu_{N}$ = $\frac{e\hbar}{2M_{p}c}$), rather
than the 0.5$\mu_{N}$ expected for a point-like spin 1/2 charged particle demonstrates that the proton posses
a structure. Similarly, the neutron's magnetic moment of -1.91$\mu_{N}$ is very different
from the 0.0$\mu_{N}$ that a point neutral particle would have. Thus, nucleons, protons
and neutrons, which were once thought to be the fundamental building blocks of matter in 
fact are complex particles that turn out to be composed into even smaller objects which we 
call quarks and gluons. The whole picture boils down to electrons (leptons), 
quarks, and gluons as the fundamental types of building blocks.
  
Several challenging questions and issues arose when physicists tried to understand 
nucleons. For example, what are these particles made off, how do 
they interact, and, most importantly, how do they bind together? These are fundamental 
and complicated questions that must be addressed, answered, and resolved by the nuclear 
and particle physics communities. The strong force is the reason why protons and neutrons 
bind together to form the nuclei. It is the theory of Quantum Chromodynamics (QCD) that 
describes the strong interaction between the quarks and gluons, which in turn make up the 
protons and neutrons.

   The strong coupling constant, $\alpha_{s}$, as defined by the theory of Quantum 
Chromodynamics \cite{halzen84}, is a measure of the strength of the strong interaction between 
quarks and gluons and is a function of the four momentum transfer squared $Q^2$:
\begin{equation} \label{eq:eps}
\alpha_{s}(Q^2) = \frac{12\pi}{(33-2N_{f})\ln(\frac{Q^2}{\Lambda^2_{QCD}})}~,
\end{equation} 
where $N_{f}$ is the number of flavors active for QCD renormalization scale 
$\Lambda_{QCD}$ \cite{brodsky73, brodsky75}.

When $Q^2$ is large (short distance scale), equation (\ref{eq:eps}) tells us that 
$\alpha_{s}$ is small. Therefore, at large momentum transfer, the quarks within a hadron 
(neutron or proton) are weakly interacting and behave almost as if they are free particles. 
On the other hand, when $Q^2$ is small (large distance scale), $\alpha_{s}$ is large and the quarks 
are strongly interacting particles and hence form hadronic matter.

\pagestyle{myheadings}
\newcommand{\sekshun}[1]
{
\section{#1}
\markboth{#1 \hfill}{#1 \hfill}
}
\section{Electron Scattering}

In order to reveal the underlying structure of the nucleon, experimental techniques 
such as electron scattering have been developed. Electron scattering in particular has 
proven to be a powerful tool for studying the structure of the nucleus. The reason 
lies in the fact that the electron is a point-like particle and has no internal structure, 
making it a clean probe of the target nucleus. In this case, the information extracted 
such as the differential cross section reflects the structure of the target without any 
contribution from the projectile. The incident electron is scattered off a nuclear target 
(single particle in the target) by exchanging a virtual photon. The electron-photon scattering 
vertex is known and understood within the theory of QED.

In electron scattering experiments, highly relativistic electrons are used.
At low energy transfers, the virtual photon interacts with the 
entire nucleus. The nucleus stays intact and the electron scatters elastically or excites
nuclear states. In this case, the virtual photon is dominantly interacting with the nuclear target by 
coupling to a vector meson resonance states or $q\bar{q}$ \cite{hohler76,gari86}.

As the energy transfer increases and becomes larger than the nuclear binding energy, 
we enter the quasi-elastic region. The virtual photon becomes more sensitive to the internal 
structure of the nuclear target and the scattering process is described in terms of 
photon-nucleon coupling rather than photon-meson coupling. In this case, the target 
is viewed by the virtual photon as a set of quasi-free nucleons. The electron scatters 
elastically from the nucleon which in turn is ejected from the nucleus.

By increasing $Q^2$ and the energy transfer further, we enter the resonance region. 
In this case, the virtual photon becomes sensitive to the internal structure of the nucleon. 
The quarks inside the nucleon absorb the virtual photons and form excited states 
which we refer to as nucleon-resonances.
 
If we increase $Q^2$ and the energy transfer even further, we enter the deep inelastic region. 
The virtual photon probes smaller distance scales and has enough energy to resolve the 
constituents of the nucleon, or partons. Perturbative QCD models (pQCD) \cite{lepage79,lepage80} seem 
to give the correct description of the scattering process at high $Q^2$.

\begin{figure}[!htbp] 
\begin{center}
\epsfig{file=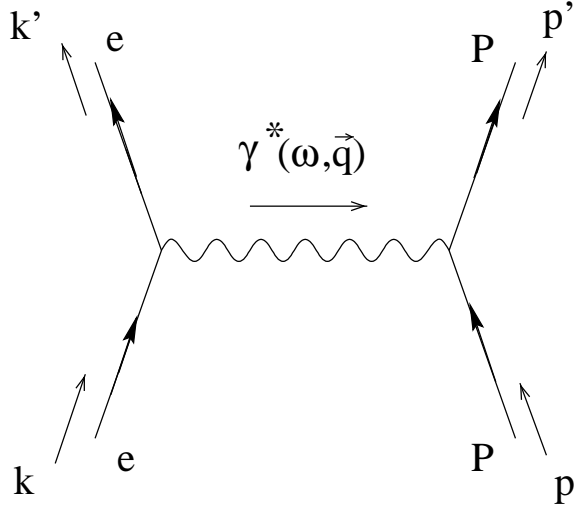,width=3in}
\end{center}
\caption[Feynman digram for electron-nucleus scattering]
{Feynman digram for electron-nucleus scattering.}
\label{fig:Feynman}
\end{figure}

In electron scattering experiments, see Figure \ref{fig:Feynman}, the initial 
electron's energy and momentum $k = (E, \vec k)$ are known. Elastic electromagnetic 
scattering of the electron can be described, to lowest
order in the electromagnetic coupling constant $\alpha$, as the exchange of a single 
virtual photon $\gamma^{*}$ of momentum $\vec{q}$ and energy $\omega$. The virtual photon 
then interacts with the target. After the scattering, the final electron's energy and 
momentum $k' = (E', \vec{k'})$ are measured. This allows us to determine the energy 
$\omega = (E-E')$ and momentum $\vec q = (\vec k - \vec{k'})$ of the virtual photon. 
Usually we refer to the energy of the virtual photon, $\omega = (E-E')$, as the energy 
transfer.

We describe the scattering process using two main variables,  
the energy transfer to the virtual photon, $\omega = E - E'$, and the square 
of the 4-momentum transfer, $Q^2 = -q_\mu q^\mu = -q^2 = | \vec k - \vec{k'} |^2- (E-E')^2$.  
We define the Bjorken variable $x = {Q^2 \over 2M_{p}\omega }$, 
where $M_{p}$ is the mass of the nucleon. The Bjorken variable $x$ varies between 0 and 1.0, 
where the $x=1.0$ case corresponds to elastic scattering, and the $x<1.0$ case corresponds 
to inelastic scattering.

The 4-momentum transfer $Q^2= - q^2$ is given in terms of the electron's kinematics 
(initial energy $E$, final energy $E'$, and the scattering angle $\theta_{e}$) as:
\begin{equation}
Q^{2} = 4EE'\sin^2 \frac{\theta_{e}}{2}~. 
\end{equation}

We refer to the scattering as an exclusive scattering if the final state is fully identified.
In this case we can also express the scattering kinematics in terms of the knocked out or recoil target (nucleon) 
rather than the electron. If the initial and the final four-momentum of the nucleon are defined as $P = (E_{p}, \vec p)$ and 
$P' = (E'_{p}, \vec{p'})$, where $E_{p}$ and $E'_{p}$ are the initial and final energy 
of the nucleon and $\vec{p}$ and $\vec{p'}$ are the initial and final 3-momentum of the 
nucleon, we can write the the 4-momentum transfer $Q^2$ as:
\begin{equation}
Q^{2} = -q^2 = -(P' - P)^2 = \Big ( (E'_{p}-E_{p})^2 -(\vec {p'} - \vec p)^2 \Big )~.
\end{equation}

For the case where the final hadronic state contains a single nucleon only, the
missing mass squared of the scattered hadron $W^{2}$ must equal the mass of the nucleon 
squared, $M_{p}^{2}$, where we define the missing mass squared as:
\begin{equation} \label{eq:missing_mass2}
W^{2} = M_{p}^2 + 2M_{p}\omega  - Q^{2}~.
\end{equation}

Examining equation (\ref{eq:missing_mass2}), we see that if the final state contains only 
a single nucleon, then $W^2 = M_{p}^2$ and $Q^{2} = 2M_{p}\omega$ yielding a value of 
$x = {Q^2 \over 2M_{p}\omega } = 1$ for the Bjorken variable. So by choosing scattering kinematics where
$x = 1$, we can isolate elastic scattering.

If the nucleus is knocked into an excited state, there is some 
additional energy transfer, and $x$ will decrease as the energy transfer increases. 
At somewhat higher energy transfer, where quasielastic scattering is the dominant process, 
the electron knocks a single nucleon out of the nucleus.  
This corresponds to scattering near $x=1$. At higher energy transfers, 
corresponding to $x<1$, the scattering is inelastic and the struck nucleon is either excited 
into a higher energy state (resonance scattering), or temporarily fragmented 
(deep inelastic scattering). At very high energy transfers, where deep
inelastic scattering dominates, the electron is primarily interacting with a
single quark via the virtual photon $\gamma^{*}$. In this case, a short wavelength virtual 
photon $\gamma^{*}$ is needed in order to resolve the quarks within the proton.

\section{Elastic Electron-Proton Scattering} \label{elastic_scattering}

The electron-photon vertex as shown in Figure \ref{fig:Feynman} 
is well understood and described by QED. On the other hand, the photon-proton
vertex is complicated and the detail of such interaction,~$\gamma^{*}+P(p) \to P(p')~$, 
cannot be calculated from first principles. This is due to the fact that the proton is 
not a point-like particle but rather a particle with internal structure. 
We introduce two $Q^2$-dependent functions that contain all the information about 
the internal structure of the proton. We refer to these two functions as the proton 
electromagnetic form factors which parameterize the internal structure of the proton.

For the single-photon exchange diagram for the electron-proton elastic scattering shown 
in Figure \ref{fig:Feynman}, the Lorentz invariant transition matrix element is given by:
\begin{equation} \label{eq:lorentz_trans_matrix}
M_{fi} = \Big (j_{\mu}(k,k') \frac{1}{q^2} J^{\mu}(p,p') \Big)~,
\end{equation}  
where $j_{\mu}$ and $J^{\mu}$ are the electromagnetic currents for the electron and
proton, respectively: 
We can express the electromagnetic currents as:
\begin{equation}
j_{\mu} = -e\bar{u}(k')\gamma_{\mu}u(k)~,
\end{equation}
\begin{equation}
J^{\mu} = e\bar{v}(p')\Gamma^{\mu}v(p)~,
\end{equation}  
where ${u}(k)$, $\bar{u}(k')$, ${v}(p)$, and $\bar{v}(p')$ are the four-component Dirac 
spinors for the initial and final electron and proton, respectively, which appear in the
plane-wave solutions for Dirac equation, and $\Gamma^{\mu}$ and $\gamma_{\mu}$ are the Dirac 4x4 matrices. 
Therefore, the Lorentz invariant transition matrix element can be expressed as:
\begin{equation}
iM_{fi} = \frac{-i}{q^2} \Big[ie\bar{v}(p')\Gamma^{\mu}(p,p')v(p)\Big] \Big[ie\bar{u}(k')\gamma_{\mu}u(k)\Big]~.
\end{equation} 

Because of the composite nature of the proton,~$\Gamma^{\mu}$ rather than $\gamma_{\mu}$ is 
used to describe the proton current since $\Gamma^{\mu}$ contains all the information about 
the internal electromagnetic structure of the proton. It is worth mentioning that if the proton 
were to be treated as a point-like particle, then $\Gamma^{\mu} \to \gamma_{\mu}$ as in the case
with the electron. The proton current $J^{\mu}$ is a Lorentz four-vector that is Lorentz invariant 
and satisfies both parity and current conservations in electromagnetic interactions, 
i.e. $\partial J^{\mu} = 0$. With this in mind, the proton current can be written as:
\begin{equation}
J^{\mu} = \bar{v}(p') \Big(F_{1}(q^2)\gamma_{\mu} + \frac{i\kappa_{p}}{2M_{p}} F_{2}(q^2)\sigma_{\mu\nu}q^{\nu} \Big) v(p)~, 
\end{equation} 
where $\kappa_{p} = 1.793 \mu_{N}$ is the proton anomalous magnetic moment and it is expressed in 
the unit of nuclear magneton $(\mu_{N})$, $M_{p}$ is the mass of the proton, $\sigma_{\mu\nu} = \frac{i}{2}[\gamma^{\mu},\gamma^{\nu}]$,
and the factor $\frac{i\kappa_{p}}{2M_{p}}$ has been included as a matter of convention. The functions $F_{1}(q^2)$ 
and $F_{2}(q^2)$ are functions of $q^{2}$ and known as Dirac and Pauli form factors, respectively. The Dirac form factor,~$F_{1}(q^2)$, 
is used to describe the helicity-conserving scattering amplitude while the Pauli form factor,~$F_{2}(q^2)$, 
describes the helicity-flip amplitude. In the limit that $q^{2} \to 0$, the structure functions
$F_{1}(q^2=0) = F_{2}(q^2=0) = 1.0$, and in this limit, the virtual photon becomes insensitive to the internal 
structure of the proton which is viewed as a point-like particle. 

The elastic differential cross section in the lab frame for the $~e(k)+P(p) \to e(k')+P'(p')~$ reaction is given by:
\begin{equation} \label{eq:ep_diff_cross_section}
\frac{d\sigma}{d\Omega} = \Bigg[\frac{|M_{fi}|^{2}}{4 \Big((k\cdot p)^{2} - m_{e}^2 M_{p}^2 \Big)} (2\pi)^{4}\delta^{4}\Big(k'+p-k-p'\Big) \frac{d^{3}k' d^{3}p'}{(2\pi)^{3}2E'(2\pi)^{3}2(M_{p}+\omega)}\Bigg]~.
\end{equation} 

By using the scattering amplitude $M_{fi}$ expressed in terms of the electron and proton currents, 
integrating out the $\delta$-function that imposes momentum conservation, and finally expressing 
the proton current in terms of the Dirac and Pauli form factors, the differential cross section for 
an unpolarized beam and target can be written as:
\begin{eqnarray} \label{eqnarray:diffrentional1}
\frac{d\sigma}{d\Omega dE'}& =& \frac{\alpha^{2}}{4E^{2}\sin^{4}({\frac{\theta_{e}}{2}})} \frac{E'}{E} \Bigg[ \Big( F_{1}^{2} + \frac{\kappa_{p}^{2}Q^2}{4M_{p}^{2}}F_{2}^{2} \Big) \cos^{2} ({\frac{\theta_{e}}{2}})
                                                                                            \nonumber\\
& & + \frac{Q^2}{4M_{p}^{2}}(F_{1}+\kappa_{p}F_{2})^{2} \sin^{2}({\frac{\theta_{e}}{2}}) \Bigg] \delta(E-E'-\frac{Q^2}{2M_{p}})~,
\end{eqnarray}
where we have averaged over initial spins and summed over final spins. It is understood that the 
$\delta$-function is used to assure elastic scattering, that is, at a given energy $E$ and angle $\theta_{e}$, the elastic differential cross section is a $\delta$-function in $E'$ at $E'= E - \frac{Q^2}{2M_{p}}$. 
If we integrate over $E'$ in equation (\ref{eqnarray:diffrentional1}) above and divide the numerator 
and denominator by $\cos^{2}({\frac{\theta_{e}}{2}})$, we can write the elastic cross section as:
\begin{equation} \label{eq:diffrentional2}
\frac{d\sigma}{d\Omega} = \sigma_{ns} \Bigg[ \Big( F_{1}^{2} + \frac{\kappa_{p}^{2}Q^2}{2M_{p}^{2}}F_{2}^{2} \Big) + \frac{Q^2}{2M_{p}^{2}}(F_{1}+\kappa_{p}F_{2})^{2}\tan^{2}({\frac{\theta_{e}}{2}}) \Bigg]~,
\end{equation}
where $\sigma_{ns}$ in equation (\ref{eq:diffrentional2}) above is known as the non-structure cross section 
and is given by:
\begin{equation}  \label{eq:nonstructure}
\sigma_{ns} = \frac{\alpha^{2} \cos^{2}({\frac{\theta_{e}}{2}})}{4E^{2}\sin^{4}({\frac{\theta_{e}}{2}})} \frac{E'}{E}~,
\end{equation}
where $\alpha$ is the fine structure constant. 

It is worth mentioning that $\sigma_{ns}$, equation (\ref{eq:nonstructure}), is nothing but the famous 
Rutherford formula for the elastic electron-proton scattering modified to account for the proton's recoil 
and the spin-orbit coupling effects:

\begin{enumerate}
\item {The recoil effect: The term $\frac{E'}{E}$, which is relativistic in nature, is due to the recoil 
of the proton. Although the protons are massive, their recoil effect can not be neglected at high momentum 
transfer squared $Q^{2}$. We can write $\frac{E'}{E}$ in terms of the electron's kinematics as:
\begin{equation}  \label{eq:recoil_effect}
\frac{E'}{E} = \frac{1}{ 1 + \frac{2E}{M_{p}} \sin^{2}({\frac{\theta_{e}}{2}})}~,
\end{equation}
where it should be clear that the term $(\frac{E'}{E}) \to 1.0$ as  $Q^{2} \to 0$.}
\item {The spin-orbit coupling: The electron is a spin-$\frac{1}{2}$ particle and has a magnetic 
moment $\mu_{e}$ which interacts with the magnetic field $B_{p}$ of the proton as it is felt
by the electron in its own frame of reference. This is referred to as the spin-orbit coupling effect 
which results in the $\cos^{2}({\frac{\theta_{e}}{2}})$ term or more precisely, 
$1 - \beta^{2}\sin^{2}({\frac{\theta_{e}}{2}})$, with $\beta = (\frac{v}{c}) \to 1.0 $ for extremely relativistic electrons.}
\end{enumerate}

We refer to the cross section without taking into account the recoil effect 
as the Mott cross section and it is given by:
\begin{equation} \label{eq:Mott}
\Big(\frac{d\sigma}{d\Omega}\Big)_{Mott}=\frac{(Z\alpha)^{2}\Big(1 - \beta^{2}\sin^{2}({\frac{\theta_{e}}{2}})\Big)}{4k^{2}\sin^{4}({\frac{\theta_{e}}{2}})}~,
\end{equation}
where $k$ is the initial momentum of the incident electron and Z is the atomic number. 
In the case of a high momentum electron scattered off a spin-1/2 point-like proton $(Z=1)$, 
we recover equation (\ref{eq:nonstructure}) 
without accounting for the recoil effect i.e. $(\frac{E'}{E} \to 1)$. 

In the non-relativistic limit where $\beta \to 0, E \to k$, and $\frac{E'}{E} \to 1$, 
equation (\ref{eq:nonstructure}) becomes:
\begin{equation} \label{eq:Rutherford}
\Big( \frac{d\sigma}{d\Omega} \Big)_{Rutherford} = \frac{\alpha^{2}}{4k^{2}\sin^{4}({\frac{\theta_{e}}{2}})}~,
\end{equation}
and this is the famous Rutherford cross section formula.

To put things in more perspective we can write:
\begin{equation} 
\sigma_{ns} = \Big(\frac{d\sigma}{d\Omega}\Big)_{Rutherford} \Big[1 - \beta^{2}\sin^{2}({\frac{\theta_{e}}{2}})\Big] \Big(\frac{E'}{E}\Big) = \Big(\frac{d\sigma}{d\Omega}\Big)_{Mott}  \Big(\frac{E'}{E}\Big)~,
\end{equation}
which describes the proton as a spin-1/2 point-like particle without any internal structure.

\section{Form Factor Interpretations}  \label{sect_interpretation}

As shown earlier the concept of the nucleus as a point-like 
particle with a static charge distribution $\varrho(\vec{r}) = Ze\delta(\vec{r})$ 
is not the correct one. This is due to the fact that the nucleus has an internal substructure 
in the form of electric charge and magnetic moment distribution. This substructure modifies the 
non-structure scattering cross section as given by equation (\ref{eq:nonstructure}) to:
\begin{equation} \label{eq:modify_nonstructure}
\frac{d\sigma}{d\Omega} = \Big( \frac{d\sigma}{d\Omega} \Big)_{ns} |F(Q^{2})|^{2}~,
\end{equation}   
where $F(Q^{2})$ is the form factor which accounts for the fact that the target nucleon 
possess an internal structure. It should be mentioned that $F$ is a function of 
$Q^{2}$ alone in the case of an elastic scattering.

It can be shown \cite{halzen84} that in the non-relativistic limit, the form factor  
$F(Q^{2})$ can be expressed in terms of the charge distribution $\varrho(\vec{r})$ of the target
nucleus as:
\begin{equation} \label{eq:form_factor}
F_{NR}(Q^{2}) = \int \varrho(\vec{r})e^{i\vec{Q}.\vec{r}} d^{3}\vec{r}~, 
\end{equation} 
where $NR$ stands for Non-Relativistic. Equation (\ref{eq:form_factor}) is the Fourier transform 
of the electric charge distribution of the target. In the case of the proton, $F_{NR}(Q^{2})$ corresponds 
to the electric form factor, $G_{Ep}(Q^{2})$. In the same way, if the target has an extended 
magnetic moment distribution, the magnetic form factor is the Fourier transform of the magnetic moment 
distribution of the target and is referred to as $G_{Mp}(Q^{2})$.

\begin{figure}[!htbp]
\begin{center}
\epsfig{file=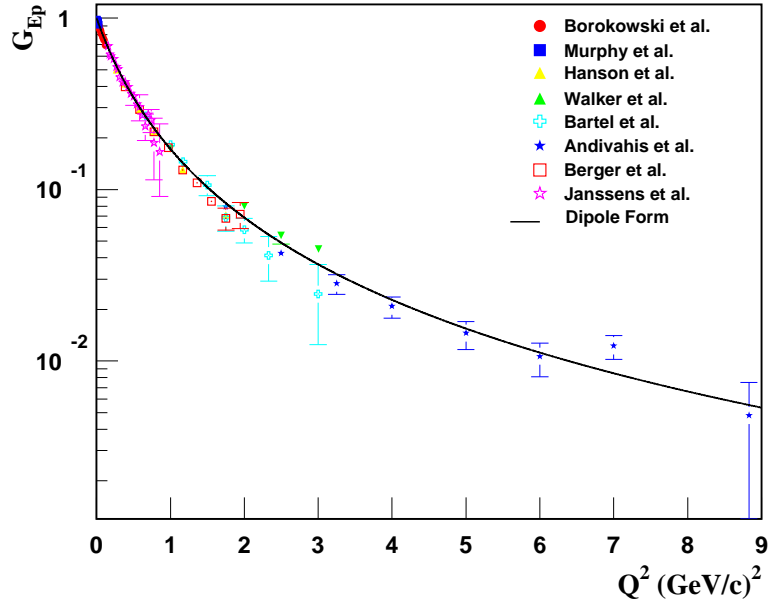,width=4in}
\end{center}
\caption[$G_{Ep}$ world data and the dipole form factor fit.]
{$G_{Ep}$ world data. The black solid line is the dipole form factor fit to data.
Data from references \cite{borkowski75,murphy74,hanson73,walker94,bartel73,andivahis94,berger71,   
janssens66}.}
\label{fig:gep}
\end{figure}

\begin{figure}[!htbp]
\begin{center}
\epsfig{file=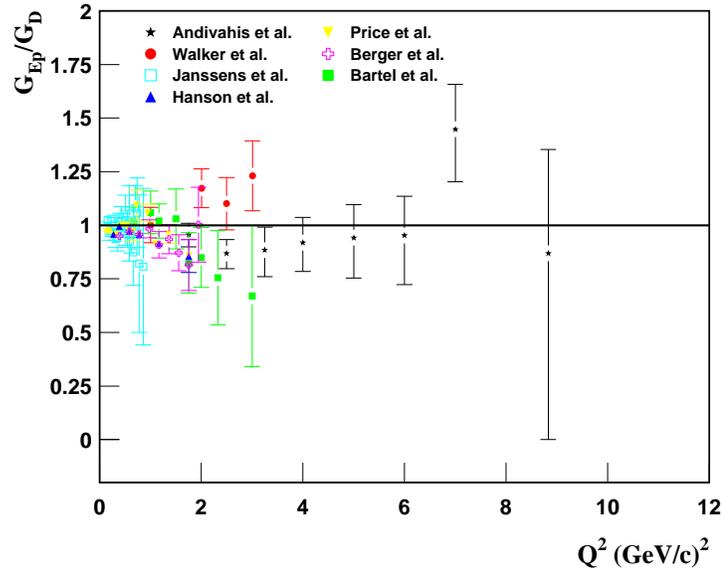,width=3.74in}
\end{center}
\caption[$G_{Ep}$ world data normalized to the dipole form factor $G_{D}$.]
{$G_{Ep}$ world data normalized to the dipole form factor $G_{D}$.
Data from references \cite{murphy74,hanson73,walker94,bartel73,andivahis94,berger71,janssens66,
price71}.}  
\label{fig:gep_gd}
\end{figure}

\begin{figure}[!htbp]
\begin{center}
\epsfig{file=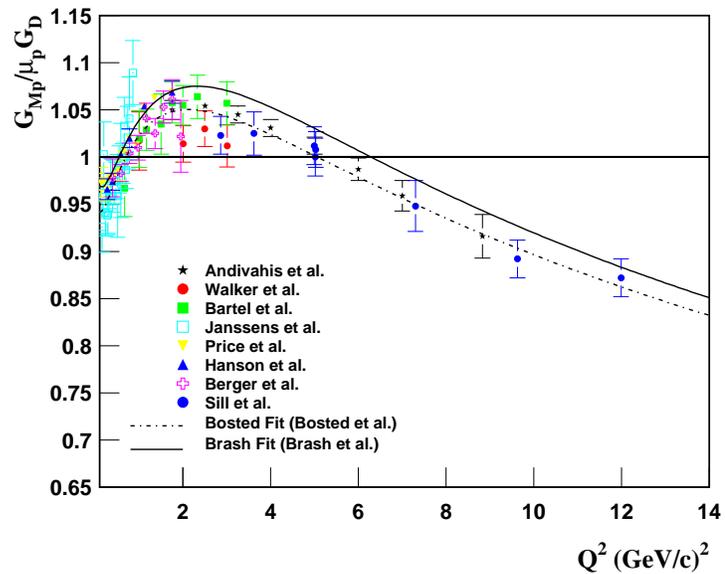,width=3.74in}
\end{center}
\caption[$G_{Mp}$ world data normalized to the dipole form factor $G_{D}$.]
{$G_{Mp}$ world data normalized to the dipole form factor $G_{D}$. The solid 
and the dotted lines are the Bosted and Brash fits, respectively. 
Data from references \cite{murphy74,hanson73,walker94,bartel73,andivahis94,berger71,janssens66,
price71,sill93}.}
\label{fig:gmp_gd}
\end{figure}

If we scatter off a point-like particle with static charge distribution 
$\varrho(\vec{r}) = Ze\delta(\vec{r})$, equation (\ref{eq:form_factor}) gives $F_{NR}(Q^{2}) = Z$, and in 
the case of an extended spatial charge distribution $\varrho(\vec{r}) = \rho_{0}e^{-(\frac{r}{a})}$, 
equation (\ref{eq:form_factor}) gives the well-known dipole form factor 
$F_{NR}(Q^{2}) =  G_D(Q^{2}) = ( 1 + a^{2}Q^{2} )^{-2}$ where $a$ is the scale of the proton radius. 
Experimentally, it was observed that both the electric and magnetic form factors could be described to 
a good approximation by the dipole form factor $G_D(Q^{2})$ with $a^2$ = $(0.71 GeV^2)^{-1}$. 
Figures \ref{fig:gep}, \ref{fig:gep_gd}, and \ref{fig:gmp_gd} show the world data on the electric and 
magnetic form factors. Figure \ref{fig:gmp_gd} also shows $\frac{G_{Mp}}{\mu{p}G_{D}}$ with the fits of 
Bosted \cite{bosted95} and Brash \cite{brash02} to the proton magnetic form factor. The Bosted and the Brash fits will 
be discussed in detail in the next chapter.

If we assume that the magnetic moment of the proton has the same spatial dependence as the charge 
distribution, then the electric and magnetic form factors (in the non-relativistic limit) are related by 
$\mu_{p} \frac{G_{Ep}}{G_{Mp}} = 1.0$ and this is referred to as form factor scaling or:
\begin{equation} \label{eq:dipole}
G_{Ep}(Q^{2}) = \frac{G_{Mp}(Q^{2})}{\mu_{p}}~.
\end{equation} 

The electric and magnetic form factors seem to approximately follow the dipole form factor as shown in
Figures \ref{fig:gep_gd} and \ref{fig:gmp_gd}. This indicates that both form factors
have the same $Q^2$ dependence although $G_{Mp}$ seems to significantly deviate from 
dipole form factor at high $Q^2$. 

For $|\vec{Q}|r<<1$, equation (\ref{eq:form_factor}) can be expanded in terms of the root-mean-square charge 
radius as \cite{perkins87}:
\begin{equation} 
F_{NR}(Q^{2}) = 1 - \frac{|Q|^{2}<r^{2}>_{r.m.s}}{6} + \frac{|Q|^{4}<r^{4}>_{r.m.s}}{120} + \cdots ~,
\end{equation} 
where the root-mean-square charge radius is given by:
\begin{equation} 
<r^{2}>_{r.m.s} = \int \varrho(\vec{r}) r^{2} d^{3}r~.
\end{equation} 

Finally, it is important to realize that the interpretation of the form factors 
(electric and magnetic) as the Fourier transform of the electric charge and magnetic 
moment distribution is only valid in the non-relativistic limit.

\section{Rosenbluth Separations Technique} \label{sect_rosenbluth}

A linear combinations of the Dirac and Pauli form factors $F_{1}(q^2)$ and $F_{2}(q^2)$ 
can be used to define the Sachs form factors \cite{sachs62}, $G_{Ep}$ and $G_{Mp}$, 
the electric and magnetic form factors of the proton:
\begin{equation} \label{eq:electric}
G_{Ep}(Q^2) = F_{1}(Q^2) - \kappa_{p} \tau F_{2}(Q^2)~,
\end{equation}
\begin{equation} \label{eq:magnetic}
G_{Mp}(Q^2) = F_{1}(Q^2) + \kappa_{p} F_{2}(Q^2)~,
\end{equation}
where $\tau = \frac{Q^2}{4M_{p}^2}$. In the limit $Q^{2} \to 0$, where the virtual 
photon becomes insensitive to the internal structure of the proton, equations 
(\ref{eq:electric}) and (\ref{eq:magnetic}) reduce to the normalization conditions for the electric
and magnetic form factors respectively:
\begin{equation}
G_{Ep}(0) = F_{1}(0) = 1~,
\end{equation} 
\begin{equation}
G_{Mp}(0) = \Big[F_{1}(0) + \kappa_{p} F_{2}(0)\Big] = \Big(1 + \kappa_{p}\Big) = \mu_{p} = 2.793\mu_{N}~,
\end{equation} 
where $\mu_{p} = 2.793$ is the proton magnetic moment in units of the nuclear magneton 
$\mu_{N}$, $\mu_{N} = \frac{e\hbar}{2M_{p}c}$.

If we express $F_{1}(Q^2)$ and $F_{2}(Q^2)$ in terms of $G_{Ep}(Q^2)$ and $G_{Mp}(Q^2)$, we can write:
\begin{equation}
F_{1}(Q^2) = \Big[\frac{G_{Ep}(Q^2) + \tau G_{Mp}(Q^2)}{1+\tau} \Big]~, 
\end{equation} 
\begin{equation}
F_{2}(Q^2) = \Big[\frac{G_{Mp}(Q^2) - G_{Ep}(Q^2)}{\kappa_{p}(1+\tau)} \Big]~. 
\end{equation}

By substituting for $F_{1}(Q^2)$ and $F_{2}(Q^2)$ in equation (\ref{eq:diffrentional2}), and 
dropping the interference term between $G_{Ep}(Q^2)$ and $G_{Mp}(Q^2)$ 
(keeping leading order terms in perturbation theory with single-photon exchange), we can express
the elastic electron-proton cross section in terms of the Sachs form factors as:
\begin{equation} \label{eq:rosenbluth1}
\frac{d\sigma}{d\Omega} = \sigma_{ns} \Bigg( \frac{G_{Ep}^{2}(Q^2)+ \tau G_{Mp}^{2}(Q^2)}{(1+\tau)} + 2\tau G_{Mp}^{2}(Q^2)\tan^{2}({\frac{\theta_{e}}{2}}) \Bigg)~. 
\end{equation}

If we express the non-structure cross section in terms of the Mott cross section, we can write
equation (\ref{eq:rosenbluth1}) as:
\begin{equation} \label{eq:rosenbluth2}
\frac{d\sigma}{d\Omega} = \Big(\frac{d\sigma}{d\Omega}\Big)_{Mott} \Big(\frac{E'}{E}\Big) \frac{1}{1+\tau} \Bigg( G_{Ep}^{2}(Q^2) + \tau \Bigg[ 1 + 2 (1+\tau) \tan^{2}({\frac{\theta_{e}}{2}})\Bigg] G_{Mp}^{2}(Q^2) \Bigg)~. 
\end{equation}

The virtual photon longitudinal polarization parameter $\varepsilon$ is defined as:
\begin{equation}
\varepsilon = \Big[ 1 + 2 (1+\tau) \tan^{2}({\frac{\theta_{e}}{2}})\Big]^{-1}~,
\end{equation}
which simplifies equation (\ref{eq:rosenbluth2}) to:
\begin{equation} \label{eq:rosenbluth3}
\frac{d\sigma}{d\Omega} = \Big(\frac{d\sigma}{d\Omega}\Big)_{Mott} \Big(\frac{E'}{E}\Big) \frac{1}{1+\tau} \Bigg( G_{Ep}^{2}(Q^2) + \frac{\tau}{\varepsilon} G_{Mp}^{2}(Q^2) \Bigg)~, 
\end{equation}
which is the Rosenbluth formula \cite{rosenbluth50}.

\begin{figure}[!htbp]
\begin{center}
\epsfig{file=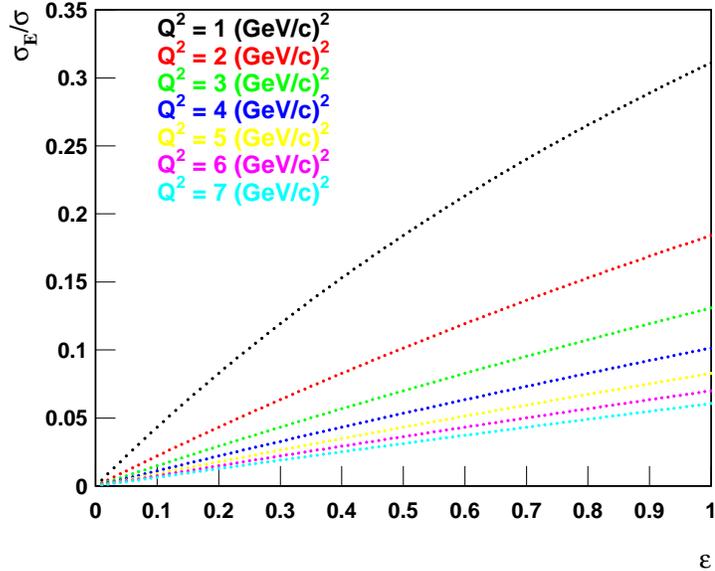,width=3.74in}
\end{center}
\caption[Fractional contribution of $G_{Ep}(Q^2)$ to the cross section.]
{Fractional contribution of $G_{Ep}(Q^2)$ to the cross section $\sigma_{E} \over \sigma$ as 
a function of the virtual photon polarization parameter $\varepsilon$ for several $Q^2$ values
assuming $\mu_{p}G_{Ep}(Q^{2}) = G_{Mp}(Q^{2})$.}
\label{fig:gep_fractional}
\end{figure}

Due to the $\tau = \frac{Q^2}{4M_{p}^2}$ factor in equation (\ref{eq:rosenbluth3})
that multiplies $G_{Mp}^{2}(Q^2)$ but not $G_{Ep}^{2}(Q^2)$, two cases of interest arise:

\begin{enumerate}
\item {Small $Q^{2}$ ($\tau<<1$): the magnetic form factor $G_{Mp}(Q^2)$ of the proton is 
suppressed and the cross section is dominated by the contribution of the electric form 
factor $G_{Ep}(Q^2)$ except as $\varepsilon$ gets close to zero.}
\item {Large $Q^{2}$ ($\tau>>1$): the electric form factor $G_{Ep}^{2}(Q^2)$ of the proton is 
suppressed and the cross section is dominated by the contribution of the magnetic form 
factor $G_{Mp}^{2}(Q^2)$. Figure \ref{fig:gep_fractional} shows the fractional contribution of 
$G_{Ep}(Q^2)$ to the cross section as a function of $\varepsilon$ for several $Q^2$ values assuming 
form factor scaling.}
\end{enumerate}

The fact that the magnetic form factor $G_{Mp}(Q^2)$ dominates at large $Q^{2}$ (small distances),
makes it difficult to extract $G_{Ep}(Q^2)$ with high accuracy from the measured cross 
section at large $Q^{2}$. On the other hand, highly accurate $G_{Mp}(Q^2)$ at small $Q^{2}$ is 
also difficult to extract. 

If we pull out the $\frac{\tau}{\varepsilon}$ factor in equation (\ref{eq:rosenbluth3}), 
we can write:
\begin{equation} \label{eq:rosenbluth4}
\frac{d\sigma}{d\Omega} = \Big(\frac{d\sigma}{d\Omega}\Big)_{Mott} \Big(\frac{E'}{E}\Big) \frac{1}{1+\tau} \frac{\tau}{\varepsilon} \Bigg(\frac{\varepsilon}{\tau} G_{Ep}^{2}(Q^2) + G_{Mp}^{2}(Q^2) \Bigg)~. 
\end{equation} 

Finally, we can define the reduced cross section $\sigma_{R}$ as:
\begin{equation} \label{eq:reduced}
\sigma_{R} = \frac{d\sigma}{d\Omega} \frac{(1+\tau)\varepsilon}{\sigma_{ns}} = \Bigg(\tau G_{Mp}^{2}(Q^2) + \varepsilon G_{Ep}^{2}(Q^2) \Bigg)~, 
\end{equation}
which is the measured cross section multiplied by a kinematic factor.
By measuring the reduced cross section $\sigma_{R}$ at several $\varepsilon$ 
points for a fixed $Q^{2}$, a linear fit of $\sigma_{R}$ to $\varepsilon$  
gives $\tau G_{Mp}^2(Q^2)$ as the intercept and $G_{Ep}^2(Q^2)$ as the slope.

\begin{figure}[!htbp]
\begin{center}
\epsfig{file=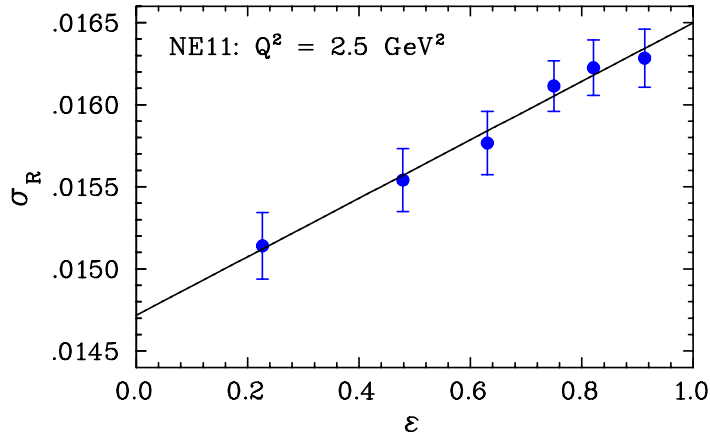,width=3.74in}
\end{center}
\caption[SLAC NE11 L-T separation at $Q^2$ = 2.5 GeV$^2$.]
{Reduced cross section $\sigma_{R}$ as a function of the virtual photon polarization parameter
$\varepsilon$ at fixed $Q^2$ = 2.5 GeV$^2$. Data is taken from the SLAC NE11 
experiment \cite{andivahis94}.}
\label{fig:NE11_LT}
\end{figure}

This is known as the Rosenbluth separations method and an example is shown in 
Figure \ref{fig:NE11_LT}. Having extracted $G_{Mp}$ and $G_{Ep}$ for a fixed $Q^{2}$, 
the ratio of the electric to magnetic form factors of the proton $\frac{\mu_{p} G_{Ep}}{G_{Mp}}$ 
can be determined for that $Q^{2}$ point.

\section{Recoil Polarization Technique} \label{sect_recoil_polarization}

In recoil polarization experiments \cite{milbrath98,jones00,gayou01,gayou02}, a longitudinally polarized beam 
of electrons is scattered by unpolarized protons. This results in a transfer of 
the polarization from the electrons to the recoil protons. It must be mentioned that any 
observed polarization of the struck proton was transfered from the electron since the 
target is unpolarized. In elastic electromagnetic scattering in the single-photon exchange approximation, 
there is no induced polarization.
\begin{figure}[!htbp]
\begin{center}
\epsfig{file=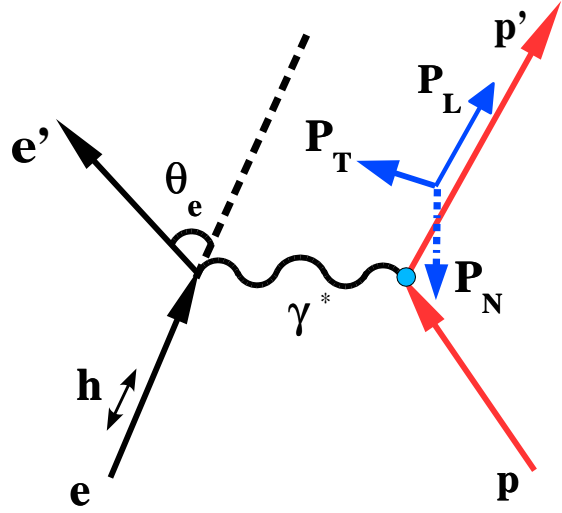,width=3in}
\end{center}
\caption[Recoil polarization and the $^{1}H(\vec{e},e',\vec{p})$ reaction spin transfer]
{Schematic diagram for the recoil polarization spin transfer in the $^{1}H(\vec{e},e',\vec{p})$ 
reaction.}
\label{fig:recoil_polarization}
\end{figure}
In the $^{1}H(\vec{e},e',\vec{p})$ reaction, see Figure \ref{fig:recoil_polarization},
the only non vanishing polarization transfer 
observables are the transverse, $P_{t}$, and the longitudinal, $P_{l}$, components of the 
transfered polarization. The normal component, $P_{n}$, does not exist in elastic scattering in 
single-photon exchange.

In the single-photon exchange approximation, it can be shown \cite{dombey69,akheizer74,arnold81} 
that the the transverse, $P_{t}$, and the longitudinal, $P_{l}$, components of the transfered
polarization are related to the Sachs form factors by:
\begin{equation} \label{eq:pl}
I_{O}P_{l} = \frac{(E+E')}{M_{p}} \sqrt{\tau(1+\tau)}G_{Mp}^{2}(Q^2) \tan^{2}({\frac{\theta_{e}}{2}})~,
\end{equation} 
\begin{equation} \label{eq:pt}
I_{O}P_{t} = -2\sqrt{\tau(1+\tau)}G_{Ep}(Q^2)G_{Mp}(Q^2) \tan({\frac{\theta_{e}}{2}})~,
\end{equation}   
where $E$, $E'$, and $\theta_{e}$ are the incident energy, final energy, and scattered angle of 
the electron, and $I_{O}$ is defined as:
\begin{equation} \label{eq:IO}
I_{O} = G_{Ep}^{2}(Q^2) + \frac{\tau}{\varepsilon} G_{Mp}^{2}(Q^2)~.
\end{equation} 

In the recoil polarization experiments, a Focal Plane Polarimeter (FPP) \cite{bimbot01} 
is used to simultaneously measure both the transverse and longitudinal polarization components
$P_{t}$ and $P_{l}$. By scattering the proton off a secondary target inside the FPP 
and measuring the azimuthal angular distribution, both $P_{t}$ and $P_{l}$ can be determined 
at the same time for a given $Q^{2}$ value.

By dividing equation (\ref{eq:pt}) by (\ref{eq:pl}) and solving for $G_{Ep} \over G_{Mp}$
we can write:
\begin{equation} \label{eq:ratio}
\frac{G_{Ep}}{G_{Mp}} = \frac{P_{t}}{P_{l}} \frac{(E+E')}{2M_{p}} \tan({\frac{\theta_{e}}{2}})~,
\end{equation}
which is the ratio of the electric to magnetic form factors of the proton $G_{Ep} \over G_{Mp}$
as extracted by direct and simultaneous measurement of the transverse and longitudinal 
polarization components of the recoiling proton. An interesting point is that while the 
Rosenbluth separations method measures absolute cross sections and then extracts 
$G_{Ep}$ and $G_{Mp}$ from these cross sections, the recoil polarization methods gives 
directly $G_{Ep} \over G_{Mp}$ without any measurements of cross sections. 
The world data for $\mu_{p}G_{Ep} \over G_{Mp}$ as extracted from the two techniques will be 
discussed in more detail in the next chapter.

\chapter{Previous Form Factor Data}\label{chap_form_factors}
\pagestyle{plain}
\section{Overview} \label{sect_overview}

Understanding the internal structure of the proton is a fundamental problem of
strong-interaction physics. It was the famous experiment of Hofstadter and
collaborators \cite{hofstadter57} at Stanford that first measured the internal structure
of the proton. Since that time the structure of hadrons has become one of the most important 
topics in nuclear physics and received the attention of experimentalists and theorists 
worldwide. The virtual photon-proton vertex cannot be calculated from first principles.
Therefore, the internal structure of the proton has been parameterized in terms of electric
$G_{Ep}(Q^2)$ and magnetic $G_{Mp}(Q^2)$ form factors. 
The electromagnetic form factors of the protons are used to describe the deviations 
of the proton from a point-like particle in elastic electron-proton scattering. 
In the non-relativistic limit, the form factors are interpreted as the Fourier transform 
of the spatial distributions of the charge and magnetic moment.

\section{Previous Measurements} \label{sect_previous_data}

Several experiments have been conducted over the last fifty years to measure 
the elastic electron-proton cross section. Some of these experiments were 
able to extract the electric and magnetic form factors of the proton, 
$G_{Ep}$ and $G_{Mp}$, using the Rosenbluth separation technique. 
The form factor ratio $\mu_{p}G_{Ep} \over G_{Mp}$ has also been measured using 
the recoil polarization technique where the elastic electron-proton cross section 
are not feasible. In this section I will briefly summarize the work done and the results 
quoted by these two techniques in a chronological order. In particular, a brief description of the 
technique used, kinematics range covered, experimental details, radiative corrections applied 
(if applicable), and uncertainties quoted in the measured cross sections and in the overall normalization 
is presented. A comparison between the results of the two techniques will be made 
and a discussion of the global analyses and fits of the world data will be presented.
Because the focus of the present work is on $\frac{\mu_{p}G_{Ep}}{G_{Mp}}$, the stress is
placed on the values of this ratio obtained in the various experiments.

\pagestyle{myheadings}

\subsection{Elastic e-p Cross Sections Measurements} \label{sect_previous_ep_data}

The following experiments measured the elastic e-p cross sections for different
$Q^2$ range. Experiments are listed under the name of the first author: 

\begin{itemize}
{\it
\item \textbf{Janssens et al, 1966 \cite{janssens66}:}
}
\end{itemize}

\begin{figure}[!htbp]
\begin{center}
\epsfig{file=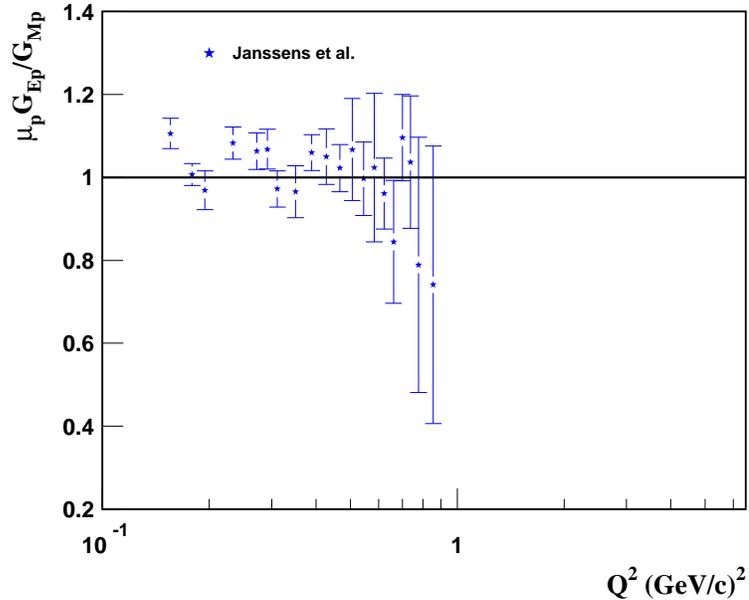,width=3.9in}
\end{center}
\caption[$\mu_{p}G_{Ep}/G_{Mp}$ by Janssens et al \cite{janssens66}.]
{The ratio of electric to magnetic form factor $\mu_{p}G_{Ep}/G_{Mp}$ by Janssens et al \cite{janssens66}.}
\label{fig:janssens}
\end{figure}

The Stanford Mark III linear accelerator was used to produce an incident electron beam 
of energies in the range of 0.25$< E_{0} <$1.0 GeV. A liquid hydrogen target 0.953 cm 
thick with 0.0254 mm stainless steel target walls was used to scatter 
electrons. The scattered electrons were detected using the 72 inch double focusing magnetic
spectrometer. Elastic e-p cross sections were measured at 25 different $Q^2$ in 
the range of 0.15$< Q^2 <$0.86 GeV$^2$ covering an angular range of 
45$^o$$< \theta <$145$^o$ with uncertainty never more than 0.08$^o$. Of the 25 $Q^2$ 
measured, 20 $Q^2$ points had enough $\varepsilon$ coverage to do an L-T separation 
(Rosenbluth separation). Typically 3-5 $\varepsilon$ points per $Q^2$ value. Measurements for a single 
$\varepsilon$ point at $Q^2$ = 1.01, 1.09, and 1.17 GeV$^2$ were made at constant spectrometer angle of 
$\theta = 145^{o}$, and that of $Q^2$ = 0.49, and 0.68 GeV$^2$ were made at constant 
spectrometer angle of $\theta=75^{o}$. The internal radiative corrections were calculated 
using the method of Tsai \cite{tsai61} and the external radiative corrections (bremsstrahlung in the target) 
were calculated using the formulas of Schwinger \cite{schwinger49} and Bethe \cite{bethe53}. 
The quoted uncertainty in the absolute elastic cross section was on the 4.0\% level with 1.6\% as an 
overall normalization uncertainty. L-T extractions of the proton form factors were performed. 
Figure \ref{fig:janssens} shows the ratio of electric to magnetic form factor of the proton from this work.
%

\begin{itemize}
{\it
\item \textbf{Litt et al, 1970 \cite{litt70}:}
}
\end{itemize}

\begin{figure}[!htbp]
\begin{center}
\epsfig{file=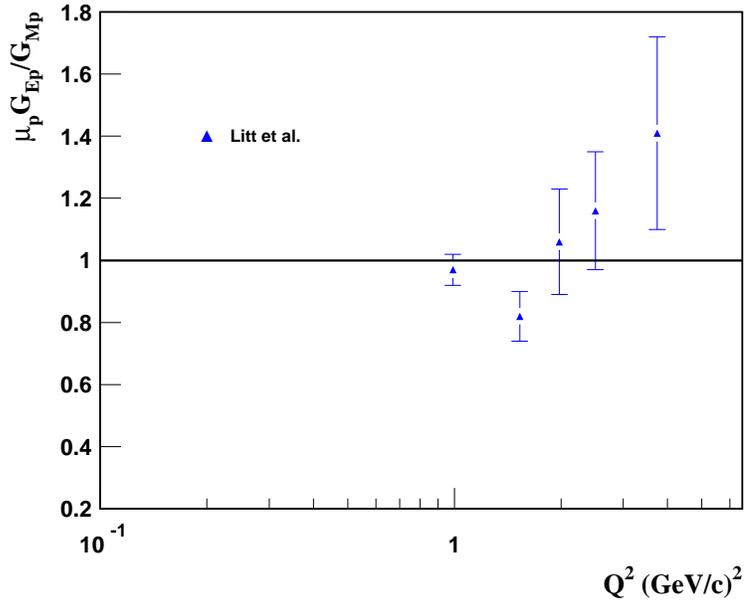,width=3.9in}
\end{center}
\caption[$\mu_{p}G_{Ep}/G_{Mp}$ by Litt et al \cite{litt70}.]
{The ratio of electric to magnetic form factor $\mu_{p}G_{Ep}/G_{Mp}$ by Litt et al \cite{litt70}.}
\label{fig:litt}
\end{figure}

An electron beam from the Stanford Linear Accelerator Center (SLAC) was produced in the range of   
4.0$< E_{0} <$10.0 GeV. A liquid-hydrogen target 23 cm in length was used to scatter electrons. 
The scattered electrons were detected using the SLAC 8-GeV/c magnetic spectrometer. 
Six $Q^2$-values in the range of 1.0$< Q^2 <$3.75 GeV$^2$, corresponding to a scattering angle 
in the range of 12.5$^o$$< \theta <$41.4$^o$, were covered. 
On the average, 3-5 $\varepsilon$ points were taken for each $Q^2$ setting allowing for an L-T 
extraction of the proton form factors. Data was corrected for radiation loss due to straggling of 
the electrons in the target using Eyges \cite{eyges49}. Internal radiative corrections 
were applied using Tsai \cite{tsai61}. Measured cross sections were determined to within
a (1.5-2.0)\% point-to-point uncertainty. An overall normalization uncertainty of 4.0\% was quoted.
Figure \ref{fig:litt} shows the ratio of electric to magnetic form factor of the proton from this work.
%

\begin{itemize}
{\it
\item \textbf{Price et al, 1971 \cite{price71}:}
}
\end{itemize}

\begin{figure}[!htbp]
\begin{center}
\epsfig{file=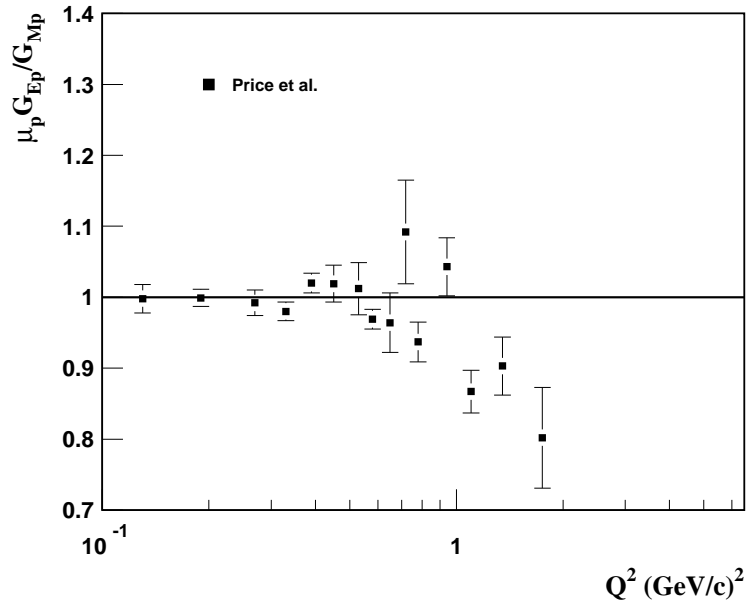,width=3.9in}
\end{center}
\caption[$\mu_{p}G_{Ep}/G_{Mp}$ by Price et al \cite{price71}.]
{The ratio of electric to magnetic form factor $\mu_{p}G_{Ep}/G_{Mp}$ by Price et al \cite{price71}.}
\label{fig:Price}
\end{figure}

This experiment is an extension of the forward angle experiment done by Goitein et al \cite{goitein70}.
Incident electron beams from the Cambridge Electron Accelerator (CEA) in the range of  
0.45$< E_{0} <$1.6 GeV were used and directed on a 3.3 cm liquid-hydrogen target in 
length. Electrons were scattered at large angle in the range of 
$80^o$$< \theta <$$90^o$ and detected using the 14\% total momentum acceptance and 0.83 msr 
solid angle magnetic spectrometer. Cross sections were measured in the range 
0.25$< Q^2 <$1.75 GeV$^2$. Radiative corrections were applied using the equivalent 
radiators method of Mo and Tsai \cite{mo69} with modifications added using Meister and Yennie \cite{meister63}. 
The cross sections from the large angle measurements were known with uncertainties of
(3.1-5.3)\% including both statistical and systematic uncertainties. There is also an overall 
normalization uncertainty of 1.9\%. It should be mentioned that the large angle data by 
itself was not sufficient to do an L-T extraction. Rather, elastic cross sections from this 
work were combined with several e-p scattering experiments and a correction for normalization 
difference between the several experiments was applied. Therefore, no estimation of the uncertainties in the cross
sections was given due to the difference in the normalization procedures used.
Global fits were performed using $G_{Ep}$ and $G_{Mp}$ as parameters of the fit. The results of combining e-p cross sections 
from several experiments showed deviations from form factor scaling. Figure \ref{fig:Price} 
shows the ratio of electric to magnetic form factor of the proton from this work.

\begin{itemize}
{\it
\item \textbf{Berger et al, 1971 \cite{berger71}:}
}
\end{itemize}

\begin{figure}[!htbp]
\begin{center}
\epsfig{file=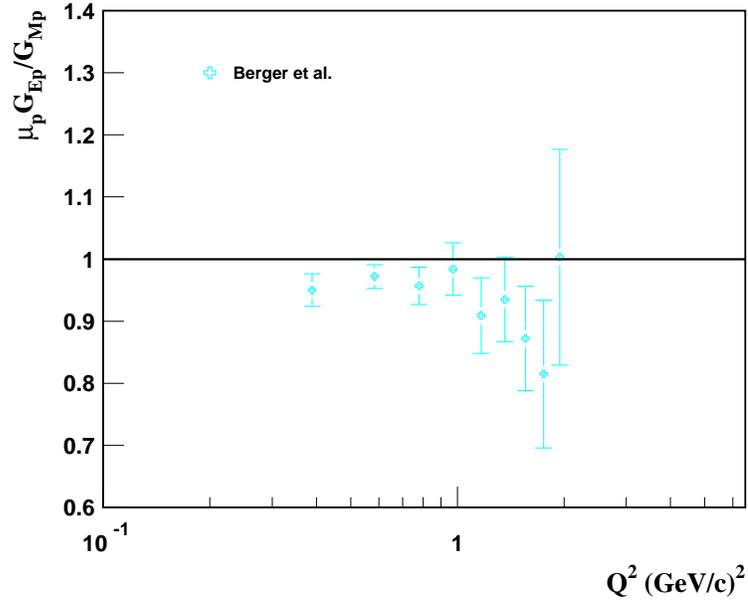,width=3.9in}
\end{center}
\caption[$\mu_{p}G_{Ep}/G_{Mp}$ by Berger et al \cite{berger71}.]
{The ratio of electric to magnetic form factor $\mu_{p}G_{Ep}/G_{Mp}$ by Berger et al \cite{berger71}.}
\label{fig:Berger}
\end{figure}

Measurements of the elastic e-p cross sections were made at the Physikalisches 
Institute at the University of Bonn in Germany. Incident electron beams of energies in the range of
0.66$< E_{0} <$1.91 GeV were directed on a liquid-hydrogen target 5 cm in diameter. 
Cross sections measurements were made for 0.10$< Q^2 <$1.95 GeV$^2$ covering angular range of 
$25^o$$< \theta <$111$^o$ with low $Q^2$ data taken at $\theta = 30^{o}$ in order to normalize
to other experiments (DESY, Bartel et al \cite{bartel66}). 
With the goal to combine the data and extract form factors, 3-14 $\varepsilon$ data points 
were taken for each $Q^2$ value. Internal radiative correction were applied
using the method of Meister and Yennie \cite{meister63}. The external bremsstrahlung corrections 
were applied using Heitler \cite{heitler54}. Cross sections were determined to within (2-6)\% with 
an overall normalization uncertainty of 4\%. It was concluded that form factor scaling was not 
valid at least in the region of 0.39$< Q^2 <$1.95 GeV$^2$. Figure \ref{fig:Berger} shows 
the ratio of electric to magnetic form factor of the proton from this work.

\begin{itemize}
{\it
\item \textbf{Kirk et al, 1972 \cite{kirk73}:}
}
\end{itemize}

Electron beams from the Stanford Linear Accelerator Center (SLAC) in the range of 
4.0$< E_{0} <$17.31 GeV were directed on a five condensation-type liquid-hydrogen target 
cells of different sizes (8-32 cm diameter with 25-75-$\mu$-thick stainless steel walls). 
The scattered electrons were detected using the SLAC 1.6-GeV/c and SLAC 8-GeV/c magnetic 
spectrometers. With the assumption that the electric form factor contribution to the elastic 
cross section is small, the experiment aimed to extract the magnetic form factors over a large $Q^2$ 
range of 1.0$< Q^2 <$25 GeV$^2$, covering three main angular regions to a maximum of 180$^{o}$. 
The 20-GeV/c spectrometer was used for small angle measurements $0^o$$< \theta <$$20^o$, 
the 8-GeV/c spectrometer was used for intermediate angle measurements $12^o$$< \theta <$$105^o$, 
and the 1.6-GeV/c spectrometer was used for backward angle measurements $25^o$$< \theta <$$165^o$. 
The main data was taken in the range of $12^o$$< \theta <$$35^o$ while the data at low $Q^2$ was 
taken to provide cross calibration with other experiments. Some of the higher $Q^2$ data 
($Q^2$ = 5 and 10 GeV$^2$) was taken to be combined with large angle measurements from different 
experiments to provide an upper limit value for $G_{Ep}$. Internal radiative corrections were applied
using Tsai \cite{tsai61}, and Eyges \cite{eyges49} for the external corrections. Cross sections were reported
with uncertainties of 2.0\% and with an overall normalization uncertainty of 4.0\%. 
An L-T extraction was impossible since there were not enough $\varepsilon$ data points taken. 
Although no plot of the ratio of electric to magnetic form factor of the proton from this 
work is possible, the cross sections measured in this experiment were combined with cross
sections from other experiments for global extractions of the proton elastic form factors.

\begin{itemize}
{\it
\item \textbf{Murphy et al, 1974 \cite{murphy74}:}
}
\end{itemize}

The University of Saskatchewan Linear Accelerator was used to provide electron beams in the
range of 0.057$< E_{0} <$0.123 GeV. The beam was directed on a gaseous-hydrogen target
in a circular cylinder 2.54 cm in radius and 3.50 cm in high. In this work, the recoil 
protons rather than the scattered electrons were detected by a double-focusing magnetic 
spectrometer. Measurements were made for 11-values of $Q^2$ in the range of
0.006$< Q^2 <$0.031 GeV$^2$ covering only two angles for the recoiled protons, 
$\theta_{p}$ = 30$^{o}$ and 45$^{o}$. Not enough $\varepsilon$ data points were covered to
perform an L-T extraction. Radiative corrections for the protons were applied using the method 
of Meister and Yennie \cite{meister63}. The experiment extracted the values of $G_{Ep}$ with total uncertainty 
in the range of (0.3-0.9)\%. No plot of the ratio of electric to magnetic form factor 
of the proton from this work is possible. Cross sections measured in this experiment were 
combined with cross sections from other experiments for global extraction of the proton 
elastic form factors.
%

\begin{itemize}
{\it
\item \textbf{Bartel et al, 1973 \cite{bartel73}:}
}
\end{itemize}

\begin{figure}[!htbp]
\begin{center}
\epsfig{file=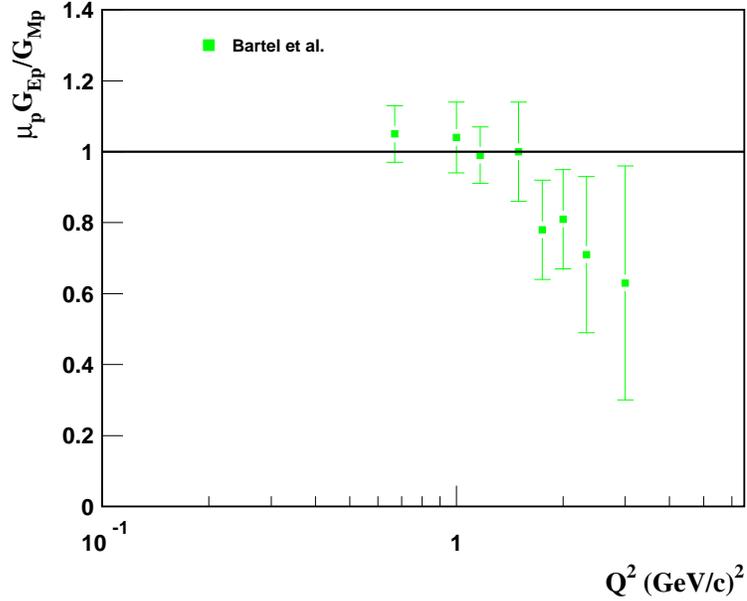,width=3.9in}
\end{center}
\caption[$\mu_{p}G_{Ep}/G_{Mp}$ by Bartel et al \cite{bartel73}.]
{The ratio of electric to magnetic form factor $\mu_{p}G_{Ep}/G_{Mp}$ by Bartel et al \cite{bartel73}.}
\label{fig:Bartel}
\end{figure}

Elastic e-p scattering cross sections were measured using the 
Deutsches Elecktronen-Synchrotron (DESY) in Hamburg Germany. An electron beams in the range 
of 0.8$< E_{0} <$3.0 GeV were directed on a cylindrical vessel 5 cm in diameter and 6 cm long 
liquid-hydrogen target. Using the high resolution magnetic spectrometer (small angle spectrometer), 
measurements of cross sections were made at electron scattering angle in the range of 
$10^o$$< \theta_{e} <$$20^o$ and by detecting protons at forward angles (corresponding to
$\theta_{e}$ = 86$^{o}$) using the recoil nucleon detector. Electrons scattered at large angle,  
$\theta_{e}$ = 86$^{o}$, were measured using the large angle spectrometer. Cross sections were 
measured at 7-values of $Q^2$ in the range of 0.67$< Q^2 <$3.0 GeV$^2$ with uncertainty on the
2-4\% level and an overall normalization uncertainty of 2.1\%. Typically 2-3 $\varepsilon$ 
points were taken per $Q^2$ value allowing for an L-T extraction. Internal radiative corrections 
were calculated using Meister and Yennie \cite{meister63}, and the external radiative contribution were 
calculated using Mo and Tsai \cite{mo69}. Electron-proton cross sections in this work and several 
other experiment were compiled and analyzed. A global fit was performed in order to extract the form factors. 
Deviations from form factor scaling were reported. Figure \ref{fig:Bartel} shows the ratio of 
electric to magnetic form factor of the proton from this work.

\begin{itemize}
{\it
\item \textbf{Stein et al, 1975 \cite{stein75}:}
}
\end{itemize}

The Stanford Linear Accelerator Center (SLAC) was used to produce an electron beams in the range of 
4.5$< E_{0} <$20.0 GeV. Elastic cross sections measurements were made in the range of 
0.004$< Q^2 <$0.07 GeV$^2$ at a scattering angle of $\theta$ = 4$^{o}$ providing one 
$\varepsilon$ data point for each $Q^2$ value measured. Clearly not enough $\varepsilon$ 
data points for an L-T extraction. The beam was directed on a vertical cylinder with 0.0076 cm 
aluminum walls liquid-hydrogen target. The scattered electrons were detected using the SLAC 
20-GeV/c spectrometer. Radiative corrections were applied to the measured cross sections 
using the procedure of Mo and Tsai \cite{mo69}. The uncertainties in the cross sections are on 
the 3.1\% level with an overall normalization uncertainty of 2.8\%. No plot of the ratio of 
electric to magnetic form factor of the proton from this work is possible.
Cross sections measured in this experiment were combined with cross sections from other experiments 
for global extractions of the proton elastic form factors.

\begin{itemize}
{\it
\item \textbf{Borkowski et al, 1975 \cite{borkowski75}:}
}
\end{itemize}

\begin{figure}[!htbp]
\begin{center}
\epsfig{file=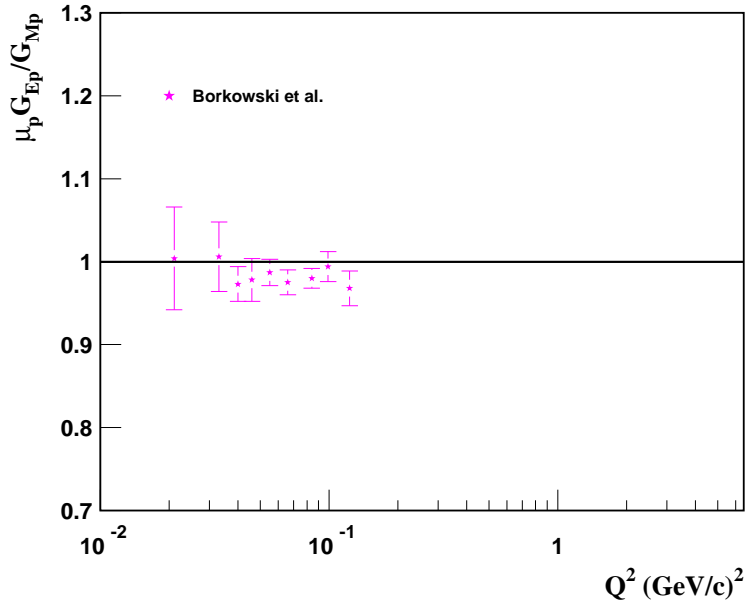,width=3.9in}
\end{center}
\caption[$\mu_{p}G_{Ep}/G_{Mp}$ by Borkowski et al \cite{borkowski75}.]
{The ratio of electric to magnetic form factor $\mu_{p}G_{Ep}/G_{Mp}$ by Borkowski et al 
\cite{borkowski75}.}
\label{fig:Borkowski}
\end{figure}

Elastic e-p scattering cross sections have been measured using the 
300 MeV Electron Linear Accelerator at Mainz. An electron beam was produced in the range of 
0.15$< E_{0} <$0.30 GeV and was directed on a 10 mm diameter thin-walled cylindrical cell
filled with liquid-hydrogen. Elastic cross sections measurements were made in the range of 
0.005$< Q^2 <$0.183 GeV$^2$ covering a scattering angle in the range of 
$28^o$$< \theta_{e} <$$150^o$. Measurements were made using two spectrometers. 
The first is the double-focusing 180$^{o}$ spectrometer and was set to different scattering angles. 
The second consisted of two quadrupoles and 12$^{o}$ bending-magnet spectrometer and was fixed at 
scattering angle of 28$^{o}$ as a monitor. Radiative corrections were applied using Mo and Tsai 
\cite{mo69}. Random uncertainties in the cross sections were reported and were on the (1.0-2.0)\% level. 
Figure \ref{fig:Borkowski} shows the ratio of electric to magnetic form factor of the proton 
from this work.

\begin{itemize}
{\it
\item \textbf{Bosted et al, 1990 \cite{bosted90}:}
}
\end{itemize}

The primary goal of this experiment was to measure the magnetic structure function $B(Q^2)$ of 
the deuteron to the largest $Q^2$-value possible. The elastic cross sections of the e-p 
scattering were measured near 180$^{o}$ and the value of the magnetic form factor $G_{Mp}$ was 
determined assuming form factor scaling with small $G_{Ep}$ contribution to the cross sections at high 
$Q^2$ points. Incident electron beam from the Stanford Linear Accelerator Center (SLAC) of energies in 
the range of 0.5$ < E_{0} <$1.3 GeV was directed on a liquid-hydrogen target with nominal lengths 
of 40, 20, 10, and 5 cm. Cross sections were measured at 11-values of $Q^2$ in the range of 
0.49$< Q^2 <$1.75 GeV$^2$ for electrons backscattered near 180$^{o}$ in coincidence with protons 
recoiling near 0$^{o}$ in a large solid-angle double-arm spectrometer. 
An L-T extraction is impossible from these measurements. The equivalent radiator approximation 
method of Tsai \cite{tsai61} was used for the internal radiative corrections. The effect of the 
bremsstrahlung and Landau straggling in the target external radiative corrections were combined with 
that of the internal corrections. Cross sections with uncertainty of 3\% were quoted with an overall 
normalization uncertainty of 1.8\%. No plot of the ratio of electric to magnetic form factor
of the proton from this work is possible. Cross sections measured in this experiment were 
combined with cross sections from other experiments for global extraction of the proton 
elastic form factors.

\begin{itemize}
{\it
\item \textbf{Rock et al, 1992 \cite{rock92}:}
}
\end{itemize}

The main goal of this experiment was to extract the elastic neutron cross sections but
elastic e-p cross sections were also measured since they were needed for the analysis. 
The Stanford Linear Accelerator Center (SLAC) electron beam with energies in the range of 
9.761$< E_{0} <$21.0 GeV was directed on a 30-cm long liquid-hydrogen cell. 
The elastic e-p cross sections were measured at 5 values of $Q^2$ in the range of 
2.5$< Q^2 <$10.0 GeV$^2$ at a fixed scattering angle of $\theta$ = $10^o$. 
An L-T extraction is impossible with this data. The scattered electrons were detected using the 
SLAC 20-GeV/c spectrometer. The cross sections were radiatively corrected using the method of Tsai 
\cite{tsai61}. The uncertainties in the cross sections were on the 1-4\% level in addition to an overall 
normalization uncertainty of 3\%. No plot of the ratio of electric to magnetic form factor of the 
proton from this work is possible. Cross sections measured in this experiment were combined with
cross sections from other experiments for global extraction of the proton elastic form factors.

\begin{itemize}
{\it
\item \textbf{Sill et al, 1993 \cite{sill93}:}
}
\end{itemize}

Elastic e-p cross sections were measured using the beam line at the Stanford 
Linear Accelerator (SLAC). An electron beams of energies in the range of 
5.0$ < E_{0} <$21.5 GeV were directed on two liquid-hydrogen targets with different 
lengths. The 25-cm target was used for 
normalization of the acceptance of the 65-cm target and to provide a test for the low $Q^2$ 
data. The 65-cm target provided higher counting rate and was used to take the majority of 
the elastic data. The scattered electrons were detected using the SLAC 8-GeV/c spectrometer.
Cross sections were measured for 13-values of $Q^2$ for the range of 2.9$< Q^2 <$31.3 
GeV$^2$ covering three-angle settings of $\theta_{e}$ = 21$^{o}$, 25$^{o}$, and 33$^{o}$.
Due to the limited angular range, an L-T extraction was impossible to perform.  
Radiative corrections were applied to the data using Mo and Tsai \cite{mo69}. 
Cross sections uncertainties were on the 3-4\% level with an overall normalization uncertainty
of 3.6\%. No plot of the ratio of electric to magnetic form factor of the proton from this 
work is possible. Cross sections measured in this experiment were combined with
cross sections from other experiments for global extraction of the proton elastic form factors.

\begin{itemize}
{\it
\item \textbf{Walker et al, 1994 \cite{walker94}:}
}
\end{itemize}

\begin{figure}[!htbp]
\begin{center}
\epsfig{file=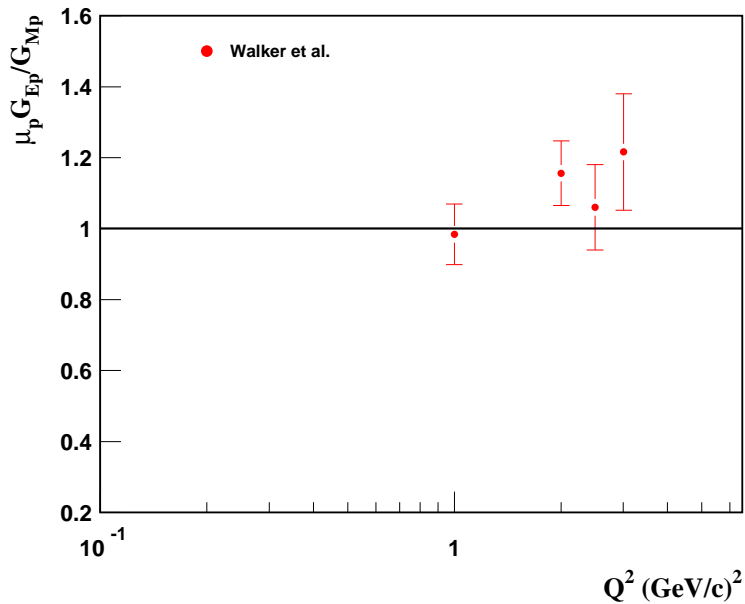,width=3.9in}
\end{center}
\caption[$\mu_{p}G_{Ep}/G_{Mp}$ by Walker et al \cite{walker94}.]
{The ratio of electric to magnetic form factor $\mu_{p}G_{Ep}/G_{Mp}$ by Walker et al \cite{walker94}.}
\label{fig:walker}
\end{figure}

The Stanford Linear Accelerator Center beam line was used to produce incident electron beams of 
energies in the range of 1.594$< E_{0} <$8.233 GeV. The beam was directed on cylindrical 
liquid-hydrogen target 20-cm in length and 5.08-cm in diameter. The scattered electrons were detected 
using the SLAC 8-GeV/c spectrometer. Elastic e-p cross sections were measured for 
4-values of $Q^2$ in the range of 1.0$< Q^2 <$3.007 GeV$^2$ covering angular range of 
$11.714^o$$< \theta <$$45.221^o$ with an average of 3-8 $\varepsilon$ points per $Q^2$ point 
allowing for an L-T extraction. Internal radiative corrections were done using Mo and Tsai \cite{mo69}. 
Also, improvements were made to the internal corrections using the equivalent radiator approximation. 
The external corrections were applied using the work of Tsai \cite{tsai61} to account for bremsstrahlung in 
the target material and the effects of the Landau tail of the ionization energy loss spectrum. 
Uncertainty in the cross sections was on the 1\% level with an overall normalization of 1.9\%. 
Figure \ref{fig:walker} shows the ratio of electric to magnetic form factor of the proton from this work. 
Cross sections from several experiments including this work were combined and a fit for a global extraction 
of the form factors was performed. Results indicate good consistency between the different data sets. 
The form factors extracted from this work supported form factor scaling. See Figure \ref{fig:combined_fit}
in section (\ref{sect_discussion}) for more detail.

\begin{itemize}
{\it
\item \textbf{Andivahis et al, 1994 \cite{andivahis94}:}
}
\end{itemize}

\begin{figure}[!htbp]
\begin{center}
\epsfig{file=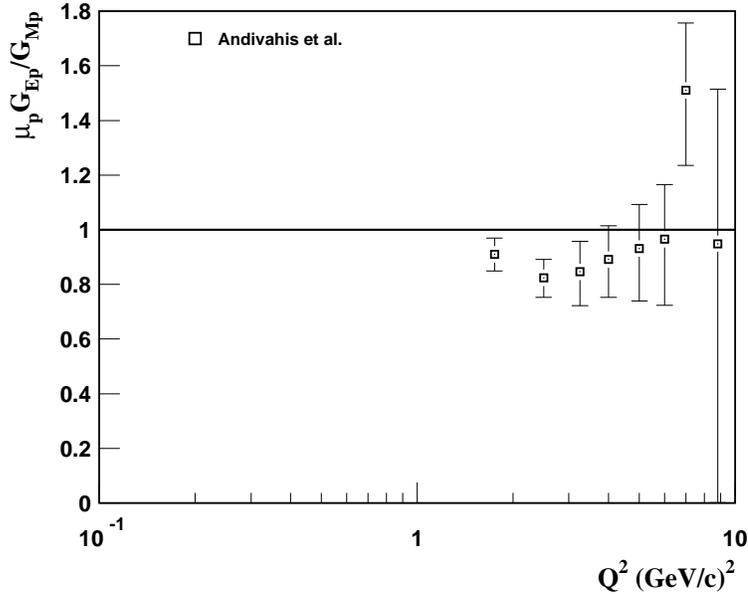,width=3.9in}
\end{center}
\caption[$\mu_{p}G_{Ep}/G_{Mp}$ by Andivahis et al \cite{andivahis94}.]
{The ratio of electric to magnetic form factor $\mu_{p}G_{Ep}/G_{Mp}$ by Andivahis et al \cite{andivahis94}.}
\label{fig:andivahis}
\end{figure}

The primary goal of this experiment was to minimize both statistical and systematic uncertainties and 
extend the measurements of the form factors to a high $Q^2$ values. The Stanford Linear Accelerator Center (SLAC) 
was used to produce electron beams with energies in the range of 1.511$< E_{0} <$9.8 GeV. The beam was directed on a 
15-cm liquid-hydrogen target. The scattered electrons were detected using both the SLAC 1.6 and 8-GeV/c spectrometers 
simultaneously. The 8-GeV/c spectrometer was used to measure 5-values of $Q^2$ in the range of 1.75$< Q^2 <$5.0 GeV$^2$ 
where an L-T extraction is possible with an average of 3-6 $\varepsilon$ points per $Q^2$. 
In addition, two extra points at $Q^2$ = 6.0 and 7.0 GeV$^2$ were measured as a single $\varepsilon$ points. 
The angular range was $13.25^o$$< \theta <$$90.066^o$. The 1.6-GeV/c spectrometer was used to measure 8-single-points of 
$\varepsilon$ at $Q^2$ = 1.75, 2.50, 3.25, 4.0, 5.0 6.0, 7.0, and 8.83 GeV$^2$ at a constant angle $\theta$ $\approx$ $90^o$ 
so data could be combined with previous forward-angle cross sections \cite{arnold86,sill93} collected at SLAC at the 
same $Q^2$. The 1.6-GeV/c cross sections were normalized to the 8-GeV/c results.
The procedure of the radiative corrections applied was similar to the procedure introduced by 
Walker et al \cite{walker94} reported above. Average uncertainties in the cross sections were less than 2\% 
with an overall normalization uncertainty of 1.77\%. Figure \ref{fig:andivahis} shows the ratio of 
electric to magnetic form factor of the proton from this work.   

\begin{itemize}
{\it
\item \textbf{Christy et al, 2004 \cite{christy04}:}
}
\end{itemize}

\begin{figure}[!htbp]
\begin{center}
\epsfig{file=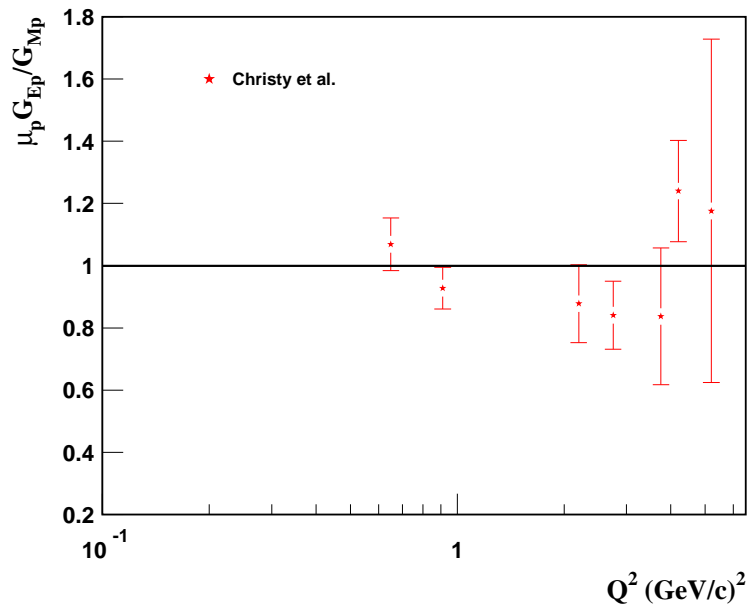,width=3.9in}
\end{center}
\caption[$\mu_{p}G_{Ep}/G_{Mp}$ by Christy et al \cite{christy04}.]
{The ratio of electric to magnetic form factor $\mu_{p}G_{Ep}/G_{Mp}$ Christy et al \cite{christy04}.}
\label{fig:christy}
\end{figure}

The focus of this experiment was to separate the longitudinal and transverse unpolarized
proton structure functions in the resonance region using Rosenbluth separation technique, but elastic 
e-p cross sections were also measured at 28-values of $Q^2$ in the range of 
0.4$< Q^2 <$5.5 GeV$^2$ covering an angular range of $12.5^o$$< \theta_{e} <$$80^o$. 
This range covered 3 $\varepsilon$ points per $Q^2$ allowing for L-T separations at 7 $Q^2$-values. 
The experiment was done at the experimental Hall C of the Thomas Jefferson National Laboratory (JLAB) 
in Newport News Virginia. An electron beam in the range of 1.148$< E_{0} <$5.494 GeV was directed 
on a tuna-can shaped cryogen liquid hydrogen cell. The cell has an inside diameter of 40.113 mm when 
warm and 39.932 mm when cold with cylindrical wall thickness of 0.125 mm. The scattered electrons were 
detected using the High Momentum Spectrometer. 
Radiative corrections were applied using the procedure described in Walker et al \cite{walker94} and based on 
the prescription of Mo and Tsai \cite{mo69}. Uncertainties in the cross sections were on the 1.96\% level 
with an overall normalization uncertainty of 1.7\%. The uncertainties in the $\mu_{p}G_{Ep} \over G_{Mp}$ 
ratio in the $Q^2$ region of interest are quite large and do not represent an improvement in accuracy over 
previous L-T determinations. However, when the polarization transfer results became available and were at 
odds with the previously known Rosenbluth results, Christy et al carefully analyzed their elastic e-p 
data and concluded that their $\mu_{p}G_{Ep} \over G_{Mp}$ were not consistent with the ratio reported by 
Jones et al. Figure \ref{fig:christy} shows the ratio of electric to magnetic form factor of the proton 
from this work.

\subsection{Polarization Measurements} \label{sect_previous_recoil_data}

The following experiments represent the proton polarization measurements (recoil polarization and polarized target
measurements) to date for different $Q^2$ range. 

\begin{itemize}
{\it
\item \textbf{Alguard et al, 1976 \cite{alguard76}:}
}
\end{itemize}

\begin{figure}[!htbp]
\begin{center}
\epsfig{file=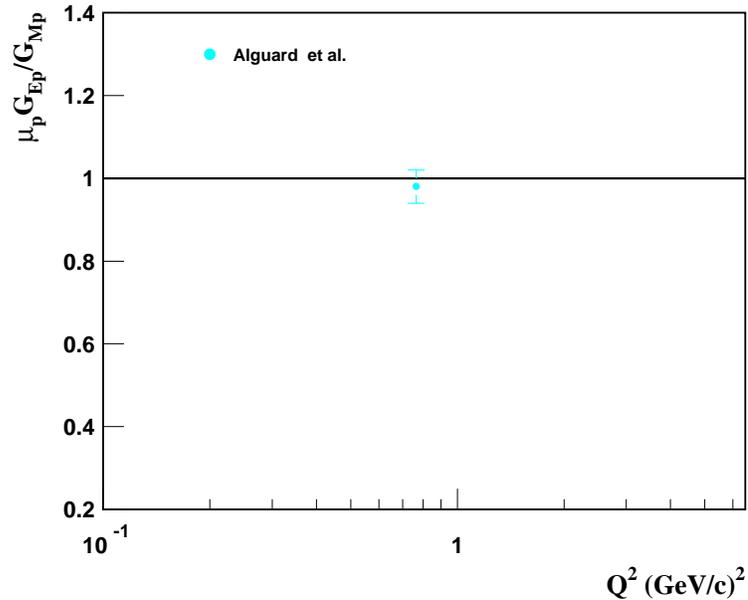,width=3.9in}
\end{center}
\caption[$\mu_{p}G_{Ep}/G_{Mp}$ by Alguard et al \cite{alguard76}.]
{The ratio of electric to magnetic form factor $\mu_{p}G_{Ep}/G_{Mp}$ by Alguard et al \cite{alguard76}.}
\label{fig:Alguard}
\end{figure}

The main idea of this experiment was to measure the antiparallel-parallel asymmetry $A$ in
the differential cross section which in turn is related to the form factors of the proton. 
The polarized electron source (PEGGY) at the 20-GeV Stanford Linear Accelerator 
Center (SLAC) was used to produce a polarized electron beam which was directed on a polarized 
proton target polarized by the method of dynamic nuclear orientation in a butanol target doped 
with 1.4\% porphyrexide. The scattered electrons were detected using the 8-GeV/c spectrometer.
Data were taken at $Q^2$ = 0.765 GeV$^2$ with incident electron energy of $E_{0}$ = 6.473 GeV 
and scattering electron angle of $\theta_{e}$ = 8.005$^{o}$. The beam and target polarization 
$P_{e}$ and $P_{p}$ were measured. The experimental asymmetry $\Delta$ was determined  
in order to solve for the antiparallel-parallel cross section asymmetry $A$ where 
$\Delta = P_{e} P_{p} F A$ and $F$ is the fraction of the elastically scattered 
electrons within the elastic missing-mass region. Figure \ref{fig:Alguard} shows the ratio of 
electric to magnetic form factor of the proton from this work.

\begin{itemize}
{\it
\item \textbf{Milbrath et al, 1998 \cite{milbrath98}:}
}
\end{itemize}

\begin{figure}[!htbp]
\begin{center}
\epsfig{file=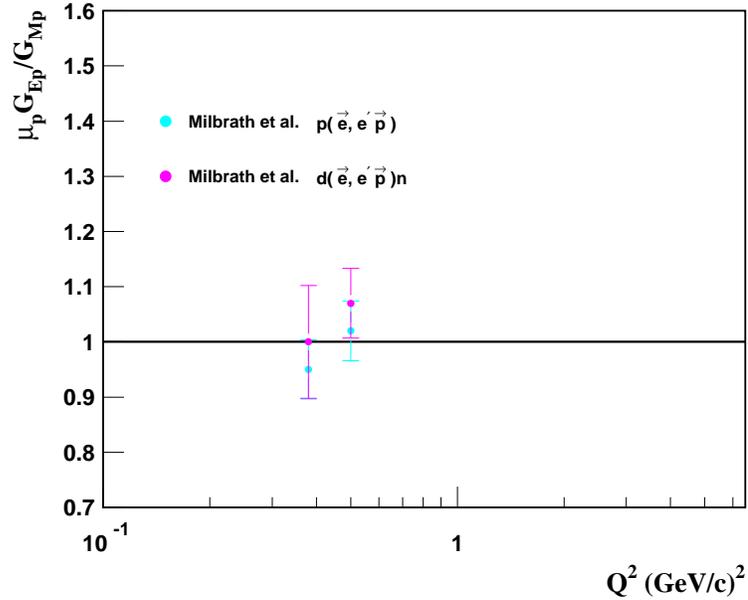,width=3.9in}
\end{center}
\caption[$\mu_{p}G_{Ep}/G_{Mp}$ by Milbrath et al \cite{milbrath98}.]
{The ratio of electric to magnetic form factor $\mu_{p}G_{Ep}/G_{Mp}$ by Milbrath et al \cite{milbrath98}.}
\label{fig:milbrath}
\end{figure}

This is the first experiment to demonstrate the feasibility of the recoil polarization as a 
technique to extract the ratio of electric to magnetic form factors of the proton.
Measurements of the recoil proton polarization observables in the reactions 
$p(\vec{e},e'\vec{p})$ and $d(\vec{e},e'\vec{p})n$ were made. The MIT-Bates Linear Accelerator Center
was used to produce a longitudinally polarized electron beam of energy $E_{0}$ = 0.58 GeV. The beam was 
directed on an unpolarized cryogenic target of liquid hydrogen and deuterium cells of 5 and 3 cm in diameter 
respectively. The scattered electrons were detected using the Medium Energy Pion Spectrometer, and the 
scattered protons were detected using One-Hundred Inch Proton Spectrometer. Two $Q^2$-values of 0.38 
and 0.5 GeV$^2$ were measured corresponding to electron-scattering angles of 
$\theta_{e}$ = $82.7^o$ and $113^o$ respectively. 
The recoil proton polarization was measured in the focal plane polarimeter (FPP). 
Figure \ref{fig:milbrath} shows the ratio of electric to magnetic form factor of the proton from this work.
%

\begin{itemize}
{\it
\item \textbf{Jones et al, 2000 \cite{jones00}:}
}
\end{itemize}

\begin{figure}[!htbp]
\begin{center}
\epsfig{file=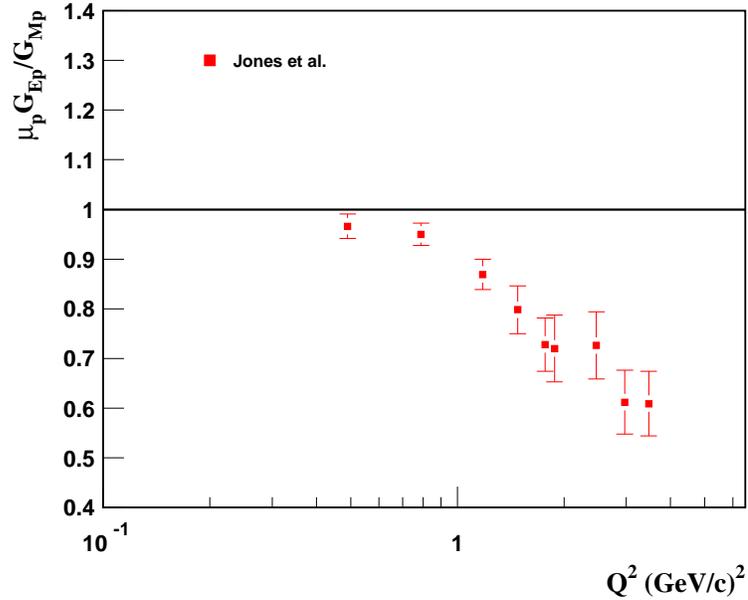,width=3.9in}
\end{center}
\caption[$\mu_{p}G_{Ep}/G_{Mp}$ by Jones et al \cite{jones00}.]
{The ratio of electric to magnetic form factor $\mu_{p}G_{Ep}/G_{Mp}$ by Jones et al \cite{jones00}.}
\label{fig:jones}
\end{figure}

Recoil proton polarization measurements were carried out at the experimental Hall A of Thomas Jefferson 
National Laboratory (JLAB) in Newport News Virginia. Polarized electron beam of energy in range
of 0.934$< E_{0} <$4.090 GeV was directed on a 15-cm-long unpolarized liquid hydrogen target.
The elastically scattered electrons and protons were detected in coincidence using the two identical
high resolution spectrometers (HRS) of Hall A. The ratio $\mu_{p} G_{EP} \over G_{MP}$ was determined at
9 $Q^2$-values in the range of 0.49$< Q^2 <$3.47 GeV$^2$ covering an angular range of 
$22.26^o$$< \theta_{e} <$$79.88^o$ for the electrons and $29.98^o$$< \theta_{p} <$$46.03^o$ for 
the protons. The recoil proton polarization was measured in the focal plane polarimeter (FPP). 
External radiative corrections were not applied. The internal radiative corrections such as the hard 
photon emission and higher order contributions were calculated \cite{afanasev01} and found 
to be on the order of a few percent and were not applied. The results of this work showed the decline of 
$\mu_{p} G_{EP} \over G_{MP}$ with increasing $Q^2$ deviating from form factor scaling and indicating
for the first time a definite difference in the spatial distribution of charge and magnetization 
currents in the proton. Figure \ref{fig:jones} shows the ratio of electric to magnetic form factor of the 
proton from this work.

\begin{itemize}
{\it
\item \textbf{Pospischil et al, 2001 \cite{pospischil01}:}
}
\end{itemize}

\begin{figure}[!htbp]
\begin{center}
\epsfig{file=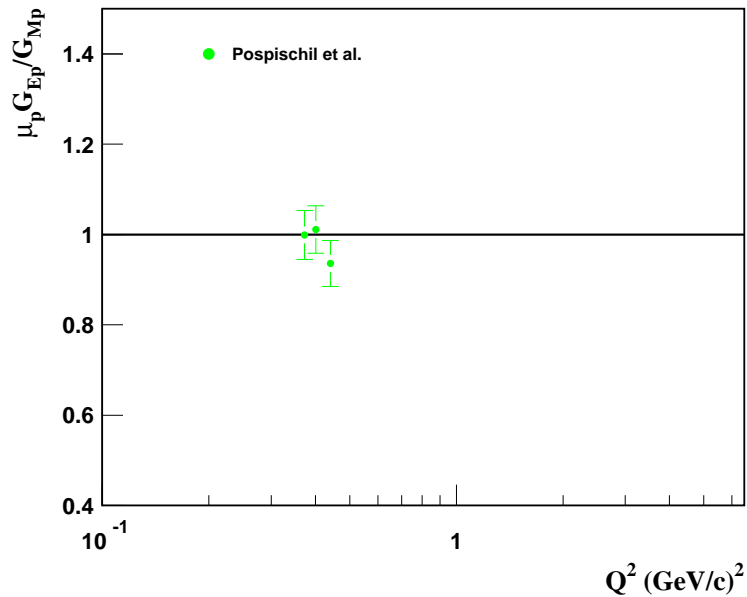,width=3.9in}
\end{center}
\caption[$\mu_{p}G_{Ep}/G_{Mp}$ by Pospischil et al \cite{pospischil01}.]
{The ratio of electric to magnetic form factor $\mu_{p}G_{Ep}/G_{Mp}$ by Pospischil et al \cite{pospischil01}.}
\label{fig:pospischil}
\end{figure}

Recoil proton polarization measurements were carried at the 3-spectrometer setup of the A1-Collaboration 
at the Mainz microtron MAMI. Longitudinally polarized electron beam of energy $E_{0}$ = 0.8544 GeV was 
directed on a 49.5-mm-long Havar cell filled with unpolarized liquid hydrogen. Data were taken at 
3 $Q^2$-values in the range of 0.373$< Q^2 <$0.441 GeV$^2$ covering an angular range of 
$48.2^o$$< \theta_{e} <$$54.4^o$ for the electrons and $45.5^o$$< \theta_{p} <$$49.5^o$ for 
the protons. The recoil proton polarization was measured in the focal plane polarimeter (FPP). 
Radiative corrections were not applied. Figure \ref{fig:pospischil} shows the ratio of electric 
to magnetic form factor of the proton from this work.  
%

\begin{itemize}
{\it
\item \textbf{Gayou et al, 2001 \cite{gayou01}:}
}
\end{itemize}

\begin{figure}[!htbp]
\begin{center}
\epsfig{file=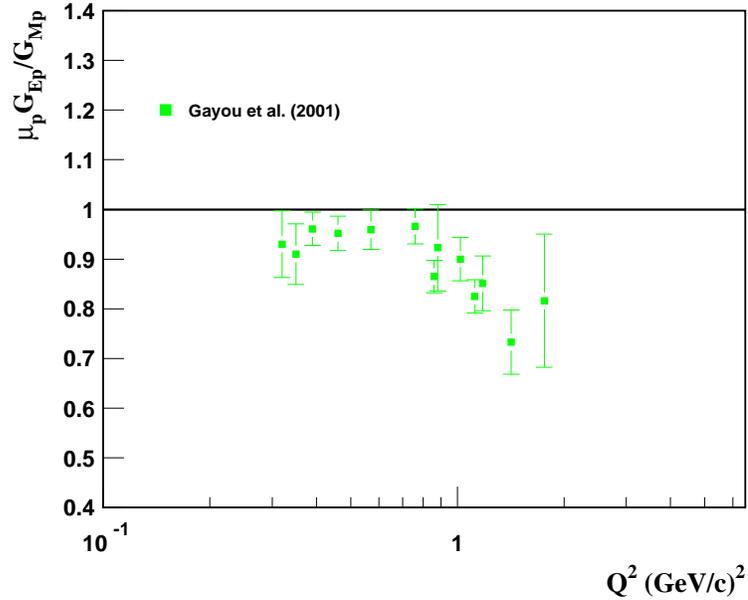,width=3.9in}
\end{center}
\caption[$\mu_{p}G_{Ep}/G_{Mp}$ by Gayou et al \cite{gayou01} (first measurements).]
{The ratio of electric to magnetic form factor $\mu_{p}G_{Ep}/G_{Mp}$ by Gayou et al \cite{gayou01} (first measurements).}
\label{fig:gayou2}
\end{figure}

Recoil proton polarization measurements were carried out at the experimental Hall A of Thomas 
Jefferson National Laboratory (JLAB) in Newport News Virginia using the same experimental
hardware used by Jones et al \cite{jones00}. Polarized electron beam of energy 
in range of 1.0$< E_{0} <$4.11 GeV was directed on a 15-cm-long unpolarized liquid hydrogen target.
The elastically scattered electrons and protons were detected in coincidence using the two identical
high resolution spectrometers (HRS) of Hall A. Although the goal of this experiment was to study the
$D(\vec\gamma,p)n$ and $H(\vec\gamma,p)\pi^{0}$ reactions, 13 measurements of coincidence 
$ep \to ep$ polarizations were performed to calibrate the focal plane polarimeter (FPP) used to 
study the reactions defined above. Data were taken in the range of  0.32$< Q^2 <$1.76 GeV$^2$ 
covering an angular range of $18.012^o$$< \theta_{e} <$$48.65^o$ for the electrons and 
$37.446^o$$< \theta_{p} <$$55.984^o$ for the protons. The recoil proton polarization was measured 
in the focal plane polarimeter. Radiative corrections were not applied.
Figure \ref{fig:gayou2} shows the ratio of electric to magnetic form factor of the proton from this work.

\begin{itemize}
{\it
\item \textbf{Gayou et al, 2002 \cite{gayou02}:}
}
\end{itemize}

\begin{figure}[!htbp]
\begin{center}
\epsfig{file=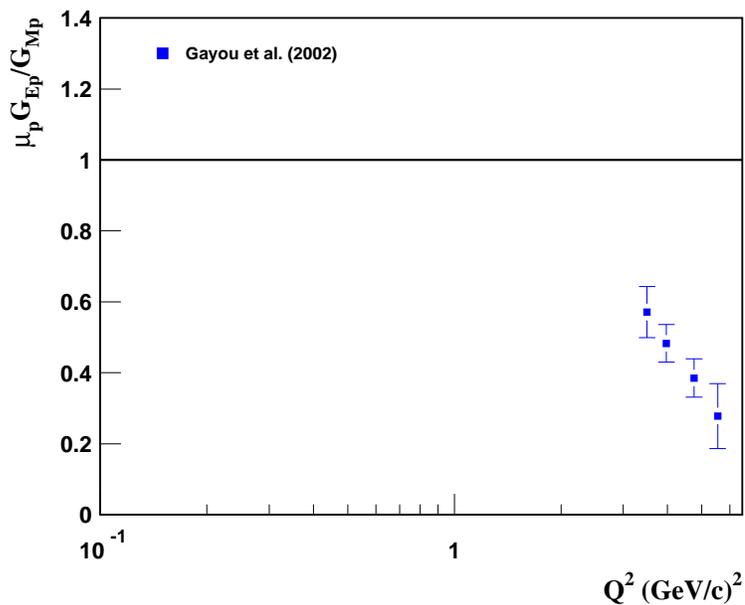,width=3.9in}
\end{center}
\caption[$\mu_{p}G_{Ep}/G_{Mp}$ by Gayou et al \cite{gayou02} (second measurements).]
{The ratio of electric to magnetic form factor $\mu_{p}G_{Ep}/G_{Mp}$ by Gayou et al 
\cite{gayou02} (second measurements).}
\label{fig:gayou1}
\end{figure}

This work in an extension of that of Jones et al \cite{jones00} where measurements were taken to higher $Q^2$ 
value of 5.54 GeV$^2$. Recoil proton polarization measurements were carried out at the experimental Hall A 
of Thomas Jefferson National Laboratory (JLAB) in Newport News Virginia also using the same experimental
hardware used by Jones et al. Polarized electron beam of energy 
in range of 4.59$< E_{0} <$4.607 GeV was directed on a 15-cm-long unpolarized liquid hydrogen target.
Data were taken in the range of  3.5$< Q^2 <$5.54 GeV$^2$. At $Q^2$ = 3.5 GeV$^2$, 
the elastically scattered  electrons and protons were detected in coincidence using the two identical 
high resolution spectrometers (HRS) of Hall A with fixed angles of $\theta_{e}$ = $30.6^o$ for the 
electrons and $\theta_{p}$ = $31.785^o$ for the protons. At higher $Q^2$ values of 3.97, 4.75, and 5.54 GeV$^2$ and at a fixed beam energy $E_{0}$ = 4.607 GeV, the electrons scattered at a higher angles than the 
protons and were detected using a calorimeter in coincidence with the protons covering an angular range 
of $19.275^o$$< \theta_{e} <$$28.587^o$ for the electrons 
and $34.5^o$$< \theta_{p} <$$51.44^o$ for the protons. The recoil proton polarization was measured using 
a focal plane polarimeter. Radiative corrections were not applied.
Figure \ref{fig:gayou1} shows the ratio of electric to magnetic form factor of the proton from this work.

\section{Summary of Previous e-p Measurements} \label{sect_previous_data_suumary}

A summary of section \ref{sect_previous_data} for the world's data on elastic e-p scattering
cross section measurements is given in Table \ref{world_ep_scattering}. 
Experiments are listed under the principal author's name, laboratory at which they were performed, 
energy, $Q^2$, number of $\varepsilon$ points measured at 
each $Q^2$ value $N_{\varepsilon}$, 
cross section uncertainty $\Delta \sigma$, and the overall normalization uncertainty in the cross 
section $\Delta \sigma_{N}$.

\begin{table}[!htbp]
\begin{center}
\begin{tabular}{||c|c|c|c|c|c|c||} \hline
\hline
Author&Laboratory&Energy      &$Q^{2}$    &$N_{\varepsilon}$&$\Delta \sigma$  &$\Delta \sigma_{N}$\\ 
      &          &(GeV)     &(GeV$^2$)    &                 &                 &               \\
      &          &            &           &                 &  \%             &    \%          \\
\hline \hline
Janssens\cite{janssens66} &Mark III    &0.250-1.00 &0.15-0.86   &3-5  &4       &1.6\\
Litt\cite{litt70}         &SLAC        &4.00-10.00 &1.00-3.75   &3-5  &1.5-2.0 &4\\
Price\cite{price71}       &CEA         &0.45-1.60  &0.25-1.75   &1    &3.1-5.3 &1.9\\
Berger\cite{berger71}     &Bonn        &0.66-1.91  &0.10-1.95   &3-14 &2-6     &4\\
Kirk\cite{kirk73}         &SLAC        &4.00-17.31 &1.00-25.00  &1    &2       &4\\
Murphy\cite{murphy74}     &Saskatchewan&0.057-0.123&0.006-0.031 &1    &-       &-\\
Bartel\cite{bartel73}     &DESY        &0.80-3.00  &0.67-3.00   &2-3  &2-4     &2.1\\ 
Stein\cite{stein75}       &SLAC        &4.50-20.00 &0.004-0.07  &1    &3.1     &2.8\\
Borkowski\cite{borkowski75}&Mainz      &0.15-0.30  &0.005-0.183 &-    &1-2     &-\\
Bosted\cite{bosted90}     &SLAC        &0.50-1.30  &0.49-1.75   &2-7  &3       &1.8\\
Rock\cite{rock92}         &SLAC        &9.761-21.00&2.50-10.00  &1    &1-4     &3\\
Sill\cite{sill93}         &SLAC        &5.00-21.50 &2.90-31.30  &1    &3-4     &3.6\\
Walker\cite{walker94}     &SLAC        &1.594-8.233&1.00-3.007  &3-8  &1       &1.9\\
Andivahis\cite{andivahis94}&SLAC       &1.511-9.80 &1.75-5.00   &3-6  &$<$ 2   &1.77\\
Christy\cite{christy04}    &JLAB       &1.148-5.50  &0.40-5.50  &3    &1.96    &1.7\\
\hline \hline 	
\end{tabular}
\caption[Summary of selected world data on e-p elastic scattering cross section measurements.]
{Summary of selected world data on e-p elastic scattering cross section measurements.}
\label{world_ep_scattering}
\end{center}
\end{table}
\section{Discussion} \label{sect_discussion}

\begin{figure}[!htbp]
\begin{center}
\epsfig{file=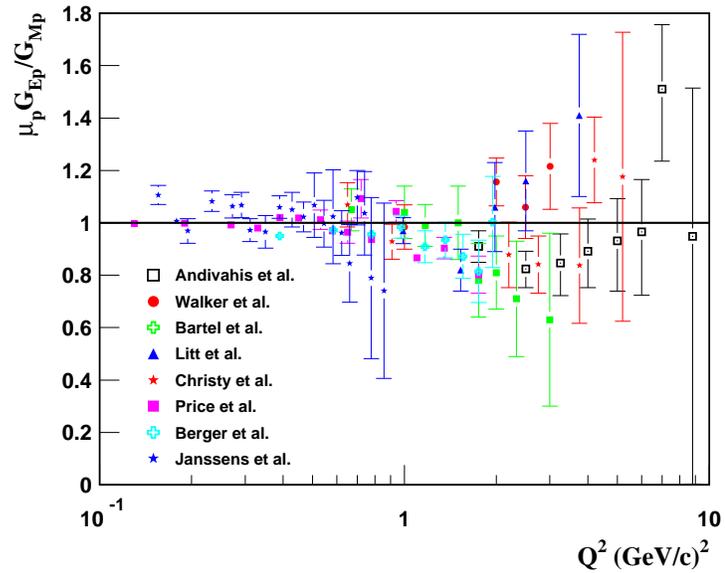,width=3.74in}
\end{center}
\caption[World data on $\mu_{p}G_{Ep}/G_{Mp}$ by Rosenbluth separation.]
{The world data on the ratio of electric to magnetic form factor $\mu_{p}G_{Ep}/G_{Mp}$ by 
Rosenbluth separation.}
\label{fig:gep_gmp_rosen}
\end{figure}

The world data on the ratio of electric to magnetic form factor of the proton as 
extracted from elastic e-p cross sections (Rosenbluth separation) and recoil 
polarization measurements in the single-photon exchange approximation have been summarized 
in section \ref{sect_previous_data}. Figure \ref{fig:gep_gmp_rosen} shows the 
$\mu_{p} G_{Ep} \over G_{Mp}$ world data as determined by the Rosenbluth separation method 
(notice the logarithmic scale for the $Q^2$ axis). With the exception of the data of 
Berger \cite{berger71} and Bartel \cite{bartel73}, which show a decrease in the ratio with increasing $Q^2$ 
especially for the $Q^2 > 1.0$ GeV$^2$, the data show approximate form factor scaling or 
$\mu_{p} G_{Ep} \over G_{Mp}$ = 1.0 with some fluctuation at high $Q^2$ points where the uncertainties become large. 
This is due to the fact that the experimental cross section is small and dominated by $G_{Mp}$ which makes it difficult 
to extract $G_{Ep}$ with high precision. The difficulty in extracting $G_{Ep}$ at high $Q^2$ values with high 
precision using the Rosenbluth separation technique was the motivation behind developing the recoil polarization 
technique, which is more sensitive to $G_{Ep}$ at large $Q^2$.

\begin{figure}[!htbp]
\begin{center}
\epsfig{file=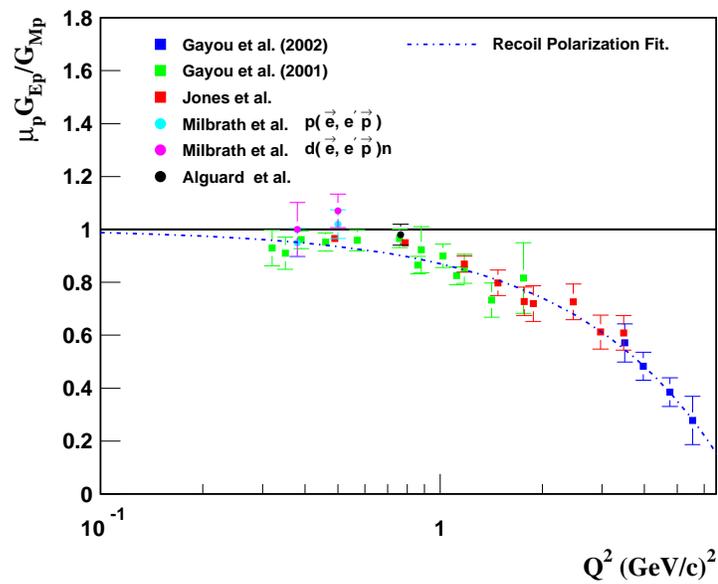,width=3.74in}
\end{center}
\caption[World data on $\mu_{p}G_{Ep}/G_{Mp}$ by recoil polarization.]
{The world data on the ratio of electric to magnetic form factors $\mu_{p}G_{Ep}/G_{Mp}$ by 
recoil polarization. The dashed line is the recoil polarization published fit.}
\label{fig:gep_gmp_recoil}
\end{figure}

Figure \ref{fig:gep_gmp_recoil} shows the world data on $\mu_{p} G_{Ep} \over G_{Mp}$ as determined
by recoil polarization technique. The data agree with form factor scaling for the region 
$Q^2 < 1.0$ GeV$^2$. However, for the region $Q^2 \ge 1.0$ GeV$^2$, the data decrease with 
increasing $Q^2$ (notice the logarithmic scale for the $Q^2$ axis) deviating significantly from form 
factor scaling as suggested by Rosenbluth separation. The data from recoil polarization 
measurements are more precise at high $Q^2$ and maybe less sensitive to systematic 
uncertainties than the Rosenbluth data. The dashed line in Figure \ref{fig:gep_gmp_recoil} 
is the recoil polarization fit to the data \cite{gayou02} or:
\begin{equation} \label{eq:recoil_fit}
\frac{\mu_{p} G_{Ep}}{G_{Mp}} = 1 - 0.13(Q^2 -0.04)~.
\end{equation}

We can see from Figures \ref{fig:gep_gmp_rosen} and \ref{fig:gep_gmp_recoil} that the two 
techniques give different results in the region $Q^2 \ge 1.0$ GeV$^2$. 
The values of $\mu_{p} G_{Ep} \over G_{Mp}$ from the two techniques differ almost by a 
factor of three at the high $Q^2$ points. This discrepancy between the results of the two techniques 
is becoming to be known as the $G^{p}_{E}$ crisis. 
This difference implies uncertainties in our knowledge of the form factors of the proton and 
raises several questions that must be answered. For example, the high precision data 
provided by the recoil polarization technique and the larger uncertainties in the L-T data
(small $G_{Ep}$ contribution at high $Q^2$) have led people to believe that the 
previous Rosenbluth extractions are inconsistent. If this is the case, then all the form factors 
extracted using Rosenbluth separation technique which are supposed to parameterize the deviation of the proton's 
structure from point-like particle are unreliable. Also, if there is a significant error in the elastic 
e-p cross section measurements ($Q^2 \ge$ 1.0 GeV$^2$), then there could be errors in all previous 
experiments that require normalization to elastic e-p cross sections, or require the use 
of the elastic cross sections \cite{dutta03,arrington04a} or form factors \cite{budd03} as an input to the analysis. 
If the cross section measurements and hence the Rosenbluth extractions are incorrect, that still will 
not solve the problem since the recoil polarization technique provides the ratio of $G_{Ep}$ to $G_{Mp}$ 
and not the actual values for the individual form factors.

At this point, the following questions are paramount:

\begin{enumerate}
{\it
\item \textbf{Why do the two techniques disagree?}
}
\begin{itemize}
{\it
\item \textbf{Is there a missing correction (radiative corrections or normalization uncertainties) or 
something wrong in the analysis of the previous Rosenbluth separations data?}
}
{\it
\item \textbf{Is there a missing correction (radiative corrections or proton spin precession
determination) or something wrong in the analysis of the previous recoil polarization data?}
}
\end{itemize}
{\it
\item \textbf{Which form factors are the correct ones to use?}
}
\begin{itemize}
{\it
\item \textbf{Is there something fundamentally wrong in the physics of one or both techniques?}
}
\end{itemize}
{\it
\item \textbf{What about all the conclusions based on the old and new theoretical models and 
calculations concerning the proton form factors?} 
}
\end{enumerate}

\begin{figure}[!htbp]
\begin{center}
\epsfig{file=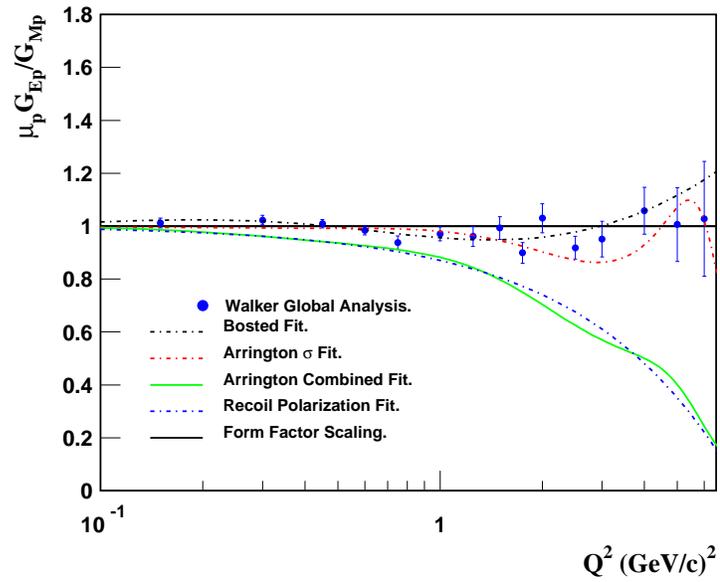,width=3.74in}
\end{center}
\caption[Fits of Arrington, Bosted, and recoil polarization along with the global analysis of Walker.]
{Global analysis of the proton form factors ratio $\mu_{p}G_{Ep}/G_{Mp}$ by Walker.
In addition, the fits of Arrington, Bosted, and recoil polarization are also shown.}
\label{fig:combined_fit}
\end{figure}

In order to provide a check on the consistency of the world's cross section measurements, several global analysis 
have been done. Walker \cite{walker94,walkerphd} combined cross sections from different experiments and performed 
a global extraction of the elastic form factors in the range of 0.1 $< Q^2 <$ 10.0 GeV$^2$. The result of the global 
fit supported the ansatz of form factor scaling as shown in Figure \ref{fig:combined_fit}.
%
%

An empirical fit to the world data of the proton form factors was made by Bosted \cite{bosted95} 
for the region 0.0 $< Q^2 <$ 30.0 GeV$^2$. The data of both $\frac{G_{Ep}}{G_{D}}$ and 
$\frac{G_{Mp}}{\mu_{p}G_{D}}$ in the region $Q^2 \le$ 7.0 GeV$^2$ are from the global analysis 
of Walker \cite{walker94}, while for $Q^2 >$ 9.0 GeV$^2$, form factor scaling was assumed to extract 
$G_{Mp}$. A good fit to the data was achieved when the form factors
were described as an inverse polynomial in $Q$:
\begin{equation} \label{eq:bosted_gep}
G_{Ep}(Q^2) = \frac{1.0}{1.0 + 0.62Q + 0.68Q^2 + 2.80Q^3 + 0.83Q^4}~,
\end{equation}
\begin{equation} \label{eq:bosted_gmp}
\frac{G_{Mp}(Q^2)}{\mu_{p}} = \frac{1.0}{1.0 + 0.35Q + 2.44Q^2 + 0.50Q^3 + 1.04Q^4 + 0.34Q^5}~.
\end{equation}

An extensive examination of the form factors extractions from cross section measurements was 
done by Arrington \cite{arrington04a,arrington03a}. A similar global analysis to that of Walker was performed. 
The fit included two extra data sets from Stein \cite{stein75} and Rock \cite{rock92} in addition to the 
data sets used in Walker's analysis. Also, the final published cross sections in the work 
of Sill \cite{sill93} and Andivahis \cite{andivahis94} were used which were not available for Walker's analysis. 
Some recent measurements of elastic scattering at Jefferson Lab by Dutta \cite{dutta03}, Niculescu \cite{niculescuphd},
and Christy \cite{christy04} were added. Also, the results of Borkowski \cite{borkowski75}, Murphy \cite{murphy74}, and
Simon \cite{simon80} were added to constrain the low $Q^2$ behavior. In addition, all the high $Q^2$ 
data up to 30 GeV$^2$ were included in the fit.

For each dataset, an overall normalization uncertainty was determined. The normalization 
uncertainties were taken either from the original published work or from Walker's global analysis. 
Independent normalization uncertainties were assigned to data taken by different detectors in the same 
experiment. In the work of Bartel \cite{bartel66}, data were taken using three different spectrometers, 
so the data were divided into three sets with different normalization uncertainty factor assigned to 
each data set. Higher order terms such as the Schwinger term \cite{schwinger49} and additional corrections for vacuum 
polarization contributions from muon and quark loops were added to the radiative corrections applied 
to the work of Janssens \cite{janssens66}, Bartel \cite{bartel66,bartel67,bartel73}, Albercht \cite{albrecht66}, 
Litt \cite{litt70}, Goitein \cite{goitein70}, Berger \cite{berger71}, and Price \cite{price71}. 
Finally, the small-angle ($<20^{o}$) data from Walker \cite{walker94} were excluded since an error was identified 
in that data. The form factors are parameterized as:
\begin{equation} \label{eq:arrington_sigma_fit}
G_{Ep}(Q^2), \frac{G_{Mp}(Q^2)}{\mu_{p}} = \frac{1.0}{1.0 + p_{2}Q^2+ p_{4}Q^4 + \ldots + p_{2N}Q^{2N}}~,
\end{equation}
where the parameters of the fit are listed in Table \ref{arrington_sigma}.
\begin{table}[!htbp]
\begin{center}
\begin{tabular}{||c|c|c||} \hline 
\hline
Parameter & $G_{Ep}$	&     $\frac{G_{Mp}}{\mu_{p}}$ \\ \hline \hline
p$_2$	&   3.226	&             3.19 \\
p$_4$	&   1.508	&             1.355 \\
p$_6$	&  -0.3773	&             0.151 \\
p$_8$	&   0.611	&           $-1.14\times10^{-2}$ \\
p$_{10}$ & -0.1853	&            $5.33\times10^{-4}$ \\
p$_{12}$ & $1.596\times10^{-2}$ &   $-9.0\times10^{-6}$ \\ \hline \hline
\end{tabular}
\caption[Arrington form factors parameters as extracted by cross sections fit.]
{Form factors parameters, equation (\ref{eq:arrington_sigma_fit}), as extracted by Arrington fit to $\sigma$.}
\label{arrington_sigma}
\end{center}
\end{table}

It was hypothesized that the discrepancy between the Rosenbluth and the polarization data was coming from 
a common systematic error in the cross section measurements and a (5-8)\% $\varepsilon$-dependent 
systematic error in the cross sections could resolve the discrepancy (see section 
\ref{two_photon_exchange_coulomb} for more detail). Therefore, a combined analysis was done by Arrington 
\cite{arrington04a} where an $\varepsilon$-dependent correction of 6\% was applied to all cross sections:
\begin{equation} \label{eq:combined_fit}
\sigma_{c} = \sigma_{o}\Big(1.0 - 0.06(\varepsilon - 1.0)\Big)~,
\end{equation}
and then the recoil polarization data were included in the fit. Here $\sigma_{c}$ and $\sigma_{o}$ 
are the corrected and uncorrected cross sections respectively. The form factors were parameterized 
using the same form as in equation (\ref{eq:arrington_sigma_fit}). The parameters of the fit are listed 
in Table \ref{arrington_combined}.
\begin{table}[!htbp]
\begin{center}
\begin{tabular}{||c|c|c||} \hline 
\hline
Parameter & $G_{Ep}$	&     $\frac{G_{Mp}}{\mu_{p}}$ \\ \hline \hline
p$_2$	&   2.94	&             3.00 \\
p$_4$	&   3.04	&             1.39 \\
p$_6$	&  -2.255	&             0.122 \\
p$_8$	&   2.002	&           $-8.34\times10^{-3}$ \\
p$_{10}$ & -0.5338	&            $4.25\times10^{-4}$ \\
p$_{12}$ & $4.875\times10^{-2}$ &   $-7.79\times10^{-6}$ \\ \hline \hline
\end{tabular}
\caption[Arrington form factors parameters as extracted from the combined fit.]
{Form factors parameters as extracted from Arrington combined fit to cross sections and recoil
polarization data.}
\label{arrington_combined}
\end{center}
\end{table}

The ratio of the electric to magnetic form factor of the proton as extracted from the 
Arrington's global fit to the world's elastic cross section data supports the results
of previous Rosenbluth extractions. The result indicated a good consistency between all 
different data sets and ruled out any possibility for a single or two bad data sets or incorrect 
normalization in the combined Rosenbluth analysis. 

An empirical fit to the proton form factors was done by Brash \cite{brash02} where most of the higher-$Q^2$
elastic e-p cross sections data were reanalyzed by using the ratio between the proton 
form factors $r$ = $\mu_{p}G_{Ep} \over G_{Mp}$ as provided by the recoil polarization data as a constraint: 
\begin{equation} \label{equation:brash_ratio}
r(Q^2) = \frac{\mu_{p}G_{Ep}}{G_{Mp}} = \begin{cases}
1 & \text{: if $Q^2 \le$ 0.04 GeV$^2$}~, \\
1 - 0.130(Q^2 - 0.04) &\text{: if 0.04 $< Q^2 <$ 7.7 GeV$^2$}~, \\
0 &\text{: if $Q^2 \geq$ 7.7 GeV$^2$}~, 
\end{cases}
\end{equation}
that way, $r(Q^2)$ fixes the ratio between the intercept ($ a = \tau G^2_{Mp}$) and the slope 
($b = G^2_{Ep}$) from the linear fit of the reduced cross section $\sigma_{R}$ to $\varepsilon$. 
The form factors are associated with the parameters of the linear fit, $\sigma_{R} = a(1+b\varepsilon)$, where 
$a =\frac{\mu^2_{p}b\tau}{r^2} = bR$ and $R = \frac{\tau\mu^2_{p}}{r^2}$.
A new parameterization of the proton magnetic form factor $G_{Mp}$ was obtained and the electric 
form factor was then calculated using the recoil polarization constrained ratio:
\begin{equation} \label{eq:brash_gmp} 
\frac{G_{Mp}(Q^2)}{\mu_{p}} = \frac{1}{1.0 + 0.116Q + 2.874Q^2 + 0.241Q^3 + 1.006Q^4 + 0.345Q^5}~,
\end{equation}
\begin{equation} \label{eq:brash_gep}
G_{Ep}(Q^2) = r(Q^2) \frac{G_{Mp}(Q^2)}{\mu_{p}}~.
\end{equation}

\begin{figure}[!htbp]
\begin{center}
\epsfig{file=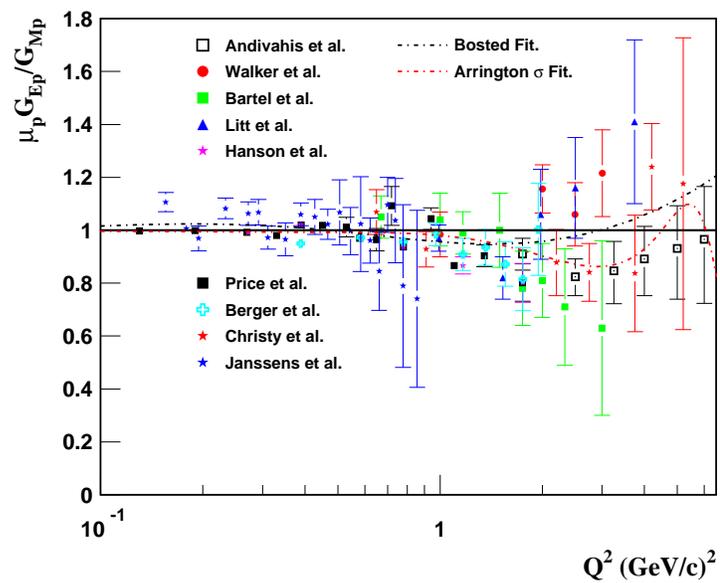,width=3.74in}
\end{center}
\caption[World data of the proton form factors by Rosenbluth separations with the fits of Arrington and Bosted.]
{Proton form factors ratio $\mu_{p}G_{Ep}/G_{Mp}$ by Rosenbluth separations.
In addition, the fits of Arrington and Bosted are also shown.}
\label{fig:fit_gepgmprosen}
\end{figure}

\begin{figure}[!htbp]
\begin{center}
\epsfig{file=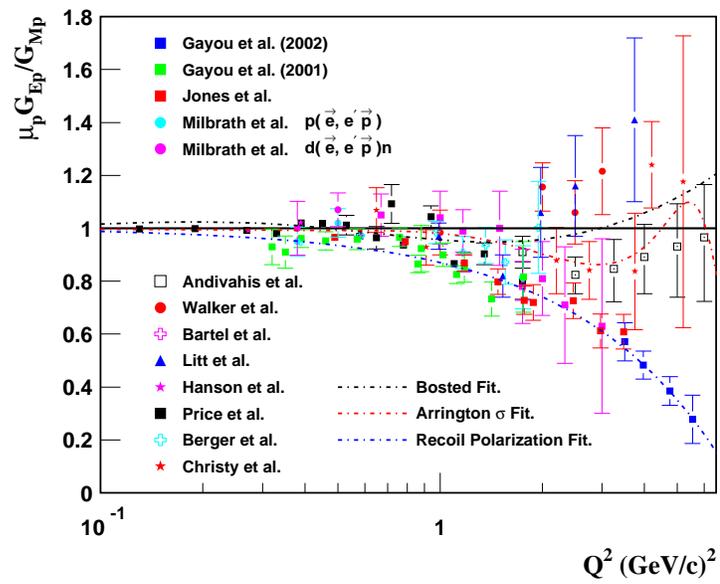,width=3.74in}
\end{center}
\caption[World data of the proton form factors by Rosenbluth separations and recoil polarization 
with the fits of Arrington, Bosted, and recoil polarization.]
{Proton form factors ratio $\mu_{p}G_{Ep}/G_{Mp}$ by Rosenbluth separations and recoil 
polarization. In addition, the fits of Arrington, Bosted, and recoil polarization are also shown.}
\label{fig:fit_gepgmprecoil}
\end{figure}

The magnetic form factor of the proton as parameterized by Brash \cite{brash02} was shown previously in Figure \ref{fig:gmp_gd} in section \ref{sect_interpretation}. In Figures \ref{fig:combined_fit}, \ref{fig:fit_gepgmprosen}, 
and \ref{fig:fit_gepgmprecoil}, the fits of Arrington, Bosted, and recoil polarization are shown 
along with the world data on proton form factors in addition to Walker's global analysis. 

\chapter{Experimental Setup}\label{chap_experimental_setup}
\pagestyle{plain}
\section{Overview} \label{overview}

Due to the inconsistency in the results of the ratio of the electric 
to magnetic form factors of the proton, $\mu_{p} G_{Ep} \over G_{Mp}$, as extracted 
from the Rosenbluth and recoil polarization techniques, and due to the fact that the
reported uncertainties in $\mu_{p} G_{Ep} \over G_{Mp}$ in the various Rosenbluth determinations
are much larger than those quoted for the polarization transfer measurements, 
a high-precision measurement of $\mu_{p} G_{Ep} \over G_{Mp}$ using the L-T 
separation technique in the $Q^2 >$ 1.0 GeV$^2$ region is important to:

\begin{itemize}
{\it
\item \textbf{Provide a comparison between the two techniques in the region 
where both can extract the value of $\mu_{p} G_{Ep} \over G_{Mp}$ with high precision.}
}
{\it
\item \textbf{Achieve uncertainties comparable to or better than the uncertainties
quoted by the recoil polarization measurements.}
}
{\it
\item \textbf{Provide a check on any possibility of additional and unaccounted for 
systematic uncertainties in the L-T or recoil polarization measurements.}
}
\end{itemize}

Experiment E01-001 sought to achieve these goals. It ran in May 2002 and was carried out in
Hall A of the Thomas Jefferson National Accelerator Facility (formerly known as Continuous 
Electron Beam Accelerator Facility, or CEBAF) which is located in Newport News Virginia in the 
U.S.A. An incident electron beam of energies in the range of 1.912$<E_{0}<$4.702 GeV
was directed on a 4-cm-long unpolarized liquid hydrogen target. High precision
measurements of the elastic e-p cross sections were made to allow for an 
L-T separation of the proton electric and magnetic form factors. Protons were detected 
simultaneously using the two identical high resolution spectrometers (HRS) or what is known by the 
left and right arm spectrometers of Hall A. The left arm spectrometer was used to measure three 
$Q^2$ points of 2.64, 3.20, and 4.10 GeV$^2$. Simultaneously, measurements at $Q^2$ = 0.5 GeV$^2$ 
were carried out using the right arm spectrometer which served as a monitor of beam charge, current, 
and target density fluctuations. 

\pagestyle{myheadings}

\begin{figure}[!htbp]
\begin{center}
\epsfig{file=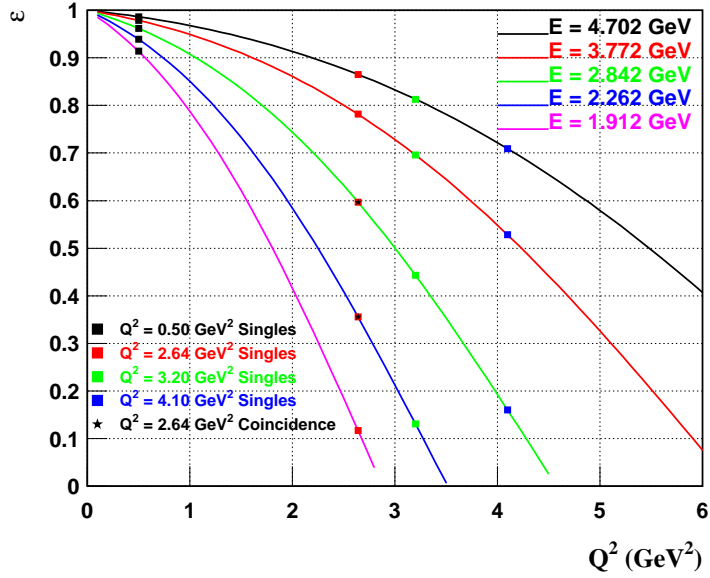,width=3.7in}
\end{center}
\caption[Plot of $\varepsilon$ vs $Q^2$ and the kinematics covered.]
{Plot of $\varepsilon$ vs $Q^2$ showing the kinematics covered.}
\label{fig:epsilon_q2}
\end{figure}

A total of 12 points (5 $\varepsilon$ points for $Q^2$ = 2.64 GeV$^2$, 4 $\varepsilon$ points for $Q^2$ = 3.20 GeV$^2$, 
and 3 $\varepsilon$ points for $Q^2$ = 4.10 GeV$^2$) were measured covering an angular range of 
12.52$^o$ $< \theta_{L} <$ 38.26$^o$ for the left arm, while the right arm was at $Q^2$ = 0.5 GeV$^2$, and used to 
simultaneously measure 5 $\varepsilon$  points covering an angular range of 58.29$^o$ $< \theta_{R} <$ 64.98$^o$. Here, 
$\theta_{L}$ and $\theta_{R}$ are the nominal angle of the struck proton with respect to the beam electron
for the left and right spectrometer, respectively. Figure \ref{fig:epsilon_q2} and Table \ref{kinematics} show and list 
the nominal kinematics covered in the E01-001 experiment and their settings. Small offsets were determined and applied 
to the energy and scattering angles. See section \ref{spect_mispoint} for details. The final kinematics used in the 
analysis are listed in Table \ref{angles_used}.

The final momentum of the scattered protons, $P_{measured}$, measured using the high resolution 
spectrometer, was compared to the final momentum of the scattered protons, $P_{calculated}(\theta_{p})$, 
calculated from two-body kinematics using the measured scattering angle of the protons $\theta_{p}$ 
(see equation (\ref{eq:scatter_angle})):
\begin{equation} \label{eq:pcal}
P_{calculated}(\theta_{p}) = \frac{2E_{beam}(M^{2}_{p}+E_{beam}M_{p})\cos\theta_{p}}
{M^{2}_{p}+E^{2}_{beam}+2E_{beam}M_{p}-E^{2}_{beam}\cos^{2}\theta_{p}}~,
\end{equation}
and the difference in momentum $\Delta P$ was then constructed:
\begin{equation} \label{eq:delta_p}
\Delta P = P_{measured} - P_{calculated}(\theta_{p})~,
\end{equation}
where $M_p$ is the mass of the proton and $E_{beam}$ is the incident electron energy.

The $\Delta P$ spectrum is made of several contributions. The main contributions come from the 
elastic peak which is due to the elastic $ep \to ep$ scattering and the radiative tail. 
In addition there are backgrounds due to quasi-elastic scattering from the aluminum target windows  
and high energy protons generated from photoreactions ($\gamma p \to \pi^0 p$ and 
$\gamma p \to \gamma p $) that contribute to the $\Delta P$ spectrum. 
For each of the kinematics covered, data were taken with the dummy target to subtract away 
the endcaps contribution from the spectrum. 
The photoproduction of $\pi^0$ or $\gamma p \to \pi^0 p$ events were simulated using a calculated
bremsstrahlung spectrum and $d \sigma \over dt$ $\propto s^{-7}$ and then subtracted away from the spectrum as well. 

\begin{table}[!htbp]
\begin{center}
\begin{tabular}{||c|c|c|c|c|c|c|c|c||} \hline
\hline
Setting & $E_{beam}$ & $\varepsilon_{L}$ & $Q^2_{L}$ & $\theta_{L}$ & $P_{L}$ & $\varepsilon_{R}$ & $\theta_{R}$ & $P_{R}$ \\ 
           & (GeV) &                   & (GeV$^2$) & ($^o$)       & (GeV/c) &        & ($^o$) & (GeV/c) \\
\hline \hline
$o$ & 1.912 & 0.117 & 2.64 & 12.631 & + 2.149 & 0.914 & 58.288 & + 0.756 \\
$a$ & 2.262 & 0.356 & 2.64 & 22.166 & + 2.149 & 0.939 & 60.075 & + 0.756 \\
$i$ & 2.842 & 0.597 & 2.64 & 29.462 & + 2.149 & 0.962 & 62.029 & + 0.756 \\
$q$ & 3.772 & 0.782 & 2.64 & 35.174 & + 2.149 & 0.979 & 63.876 & + 0.756 \\
$l$ & 4.702 & 0.865 & 2.64 & 38.261 & + 2.149 & 0.986 & 64.978 & + 0.756 \\
\hline \hline 
$b$ & 2.262 & 0.131 & 3.20 & 12.525 & + 2.471 & 0.939 & 60.075 & + 0.756 \\
$j$ & 2.842 & 0.443 & 3.20 & 23.395 & + 2.471 & 0.962 & 62.029 & + 0.756 \\
$p$ & 3.772 & 0.696 & 3.20 & 30.501 & + 2.471 & 0.979 & 63.876 & + 0.756 \\
$m$ & 4.702 & 0.813 & 3.20 & 34.139 & + 2.471 & 0.986 & 64.978 & + 0.756 \\
\hline \hline
$k$ & 2.842 & 0.160 & 4.10 & 12.682 & + 2.979 & 0.962 & 62.029 & + 0.756 \\
$r$ & 3.772 & 0.528 & 4.10 & 23.665 & + 2.979 & 0.979 & 63.876 & + 0.756 \\
$n$ & 4.702 & 0.709 & 4.10 & 28.380 & + 2.979 & 0.986 & 64.978 & + 0.756 \\ 
\hline \hline
$coin1$ & 2.262 & 0.356 & 2.64 & 22.166 & + 2.149 & 0.356 & 71.481 & - 0.855 \\
$coin2$ & 2.842 & 0.597 & 2.64 & 29.462 & + 2.149 & 0.597 & 47.439 & - 1.435 \\
$coin3$ & 3.362 & 0.398 & 4.10 & 20.257 & + 2.185 & 0.398 & 61.184 & - 1.177 \\
\hline \hline
\end{tabular}
\caption[The nominal kinematics setting for the E01-001 experiment.]
{The nominal kinematics setting for the E01-001 experiment. For each setting, $E_{beam}$ is the nominal electron beam energy,
$\varepsilon_{L}$ ($\varepsilon_{R}$) is the virtual photon polarization parameter for the
left (right) arm spectrometer, $Q^2_{L}$ is the four-momentum transfer squared for the left arm spectrometer, 
$\theta_{L}$ ($\theta_{R}$) is the spectrometer nominal scattering angle for the left (right) arm spectrometer, 
and $P_{L}$ ($P_{R}$) is the central momentum for the left (right) arm spectrometer. Note that the four-momentum transfer 
squared for the right arm $Q^2_{R}$ (not listed) was at 0.50 GeV$^2$ for all the settings except for the coincidence kinematics
settings $coin1$, $coin2$, and $coin3$ where $Q^2_{R}$ = $Q^2_{L}$. Small offsets were determined and applied to the energy and scattering
angles. See section \ref{spect_mispoint} for details. The final kinematics used in the analysis are listed in table \ref{angles_used}.}
\label{kinematics}
\end{center}
\end{table}

The net number of elastic events from data is then compared to the number of elastic events in the 
e-p peak as simulated using the Monte Carlo simulation program SIMC \cite{naomithesis,tomthesis}, under the same 
conditions (cuts), for a given narrow window cut on the $\Delta P$ spectrum. 
The ratio of the number of events from the data to that of the simulation normalized to input e-p 
cross section is then determined for that $\Delta P$ window cut.     

In addition, three coincidence kinematics were taken in order to allow for a separation of the elastic 
events from background events. Such separations helped us test our calculations of the lineshapes of the $\Delta P$ 
spectrum. The left arm was used to detect protons while electrons scattered in coincidence with protons were 
detected using the right arm spectrometer. The coincidence data were also used to provide a check on the 
scattering kinematics and to measure the proton detection efficiency and absorption. 

\section{Why Detect Protons?} \label{proton_vs_electron}

The majority of the previous elastic e-p cross section measurements
were made by detecting electrons rather than protons. However, proton
detection has several advantages over electron detection:
\begin{enumerate}
{\it
\item {Protons at moderately large angles correspond to electrons scattered at small angles.
Detecting protons allows us to go to lower values of electron scattering angles 
(down to $\sim$ 7$^o$) than would normally be possible.}
}
{\it
\item {Detecting protons reduces the variation of the cross section with the scattering angle. 
The cross section for the forward angle electrons varies rapidly with the scattering angle, 
while such variation for the corresponding protons is much smaller.
The reverse is true for the backwards angle electrons where the variation of the cross 
section with the scattering angle is greater for the corresponding forward angle protons,
however, it is still a smaller effect than that for the forward angle electrons.
}
}
{\it
\item {Detecting protons reduces the variation of the cross section with the beam energy.}
}
{\it
\item {Detecting backwards angle electrons results in a reduced cross section within the 
angular acceptance of the spectrometer. On the other hand, the corresponding protons fall
within a narrow angular window resulting in higher counting rates in less running time. That is, electrons
are cross section limited at large angles (small $\varepsilon$), while protons cross section is 10-20 times 
larger for small $\varepsilon$.}
}
{\it
\item {The proton momentum is constant for all $\varepsilon$ values for a given $Q^2$.}
}
{\it
\item {The linear $\varepsilon$ dependence of the radiative corrections is smaller for the protons
than for the electrons.}
} 
\end{enumerate}

\begin{figure}[!htbp]
\begin{center}
\epsfig{file=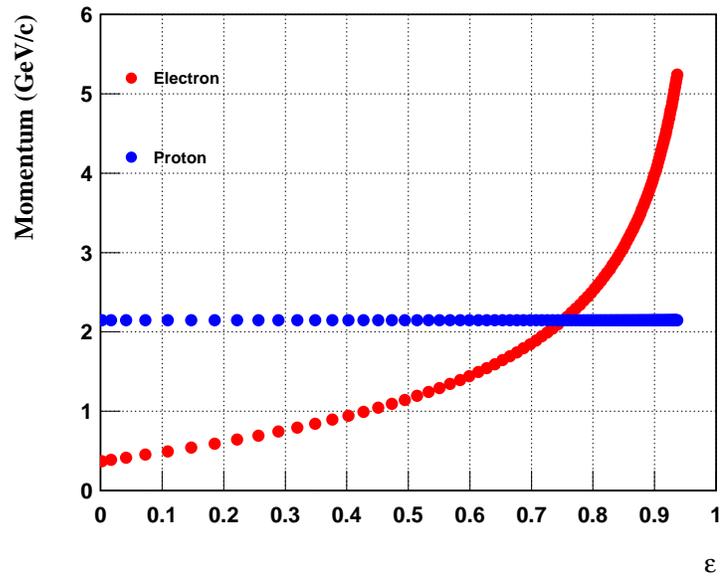,width=3.74in}
\end{center}
\caption[Proton and electron momentum as a function of $\varepsilon$ at $Q^2$ = 2.64 GeV$^2$.]
{Plot of the proton and electron momentum as a function of $\varepsilon$ at $Q^2$ = 2.64 GeV$^2$.}
\label{fig:momentum_epsilon}
\end{figure}

Because of these advantages, detecting protons greatly reduces the $\varepsilon$-dependent systematic 
corrections and associated uncertainties applied to the measured cross sections as compared to detecting 
electrons. Figure \ref{fig:momentum_epsilon} shows the momentum of the proton and electron as 
a function of the virtual photon polarization parameter $\varepsilon$ for $Q^2$ = 2.64 GeV$^2$.
The momentum is constant for the proton, while it varies by a factor of $\sim$ 20 for the electron. 
The fact that the proton momentum is the same for all $\varepsilon$ values at 
a given $Q^2$ point means that there is no $\varepsilon$ dependence due to any momentum-dependent corrections 
due to detector efficiency, particle identification, and multiple scattering. It should be
mentioned that any momentum-dependent correction will introduce an uncertainty in the reduced cross
sections at a given $Q^2$ point which in turn introduces an uncertainty in both $G_{Ep}$ and 
$G_{Mp}$ but not in the ratio.

\begin{figure}[!htbp]
\begin{center}
\epsfig{file=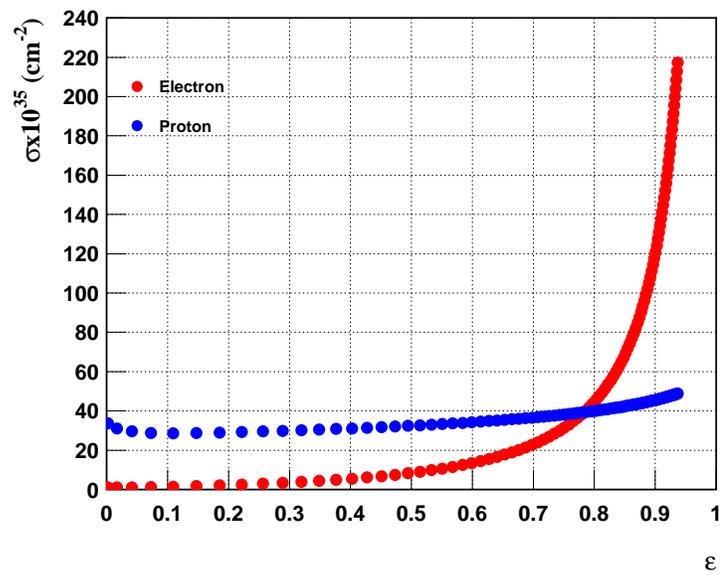,width=3.74in}
\end{center}
\caption[Cross section of the proton and electron as a function of $\varepsilon$ at $Q^2$ = 2.64 GeV$^2$.]
{Plot of the proton and electron cross section ($\sigma \times 10^{35}$) (cm$^{-2}$) as a function of 
$\varepsilon$ at $Q^2$ = 2.64 GeV$^2$.}
\label{fig:sigma_epsilon}
\end{figure}

Figure \ref{fig:sigma_epsilon} shows the cross section for both the proton, $\frac{d\sigma}{d\Omega_{p}}$, 
and electron, $\frac{d\sigma}{d\Omega_{e}}$, as a function of $\varepsilon$ for $Q^2$ = 2.64 GeV$^2$. 
The cross section, and thus rate for fixed conditions, is nearly constant for the 
proton and that will reduce the effect of rate-dependent uncertainties dramatically. Also, the low $\varepsilon$  
cross sections are no longer rate limited for the proton as it is the case for the electron which will require
more running time for comparable statistics. 

\begin{figure}[!htbp]
\begin{center}
\epsfig{file=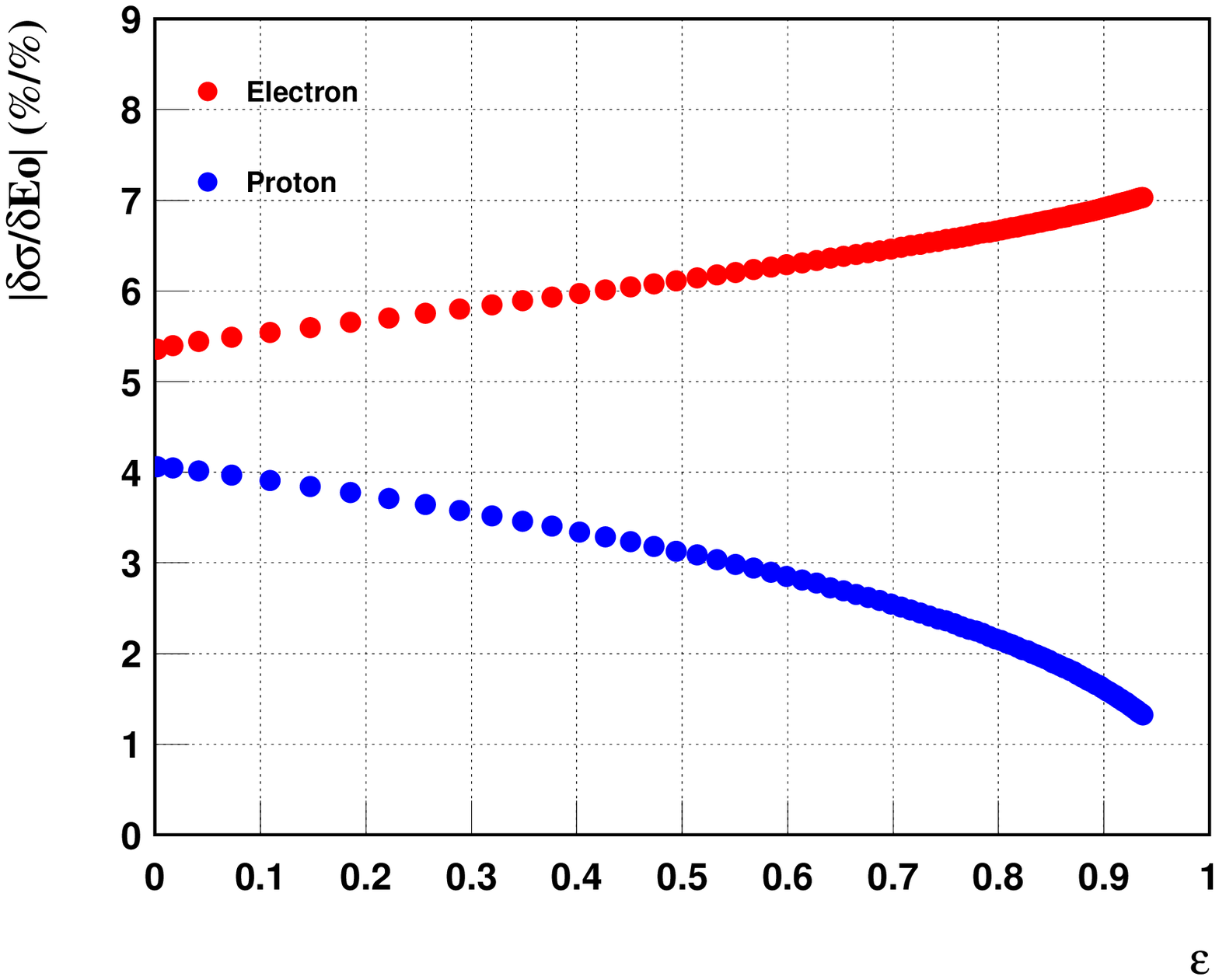,width=3.74in}
\end{center}
\caption[The absolute value of the proton and electron cross section sensitivity to the incident 
electron beam energy offset as a function of $\varepsilon$ at $Q^2$ = 2.64 GeV$^2$.]
{Plot of the proton and electron cross section sensitivity to the electron beam 
energy offset ($\delta \sigma \over \delta E_{o}$) (\%/\%) as a function of $\varepsilon$ at $Q^2$ = 2.64 
GeV$^2$.}
\label{fig:dsigma_de_epsilon}
\end{figure}

\begin{figure}[!htbp]
\begin{center}
\epsfig{file=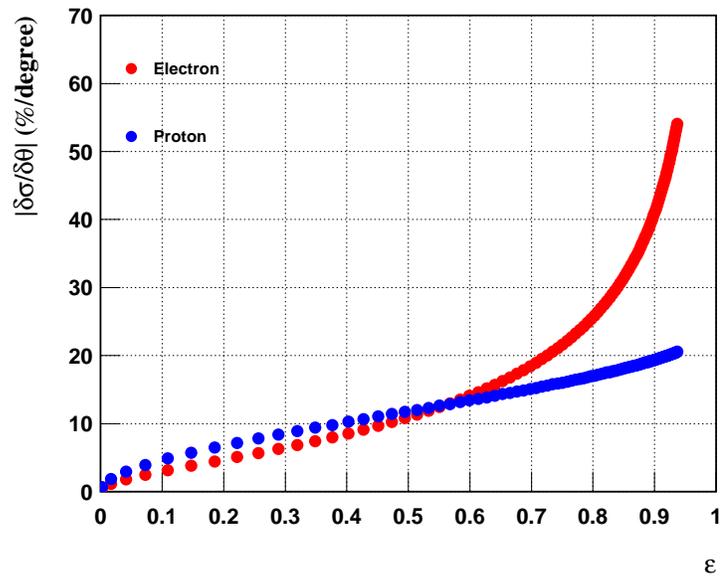,width=3.74in}
\end{center}
\caption[The proton and electron cross section sensitivity to the scattering angle 
offset as a function of $\varepsilon$ at $Q^2$ = 2.64 GeV$^2$.]
{Plot of the proton and electron cross section sensitivity to the scattering angle  
offset ($\delta \sigma \over \delta \theta$) (\%/degree) as a function of $\varepsilon$ at $Q^2$ = 2.64 
GeV$^2$.}
\label{fig:dsigma_dtheta_epsilon}
\end{figure}

Figures \ref{fig:dsigma_de_epsilon} and \ref{fig:dsigma_dtheta_epsilon} show the sensitivity
of the cross section to the beam energy (percentage change in the cross section
per one percent change in the beam energy) and scattering angle 
(percentage change in the cross section per one degree change in the scattering angle) for both 
particles as a function of $\varepsilon$ for $Q^2$ = 2.64 GeV$^2$. The variation of the 
cross section with beam energy and especially with scattering angle as a function of $\varepsilon$ is less 
for the proton than that of the electron. 

Figure \ref{fig:dsigma_de_epsilon} shows the sensitivity of the electron and proton 
cross section to a 1\% change in the beam energy. For the electron, the maximum variation 
of the cross section is $\sim$ 7\% and for the proton it is 4\%. If a 0.5\% measurement in the cross 
section is desired, then the beam energy must be known with a precision of 7.14$\times 10^{-4}$ for the electron and 
1.25$\times 10^{-3}$ for the proton. Such high precision is achievable for both particles knowing that 
energy measurements at JLAB can performed with precision as high 
as $\delta E_{o} \over E_{o}$ = 2$\times 10^{-4}$. 

Looking at Figure \ref{fig:dsigma_dtheta_epsilon} it can be seen that the biggest variation of the cross section of 
the electron for a one degree change in the scattering angle is $\sim$ 55\%. If we want to achieve a 0.5\% 
measurement in the cross section, we need to know the scattering angle to within $\sim$ 0.158 mrad which
is very difficult to achieve. On the other hand, for the proton, over the same $\varepsilon$ range, the variation is 
a factor of 3 less and the scattering angle need only to be known to within $\sim$ 0.476 mrad which is 
much more readily achievable.

\begin{figure}[!htbp]
\begin{center}
\epsfig{file=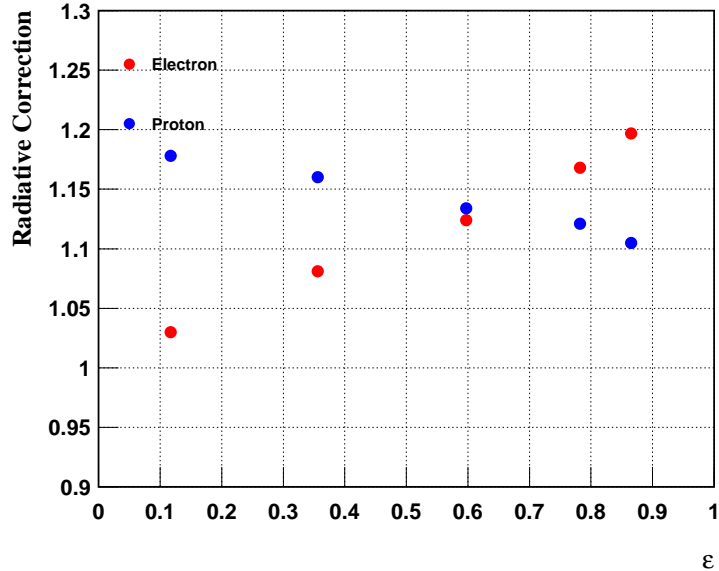,width=3.74in}
\end{center}
\caption[Radiative correction factor (internal corrections only) for the proton and electron as a function of $\varepsilon$ at 
$Q^2$ = 2.64 GeV$^2$.]
{Plot of the proton and electron radiative correction factor (internal corrections only) as a function of $\varepsilon$ at 
$Q^2$ = 2.64 GeV$^2$.}
\label{fig:rad_corr_epsilon}
\end{figure}

Finally, Figure \ref{fig:rad_corr_epsilon} shows the radiative correction factor (internal corrections only) as a function of 
$\varepsilon$ for $Q^2$ = 2.64 GeV$^2$ based on calculations done by Afanasev et al \cite{afanasev01}. While the magnitude of the 
corrections is similar and both show an approximately linear dependence on $\varepsilon$ the dependence is much smaller for 
protons, $\sim-$8\%, than for the electron $\sim$ 17\%. Note that the radiative corrections have an $\varepsilon$ 
dependence that is comparable to the slope brought about by the form factors. It is therefore extremely important that the 
radiative corrections (section \ref{rad_corrections}) are correctly handled. 

%

%

\section{The Continuous Electron Beam Accelerator Facility (CEBAF)}

\begin{figure}[!htbp]
\begin{center}
\epsfig{file=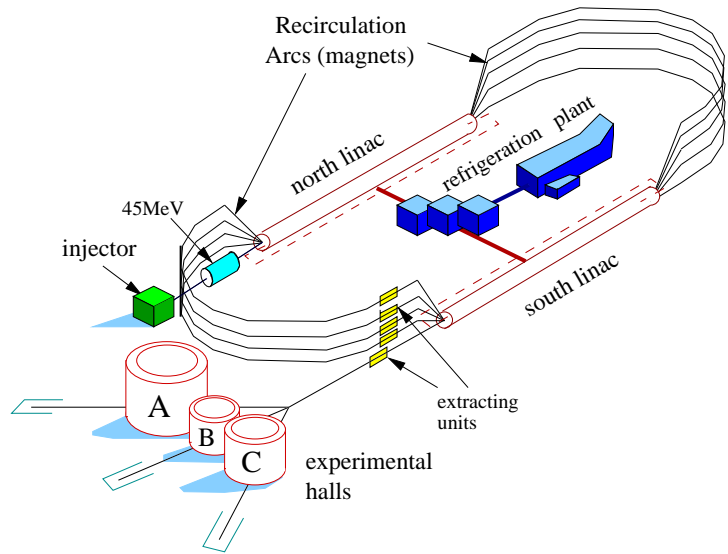,width=4in}
\end{center}
\caption[Thomas Jefferson National Accelerator Facility]
{Schematic layout of the Thomas Jefferson National Accelerator Facility.}
\label{fig:accelerator}
\end{figure}

During the run of the E01-001 experiment, the Continuous Electron Beam
Accelerator Facility of the Thomas Jefferson National Accelerator Facility \cite{cebaf} provided 
an unpolarized electron beam in the range of 1.912$< E_{0} <$4.702 GeV with beam
currents up to 70 $\micro$A. Figure \ref{fig:accelerator} shows the layout 
of the Jefferson Lab accelerator.
  
Using a state-of-the-art strained GaAs photocathode gun system with maximum current of
($I\sim$ 200 $\micro$A) and polarization above 70\%, continuous-wave (CW) beams 
of high current and polarization are delivered to both Hall A and Hall C ($I\sim$ 100 $\micro$A). 
Meanwhile, a high polarization and low current beam is delivered to Hall B 
($I\sim$ 100 nA). 

First, electrons from the photocathode gun is accelerated
to 50 MeV, and then injected into the north-linac. 
The north-linac consists of 20 Radio Frequency (RF) cryomodules. Each cryomodule has eight accelerating 
superconducting niobium cavities kept at a temperature of 2 K using liquid helium coolant form the 
Central Helium Liquefier. By the time the electrons reach the end of the north-linac, they 
could have been accelerated up to 600 MeV by the 160 cavities. At the end of the north-linac, 
180$^{o}$ bending arcs (east arc) with a radius of 80 meters join the north-linac to the 
identical and antiparallel superconducting south-linac forming a recirculating beamline. 
The beam through each arc is focused and steered using quadrupole and dipole magnets located inside 
each arc. The east arc has a total of 5 arcs on top of each other each with different bending field. 
The beam is steered through the east arc and passed on to the south-linac where it gets accelerated 
again and gain up to 600 MeV. 
At the end of the south-linac, the beam can be sent through the west arc for another pass or 
it can be sent to the Beam Switch Yard with a microstructure that consists of short (1.67 ps) 
bursts of beam coming at 1497 MHz. Each hall receives one third of these bursts, giving a pulse 
train of 499 MHz in each hall.

If another beam pass is desired for higher energy, the beam can be sent through the west recirculating 
linac for additional acceleration in the linacs, up to 5 passes through the accelerator. 
At the west recirculating linac, there are 4 different arcs on top of each other each with different 
bending field. It must be mentioned that the energy of the extracted beam is always a multiple of the 
combined linac energies $E_{linacs}$ = $E_{north~linac} + E_{south~linac}$, plus the initial injector 
energy $E_{injector}$ or ($E_{injector} + nE_{linacs}$) where $n$ is the number of passes.

\section{Hall A Beam Energy Measurements} \label{beam_energy_measurements}

Precise measurements of the beam energy are required in order to extract the form factors of the 
protons from the elastic e-p cross sections. There are two different measurements that
can be performed to determine the energy of incident electron beam with precision as high as 
$\delta E_{o} \over E_{o}$ = 2$\times 10^{-4}$. These measurements are known as the arc and ep
measurements.

\subsection{The Arc Beam Energy Measurement} \label{arc_beam_energy}

In the arc measurement \cite{marchandthesis,alcorn04,hallabeam}, the momentum $p$ of the incident electrons 
is determined by knowing the net bend angle $\theta$ of the electrons and the integral of the magnetic
field along the arc section (electron path) of the beam line:
\begin{equation} \label{eq:arc}
p = c \frac{\int Bdl}{\theta}~, 
\end{equation}
where $c$ is the speed of light. Equation (\ref{eq:arc}) is derived based on the fact that when an 
electron moves in a circular motion with velocity $\vec{v}$ in region of magnetic field $\vec{B}$, 
where the magnetic force on the electron $\frac{e}{c} \vec{v} \times \vec{B}$ provides the central 
force $\frac{mv^{2}}{r}$. Here, $e$ is the charge of the electron, $m$ is the mass of the electron, 
and $r$ is the radius of circular path which is related to the net bend angle and linear electron path 
$l$ as $l = r\theta$.

In the arc measurements, simultaneous measurements of the bend angle and the magnetic
field integral of the 8 bending dipoles (based on the measurement of the 9$^{th}$ dipole as a reference) 
in the arc section of the beamline are made. The bend angle is determined by measuring the beam position 
and profile at the entrance and exist of the arc using four wire scanners or superharps.
During the bend angle measurement, the quadruples are turned off (dispersive mode). 
The nominal bend angle of the beam in the arc section of the beamline is $\phi$ = 34.3$^{o}$. 

\begin{figure}[!htbp]
\begin{center}
\epsfig{file=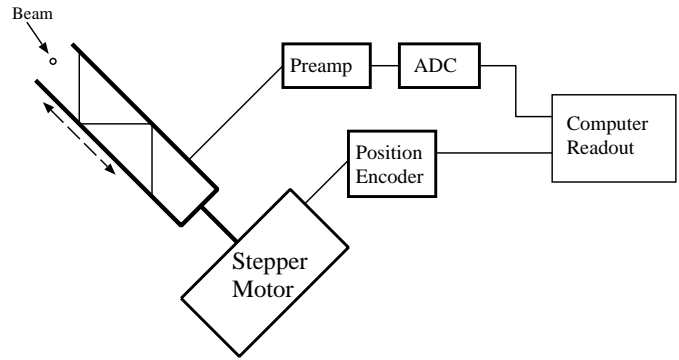,width=3.5in}
\end{center}
\caption[Schematic of the Superharp Systems]
{Schematic of the superharp system.}
\label{fig:harp}
\end{figure}

\begin{figure}[!htbp]
\begin{center}
\epsfig{file=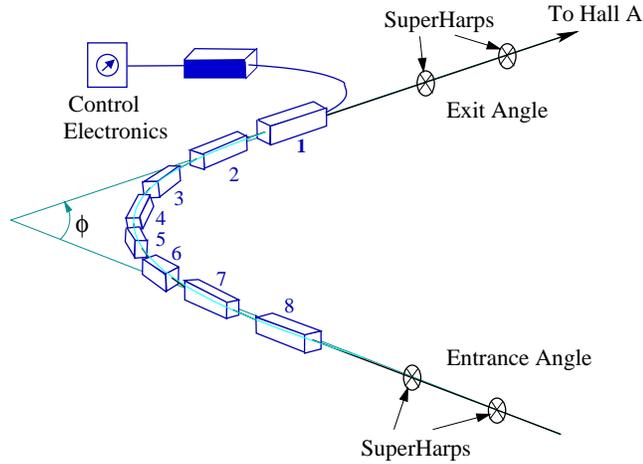,width=3.5in}
\end{center}
\caption[Schematic Layout of the Arc Section of the Beamline]
{Schematic Layout of the Arc Section of the Beamline.}
\label{fig:arc}
\end{figure}

A superharp consists of three wires, two vertical wires for the  horizontal beam profile measurement 
and one horizontal wire for the vertical beam profile. The signal from the wires gets picked up by an 
analog-to-digital converter (ADC). In addition, a position encoder measures the position of the ladder 
as the wires pass through the beam. The signal from the ADC and the position of the ladder determines 
the position and profile of the beam. Figures \ref{fig:harp} and \ref{fig:arc} show the 
superharp system and the arc section of the beamline. 
\subsection{The ep Beam Energy Measurement} \label{ep_beam_energy}

\begin{figure}[!htbp]
\begin{center}
\epsfig{file=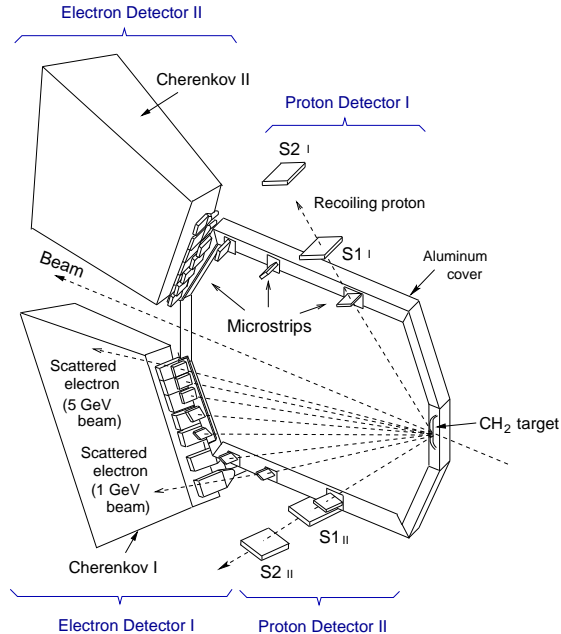,width=3in}
\end{center}
\caption[Schematic Layout of the ep energy measuring system.]
{Schematic Layout of the ep energy measuring system.}
\label{fig:ep_device}
\end{figure}

The second method of measuring the beam energy is the ep measurement \cite{alcorn04,ravelthesis,hallabeam}. 
A stand-alone device along the beamline located 17 m upstream of the target is used for the ep measurement. 
Based on the two-body kinematics in the elastic $^{1}H(\vec{e},e'\vec{p})$ reaction, an incident 
electron is scattered by a CH$_{2}$ film target enclosed by an aluminum cover. The scattered electron 
and recoiled proton are detected using two identical arms, each of which has a detector package as shown 
in Figure \ref{fig:ep_device}. The angles of the scattered electron $\theta_{e}$ and recoiled proton 
$\theta_{p}$ are measured, and the incident electron energy $E_{o}$ is then determined from the two-body 
kinematic equation:
\begin{equation} \label{eq:ep_energy}
E_{o} = M_{p} \Big (\frac{\cos{\theta_{e}} + (\sin{\theta_{e}}/\tan{\theta_{p}}) -1}{1 - cos{\theta_{e}}} \Big ) + O(\frac{m^2_{e}}{E})~,
\end{equation}
where $M_{p}$ is the mass of the proton, $m_{e}$ is the mass of the electron, and $E$ is the
final energy of the scattered electron. The energy measurements by the arc and ep methods
show an excellent agreement with each other within an experimental uncertainty of 
$\leq 3 \times 10^{-4}$. During the run of the E01-001 experiment, three arc and two ep 
measurements were made to determine the beam energy. 

Table \ref{beam_energy} lists all of the beam energy measurements for the E01-001 experiment. 
The Tiefenback energy measurement is similar to the arc measurement but is determined using Hall 
A arc beam position monitors (BPMs) instead of the superharps to calculate the beam energy. 
The Tiefenback value for the beam energy, which is consistent with the arc measurements, is used in the 
analysis of the E01-001 experiment. 
\begin{table}[!htbp]
\begin{center}
\begin{tabular}{||c|c|c|c|c|c|c||} \hline \hline
Pass & $E$(Tiefenback) & $\Delta E$ & Arc   & ep     & Arc/Tiefenback& ep/Arc \\
     & (MeV)           & (MeV)      &(MeV)  & (MeV)&               &           \\
\hline \hline
2    & 1912.94         & 0.69       & -      & -      & -             & -         \\
2    & 2260.00         & 0.81       & 2260.20& 2260.83& 1.000088      & 1.000279   \\
3    & 2844.71         & 1.03       & -      & -      & -             & -           \\
4    & 3772.80         & 1.36       & 3773.10& 3775.23& 1.000080      & 1.000565     \\
5    & 4702.52         & 1.70       & -      & -      & -             & -             \\
5    & 5554.60         & 2.00       & 5555.17& -      & 1.000103      & -              \\
\hline \hline
\end{tabular}
\caption[Tiefenback, arc, and ep beam energy measurements of the E01-001 experiment.]
{Tiefenback, Tiefenback quoted uncertainty $\Delta E$, arc, and ep beam energy measurements of the 
E01-001 experiment. Tiefenback energy was used for the analysis with final uncertainty of 0.03\% 
offset combined with 0.02\% point-to-point uncertainty.}
\label{beam_energy}
\end{center}
\end{table}

Random (point-to-point) and scale uncertainties of 0.01\% and 0.05\% have been reported on the 
non-invasive arc measurements (the Tiefenback energy), respectively, and a 0.02\% random uncertainty 
in the full invasive arc measurement or the ep measurement \cite{alcorn04}. 
Since the Tiefenback results were consistent with the full arc and ep measurements where we took them, 
we can assume that the absolute uncertainty in the Tiefenback is closer to the 0.02\%. Therefore, an 
overall scale and random uncertainties of 0.03\% and 0.02\% are estimated on Tiefenback energy, 
respectively. 

\begin{figure}[!htbp]
\begin{center}
\epsfig{file=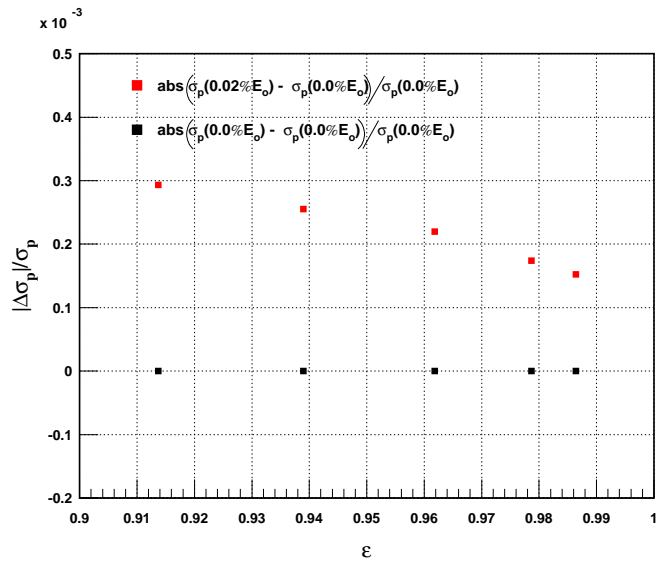,width=3.4in}
\end{center}
\caption[The relative difference between the right arm nominal cross sections and  
cross sections with a 0.02\% beam energy offset as a function of $\varepsilon$.] 
{The relative difference between the right arm nominal cross sections and  
cross sections with a 0.02\% beam energy offset (red squares) as a function
of $\varepsilon$ at all 5 incident energies. The black squares are the nominal cross sections 
relative to themselves.} 
\label{fig:right_proton_sigma0.02e0_epsilon}
\end{figure}

\begin{figure}[!htbp]
\begin{center}
\epsfig{file=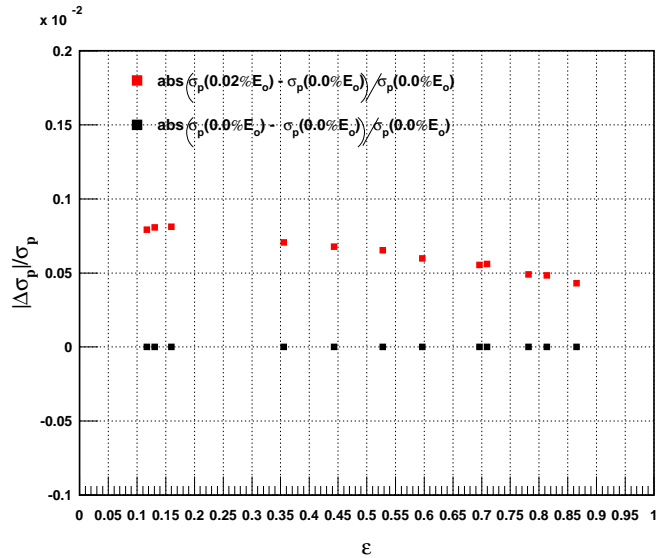,width=3.4in}
\end{center}
\caption[The relative difference between the left arm nominal cross sections and  
cross sections with a 0.02\% beam energy offset as a function of $\varepsilon$.] 
{The relative difference between the left arm nominal cross sections and  
cross sections with a 0.02\% beam energy offset (red squares) as a function
of $\varepsilon$. The black squares are the nominal cross sections relative to themselves.} 
\label{fig:left_proton_sigma0.02e0_epsilon}
\end{figure}

The classification of the uncertainty into a slope, random, and scale will be discussed in 
detail in section \ref{qeff_intro}.
The sensitivity of the cross sections to a 0.02\% beam energy offset has been studied. 
Figures \ref{fig:right_proton_sigma0.02e0_epsilon} and \ref{fig:left_proton_sigma0.02e0_epsilon} show 
the relative difference between the nominal cross sections, determined at the nominal 
scattering angle and energy, and cross sections with a 0.02\% shift in the beam energy
for the right and left arms and at all kinematics. A 0.02\% energy fluctuation changes
the cross section by 0.01-0.03\% for the right arm and 0.04-0.08\% for the left arm.
Such change in the cross sections was applied as a random uncertainty to each $\varepsilon$ point.

\begin{figure}[!htbp]
\begin{center}
\epsfig{file=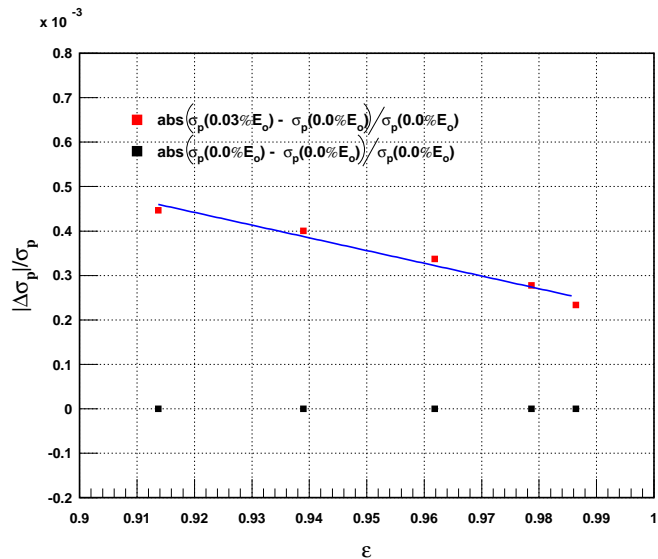,width=3.4in}
\end{center}
\caption[The relative difference between the right arm nominal cross sections and  
cross sections with a 0.03\% beam energy offset as a function of $\varepsilon$.] 
{The relative difference between the right arm nominal cross sections and  
cross sections with a 0.03\% beam energy offset (red squares) as a function
of $\varepsilon$ at all 5 incident energies. The solid blue line is a linear fit to the data.}
\label{fig:right_proton_sigma0.03e0_epsilon}
\end{figure}

\begin{figure}[!htbp]
\begin{center}
\epsfig{file=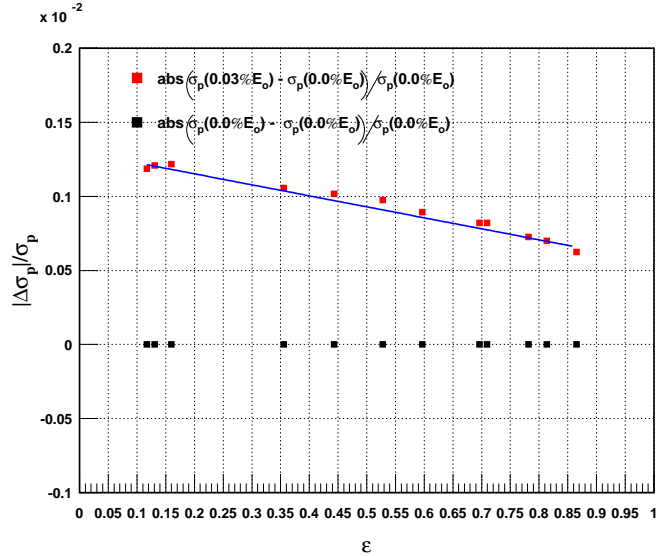,width=3.4in}
\end{center}
\caption[The relative difference between the left arm nominal cross sections and  
cross sections with a 0.03\% beam energy offset as a function of $\varepsilon$.] 
{The relative difference between the left arm nominal cross sections and  
cross sections with a 0.03\% beam energy offset (red squares) as a function
of $\varepsilon$. The solid blue line is a linear fit to the data.}
\label{fig:left_proton_sigma0.03e0_epsilon}
\end{figure}

Similarly, the sensitivity of the cross sections to an overall 0.03\% beam energy offset has been studied.
Figures \ref{fig:right_proton_sigma0.03e0_epsilon} and 
\ref{fig:left_proton_sigma0.03e0_epsilon} show the relative difference between the nominal cross 
sections and cross sections with a 0.03\% beam energy offset for the right and left arms and at all kinematics. 
In order to determine the slope uncertainty in the cross sections due to a 0.03\% beam energy offset, a fit of the 
relative difference in the cross sections to a straight 
line was performed at each $Q^2$ value. The average slope was used as the overall slope uncertainty. 
On the other hand, the deviation of the relative difference in the cross sections from the data and fit
was then used as a measure of the random uncertainty in each $\varepsilon$ point, while the average of 
the relative difference in the cross sections was used as an overall scale uncertainty.
The right arm results show an average scale, random, and slope uncertainties of 
0.034\%, 0.01\%, and 0.29\%, respectively. The left arm results show an average scale, 
random, and slope uncertainties of 0.13\%, 0.02\%, and 0.073\%, respectively.
It should be mentioned that the same procedure of estimating the scale, random, and slope
uncertainties in the sensitivity of the cross sections to a 0.10 mrad and 0.18 mrad 
angle offset will be used in section \ref{spect_mispoint}.

%

\section{Beam Position Measurements} \label{beam_position}

Knowing the position and direction of the beam on the target is crucial since any 
beam offset could be translated into an uncertainty in the effective target length 
used in the analysis. During the E01-001 experiment, the beam first was rastered using 
a rectangular raster \cite{alcorn04} producing a $\sim$ 2 mm $\times$ 2 mm spot size 
at the target to prevent any damage to the target by overheating it or local density fluctuations. 

In order to determine the beam position and direction at the target, two beam position
monitors (BPMs), BPMA and BPMB, \cite{alcorn04,hallabeam,xzhengthesis} located 7.516 and 2.378 m upstream from the target 
were used. The BPM is a cavity with 4-wire antenna in one plane with frequency 
tuned to match the RF frequency of the beam (1497 MHz). The absolute position of the beam is
determined by the BPMs by calibrating the BPMs with respect to wire scanners (superharps)
located adjacent to each BPM at 7.345 and 2.214 m upstream of the target.

The standard difference-over-sum method is used to determine the relative position of the 
beam to the wires. The position of the beam averaged over 0.3 sec, as read from the BPMs, 
is written into a data stream system on a event-by-event basis. The actual beam position and angle
at the target can be reconstructed as:
\begin{equation} \label{eq:beam_x}
x_{beam} = \frac{1}{a_3} \Big (x_{bpma}z_{bpmb} - x_{bpmb}z_{bpma} \Big )~,
\end{equation}
\begin{equation} \label{eq:beam_y}
y_{beam} = \frac{1}{a_3} \Big (y_{bpma}z_{bpmb} - y_{bpmb}z_{bpma} \Big )~,
\end{equation}
\begin{equation} \label{eq:beam_theta} 
\theta_{beam} = \frac{a_1}{a_3}~,
\end{equation}  
\begin{equation} \label{eq:beam_phi} 
\phi_{beam} = \frac{a_2}{\sqrt{a_{1}^2 + a_{3}^2}}~,
\end{equation}  
where $a_1$ = $x_{bpmb} - x_{bpma}$, $a_2$ = $y_{bpmb} - y_{bpma}$, and $a_3$ = $z_{bpmb} - z_{bpma}$.
Also, $x_{bpma}$, $y_{bpma}$, $x_{bpmb}$, and $y_{bpmb}$ are the $x$ and $y$ coordinates of the beam
as determined by beam position monitor a and b (BPMA and BPMB). Finally, $z_{bpma}$ and $z_{bpmb}$
are the locations of the superharps for both BPMA and BPMB and have a value of -2.214 and -7.345 m,
respectively. The values of $z_{bpma}$ and $z_{bpmb}$ are negative because the superharps 
are located upstream with respect to the target. 

\begin{figure}[!htbp]
\begin{center}
\epsfig{file=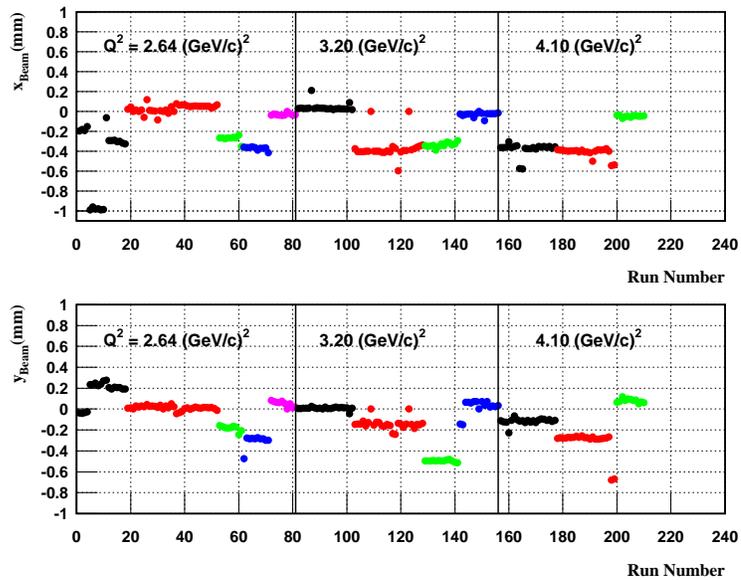,width=3.94in}
\end{center}
\caption[The x and y coordinates of the beam at the target.]
{The top(bottom) plot shows the x(y) coordinate of the beam at the target.}
\label{fig:beam_position}
\end{figure}

Using the above equations, the position of the beam on the target was reconstructed.
Figure \ref{fig:beam_position} top(bottom) shows the $x$($y$) coordinate of the beam at the target
for all the runs in order of increasing $\varepsilon$ and $Q^2$. For example, the point 
$Q^2$ = 2.64 GeV$^2$ has 5 $\varepsilon$ points and a different color is given for each 
$\varepsilon$ value starting with black for the lowest $\varepsilon$ point and then red for the 
next and higher $\varepsilon$ point and so on. See Table \ref{kinematics} for kinematics 
description. The beam was well focused on the target with an average (x,y) position of 
(-0.20,-0.10) mm, and an average beam drift of $\pm$0.30 mm. The position uncertainties of 0.30 mm
in two BPMs a $\sim$ 5 m apart yield a $\sim$ 0.07 mrad angle uncertainty in the beam angle.

\section{Beam Current Measurements} \label{beam_current}

\begin{figure}[!htbp]
\begin{center}
\epsfig{file=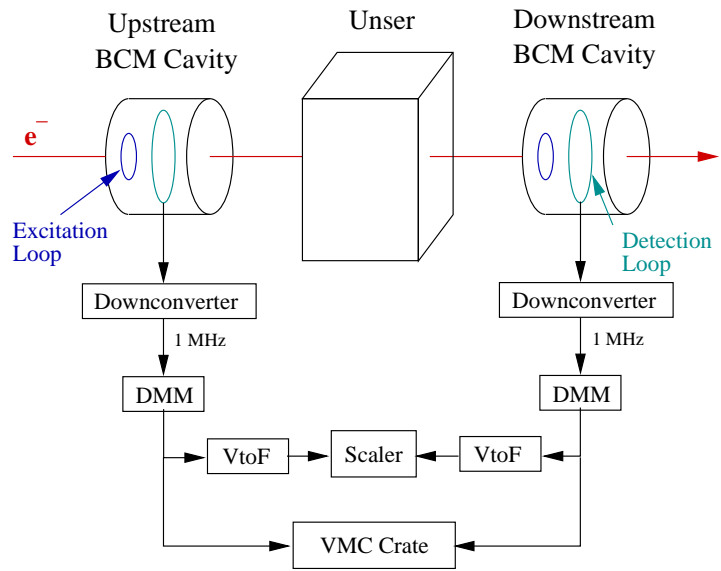,width=4.0in}
\end{center}
\caption[Schematic Layout of the Hall A Beam Current Monitors.]
{Schematic layout of the Hall A Beam Current Monitors readout block.}
\label{fig:BCM}
\end{figure}

In order to measure the current used and therefore the total charge 
accumulated during each run in each kinematics, two identical beam current 
monitors (BCMs) located $\sim$ 25 m upstream from the target are used. The two BCMs
are calibrated at several beam currents relative to a parametric current transformer or 
Unser monitor. The Unser monitor is located halfway between the two BCMs and is  
calibrated by passing a precisely known current through it. 
Hall A beam current monitors \cite{alcorn04} are designed to provide a non-invasive 
continuous measurements for the beam current. The Hall A BCM is a stainless
steel cylindrical cavity 15.24 cm in length and 15.48 cm in diameter.
The cylindrical axis of the cavity coincides with the beam direction. When the 
electron beam enters the cavity, it excites the transverse magnetic mode 
of the cavity. The resonant frequency of the cavity is tuned to the frequency of the electron
beam (1497 MHz) and hence an output signal from one of the two antennas inside the cavity
is produced. The output signal is then amplified and split into two components, 
one of which is sent to an AC Voltmeter which measures the beam current averaged over 
1.0 sec period. The second signal is then converted into an analog DC voltage level
by using an RMS-to-DC converter. The analog DC voltage level is then converted to a frequency 
signal using a voltage-to-frequency converter. Scalers gated by the start and end of each run receive the
frequency signal and then provide a measurement of the accumulated charge during the runs.
Figure \ref{fig:BCM} shows the schematic layout of the BCM readout block.

\section{Hall A Target System}

The target system of Hall A \cite{alcorn04,hallatarget} consists of the scattering vacuum chamber 
and the cryogenic targets. The 4-cm long liquid hydrogen target 
was used in this experiment.
  
\subsection{The Scattering Chamber}

The scattering vacuum chamber of Hall A consists of three main sections or rings.
The first section, the base ring, is fixed on the pivot of the hall and made of 
stainless-steel. It contains a vacuum pump-out port and several viewing and 
electrical ports. The middle section, middle ring, is located at the beam height and
is made of aluminum. It has a 104 cm inner diameter and 5 cm thick aluminum wall. 
Also a 15.2 cm vertical cutout on each of the beam side over the whole angular range of 
12.5$^o$$< \theta <$167.5$^o$ has been introduced. A 0.38 mm thin aluminum foil 
covers the vertical cutout on both side of the beam. The middle section has beam 
entrance and exit ports which are vacuum coupled to the electron beamline to prevent the beam 
from interacting with any material except the target. The third section, the upper ring, houses 
the cryogenic targets system.

\subsection{The Cryogenic Target}

The cryogenic target system of Hall A is mounted on a ladder inside the scattering vacuum
chamber. The ladder contains sub-systems for cooling, gas handling, 
temperature and pressure monitoring, and target control and motion. The ladder also 
contains a selection of solid targets such as dummy target for background measurements and BeO, 
$^{12}$C, and optics targets for beam viewing and calibration. 
The desired target can be selected from the control room (counting house) by moving the ladder 
vertically up and down until the target is aligned with the beam.

The cryogenic target has three independent loops. Two loops were configured to hold liquid hydrogen
(LH$_2$) or liquid deuterium (LD$_2$), and one for a gaseous helium.
Fans are used to circulate the liquid or the gas through each loop.
Each of the LH$_2$ and LD$_2$ loops has two aluminum cylindrical target cells of 
either 4 or 15 cm in length and 6.35 cm in diameter. The sidewalls of 
the cells are 0.178 mm thick with entrance (upstream) and exit (downstream) windows 
of 0.071 and 0.102 mm thick, respectively.

During the E01-001 experiment, the 4-cm LH$_2$ target was used and operated at a constant
temperature of 19 K and pressure of 25 psi with a density of 0.0723 g/cm$^3$. The 
temperature of the target was stabilized using a high-power heater to compensate for
the effect of any beam intensity variation on the target's temperature. 
The target was cooled using a target coolant (liquid helium) at 15 K supplied by the 
End Station Refrigerator (ESR). Because of the small spot size of the beam, the beam was rastered 
using a rectangular raster producing a $\sim$ 2 mm $\times$ 2 mm spot size at the target 
to prevent any damage to the target by overheating it.

\section{Hall A High Resolution Spectrometers}

Protons were detected simultaneously using the two identical high resolution spectrometers (HRS),
also called the left and right arm spectrometers of Hall A. The left arm spectrometer was 
used to measure three $Q^2$ points of 2.64, 3.20, and 4.10 GeV$^2$. Simultaneously, measurements at 
$Q^2$ = 0.5 GeV$^2$ were carried out using the right arm spectrometer which served as a luminosity 
monitor to remove any uncertainties due to beam charge, current, and target density fluctuations. 

\begin{figure}[!htbp]
\begin{center}
\epsfig{file=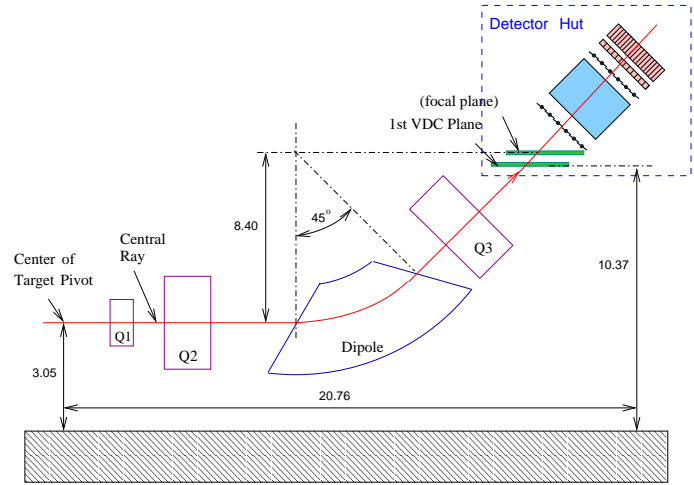,width=3.74in}
\end{center}
\caption[Schematic layout of the Hall A high resolution spectrometer (HRS) and the detector hut. 
Dimension are in meters.]
{Schematic layout of the Hall A high resolution spectrometer (HRS) and the detector hut. Dimension are in meters.}
\label{fig:HRS}
\end{figure}
\begin{table}[!htbp]
\begin{center}
\begin{tabular}{||c|c||} \hline \hline 
Configuration                                     &  $QQDQ$            \\
Momentum Range $p$ (GeV/c)                        &  0.30 - 4.0         \\
Bend Angle ($^o$)                                 &  45                  \\
Optical Length (m)                                &  23.4                 \\
Momentum Acceptance $\delta p/p$ (\%)             &  $\pm$ 4.50            \\
Dispersion (D) (cm/\%)                            &  12.4                   \\
Radial Linear Magnification (M)                   &  2.5                     \\
D/M                                               &  5.0                      \\
Momentum Resolution (FWHM) $\delta p/p$           &  1$\times$10$^{-4}$        \\
Angular Acceptance (Horizontal) (mrad)            &  $\pm$ 28                   \\
Angular Acceptance (Vertical) (mrad)              &  $\pm$ 60                    \\
Solid Angle $\Delta \Omega$ (msr)                 &  $\sim$ 6.7                   \\
Angular Resolution (FWHM) Horizontal $\phi$ (mrad)&  0.6                           \\
Angular Resolution (FWHM) Vertical $\theta$ (mrad)&  2.0                            \\
Transverse Length Acceptance (cm)                 &  $\pm$ 5.0                       \\
Transverse Position Resolution (FWHM) (mm)        &  1.50                             \\
Spectrometer Angle Determination Accuracy (mrad)  &  0.10                              \\
\hline \hline 
\end{tabular}
\caption[Characteristics of the Hall A high resolution spectrometer.]  
{Characteristics of the Hall A high resolution spectrometer.}
\label{HRS_character}
\end{center}
\end{table}

The two spectrometers are identical in their design \cite{alcorn04}. 
They can provide a maximum central momentum of $\sim$ 4 GeV/c with momentum resolution better than 
2$\times$10$^{-4}$ and a horizontal angular resolution better than 2 mrad. Figure \ref{fig:HRS} shows 
the basic layout of the high resolution spectrometer, as well as the detector hut which will be discussed 
later on in more detail. Table \ref{HRS_character} lists some of the characteristics of the Hall A high resolution 
spectrometer. A detailed description of the spectrometer design can be found in \cite{alcorn04,hallaspectrometer}
and references therein. Each high resolution spectrometer uses a $QQDQ$ configuration of super-conducting 
magnets to focus charged particles onto their focal planes. Each spectrometer consists of two 
super-conducting quadrupoles followed by a 6.6 m long indexed dipole magnet for bending with focusing 
entrance and exit windows. Due to the trapezoidal cross sectional shape of the dipole, the bending field 
inside the dipole is not radially uniform. The first quadrupole is used to focus in the vertical plane, 
while the second and the third quadrupoles are used to focus in the horizontal plane. The net effect of 
the $QQDQ$ configuration is to provide a vertical bending with nominal bending angle of 45$^o$.

\section{Detector Package}

The detector package of each HRS is located in a large steel and concrete 
detector hut after the magnet system. The purpose of these detectors is
to select and identify the charged particles coming through the spectrometers.
During the E01-001 experiment, the two spectrometers (arms) included similar detector 
packages with a slight difference between the two arms. A detailed description of the
detector package on each arm can be found in \cite{alcorn04,halladetectors} and references therein. 
The following detectors were used on each arm:

\begin{itemize}
\item \textbf{A set of two Vertical Drift Chambers (VDC's) for tracking purposes.}
\item \textbf{Two scintillator planes, $S_1$ and $S_2$, for trigger activation and 
time-of-flight determination. In addition, an extra plane $S_0$ was added to the left arm
for efficiency determination.}
\item \textbf{Aerogel Cerenkov detectors $A_1$ for the left arm and $A_2$ for
the right arm for particle identification ($p$ and $\pi^{+}$ separation).} 
\item \textbf{A gas Cerenkov detector for particle identification. The right arm gas Cerenkov detector
was used during the coincidence runs for $e^-$ and $\pi^{-}$ separation.}
\end{itemize}

\begin{figure}[!htbp]
\begin{center}
\epsfig{file=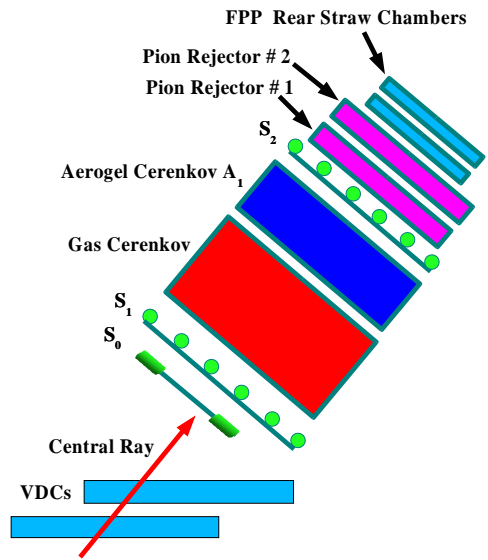,width=2.5in}
\end{center}
\caption[Schematic layout of the left arm detector package used during the E01-001 experiment.]
{Schematic layout of the left arm detector package used during the E01-001 experiment.}
\label{fig:left_detector}
\end{figure}

\begin{figure}[!htbp]
\begin{center}
\epsfig{file=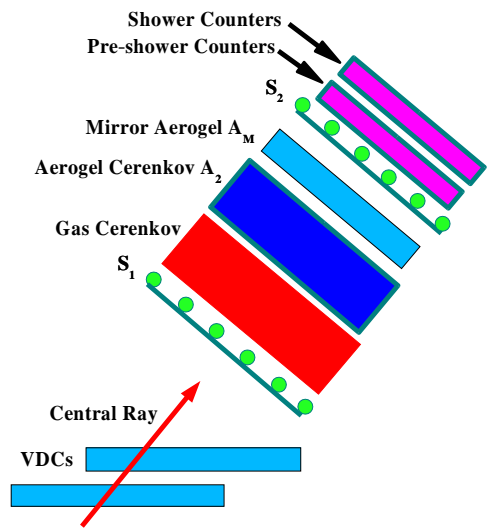,width=2.5in}
\end{center}
\caption[Schematic layout of the right arm detector package used during the E01-001 experiment.]
{Schematic layout of the right arm detector package used during the E01-001 experiment.}
\label{fig:right_detector}
\end{figure}

In addition, the right and left arm detector packages included some extra detectors that were not
used during the E01-001. The right arm had a mirror aerogel detector and calorimeter. 
The left arm included a gas Cerenkov, pion rejector (two layers), and focal plane polarimeter 
(FPP) (rear straw chambers). Although these extra detectors were not used during the E01-001 experiment, 
they do contribute to proton absorption (absorption in the material of each detector) if they were 
located before the scintillator plane $S_2$. The mirror aerogel detector and the gas
Cerenkov detector are the most relevant in this case and their contribution to proton 
absorption should be taken into an account. On the other hand, both layers of the pion rejector and 
the two rear straw chambers of the FPP were located behind the $S_2$ scintillator and were not 
relevant for the proton absorption determination. Figures \ref{fig:left_detector} and 
\ref{fig:right_detector} show a schematic layout of the detector package used with each arm 
during the E01-001 experiment. 

\subsection{Vertical Drift Chambers} \label{VDCs}

\begin{figure}[!htbp]
\begin{center}
\epsfig{file=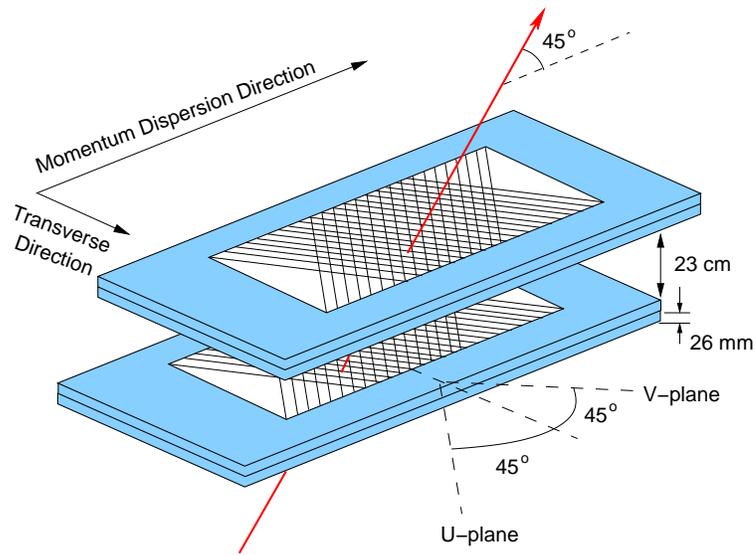,width=4.0in}
\end{center}
\caption[Schematic layout of the Hall A vertical drift chambers.]
{Schematic layout of the Hall A vertical drift chambers.}
\label{fig:VDCs}
\end{figure}

\begin{figure}[!htbp]
\begin{center}
\epsfig{file=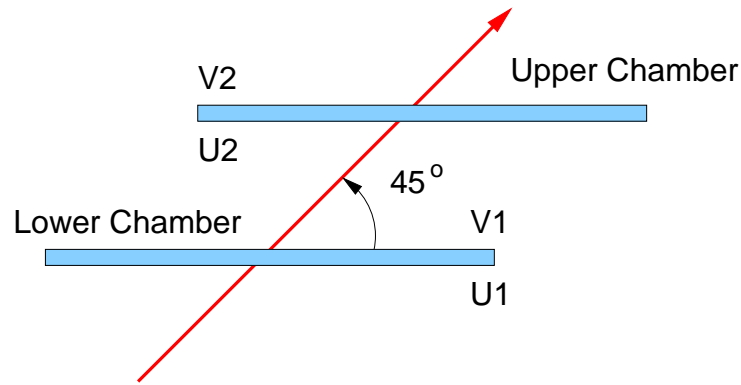,width=4.0in}
\end{center}
\caption[Side view of the Hall A vertical drift chambers planes. The red arrow is the
nominal 45$^o$ particle trajectory crosses the planes of the VDCs.]
{Side view of the Hall A vertical drift chambers planes. The red arrow is the
nominal 45$^o$ particle trajectory crosses the planes of the VDCs.}
\label{fig:VDC_sideview}
\end{figure}

Vertical drift chambers (VDCs) \cite{alcorn04,fissum01} are used for tracking (position and slope 
of the particle trajectory) of the scattered particles. Each spectrometer of Hall A has two vertical drift chambers. 
The first VDC is located at the focal plane of the spectrometer, and the second VDC is positioned 
parallel to and 23 cm away from the first VDC. Both VDCs intersect the spectrometer central ray at an 
angle of 45$^o$ which is the approximate nominal angle at which the particle trajectory 
crosses the wire planes of each VDC. Each VDC has two wire planes known as the U and V 
planes. Each wire plane contains 368 parallel 20$\micro$m-diameter 
gold-coated tungsten wires sandwiched in between gold-coated mylar planes. 
The wires of the U plane are perpendicular to that of the V plane and they are oriented 
with an angle of 45$^o$ (-45$^o$) with respect to the dispersive 
(transverse) direction. We refer to the U and V planes as U$_1$ and V$_1$ planes
in the first VDC and U$_2$ and V$_2$ planes in the second. Figures \ref{fig:VDCs} and 
\ref{fig:VDC_sideview} show a pair of vertical drift chambers and a side view of a 
nominal 45$^o$ particle trajectory crosses the planes of the VDCs.
To operate the VDCs, a gas mixture of argon (62\%) and ethane (38\%) (C$_2$H$_6$) fills 
the area between the mylar planes. The mylar planes are kept at negative potential of 
4KV, while the tungsten wires are grounded. When a charged particle crosses the VDC planes, 
it ionizes the atoms in the gas mixture. The released electrons are accelerated (drift) due 
to the potential difference between the mylar planes and the tungsten wires taking the path of 
least time (also known as the geodetic path). Once the electrons are close to the sense wires 
where the electric field is the strongest, the drifting electrons create an electron avalanche. 
This avalanche of electrons hits the sense wire and then generates a signal 
which gets amplified, discriminated, and then sent to a multihit time-to-digital 
converter (TDC) to measure the time for the least path defined above. The TDC is started by the 
signal from the sense wire and stopped by the event trigger supervisor. 

\begin{figure}[!htbp]
\begin{center}
\epsfig{file=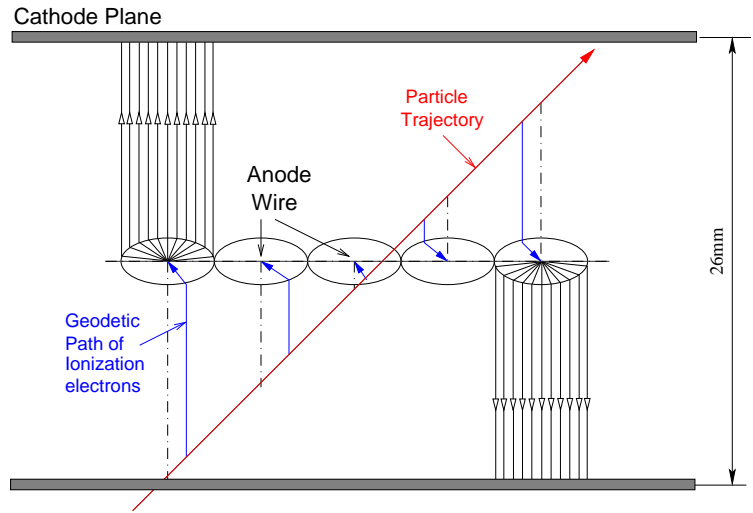,width=4.0in}
\end{center}
\caption[The reconstruction of the charged particle track in the VDC planes.]
{The avalanche process in the VDC and the reconstruction of the charged particle 
track.}
\label{fig:vdc_wire}
\end{figure}

In order to get the tracking information or the drift distance of the charged particle
from each fired wire, we use the drift velocity of the ionized electrons in the C$_2$H$_6$
gas mixture (known to be 50 $\micro$m/ns), and the output of the TDC signal (time). 
Having calculated the drift distance of the charged particle from each fired wire, 
the trajectory (track) of the charged particle can be reconstructed. A 45$^o$ nominal trajectory 
(track) typically triggers five wires. Figure \ref{fig:vdc_wire} shows the reconstruction of the charged 
particle track in the VDC planes.

\subsection{Scintillators and Triggers} \label{scintillators}

For most of the data taking, proton singles events were detected. In addition, coincidence runs 
at Q$^2$ = 2.64 and 4.10 GeV$^2$ were taken. Electrons in coincidence with protons were detected using the right 
arm spectrometer while the protons were detected using the left arm. Triggering of such events was made 
possible by the use of the Hall A scintillators and trigger system \cite{alcorn04,hallaacquisition}. 

\begin{figure}[!htbp]
\begin{center}
\epsfig{file=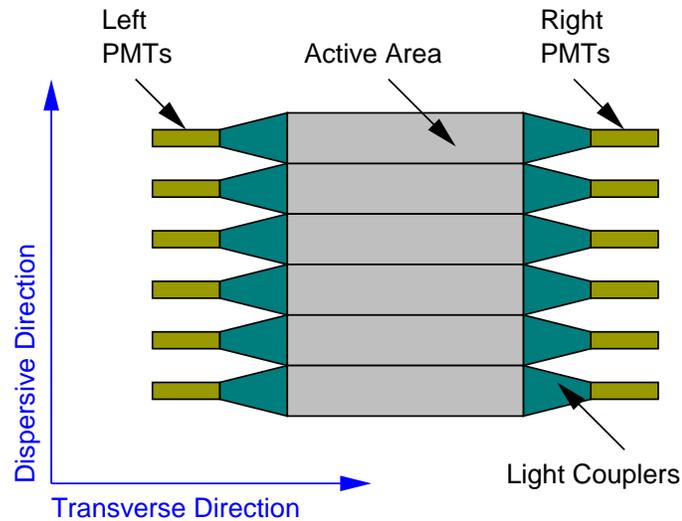,width=4.0in}
\end{center}
\caption[Scintillator plane configuration. Each scintillator plane is perpendicular
to the nominal central ray of the spectrometer.]
{Scintillator plane configuration. Each scintillator plane is perpendicular
to the nominal central ray of the spectrometer.}
\label{fig:scintillator}
\end{figure}

For each spectrometer in Hall A, there are two scintillators planes labeled S$_1$ and 
S$_2$. The two planes are parallel to each other and both are perpendicular to the 
nominal central ray of the spectrometer. The two scintillators planes are $\sim$ 2 
meters apart. Each scintillator plane has 6 identical overlapping scintillators 
paddles made of thin plastic (0.5 cm thick BICON 408) to minimize hadron absorption. 
A photo-multiplier tube (PMT) is attached to each end of each scintillator paddle to collect 
the photons produced by particles passing through scintillator. We refer to these PMTs as the left 
and right PMT of the scintillator paddle. Figure \ref{fig:scintillator} shows the scintillator 
configuration.

The active area of S$_1$ is $\sim$ 170 cm $\times$ 36 cm 
(30 cm $\times$ 36 cm for each paddle) and $\sim$ 220 cm $\times$ 60 cm 
(37 cm $\times$ 60 cm for each paddle) for S$_2$. The time resolution for each scintillator 
plane is $\sim$ 0.30 ns. For the E01-001 experiment, an additional scintillator counter
, S$_0$, was added to the left arm trigger system for a more accurate measurement of the efficiency.
S$_0$ was installed before the S$_1$ scintillator plane and it is a 1 cm thick scintillator
paddle with an active area of $\sim$ 190 cm $\times$ 40 cm. The S$_0$ paddle has 2 PMTs 
labeled top and bottom.
The analog signals from the PMTs are first sent to a discriminator (LeCroy model 4413/200)
providing both digitized and analog outputs. The analog signals are sent to ADCs and the
digitized signal are split into three signals. One signal is sent to TDCs, the second is
sent to a scalers gated by the start and the end of each run, and the third is sent
to a logical unit called the AND unit to make a coincidence between the pairs of PMTs viewing
the same paddle. For each of the spectrometers, the 12 outputs of the logical AND unit are
fed into the Memory Lookup unit (MLU) which takes in a combination of logical signals
at its input and gives out a combination of logical signals at its output.

There are five basic triggers (event types) generated from the scintillators timing 
information. They are classified as main physics triggers and loose (auxiliary) physics 
triggers. The conventional triggers types used for the left arm are the main singles trigger 
T$_3$ and the loose singles trigger T$_4$. For the right arm, we refer to the main 
singles trigger by T$_1$ and the loose singles trigger by T$_2$. Finally, there is 
the coincidence trigger T$_5$ which is the coincidence of T$_1$ and T$_3$ implying that 
the two events T$_1$ and T$_3$ were produced at the target simultaneously. The loose
triggers are usually used to determine the efficiencies of the main triggers and the
scintillators planes in general. 

The output of each ADC is proportional to the number of photons produced inside the 
scintillator and in turn represents the ionization energy loss of the particle, 
or equivalently, the energy deposition in the scintillator. For low momenta, the heavier the 
particle, the more energy it deposits in the scintillator. The TDC signal provides 
timing information for the different type triggers used. By determining the 
time-of-flight of the particle between the scintillator planes form the TDC signals, 
and by knowing the distance between the two scintillator planes, the velocity of the particle $v$ 
or $\beta$ = $v/c$ can be determined. Usually $\beta$ is used for particle identification (PID) 
since different charged particles of a known momentum can be separated by knowing their velocity 
and hence $\beta$. 
Figure \ref{fig:adc_tdc_signals} shows the ADC and TDC spectrum from one of the PMTs from the 
second scintillator plane S$_2$ for the low-momentum setting spectrometer (right arm). 
A $\beta$ spectrum where different particles are clearly separated is also shown.

The left and right arm MLUs were programmed to generate a logical signal which defines and assigns 
the trigger (event type) T$_1$(T$_3$) to an event (particle) if the event satisfies the following:

\begin{figure}[!htbp]
\begin{center}
\epsfig{file=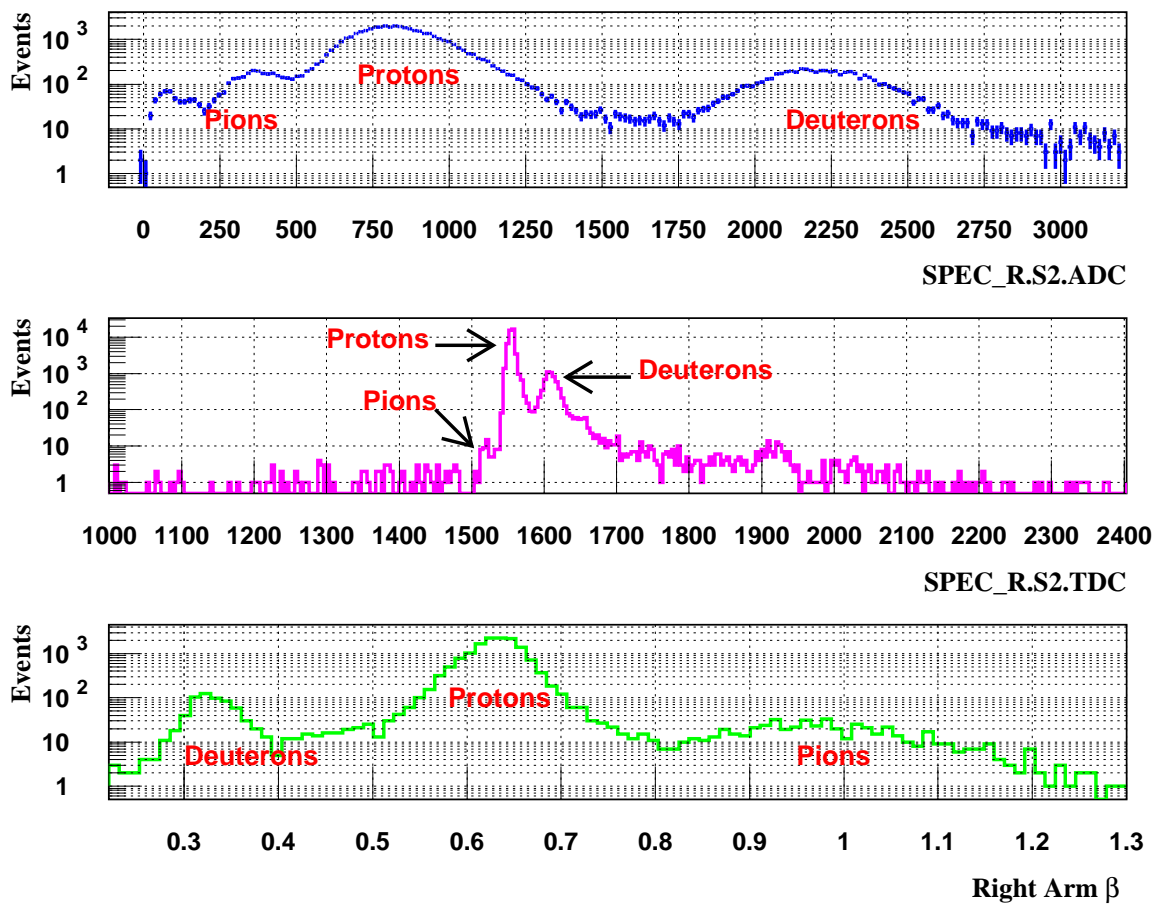,width=6.0in}
\end{center}
\caption[Typical ADC and TDC spectra from a scintillator PMT (upper two plots). Different 
particles have been identified using the ADC spectrum based on the amount of the energy deposited 
in the scintillator. The calculated $\beta$ spectrum based on the TDC timing information and 
distance between the two scintillators planes is also shown (lower plot). The $\beta$ spectrum
can be used for particle identification.] 
{Typical ADC and TDC spectra from a scintillator PMT (upper two plots) from the 
low-momentum setting spectrometer. Different particles have been identified using the ADC spectrum based 
on the amount of the energy deposited in the scintillator. Pions show first (time wise) in the TDC 
spectrum since they are lighter than the protons and therefore have smaller time-of-flight 
(larger $\beta$). The calculated $\beta$ spectrum based on the TDC timing information and 
distance between the two scintillators planes is also shown (lower plot). The $\beta$ spectrum
is used for particle identification.}
\label{fig:adc_tdc_signals}
\end{figure}

\begin{itemize}
\item {The main T$_1$(T$_3$) event type is generated if:}
\begin{enumerate}
\item {A scintillator paddle is said to have fired if and only if a signal from both
the right and left PMT in that paddle is generated.
The logic AND is then formed between the right and left PMT's signal to generate
that paddle's signal.}
\item {Each scintillator plane will have 6 signals (one signal from each paddle in each plane).}
\item {The logic OR of the 6 signals in that plane generate the main signal 
for that plane (one main signal for each scintillator plane).}
\item{The right arm main trigger (T$_1$) is defined as:

\hspace{1cm}(S$_1$ main signal) AND (S$_2$ main signal).}
\item{A third scintillator S$_0$ was added to the left arm only. Therefore, the left arm 
main trigger (T$_3$) is defined as:

\hspace{1cm}(S$_0$ main signal) AND (S$_1$ main signal) AND 
(S$_2$ main signal).}
\end{enumerate}   
\item {The loose T$_2$(T$_4$) event type is generated if:}
\begin{enumerate}
\item{T$_2$ is the same as T$_1$ defined above except the AND in point number 4 is
replaced with an OR:

\hspace{1cm}(S$_1$ main signal) OR (S$_2$ main signal).}
\item{T$_4$ is the same as T$_3$ defined above except the AND between S$_1$ and S$_2$ in point number 5 is replaced with an OR: 

\hspace{1cm}(S$_0$ main signal) AND \Big((S$_1$ main signal) OR (S$_2$ main signal)\Big).}
\end{enumerate}
\item {The coincidence trigger T$_5$ is generated as the coincidence between T$_1$ and T$_3$.}
\end{itemize}

\subsection{Gas Cerenkov}

Coincidence kinematics were taken at Q$^2$ = 2.64 and 4.10 GeV$^2$. Electrons were detected using 
the right arm and protons using the left arm. See Table \ref{kinematics} for more 
details. In order to separate electrons from background particles such as the negatively charged 
pions, a Cerenkov detector was used in the right arm for particle identification. 

The gas Cerenkov detector used in Hall A \cite{alcorn04,iodice98} was installed between the
scintillator S$_1$ and S$_2$ planes. It is a rectangular chamber filled with CO$_2$ 
gas at atmospheric pressure. When a high energy charged particle travels through a material 
of index of refraction $n$ with velocity $v$ larger than the velocity of light in that material 
($c/n$) where $c$ is the speed of light, electromagnetic radiation (light) is emitted. 
This is known as the Cerenkov effect. For the Hall A Cerenkov detector, 
10 spherical mirrors at the back wall focus the Cerenkov radiation on 10 PMT photocathodes. 

A charged particle can emit Cerenkov radiation only when its velocity is larger 
than a threshold velocity set by the index of refraction of that material:
\begin{equation}
v_{th} = \frac{c}{n}~,
\end{equation}  
and therefore a threshold momentum:
\begin{equation} \label{eq:p_th}
p_{th} = \frac{mv_{th}}{\sqrt{1 -\frac{v^2_{th}}{c^2}}}~.
\end{equation} 

\begin{figure}[!htbp]
\begin{center}
\epsfig{file=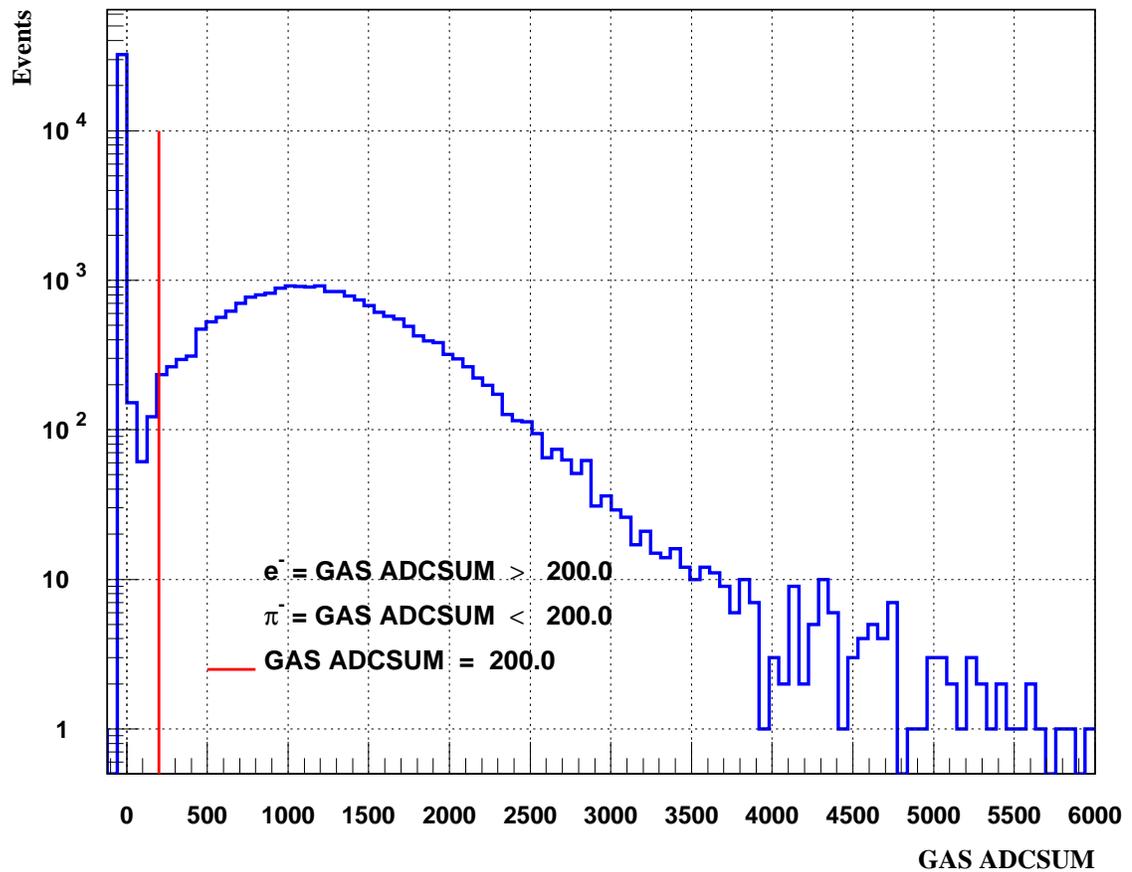,width=6in}
\end{center}
\caption[Gas Cerenkov ADCs sum signal. The peak at GAS ADCSUM = zero represents
the $\pi^{-}$ which did not fire the gas Cerenkov and it extends up to 
GAS ADCSUM = 200. GAS ADCSUM $>$ 200 represents the $e^{-}$ signal which fired the 
gas Cerenkov. The red solid line is the boundary between the $\pi^{-}$ region 
(GAS ADCSUM $<$ 200) and $e^{-}$ region (GAS ADCSUM $>$ 200).]
{Gas Cerenkov ADCs sum signal. The peak at GAS ADCSUM = zero represents
the $\pi^{-}$ which did not fire the gas Cerenkov. GAS ADCSUM $>$ 200 represents 
the $e^{-}$ signal which fired the gas Cerenkov. The red solid line is the boundary between the $\pi^{-}$ region 
(GAS ADCSUM $<$ 200) and $e^{-}$ region (GAS ADCSUM $>$ 200).}
\label{fig:cersum}
\end{figure}

For the CO$_2$ gas used, the index of refraction is $n$ = 1.00041. Equation (\ref{eq:p_th}) then 
gives a momentum threshold of $p_{th}$ $\sim$ 17 MeV/c for electrons and $p_{th}$ $\sim$ 
4.8 GeV/c for pions which is larger than the momentum range of the Hall A HRS which is 
0.30$-$4.0 GeV/c. So, only electrons give a signal in the analog-to-digital (ADC) of 
the Cerenkov detector. Figure \ref{fig:cersum} shows the distribution of the sum of the 
10 ADCs signals of the gas Cerenkov PMTs. 

\subsection{Aerogel Cerenkov}

Two Aerogel Cerenkov detectors A$_1$ and A$_2$ \cite{alcorn04,comanthesis,bwojtsekhowski} 
were used for particle identification where protons and positively charged pions could be separated 
based on the Cerenkov radiation effect discussed above. The A$_2$ aerogel detector was used with the 
right arm, while the A$_1$ aerogel was used with the left arm. In addition, a mirror aerogel, A$_M$, 
was installed on the right arm between the S$_1$ and S$_2$ scintillators but was never used. 
The A$_M$ Aerogel contributes to the proton absorption in the right arm.

The two aerogel detectors A$_1$ and A$_2$ are designed in the same way. The A$_1$
aerogel has a 9 cm aerogel radiator with index of refraction $n_{A_1}$ = 1.015 
while A$_2$ has a 5 cm aerogel radiator with index of refraction $n_{A_2}$ = 1.055. 
There are 24 PMTs installed in A$_1$ and 26 PMTs in A$_2$. 

\begin{figure}[!htbp]
\begin{center}
\epsfig{file=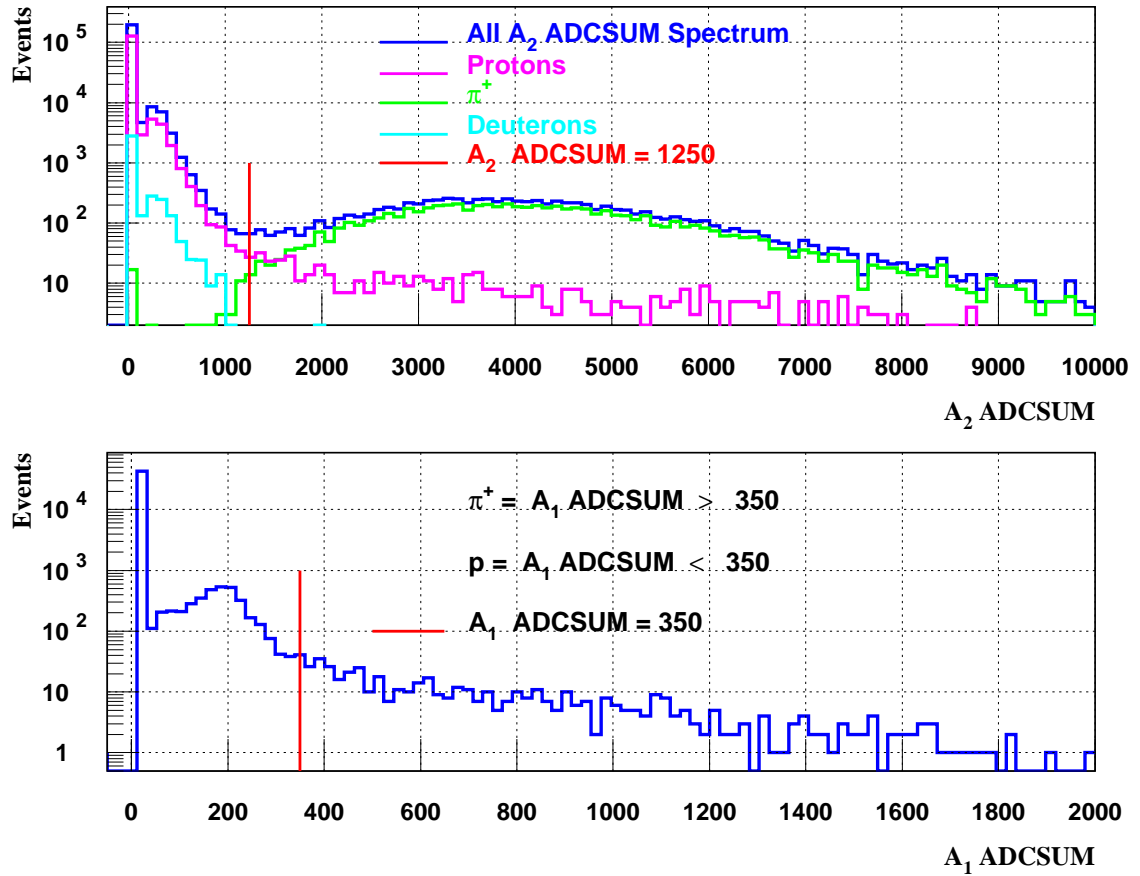,width=6.0in}
\end{center}
\caption[A$_2$(A$_1$) aerogel Cerenkov ADCs sum signal.] 
{A$_2$(A$_1$) aerogel Cerenkov ADCs sum signal top(bottom) in blue.
The peak at A$_2$(A$_1$) ADCSUM = zero represents the protons which did not fire the aerogel  
Cerenkov. The contribution of protons, $\pi^{+}$, and deuterons to A$_2$ ADCSUM (selected
by cuts on the time-of-flight and amount of energy deposited in the S$_1$ scintillator) is shown. 
The contribution of protons, $\pi^{+}$, and deuterons to A$_1$ ADCSUM is not shown since they cannot
be separated using any other detector and it will be discussed in the next chapter in more detail. 
The A$_2$(A$_1$) ADCSUM $>$ 1250(350) represents the cut we use to separate the $\pi^{+}$ signal which 
fired the aerogel Cerenkov. Notice the leakage of protons into the $\pi^{+}$ area and the small 
contamination of the protons area with $\pi^{+}$ and deuterons as can be seen form A$_2$ ADCSUM spectrum.} 
\label{fig:a2a1sum}
\end{figure}

For the A$_1$ aerogel ($n_{A_1}$ = 1.015), equation (\ref{eq:p_th}) gives a momentum threshold 
of $p_{th}$ $\sim$ 5.4 GeV/c for protons and $p_{th}$ $\sim$ 0.8 GeV/c for positive pions,
while, for the A$_2$ aerogel ($n_{A_1}$ = 1.055), equation (\ref{eq:p_th}) gives a momentum 
threshold of $p_{th}$ $\sim$ 2.8 GeV/c for protons and $p_{th}$ $\sim$ 0.40 GeV/c for 
positive pions. For the A$_1$ aerogel (left arm), the momentum threshold for the 
protons is larger even than the momentum range of the Hall A HRS (0.30$-$4.0 GeV/c) and 
that makes it impossible for the protons to fire the aerogel Cerenkov detector. 
On the other hand, the momentum threshold for the positive pions is within the momentum range of 
the Hall A HRS and hence pions above 0.8 GeV/c should produce a signal in the analog-to-digital 
(ADC) of the A$_1$ aerogel Cerenkov detector. For the right arm where A$_2$ aerogel was used, 
only positive pions could trigger the analog-to-digital and fire the aerogel Cerenkov detector 
since the central momentum for the right arm spectrometer was always set to 0.756 GeV/c which is 
less than the threshold momentum needed for the protons to trigger the aerogel Cerenkov detector. 
Figure \ref{fig:a2a1sum} shows the distribution of the sum of the ADCs signals of the 
A$_2$(A$_1$) aerogel Cerenkov PMTs.
 
\section{Data Acquisition System}

The data acquisition system in Hall A (DAQ) was used during the run of experiment 
E01-001. The data acquisition system is controlled by CEBAF On-line Data Acquisition 
system (CODA) \cite{alcorn04,coda} which is developed by the JLAB data acquisition group and 
designed for nuclear physics experiments. CODA is a toolkit that is composed of a set 
of software and hardware packages from which DAQ can be built to manage the acquisition, 
monitoring, and storage of data.

Data acquisition in Hall A takes the following steps:
\begin{itemize}
\item{The data is read out from Read-Out Controllers (ROCs). The ROCs are CPUs in 
Fastbus and VME crates in the hall and in the electronics room.  These crates contain
the ADCs, TDCs, and scalers that contain the event information.}
\item{The Trigger Supervisor controls the state of the run, and generates the 
triggers that cause the ROCs to be read out.}
\item{The Event Builder subsystem (EB) is the part of CODA that reads in the data 
fragments from the ROCs and puts the data together into an event, incorporating all 
of the necessary CODA header information needed to describe and label the event and the 
data fragments.}
\item{CODA manages the data acquisition system, and takes care of handling the data
from the events.}
\item{After the event is built by the EB, it is placed into a buffer, after which it can 
be tested and recorded using the event recorder or rejected if desired.}
\item{The accepted events (data) are written to a local disk and then transferred to the 
Mass Storage System.}
\item{Data from various control systems and scalers are injected into
the data stream every few seconds using the event transfer.}
\item{In addition to running the data acquisition, CODA also includes a graphical user 
interface (RunControl) which allows the user to start and stop runs, as well as define 
run parameters.}
\item{Events are classified as physics events which come from the spectrometers (detectors)
and beamline information (Beam position monitors, beam loss monitors, and beam raster readback 
values recorded for each event), or EPICS events \cite{epics} such as the readout of hardware 
(spectrometers magnets settings, angles, and target controls).}
\end{itemize}

\chapter{Data Analysis I: Efficiencies and Corrections} \label{chap_corrections}
\pagestyle{plain}
\section{Analysis Introduction} \label{qeff_intro}

In this chapter a description of the event reconstruction procedure
and the corrections and efficiencies applied to the measured beam charge, $Q$, will be presented. 
These corrections and efficiencies include the computer and electronics 
livetimes , $CLT$, and $ELT$, respectively, VDCs tracking efficiency, $\epsilon_{VDC}$, VDCs hardware cuts
efficiency $\epsilon_{VDCH}$, scintillators efficiency (product of the two scintillators efficiency  
or $\epsilon_{S_1} \times \epsilon_{S_2}$), particle identification (PID) efficiency, 
$\epsilon_{PID}$, proton absorption correction, $C_{Absorption}$, target boiling 
correction, $C_{TB}$, and target length correction, $C_{TL}$. 
In addition, a brief description of the spectrometer optics calibration and spectrometer
mispointing measurements will be given. 
Having applied these corrections and efficiencies to the beam 
charge, we refer to the corrected charge as the effective charge or $Q_{eff}$ and
it is defined as:
%
\begin{eqnarray} \label{eq:qeff}
Q_{eff} = \frac{1}{ps} \Big(Q \times ELT \times CLT \times \epsilon_{VDC} \times \epsilon_{VDCH} 
\times \epsilon_{S_1} \times \epsilon_{S_2} \times \epsilon_{PID} 
\times C_{Absorption} {}
\nonumber\\
\times C_{TB} \times C_{TL}\Big)~,~~~~~~~~~~~~~~~~~~~~~~~~~~~~~~~~~~~~~~~~~~~~~~~~~~~~~~~~~~~~~~~~~~{}
\end{eqnarray}
%
where $ps$ is the prescale factor. The prescale factor $n$ for the trigger type T$_i$ 
($i = 1,\cdots,5$) means that the Trigger Supervisor will read out every $n$th event of 
type T$_i$.

The advantages of detecting protons as against electrons were discussed in section \ref{proton_vs_electron}.
Because of these advantages, the E01-001 experiment was able to greatly reduce any 
$\varepsilon$-dependent systematic corrections and associated uncertainties applied to the measured cross 
sections. In addition, measurements at $Q^2$ = 0.5 GeV$^2$ using the right arm spectrometer which
served as a luminosity monitor will check the uncertainties due to beam charge, current, and target 
density fluctuations.

\pagestyle{myheadings}

The systematic uncertainties in these efficiencies and corrections will be discussed.
These systematic uncertainties are a measure of how accurately we know the various efficiencies and 
corrections. The systematic uncertainties are classified into three types, based on how they contribute 
to the quantities we want to extract:
\begin{itemize}
\item {Scale Uncertainty: Sometimes referred to as normalization uncertainty as 
it has the same effect on all $\varepsilon$ points at a given Q$^2$. An example would be the 
uncertainty in the target length. Such uncertainty would affect both $G_{Ep}$ and $G_{Mp}$ the 
same way but not the ratio.}  
\item {Random Uncertainty: Sometimes referred to as point-to-point uncertainty. It has 
an uncorrelated effect on each $\varepsilon$ point at a given Q$^2$. Two examples are  
the statistical uncertainty and the uncorrelated shifts in the beam energy and scattering angle for different
kinematics. Such uncertainties affect the extraction of $G_{Ep}$, $G_{Mp}$, and the ratio.}
\item {Slope Uncertainty: Refers to a correlated uncertainty in a correction that varies linearly 
with $\varepsilon$. An example would be the effect of a fixed scattering angle or 
beam energy offset for all kinematics which also leads to a scale uncertainty. Such an uncertainty
will modify the slope of the reduced cross section verses $\varepsilon$, and thus $G_{Ep}$ and $G_{Ep} \over G_{Mp}$, but will not 
spoil the expected linearity.}
\end{itemize}

It must be mentioned that in some cases it is unclear whether the uncertainty is scale, random, or 
slope. In these cases, the worst case of the three uncertainties is assumed. 
Table \ref{uncert_contrib} indicates whether the different types of uncertainties effect 
the individual form factors, their ratio, and the linearity of the L-T plots. Note that the slope uncertainty is 
the change in slope between $\varepsilon$ = 0.0 and $\varepsilon$ = 1.0. So the value for the slope uncertainty is the 
change over the actual $\varepsilon$ range in data divided by $\Delta \varepsilon$. 
Since $\Delta \varepsilon$ $\sim$ 0.70 for the left arm and $\Delta \varepsilon$ $\sim$ 0.07 for the right arm, the
slope uncertainty is much larger for the right arm. 

  
\begin{table}[!htbp]
\begin{center}
\begin{tabular}{||c|c|c|c|c||} \hline \hline
Uncertainty Type & $ G_{Ep}$ & $G_{Mp}$ & $\mu_{p}G_{Ep}$/$G_{Mp}$ & Linearity \\
\hline \hline
Scale       & Yes       &  Yes     & No                    & No              \\
Random      & Yes       &  Yes     & Yes                   & Yes              \\
Slope       & Yes       &  Yes     & Yes                   & No                \\
\hline \hline
\end{tabular}
\caption[The effect of the scale, random, and scale uncertainties on the individual form
factors, their ratio, and the linearity of the L-T plots.]
{The effect of the scale, random, and scale uncertainties on the individual form
factors, their ratio, and the linearity of the L-T plots.} 
\label{uncert_contrib}
\end{center}
\end{table}

\section{Event Reconstruction} \label{event_reconstruction}

The raw data files imported from the DAQ system were replayed using the 
standard Hall A event processing software ESPACE (Event Scanning Program
for Hall A Collaboration Experiments) \cite{offermanespace}. 
The multipurpose ESPACE reads the raw events and then decodes, filters, 
calibrates, and reconstructs physical variables of interest for the analysis on 
an event-by-event basis. These physical variables can be detector hits, tracks and particle 
identification (PID) signals, or the coordinates of the reaction vertex in the target. 
The outputs of ESPACE are histograms and ntuples (multi-dimensional matrices of the physical 
quantities of interest) saved in HBOOK formated files \cite{cernhbook}. An HBOOK file is generated
for each production run. These HBOOKs are read using the Physics Analysis Workstation (PAW) 
software \cite{paw} where cuts can be applied on these physical quantities.

\section{Optics} \label{e01001_optics}

\begin{figure}[!htbp]
\begin{center}
\epsfig{file=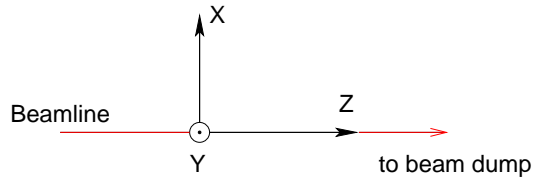,width=3in}
\end{center}
\caption[The hall coordinate system.]
{The hall coordinate system viewed from above: Z is the beam direction, Y is vertically 
pointing up, and X is to the left of the beam direction and perpendicular to both Z and Y.}
\label{fig:halla_coordinate}
\end{figure}

\begin{figure}[!htbp]
\begin{center}
\epsfig{file=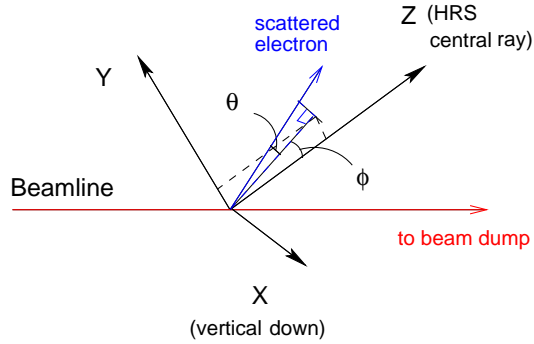,width=3in}
\end{center}
\caption[The target (spectrometer) coordinate system.]
{The target coordinate system: Z is the spectrometer central ray direction, 
X is pointing down (dispersive direction), and Y is perpendicular to both Z and X
(transverse direction), $\phi$ and $\theta$ (usually called $\phi_{tg}$ and $\theta_{tg}$ 
where $tg$ is short for target) are the in-plane and out-of-plane scattering
angles, respectively, as measured with respect to the spectrometer central ray.}  
\label{fig:target_coordinate}
\end{figure}

There are different coordinate systems used in Jefferson Lab Hall A. These
coordinate systems are described in detail in \cite{offermanespace,nilanga} and can 
be classified as: the hall coordinate system, the target coordinate system 
(spectrometer reconstructed coordinate system), the detector coordinate system, 
the transport coordinate system, and finally the focal plane coordinate system. 
The hall and target coordinate systems are commonly used to define and calculate the
physical quantities of interest for data analysis (see Figures \ref{fig:halla_coordinate} 
and \ref{fig:target_coordinate}), while the detector, transport, and focal plane 
coordinate systems are used for Monte Carlo simulations and optics calibrations. 

For event reconstruction, ESPACE starts by reconstructing the event at the focal plane of 
the spectrometer (passing through the center of the $U_{1}$ plane of the first VDC). 
That includes determination of the electrons drift times in the wire chambers and the drift 
distances (see section \ref{VDCs}), reconstruction of the particle trajectories, and calculation 
of $\beta = v/c$ and positions and angles of the track in the focal plane coordinate system. 
The detector coordinate system, the transport coordinate system, and the focal plane coordinate system all 
share the same origin defined as the intersection of central wire 184 of the VDC $U_{1}$ plane and the central wire projection of the VDC $V_{1}$ plane. 

Having reconstructed the focal plane variables, the reconstruction of the target variables
or reconstruction of the trajectories at the target can be done by using the optics database
or transformation matrix between the focal plane and target variables \cite{nilanga}. 
These reconstructed target variables are the in-plane and out-of-plane scattering angles 
$\phi$ or $\phi_{tg}$, and $\theta$ or $\theta_{tg}$ where $tg$ is short for target, 
the y-coordinate of the extended target length $y_{tg}$, and the deviation from the central 
momentum $\delta = \frac{p-p_{o}}{p_{o}}$. 

\begin{figure}[!htbp]
\begin{center}
\epsfig{file=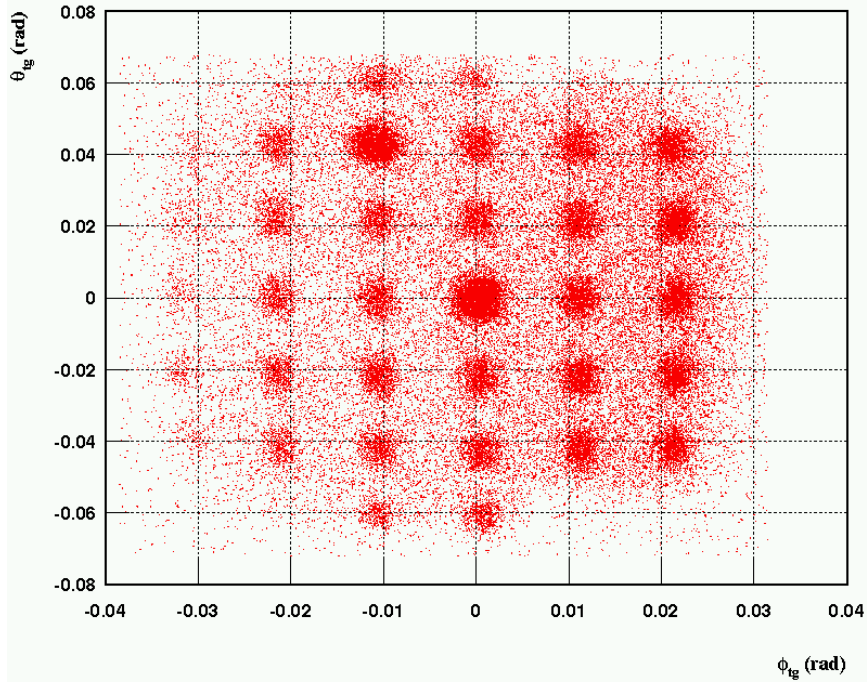,width=4.5in}
\end{center}
\caption[The reconstruction of the sieve slit image into the right arm spectrometer due to electrons 
scattered from a thin $^{12}$C target.]
{The reconstruction of the sieve slit image into the right arm spectrometer due to electrons 
scattered from a thin $^{12}$C target.}
\label{fig:right_thetatg_phitg}
\end{figure}

\begin{figure}[!htbp]
\begin{center}
\epsfig{file=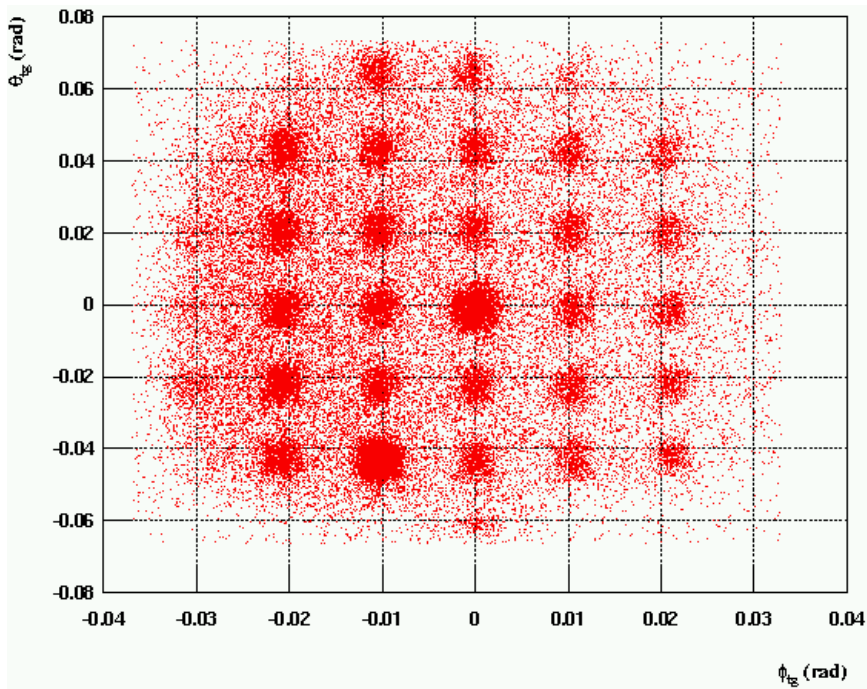,width=4.5in}
\end{center}
\caption[The reconstruction of the sieve slit image into the left arm spectrometer due to electrons 
scattered from a thin $^{12}$C target.]
{The reconstruction of the sieve slit image into the left arm spectrometer due to electrons 
scattered from a thin $^{12}$C target.}
\label{fig:left_thetatg_phitg}
\end{figure}

In order to study the optical properties of the spectrometer, sieve slits collimators \cite{alcorn04,collimation} 
positioned 1.184$\pm$0.005 and 1.176$\pm$0.005 m from the target on the left and right arm 
spectrometers, respectively, are used. The sieve slit is a 5 mm thick stainless steel sheet with 
a pattern of 49 holes (7 x 7), spaced 12.5 mm apart horizontally and 25 mm apart vertically. 
Two of the holes, one in the center and one displaced two rows vertically and one horizontally, 
are 4 mm in diameter, while the rest are 2 mm in diameter.
Figures \ref{fig:right_thetatg_phitg} and \ref{fig:left_thetatg_phitg} show the reconstruction 
of the sieve slit image due to electrons scattering from a thin $^{12}$C target as seen by 
the right and left arm spectrometers, respectively. 
The fact that the optics database used by ESPACE during the E01-001 analysis produces the correct 
position of the holes in the sieve slit for both spectrometers indicates that the optics 
database is well calibrated and no further calibration is needed. 
The calibration procedure is documented in detail in \cite{nilanga}. 

During the analysis of the E01-001 experiment, a 1.6 msr solid angle software cut was applied.
This cut corresponds to an out-of-plane angle of -40.0$<\theta_{tg}<$40.0 mrad and in-plane
angle of -10.0$<\phi_{tg}<$10.0 mrad. See section \ref{recon} and Table \ref{recon_cuts} for a full 
description of the cuts applied during the analysis. We estimate that we know the edge of the
$\theta_{tg}$ and $\phi_{tg}$ cuts from looking at the adjacent hole separation in the vertical 
and horizontal directions to within 0.20 mrad.
This translates into a 2.0\% scale uncertainty in the in-plane angle (0.2 mrad/10 mrad),
and 0.5\% scale uncertainty in the out-of-plane angle (0.2 mrad/40 mrad). The sum in 
quadrature of the two angle scale uncertainties was used to determine the final estimated
scale uncertainty in the 1.6 msr solid angle cut of 2.06\%. Because the 1.6 msr solid angle 
cut is identical for all kinematics, the uncertainty in the solid angle only contributes
to the scale uncertainty. 

\section{Spectrometer Mispointing} \label{spect_mispoint}

Experiment E01-001 requires a precise knowledge of the e-p reduced cross sections.
In order to achieve that, an accurate knowledge of the scattering angle
is required. Knowing the scattering angle accurately requires accurate 
knowledge of the spectrometer optics and offsets, target position, and 
beam position. Figure \ref{fig:target_coordinate} shows the coordinate system of the 
spectrometer. Ideally, the target center coincides with the hall center. 
Due to translational movements of the spectrometer around the hall center, 
the central ray of the spectrometer can miss the hall center in both the horizontal and 
vertical directions. The horizontal offset or the horizontal distance between the hall center 
and the central ray of the spectrometer is referred to as the spectrometer mispointing. 

There are two different and reliable methods by which the horizontal
and vertical offsets can be measured and hence the spectrometer angle: 
the survey method and the carbon-pointing method.
The survey method is the most precise. The two spectrometers used during the 
E01-001 experiment were surveyed at several kinematics settings by the JLAB
survey group. 

For the right arm spectrometer, the spectrometer angle was surveyed 
and determined at all 5 $\varepsilon$ points, and the survey angles were used
in the analysis. For the left arm, there were several spectrometer settings 
where a survey was not performed, and so the carbon-pointing method was used. 
In the carbon pointing method, the spectrometer mispointing, $\Delta h$, and spectrometer 
angle, $\theta_{s}$, are determined using carbon-pointing runs where electrons are scattered by
a thin carbon foil. By knowing the spectrometer central angle, $\theta_{o}$, (defined as the 
spectrometer angle with no mispointing), 
as determined from the hall floor marks, target position as reconstructed by the spectrometer, 
$y_{tg}$, and the target offset along the beam direction, $z_{off}$, as measured by the target 
survey group, $\Delta h$ can be determined as:
\begin{equation} \label{eq:offset}
\Delta h = \pm y_{tg} + z_{off} \sin(\theta_{o})~,
\end{equation}
and hence the spectrometer scattering angle (angle setting) can be calculated as:
\begin{equation} \label{eq:spec_angle}
\theta_{s} = \theta_{o} +  \frac{\Delta h}{L}~,
\end{equation}  
where the plus(minus) sign in front of $y_{tg}$ in equation (\ref{eq:offset}) is used 
with the right(left) arm and $L$ is the distance between the hall center and the floor marks 
where the angles are scripted and has a value of 8.458 m. Note that $\frac{\Delta h}{L} = \Delta\theta_{o}$ in equation 
(\ref{eq:offset}) above represents the correction to the central scattering angle 
of the spectrometer, $\theta_{s} = \theta_{o} \pm \Delta\theta_{o}$. 
The spectrometer is said to be mispointed downstream(upstream) if $\Delta h$ is 
positive(negative).

\begin{figure}[!htbp]
\begin{center}
\epsfig{file=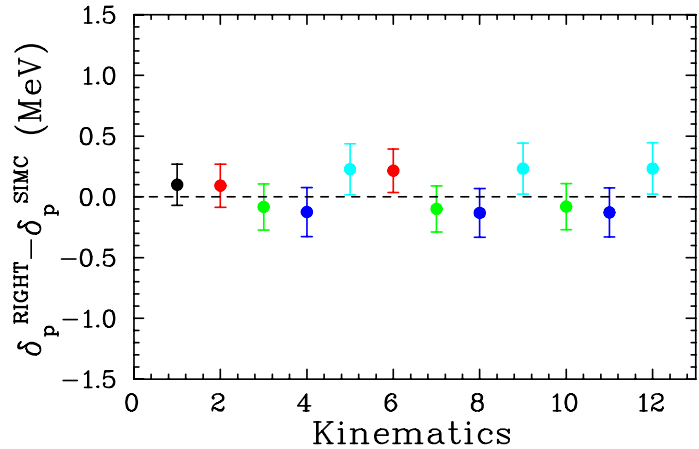,width=3.6in}
\end{center}
\caption[The right arm difference in the elastic peak position from data and that of 
simulations after applying the angular offset.] 
{The right arm difference in the elastic peak position from data ($\delta_{P}$ RIGHT) and that of 
simulations ($\delta_{P}$ SIMC) after applying the angular offset. The points are sorted according to the 
kinematics of the left arm. Kinematics 1-5 correspond to $Q^2$ = 2.64 GeV$^2$, while kinematics 
6-9 (10-12) correspond to 3.20 (4.10) GeV$^2$. For each $Q^2$ value, the points are sorted by 
$\varepsilon$ (low to high). Note that for the right arm only the first five measurements are truly independent. 
The later points are repeats at the same kinematics.}
\label{fig:rdeltap_angle_offset}
\end{figure}

\begin{figure}[!htbp]
\begin{center}
\epsfig{file=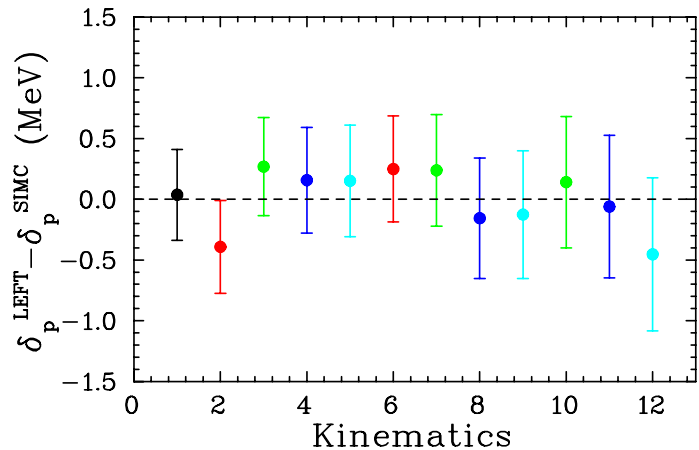,width=3.6in}
\end{center}
\caption[The left arm difference in the elastic peak position from data and that of 
simulations after applying the angular offset.] 
{The left arm difference in the elastic peak position from data ($\delta_{P}$ LEFT) and that of 
simulations ($\delta_{P}$ SIMC) after applying the angular offset. Kinematics 1-5 correspond to 
$Q^2$ = 2.64 GeV$^2$, while kinematics 6-9 (10-12) correspond to 3.20 (4.10) GeV$^2$. For each $Q^2$ 
value, the points are sorted by $\varepsilon$ (low to high).}
\label{fig:ldeltap_angle_offset}
\end{figure}

We can test the nominal kinematics by looking at the reconstructed kinematics for elastic scattering.
The reconstruction of the $\Delta P$ spectrum is discussed in section \ref{recon}. 
The elastic peak position should be near $\Delta P$ = 0.0 MeV, but there are small corrections due to energy loss and
radiative corrections, which are modeled in the Monte Carlo simulation program SIMC. See section \ref{ep_simc} for details. 
The elastic peak position in the $\Delta P$ spectrum from data, $\delta_{P}$ RIGHT(LEFT), 
is compared to that in simulations $\delta_{P}$ SIMC. 
This comparison was done for the right and left arms and at all kinematics. For the left arm, an overall angular offset of 
0.19 mrad was applied to the pointing angles to best center the elastic peak position from data to that of simulations
at each kinematics. Note that an offset of 0.28 mrad is needed to match the carbon angles to the survey ones. The two offsets 
are consistent taking into account the 0.18 mrad scale uncertainty assigned in the scattering angle (discussed below).  
For the right arm, the survey angles are used and yielded a good $\Delta P$ peak position. Therefore, no additional offset 
was needed. Figures \ref{fig:rdeltap_angle_offset} and \ref{fig:ldeltap_angle_offset} show the difference in the elastic peak 
position from data and that of simulations after applying the angular offset and for both arms. The error bars assume random 
uncertainties of 0.10 mrad for the angle (see discussion below) and 0.02\% for the beam energy 
(see section \ref{beam_energy_measurements}). With these uncertainties, the values of $\delta_{P}$ from data 
are in good agreement with those of simulations. In fact, the difference is zero to better than the error
bars, indicating that we might be slightly overestimating the random uncertainties, but we will stick 
with these uncertainties to be conservative.

A 0.10 mrad random uncertainty in the angular offset is assigned. This is 
based on the following contributions combined in quadrature: 0.07 mrad due to drifts in the beam angle
as determined from the BPMs, $\sim$ 0.05 mrad uncertainty that comes from the $\sim$ 0.1 MeV 
uncertainties in determining the peak position, and the $\sim$ 0.05 mrad uncertainty from the pointing determination 
(run to run scatter and uncertainty in determining the target position). Table \ref{angles_used} shows the Tiefenback energy 
and the spectrometer settings used for the left and right arms. 

\begin{table}[!htbp]
\begin{center}
\begin{tabular}{||c|c|c|c||} \hline
\hline
Kinematics & Tiefenback Energy & $\theta_{L}$ & $\theta_{R}$ \\ 
           & (MeV)             & ($^o$)       & ($^o$)\\
\hline \hline
$o$        & 1912.94           & 12.6311     & 58.3090 \\
$a$        & 2260.00           & 22.1592     & 60.0700 \\
$i$        & 2844.71           & 29.4590     & 62.0380 \\
$q$        & 3772.80           & 35.1512     & 63.8710 \\
$l$        & 4702.52           & 38.2512     & 64.9810 \\
\hline \hline
$b$        & 2260.00           & 12.5226     & 60.0700 \\
$j$        & 2844.71           & 23.3896     & 62.0380 \\
$p$        & 3772.80           & 30.4802     & 63.8710 \\
$m$        & 4702.52           & 34.1225     & 64.9810 \\
\hline \hline
$k$        & 2844.71           & 12.6807     & 62.0380 \\
$r$        & 3772.80           & 23.6586     & 63.8710 \\
$n$        & 4702.52           & 28.3735     & 64.9810 \\ 
\hline \hline
\end{tabular}
\caption[The Tiefenback energy and spectrometer settings used in the actual analysis of the E01-001 
kinematics.]
{The Tiefenback energy and spectrometer settings 
used in the actual analysis of the E01-001 kinematics. The right arm spectrometer settings 
$\theta_{R}$ are the survey angles, while the left arm spectrometer settings $\theta_{L}$ are the angles as determined 
by carbon-pointing runs measurements corrected by 0.19 mrad to center the elastic peak.} 
\label{angles_used}
\end{center}
\end{table}

A 0.18 mrad scale uncertainty in the angular offset is assigned. The scale uncertainty is determined 
based on the following contributions combined in quadrature: 0.13 mrad uncertainty due to uncertainty in modeling the 
energy loss and radiative effects on the elastic peak (if we shift the peak centers by 0.2 MeV due to error in SIMC 
energy loss, or small errors in the smearing we use, the angle offset varies by 0.13 mrad), 0.07 mrad uncertainty for 
possible beam angle offset, and 0.10 mrad uncertainty coming from the fact that a small shift in the beam energy can give 
a solution with the wrong angle offset (if we shift the beam energy by 0.03\%, we still get decent elastic peak 
positions by shifting the angle by 0.1 mrad. The $\chi^2$ is worse, but not totally unreasonable).

\begin{figure}[!htbp]
\begin{center}
\epsfig{file=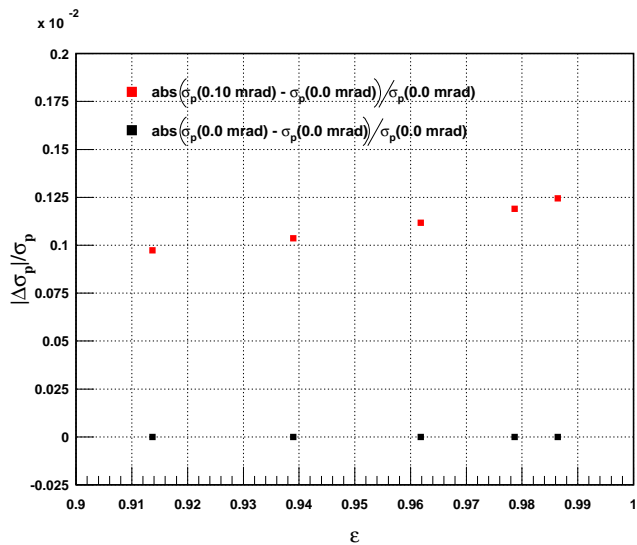,width=3.3in}
\end{center}
\caption[The relative difference between the right arm nominal cross sections and  
cross sections with a 0.10 mrad offset in the scattering angle as a function
of $\varepsilon$.] 
{The relative difference between the right arm nominal cross sections and  
cross sections with a 0.10 mrad offset in the scattering angle (red squares) as a function
of $\varepsilon$ at all 5 incident energies. The black squares are the nominal cross sections 
relative to themselves.} 
\label{fig:right_proton_sigma0.10mrad_epsilon}
\end{figure}
\begin{figure}[!htbp]
\begin{center}
\epsfig{file=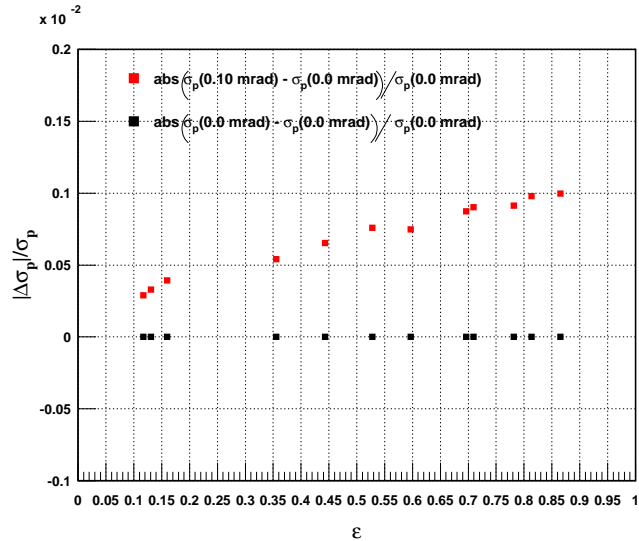,width=3.3in}
\end{center}
\caption[The relative difference between the left arm nominal cross sections and  
cross sections with a 0.10 mrad offset in the scattering angle as a function
of $\varepsilon$.] 
{The relative difference between the left arm nominal cross sections and  
cross sections with a 0.10 mrad offset in the scattering angle (red squares) as a function
of $\varepsilon$.} 
\label{fig:left_proton_sigma0.10mrad_epsilon}
\end{figure}

In estimating the scale, random, and slope uncertainties in the sensitivity of the 
cross sections to a 0.10 mrad and 0.18 mrad angle offset, the same procedure described in 
section \ref{ep_beam_energy} is used.
The sensitivity of the cross sections to a 0.10 mrad offset in the scattering angle 
has been studied. Figures \ref{fig:right_proton_sigma0.10mrad_epsilon} and 
\ref{fig:left_proton_sigma0.10mrad_epsilon} show the relative difference between the nominal cross 
sections, determined at the nominal scattering angle and energy, and the cross sections with a 
0.10 mrad change in the scattering angle for the right and left arms at all kinematics. 
A 0.10 mrad angle fluctuation changes the cross section by (0.10-0.12)\% for the right arm 
and (0.02-0.10)\% for the left arm. Such change in the cross sections was applied as a random 
uncertainty to each $\varepsilon$ point.

\begin{figure}[!htbp]
\begin{center}
\epsfig{file=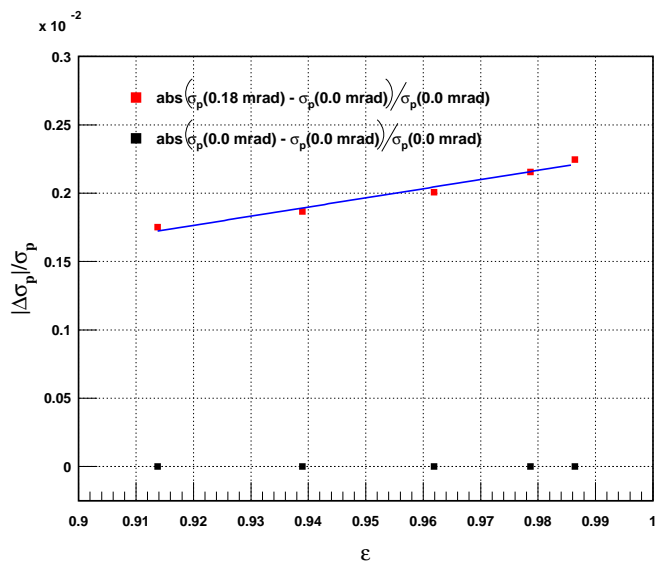,width=3.4in}
\end{center}
\caption[The relative difference between the right arm nominal cross sections and  
cross sections with a 0.18 mrad offset in the scattering angle as a function
of $\varepsilon$.] 
{The relative difference between the right arm nominal cross sections and  
cross sections with a 0.18 mrad offset in the scattering angle (red squares) as a function
of $\varepsilon$ at all 5 incident energies. The solid blue line is a linear fit to the data.} 
\label{fig:right_proton_sigma0.18mrad_epsilon}
\end{figure}

\begin{figure}[!htbp]
\begin{center}
\epsfig{file=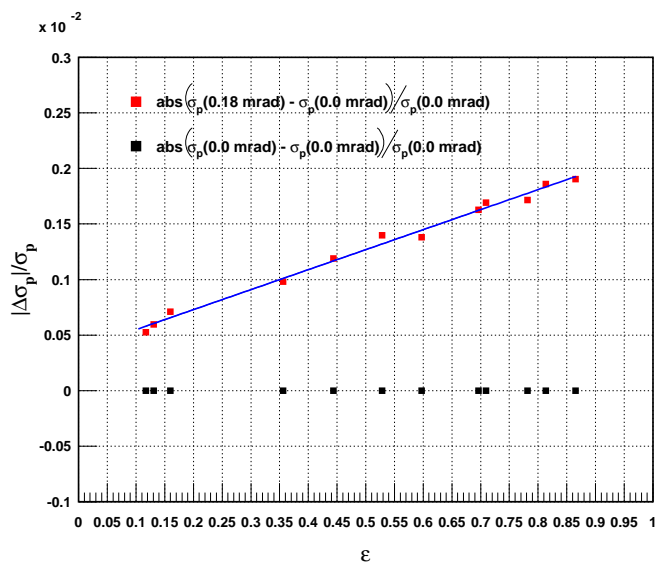,width=3.4in}
\end{center}
\caption[The relative difference between the left arm nominal cross sections and  
cross sections with a 0.18 mrad offset in the scattering angle.] 
{The relative difference between the left arm nominal cross sections and  
cross sections with a 0.18 mrad offset in the scattering angle (red squares) as a function
of $\varepsilon$. The solid blue line is a linear fit to the data.} 
\label{fig:left_proton_sigma0.18mrad_epsilon}
\end{figure}

Similarly, the sensitivity of the cross sections to a 0.18 mrad offset in the scattering angle has 
been studied. Figures \ref{fig:right_proton_sigma0.18mrad_epsilon} and 
\ref{fig:left_proton_sigma0.18mrad_epsilon} show the relative difference between the nominal cross 
sections and the cross sections with a 0.18 mrad angle offset for the right and left arms and 
at all kinematics. The right arm results show an average scale, random, and slope uncertainties of 
0.20\%, 0.02\%, and 0.67\%, respectively. While, the left arm results show an average scale, 
random, and slope uncertainties of 0.13\%, 0.01\%, and 0.18\%, respectively.
%

\section{VDC Tracking Efficiency} \label{vdcs_track_eff}

The VDC tracking inefficiency is defined as the fraction of good events where we do not reconstruct 
a track in the VDCs. It is determined by taking the fraction of the zero- and multiple-track 
events that passed through both planes of the two VDCs and caused a trigger in the scintillators 
planes: $(N_{zero}+N_{multiple})/N_{total}$ where $N_{zero}$ and $N_{multiple}$ are the number 
of events with zero- and multiple-track, respectively, and $N_{total}$ is the total number of good events. 
That in turn defines the VDC tracking efficiency $\epsilon_{VDC}$ as the fraction of one track 
events $N_{one}/N_{total}$. In the analysis of E01-001, we require only one-track events. Therefore, 
we reject multi-track events and correct for lost events.
A good-track event is defined as a one-track event that fired both VDCs planes and was in the the fiducial area, i.e, hit the
central paddles of the $S_{1}$ and $S_{2}$ scintillator planes. That requires a hit in paddle 3 or 4 of each scintillator plane. 
In addition, tracks with zero or many hits (hits per plane $>$30) are excluded. In addition on both arms, particle identification 
(PID) cuts using $A_{1}$ and $A_{2}$ aerogels were applied to exclude $\pi^{+}$. See section \ref{pid_cuts} for a
more detailed description of the PID cuts.
\begin{figure}[!htbp]
\begin{center}
\epsfig{file=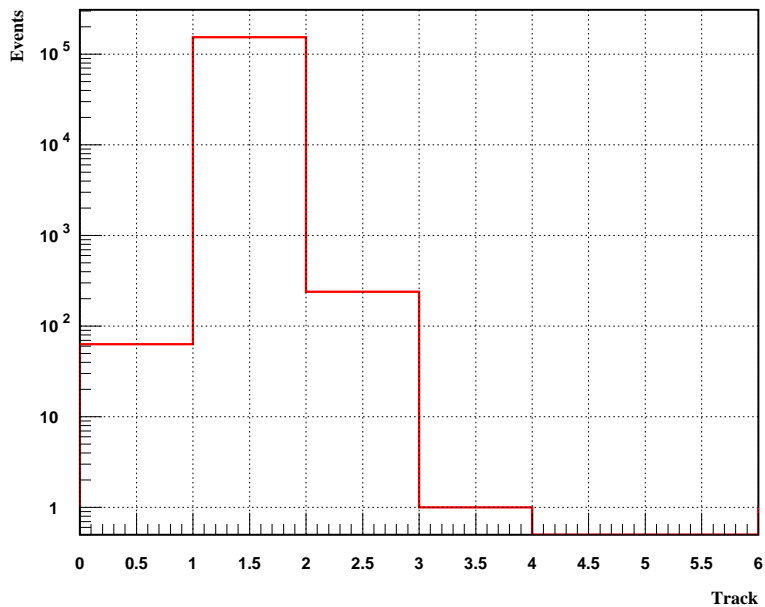,width=4in}
\end{center}
\caption[Track multiplicity in the left arm spectrometer for run 1597, kinematics $i$.]
{Track multiplicity in the left arm spectrometer for run 1597, kinematics $i$.}
\label{fig:left_track}
\end{figure}
Figure \ref{fig:left_track} shows the distribution of the number of reconstructed
tracks that satisfy the good-track event condition defined above for the  
left arm spectrometer. Although events with one track are desired, events with zero- or 
multiple-track are also present. Table \ref{vdc_effic} shows the number of zero-, one-, multiple-track 
events, and the VDC tracking efficiency for the right and left arm spectrometers as determined for 
a single run. The efficiency is calculated and applied to each run separately. 
The zero- and multiple-track inefficiencies for the left arm are $\leq$ 0.21\% and
$\leq$ 0.38\%, respectively, and $\leq$ 0.017\% and $\leq$ 1.26\% for the right arm, respectively.

\begin{figure}[!htbp]
\begin{center}
\epsfig{file=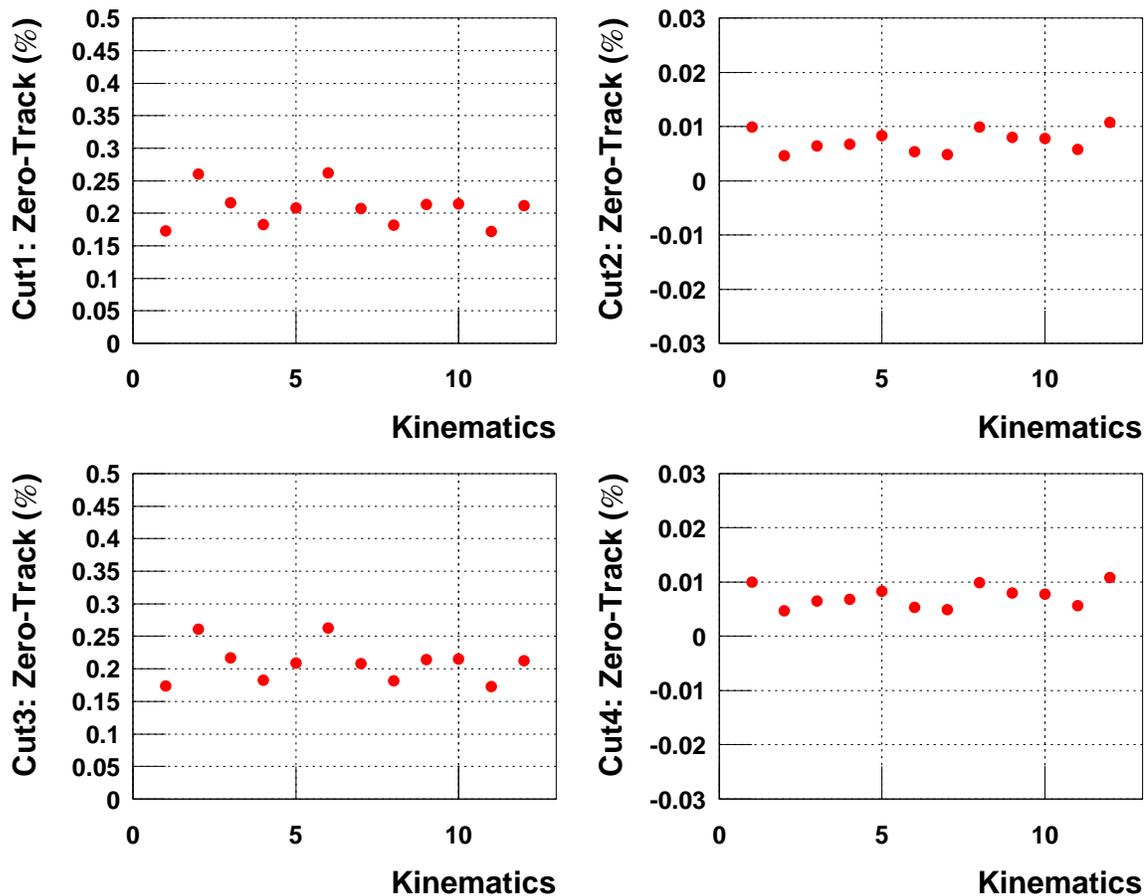,width=6in}
\end{center}
\caption[The average zero-track inefficiency for the right arm with several cuts applied.]
{The average zero-track inefficiency for the right arm with several cuts applied plotted as
a function of kinematics. The points are sorted according to the kinematics of the left arm.
Kinematics 1-5 correspond to $Q^2$ = 2.64 GeV$^2$, while kinematics 6-9 (10-12) correspond
to 3.20 (4.10) GeV$^2$. For each $Q^2$ value, the points are sorted by $\varepsilon$ (low to
high). See Table \ref{kinematics} for detail. Note that cut4 was used in the analysis.}
\label{fig:rzero_track_fraction}
\end{figure}

\begin{figure}[!htbp]
\begin{center}
\epsfig{file=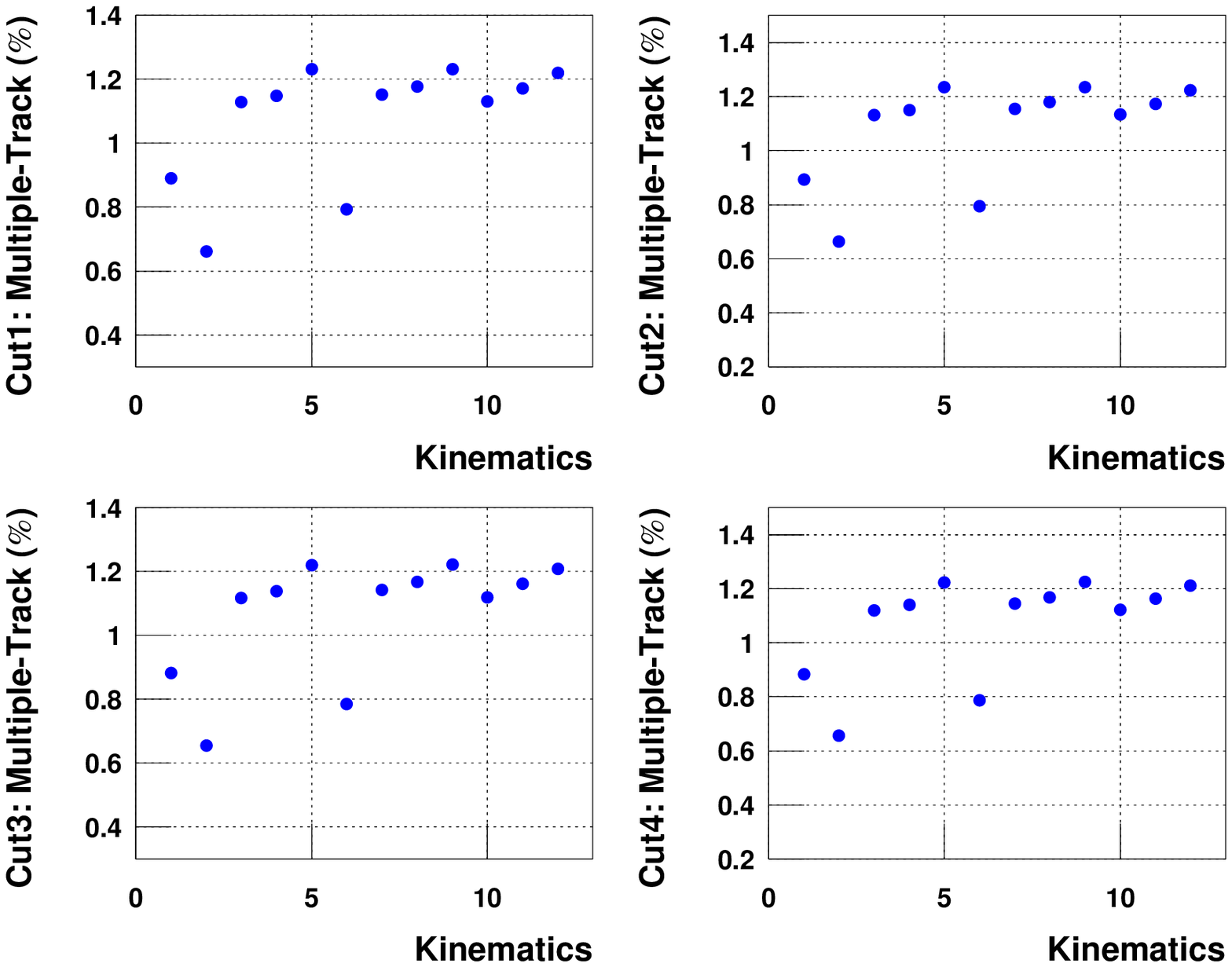,width=6in}
\end{center}
\caption[The average multiple-track inefficiency for the right arm with several cuts applied.]
{The average multiple-track inefficiency for the right arm with several cuts applied plotted as
a function of kinematics. The points are sorted according to the kinematics of the left arm.
Kinematics 1-5 correspond to $Q^2$ = 2.64 GeV$^2$, while kinematics 6-9 (10-12) correspond
to 3.20 (4.10) GeV$^2$. For each $Q^2$ value, the points are sorted by $\varepsilon$ (low to
high). See Table \ref{kinematics} for detail. Note that cut4 was used in the analysis.}
\label{fig:rmultiple_track_fraction}
\end{figure}

\begin{figure}[!htbp]
\begin{center}
\epsfig{file=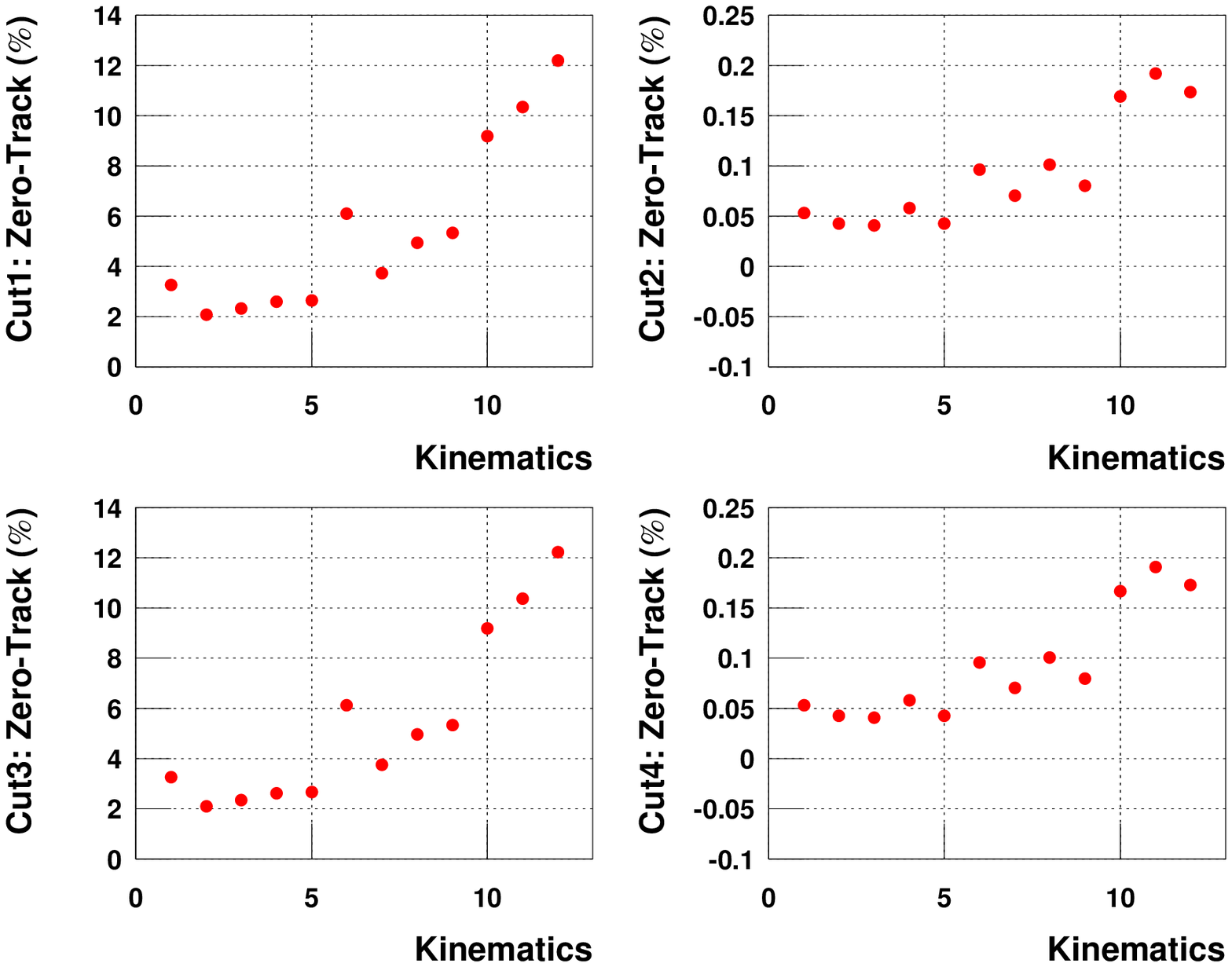,width=6in}
\end{center}
\caption[The average zero-track inefficiency for the left arm with several cuts applied.]
{The average zero-track inefficiency for the left arm with several cuts applied plotted as
a function of kinematics. Kinematics 1-5 correspond to $Q^2$ = 2.64 GeV$^2$, while kinematics 6-9 
(10-12) correspond to 3.20 (4.10) GeV$^2$. For each $Q^2$ value, the points are sorted by 
$\varepsilon$ (low to high). See Table \ref{kinematics} for detail. Note that cut4 was used in the analysis.}
\label{fig:lzero_track_fraction}
\end{figure}

\begin{figure}[!htbp]
\begin{center}
\epsfig{file=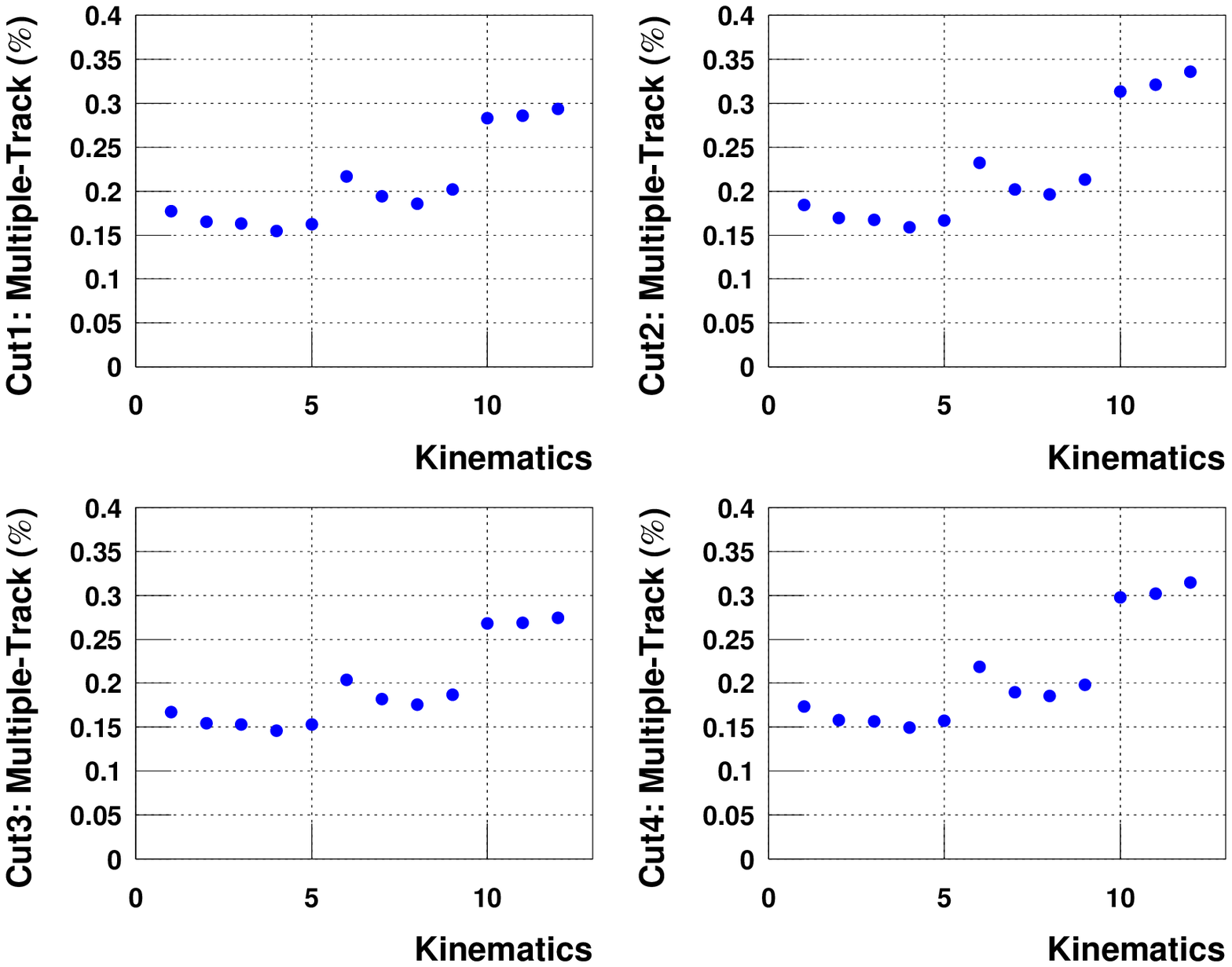,width=6in}
\end{center}
\caption[The average multiple-track inefficiency for the left arm with several cuts applied.]
{The average multiple-track inefficiency for the left arm with several cuts applied plotted as
a function of kinematics. Kinematics 1-5 correspond to $Q^2$ = 2.64 GeV$^2$, while kinematics 6-9 
(10-12) correspond to 3.20 (4.10) GeV$^2$. For each $Q^2$ value, the points are sorted by 
$\varepsilon$ (low to high). See Table \ref{kinematics} for detail. Note that cut4 was used in the analysis.}
\label{fig:lmultiple_track_fraction}
\end{figure}

In order to determine the uncertainty in the VDC tracking efficiency, the average zero- and 
multiple-track inefficiencies for all the runs in all the kinematics were determined by applying several
cuts. These cuts include: cut1 = fiducial cut, cut2 = fiducial plus no zero-hit cut, 
cut3 = fiducial plus no multi-hit cut, and cut4 = fiducial plus no zero-hit plus no multi-hit cut. 
Figures \ref{fig:rzero_track_fraction}, \ref{fig:rmultiple_track_fraction}, \ref{fig:lzero_track_fraction}, 
and \ref{fig:lmultiple_track_fraction} show the average zero- and multiple-track inefficiencies as 
determined under these cuts for the right and left arms, respectively. 
For the zero-track inefficiency, Figures \ref{fig:rzero_track_fraction} and \ref{fig:lzero_track_fraction}
show that cut1 and cut3 produce similar results as do cut2 and cut4. Clearly, exclusion of the no 
multi-hit cut does not alter the results and suggests the importance of the fiducial and no zero-hit cut 
applied only. Therefore, in the analysis of the E01-001, cut4 was used to determine the zero- and 
multiple-track inefficiencies. Clearly the zero-track inefficiency for the left arm is larger than that
for the right arm and has a range of (0.05-0.20)\% with $Q^2$ dependence but not any $\varepsilon$ dependence.
There is an estimated random or point-to-point uncertainty (fluctuation) of $\le$ 0.01\% in the average zero-track 
inefficiency for both arms. Therefore, the random uncertainty will be set to 0.0\%. The average zero-track inefficiency 
on the right arm does not have any significant $\varepsilon$ dependence or scale offset. Therefore, a scale and slope 
uncertainties of 0.0\% are assigned. On the other hand, a scale and slope uncertainties of 0.1\% and 0.0\% are assigned 
for the left arm.

\begin{figure}[!htbp]
\begin{center}
\epsfig{file=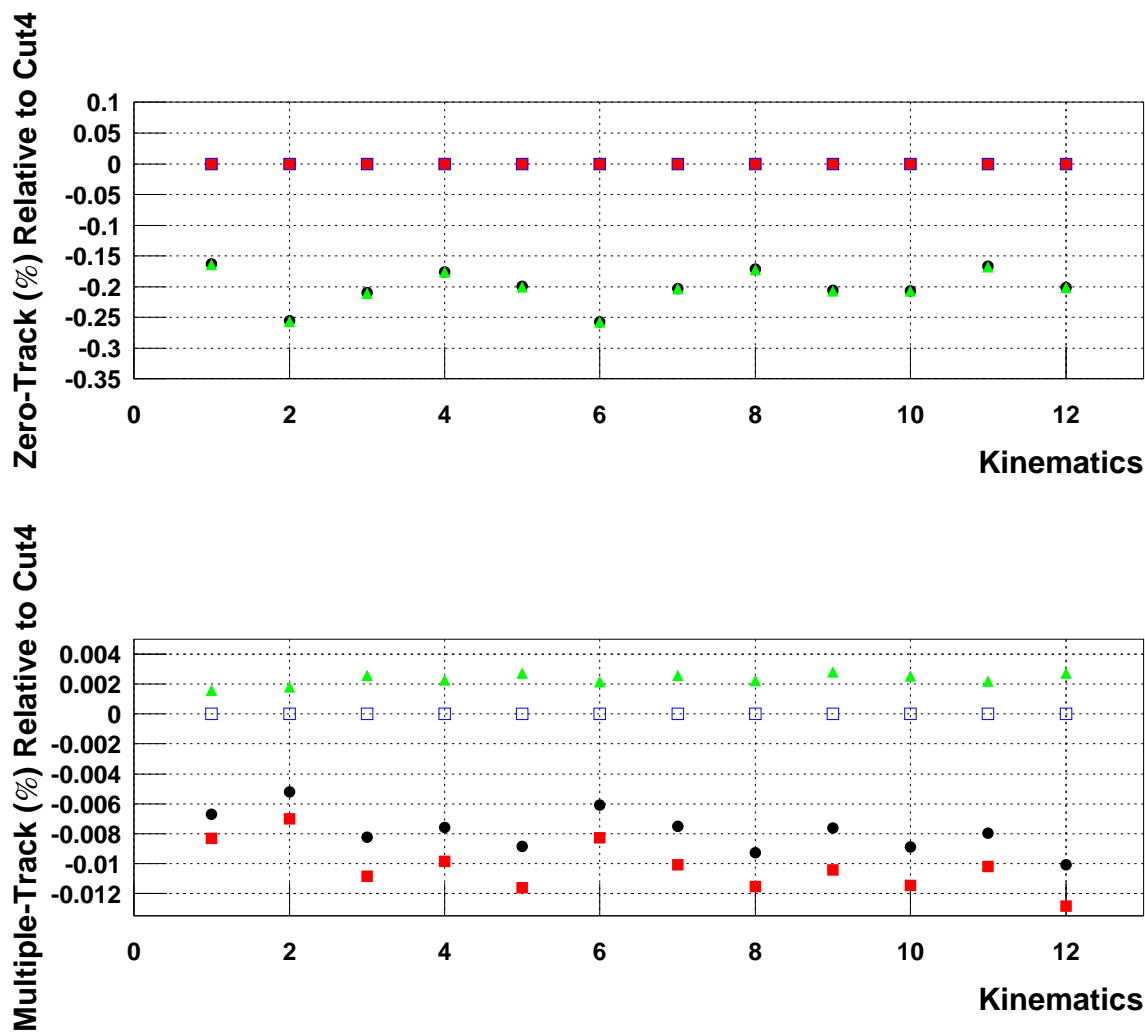,width=6in}
\end{center}
\caption[The right arm average zero- and multiple-track inefficiency with several cuts applied 
plotted relative to cut number 4.] 
{Top: The right arm average zero-track inefficiency with several cuts applied (cut1 = solid black circles, 
cut2 = solid red squares, cut3 = solid green triangles) relative to cut4 (open blue squares) plotted  
as a function of kinematics. Note that the open blue squares are the average zero-track inefficiency 
from cut4 relative to itself. The points are sorted according to the kinematics of the left arm.
Kinematics 1-5 correspond to $Q^2$ = 2.64 GeV$^2$, while kinematics 6-9 
(10-12) correspond to 3.20 (4.10) GeV$^2$. For each $Q^2$ value, the points are sorted by 
$\varepsilon$ (low to high). See Table \ref{kinematics} for detail.
Bottom: The right arm average multiple-track inefficiency with several cuts applied 
(cut1 = solid black circles, cut2 = solid red squares, cut3 = solid green triangles) 
relative to cut4 (open blue squares) plotted as a function of kinematics.  
The points are sorted according to the kinematics of the left arm as discussed in top caption
above.}
\label{fig:rrelat_zero_multiple_fraction}
\end{figure}

\begin{figure}[!htbp]
\begin{center}
\epsfig{file=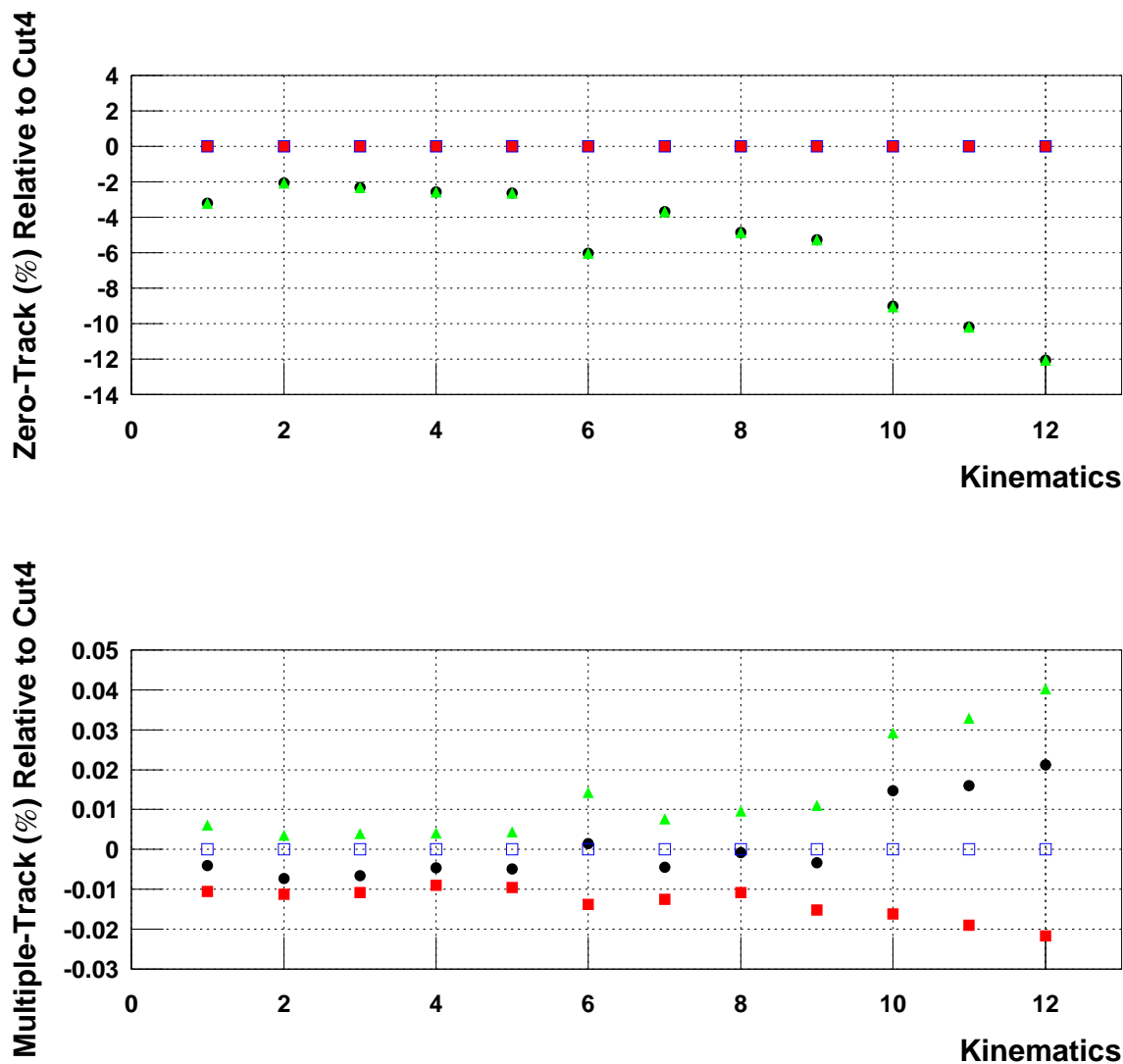,width=6in}
\end{center}
\caption[The left arm average zero- and multiple-track inefficiency with several cuts applied 
plotted relative to cut number 4.] 
{Top: The left arm average zero-track inefficiency with several cuts applied (cut1 = solid black circles, cut2 = 
solid red squares, cut3 = solid green triangles) relative to cut4 (open blue squares) plotted  
as a function of kinematics. Note that the open blue squares are the average zero-track inefficiency 
from cut4 relative to itself. 
Kinematics 1-5 correspond to $Q^2$ = 2.64 GeV$^2$, while kinematics 6-9 
(10-12) correspond to 3.20 (4.10) GeV$^2$. For each $Q^2$ value, the points are sorted by 
$\varepsilon$ (low to high). See Table \ref{kinematics} for detail.
Bottom: The left arm average multiple-track inefficiency with several cuts applied (cut1 = solid black circles, cut2 = solid red squares, cut3 = solid green triangles) 
relative to cut4 (open blue squares) plotted as a function of kinematics. The points are sorted according to the kinematics of the left arm as discussed in top caption above.}
\label{fig:lrelat_zero_multiple_fraction}
\end{figure}

As for the multiple-track inefficiency, Figures \ref{fig:rmultiple_track_fraction} and  
\ref{fig:lmultiple_track_fraction} show that the different cuts seem to produce relatively close results. 
The multiple-track inefficiency for the right arm has a range of (0.60-1.30)\% mainly rate dependence with an average 
inefficiency of $\sim$ 1.0\% and shows an $\varepsilon$ dependence of 0.05\%. If a 10\% measurement is assumed, this will 
yield a 0.10\% scale uncertainty. A slope uncertainty of 0.05\%/0.07 $\approx$ 0.70\% is assigned when
we consider the $\Delta \varepsilon$ range of 0.07. On the other hand, the multiple-track inefficiency for left arm has a range 
of (0.15-0.30)\% which varies with $Q^2$. A scale uncertainty of $\sim$ 0.10\% will be assigned as well. The multiple-track 
inefficiency shows insignificant $\varepsilon$ dependence and a slope uncertainty of 0.0\% is assigned. 
In order to determine the random uncertainty in the multiple-track inefficiency for both arms, the average multiple-track 
inefficiency from all the cuts (cut1-cut4) were plotted relative to cut4 and in order of increasing
$\varepsilon$. This was done for the average zero-track inefficiency as well. Figure \ref{fig:rrelat_zero_multiple_fraction} 
and \ref{fig:lrelat_zero_multiple_fraction} show the results. Again, the multiple-track inefficiency for the left arm is larger 
than that for the right arm and has some $Q^2$ dependence as well.
The average multiple-track inefficiency for the right arm has on the average a random fluctuation of 
0.01\% (set to 0.0\%), while the left arm average multiple-track inefficiency shows a random fluctuation of 
0.02\%.
%
%
\begin{table}[!htbp]
\begin{center}
\begin{tabular}{||c|c|c||} 
\hline \hline
Number of Tracks & Right Arm  & Left Arm \\ 
\hline \hline
Zero             & 10         & 63         \\
One              & 198563     & 154028      \\
Two              & 2231       & 240          \\
Three            & 16         & 1             \\
Four             & 0          & 0              \\
Five             & 0          & 0               \\
\hline 
Total            & 200820     & 154332            \\
\hline 
Zero-track fraction (\%) & 0.00498 & 0.0408     \\
\hline
Multiple-track fraction (\%) &1.119 &0.156      \\
\hline
$\epsilon_{VDC}$  & 0.988  & 0.998 \\ 
\hline \hline 
\end{tabular}
\caption[The number of zero-, one-, multiple-track events, and the VDC tracking efficiency 
for the right and left arm spectrometers.]
{The number of zero-, one-, multiple-track events, and the VDC tracking efficiency 
for the right and left arm spectrometers for run number 1597, kinematics $i$.}
\label{vdc_effic}
\end{center}
\end{table}

\newpage
\subsection{VDC Hardware Cuts Inefficiency} \label{vdcs_hardware_eff}

In the previous section we discussed the VDC tracking efficiency. Even after selecting single-track event,
not every reconstructed track corresponds to the true particle trajectory .
A typical single-track event making an angle of 45$^o$ with the VDC surface has a multiplicity, 
number of hit wires in the cluster that were fired by the event, of 4-6 \cite{offermanespace,fissum01}. 
Some of the good single-track events do not have enough hits in the VDC cluster or have noise hits not associated 
with the true particles. This can lead to tracks that do not reproduce the true particle trajectory.  
Such tracks lead to a long tails in the distribution of the reconstructed physical quantities. 

In order to eliminate these long tails, we apply a VDC hardware cuts. These cuts require that the number of 
clusters for each VDC plane = 1, minimum number of hits per cluster in each VDC plane = 3, maximum number 
of hits per cluster in each VDC plane = 6. However, removing these long tails comes at the expense of loosing 
good events which must be accounted for. We account for such loss by multiplying the beam charge $Q$ with
the efficiency of the VDC hardware cuts $\epsilon_{VDCH}$. 

\begin{figure}[!htbp]
\begin{center}
\epsfig{file=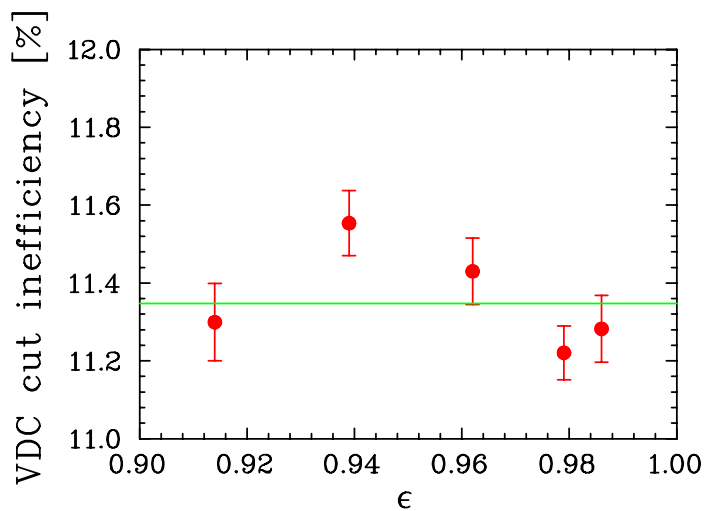,width=3.6in}
\end{center}
\caption[The right arm VDCs hardware cuts inefficiency.]
{The right arm VDCs hardware cuts inefficiency as a function of $\varepsilon$ at all
5 incident energies (solid red circles). The solid green line is the average inefficiency of 11.35\%
used in the analysis.}  
\label{fig:right_vdc_hardware}
\end{figure}

Figure \ref{fig:right_vdc_hardware} shows the inefficiency of the VDC hardware cuts 
applied for the right arm as a function of $\varepsilon$ for all 5 incidents energies. 
This inefficiency represents the fraction of good events that were lost after applying the VDC 
hardware cuts to remove the long tails seen in the distribution of the reconstructed physical 
quantities. In determining the inefficiency, we applied a set of physics/data quality cuts to get rid of 
junk events (see section \ref{recon} for a full list of cuts applied) and the results have some cut 
dependence. The results show an average inefficiency of 11.35\% giving $\epsilon_{VDCH}$ = 0.8865. 
We do not expect this efficiency to depend on $\varepsilon$ and the results are consistent with no slope. 
Therefore a slope uncertainty of 0.0\% is assigned. The results also suggest a scale and random 
uncertainties of 0.5\% and 0.1\%, respectively, due to the dependence on the quality cuts applied.

\begin{figure}[!htbp]
\begin{center}
\epsfig{file=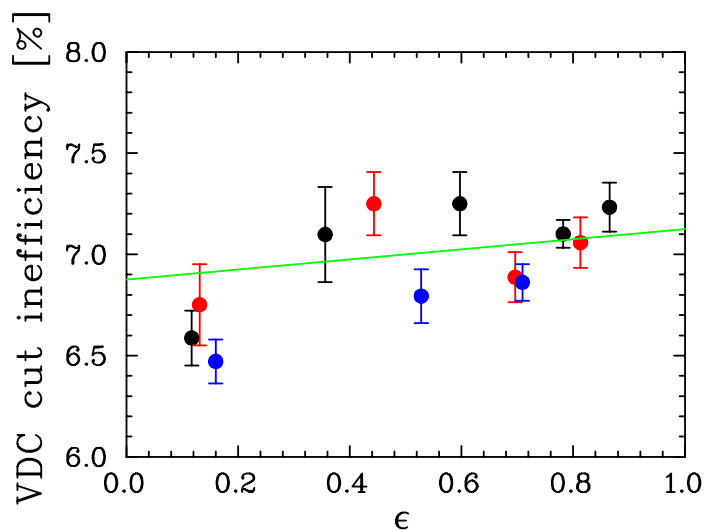,width=3.6in}
\end{center}
\caption[The left arm VDCs hardware cuts inefficiency.]
{The left arm VDCs hardware cuts inefficiency as a function of $\varepsilon$ for all kinematics. 
The $Q^2$ = 2.64 GeV$^2$ kinematics are shown as solid black circles, $Q^2$ = 3.20 GeV$^2$ as solid red 
circles, and $Q^2$ = 4.10 GeV$^2$ as solid blue circles. For each $Q^2$ value, the points are sorted by 
$\varepsilon$ (low to high). The solid green line is the inefficiency we use in the analysis as given by
7.0\%+0.25\%($\varepsilon$-0.50). See text for details.}
\label{fig:left_vdc_hardware}
\end{figure}

Figure \ref{fig:left_vdc_hardware} shows the inefficiency of the VDCs hardware 
cuts applied as a function of $\varepsilon$ for all kinematics for the left arm. Each of the three $Q^2$ values 
gives a small slope. If we average the three slopes we get an average slope of (0.60 $\pm$ 0.14)\%. 
However, most of the slope comes from the lowest $\varepsilon$ point. If we remove the lowest point for 
each $Q^2$, the result is consistent with zero slope. 
We do not know of any reason why the efficiency of the VDCs hardware cuts should depend on $\varepsilon$. 
However, we do know of reasons why our estimate of the efficiency might have an $\varepsilon$ 
dependence and that is because of the overlap of the VDCs cuts with some of our physics/data quality
cuts. So we take our best set of cuts, and apply a correction that has a large enough uncertainty 
to be consistent with no $\varepsilon$ dependence. 
Therefore, we apply a small slope with a large uncertainty of (0.25 $\pm$ 0.25)\% to try and cover 
both possibilities. The green line shown in Figure \ref{fig:left_vdc_hardware} is the inefficiency or 
correction we apply to each kinematics and is given by: 7.0\% + 0.25\%($\varepsilon$-0.5). 
The results also suggest a scale and random uncertainties of 0.5\% and 0.1\%, respectively.

\section{Scintillator Efficiency} \label{scint_effic}

The scintillator planes and trigger system are described in detail in 
section \ref{scintillators}. Too little or no energy deposited by the charged particles 
in the scintillator paddles, the inefficient transmission of light emitted by the charged particle 
inside the scintillators paddles to the PMTs, or inefficiencies of the PMTs used, will result in 
scintillator inefficiency and hence trigger inefficiency. Trigger inefficiency is estimated by taking 
the fraction of good events that were not counted by the main physics trigger $T_{1}$($T_{3}$) in the 
right(left) arm but still caused a trigger as a $T_{2}$($T_{4}$) event type in the right(left) arm.

In calculating the scintillator efficiency, single-track events that fell inside the scintillator
boundaries were kept by projecting the track to the scintillator plane and excluding events that missed
the detectors. This is done by applying cuts on the VDCs planes multiplicity and focal plane x and y axes, 
$xfp$ and $yfp$, and angles, $xpfp$ and $ypfp$. PID cuts were used to separate $\pi^{+}$ background from protons. 
Photomultiplier hits in the left and right side of each scintillators plane were required and software solid angle cut 
of 1.6 msr was applied to assure that these protons within the desired spectrometer's angular acceptance. 

Each scintillator plane efficiency $\epsilon_{S1,S2}$ was calculated using:
\begin{equation} \label{eq:scinteff}
\epsilon_{S1,S2} = \frac{N_{1(3)} + N_{5}}{N_{1(3)} + N_{5} + N_{2(4)}}~,
\end{equation} 
where $N_{i}$ ($i = 1,\cdots,5$) is the number of events of trigger type $i$ 
corrected for prescaling factor and electronic and computer deadtimes that fell 
inside the scintillators boundaries as defined by Table \ref{focal_plane_cuts}. 
It must be clear that event type $T_{2(4)}$ requires a hit in $S_{1}$ or $S_{2}$
but not both. In determining the efficiency of $S_{1}$, $N_{2(4)}$ in equation (\ref{eq:scinteff}) 
is the number of events that fired $S_{2}$ scintillator plane but not necessarily $S_{1}$. Similarly, 
if the efficiency of $S_{2}$ is desired, $N_{2(4)}$ will be the number of events that fired $S_{1}$ 
scintillator plane but not necessarily $S_{2}$. 
\begin{table}[!htbp]
\begin{center}
\begin{tabular}{||c|c|c||} 
\hline \hline
Arm              & S$_1$ plane boundary  & S$_2$ plane boundary \\
\hline \hline
Right            & $-1.05<xfp+1.381xpfp<0.90$ & $-1.30<xfp+3.314xpfp<1.00$ \\

                 & $-0.18<yfp+1.381ypfp<0.18$ & $-0.32<yfp+3.314ypfp<0.32$ \\
\hline \hline
Left             & $-1.05<xfp+1.287xpfp<0.90$ & $-1.30<xfp+3.141xpfp<1.00$ \\
                 & $-0.18<yfp+1.287ypfp<0.18$ & $-0.32<yfp+3.141ypfp<0.32$ \\
\hline \hline
\end{tabular}
\caption[Track projected from the focal plane to the location of $S_{1}$ and $S_{2}$ scintillators
and required to be inside of the scintillator plane.] 
{Track projected from the focal plane to the location of $S_{1}$ and $S_{2}$ scintillators
and required to be inside of the scintillator plane. The focal plane variables used are the x and y 
axis coordinates, $xfp$ and $yfp$ (measured in meter), and angles, $xpfp$ and $ypfp$.} 
\label{focal_plane_cuts}
\end{center}
\end{table}
\begin{figure}[!htbp]
\begin{center}
\epsfig{file=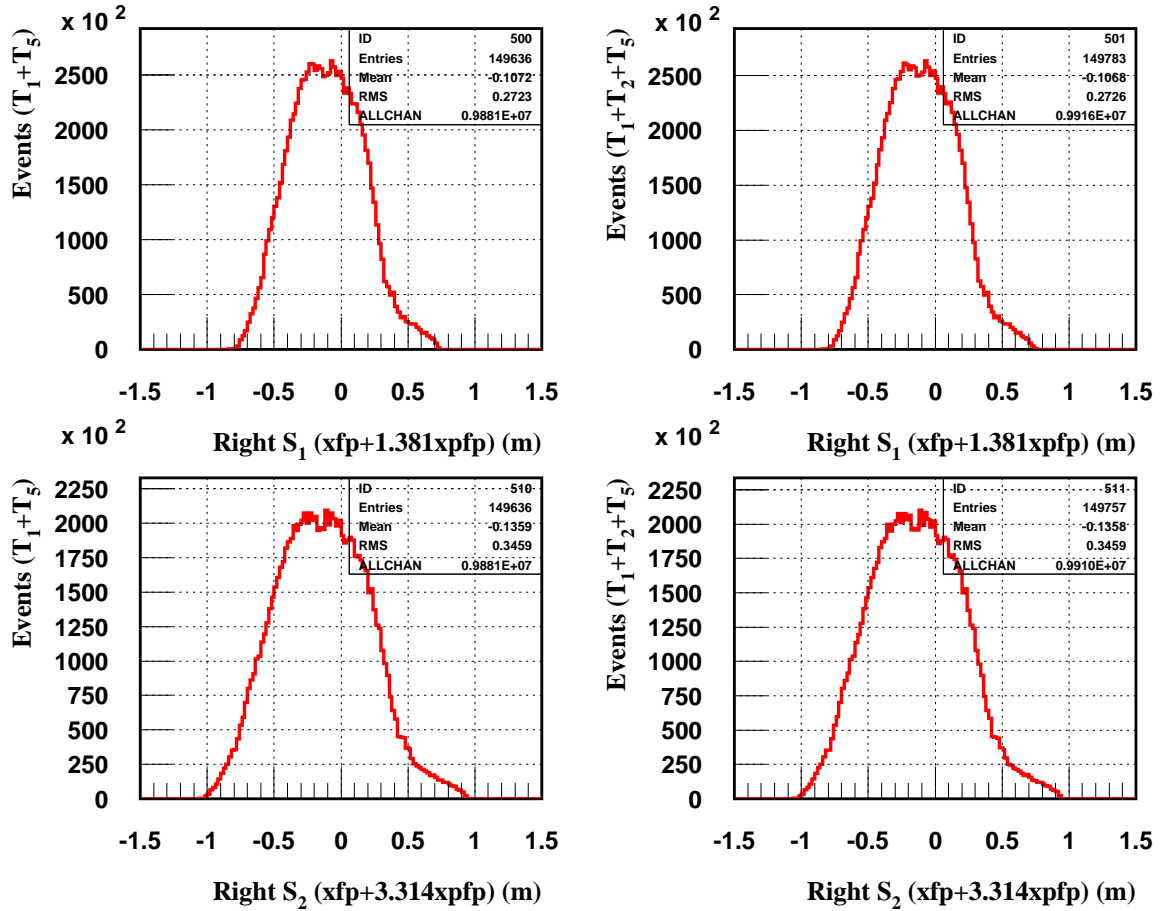,width=6in}
\end{center}
\caption[Events that fell within the scintillators boundaries as defined by Table \ref{focal_plane_cuts}
and needed for determination of $S_{1}$ and $S_{2}$ scintillators efficiencies.]
{Events that fell within the scintillators boundaries as defined by Table \ref{focal_plane_cuts} and
needed for determination of $S_{1}$ and $S_{2}$ scintillators efficiencies for run 1597, kinematic 
$i$. Top left: Events of type ($T_1+T_5$) that fell within the right arm S$_1$ scintillator boundary as 
projected from the focal plane (xfp+1.381xpfp). Top right: Events of type ($T_1+T_2+T_5$) that fell 
within the right arm S$_1$ scintillator boundary as projected from the focal plane (xfp+1.381xpfp). 
The $S_{1}$ scintillator efficiency is ratio of events in the two plots.
Bottom left: Events of type ($T_1+T_5$) that fell within the right arm S$_2$ scintillator boundary as 
projected from the focal plane (xfp+3.314xpfp). Bottom right: Events of type ($T_1+T_2+T_5$) that fell 
within the right arm S$_2$ scintillator boundary as projected from the focal plane (xfp+3.314xpfp).
The $S_{2}$ scintillator efficiency is ratio of events in the two plots. 
Similar plots can be generated for the left arm scintillators with $T_{1(2)} \to T_{3(4)}$.} 
\label{fig:right_s1s2effic}
\end{figure}

Figure \ref{fig:right_s1s2effic} shows how the efficiency for the two scintillators 
planes $\epsilon_{S_1}$ and $\epsilon_{S_2}$, as defined by equation (\ref{eq:scinteff}), for the right arm 
spectrometer is calculated. Similar plots (not shown) were generated for the left arm spectrometer and for 
each run. The final and total scintillators efficiency for any run is the product of the two scintillators 
efficiencies or $\epsilon_{S_1} \epsilon_{S_2}$. Table \ref{scint_effic_table} lists the results for 
$\epsilon_{S_1}$, $\epsilon_{S_2}$, and $\epsilon_{S_1} \epsilon_{S_2}$ for both the right and left 
arm spectrometers for a selected run. The scintillators efficiencies were 
typically $\geq$ 99.5\% and $\geq$ 99.6\% for S$_1$ and S$_2$, respectively, and were calculated and
applied to each run separately. 
\begin{table}[!htbp]
\begin{center}
\begin{tabular}{||c|c|c|c||} 
\hline \hline
Arm              & $\epsilon_{S1}$  & $\epsilon_{S2}$ & $\epsilon_{S_1} \epsilon_{S_2}$ \\
\hline \hline
Right            & 0.996         &  0.997       & 0.994                            \\
Left             & 0.999         &  0.999       & 0.998                             \\
\hline \hline
\end{tabular}
\caption[The S$_1$ and S$_2$ efficiencies, $\epsilon_{S_1}$ and $\epsilon_{S_2}$, and total scintillators efficiency, 
$\epsilon_{S_1} \epsilon_{S_2}$, for the right and left arm spectrometers as determined for run 1597, kinematics i.]
{The S$_1$ and S$_2$ efficiencies, $\epsilon_{S_1}$ and $\epsilon_{S_2}$, and total scintillators efficiency, 
$\epsilon_{S_1} \epsilon_{S_2}$, for the right and left arm spectrometers as determined for run 1597, kinematics i.}
\label{scint_effic_table}
\end{center}
\end{table}
\begin{figure}[!htbp]
\begin{center}
\epsfig{file=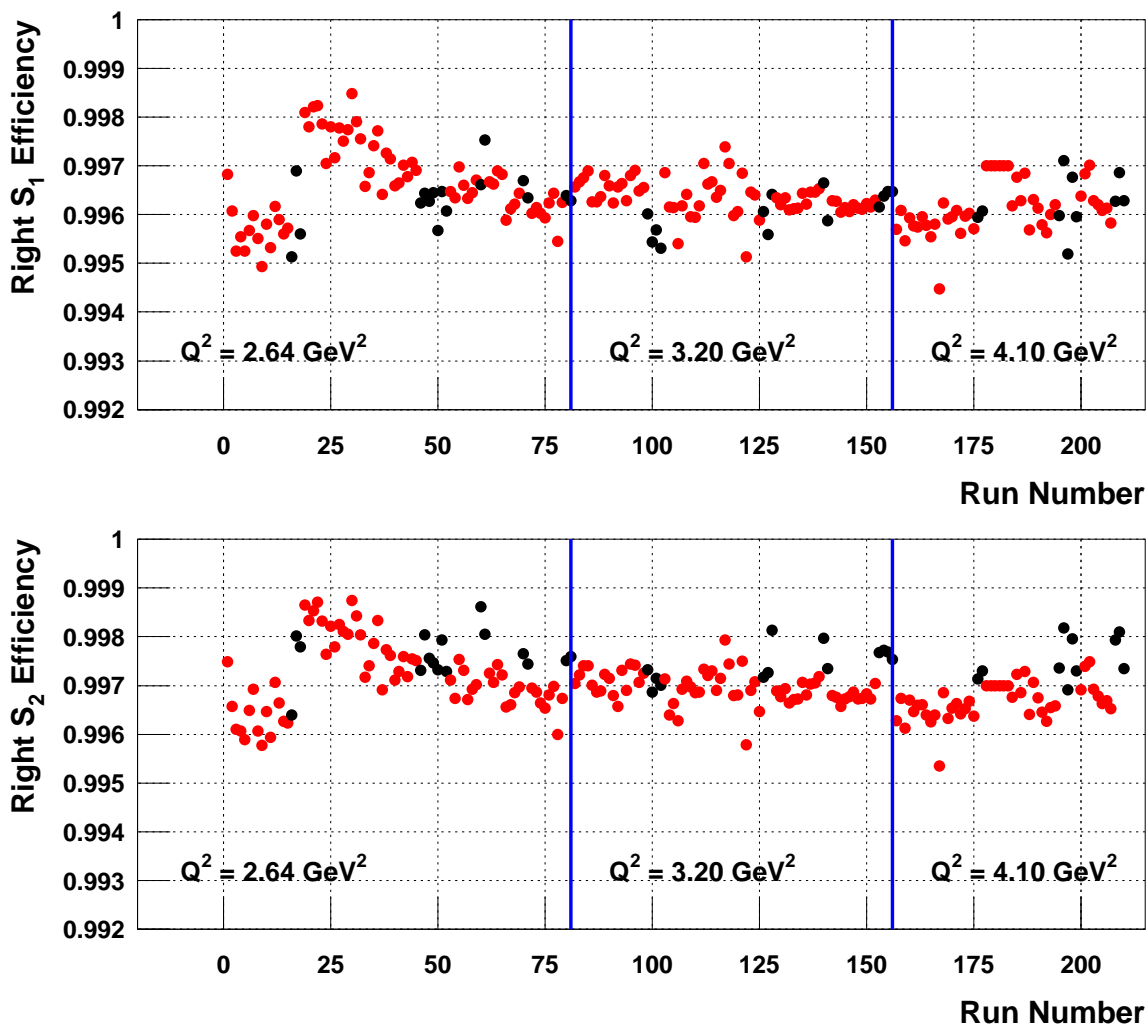,width=6in}
\end{center}
\caption[The right arm S$_1$ and S$_2$ scintillators efficiency for both LH$_2$ and dummy data.]
{The right arm S$_1$ and S$_2$ scintillators efficiency for both LH$_2$ (red) and dummy (black) for all of the 
elastic kinematics. The efficiency is plotted as a function of run number. The run number is sorted according to the 
kinematics of the left arm and by increasing $Q^2$. For each $Q^2$, the kinematics are sorted by $\varepsilon$ (low to high). 
See Table \ref{kinematics} for details.}
\label{fig:right_s1s2_scint_effic}
\end{figure}
\begin{figure}[!htbp]
\begin{center}
\epsfig{file=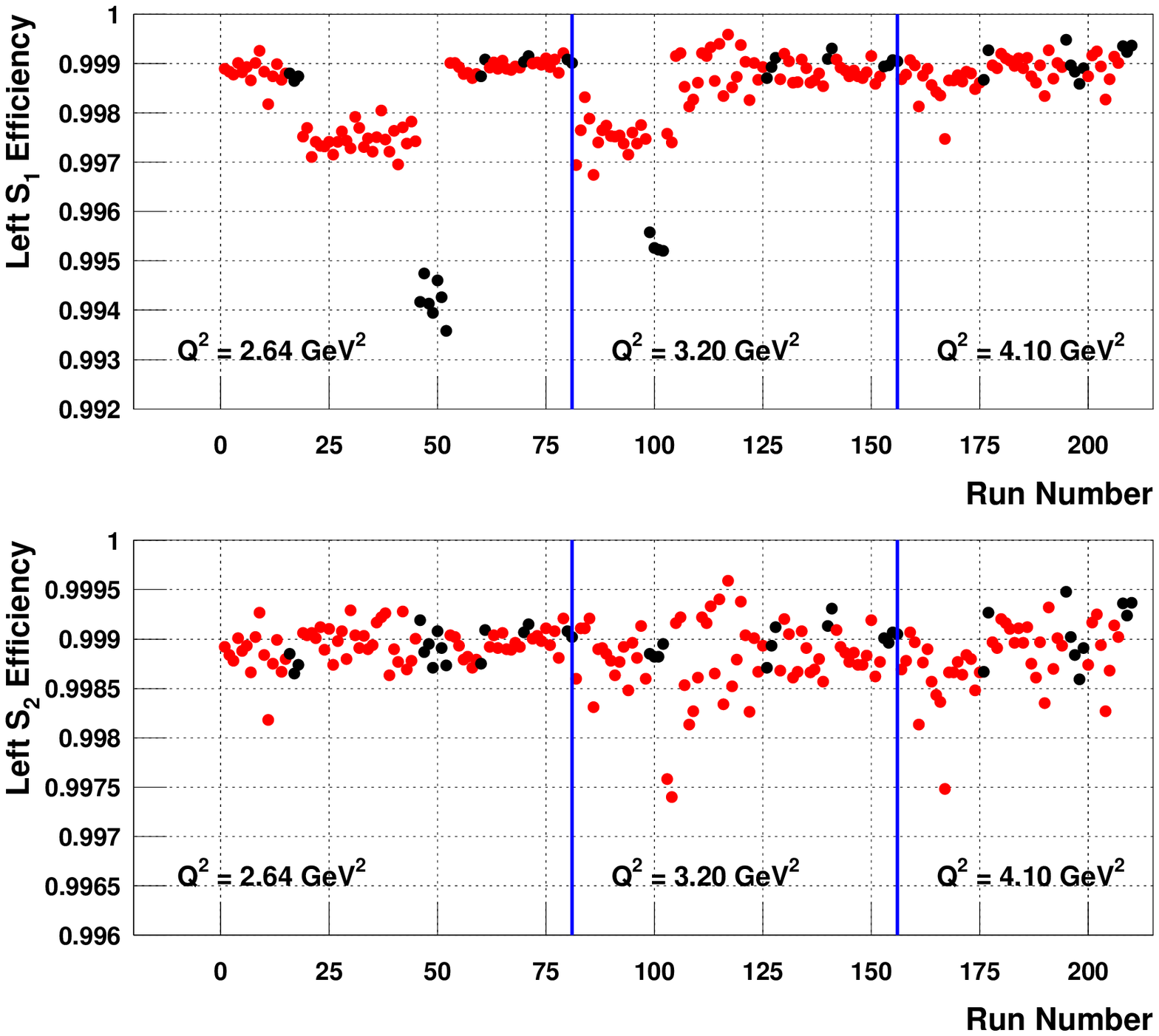,width=6in}
\end{center}
\caption[The left arm S$_1$ and S$_2$ scintillators efficiency for both LH$_2$ and dummy data.]
{The left arm S$_1$ and S$_2$ scintillators efficiency for both LH$_2$ (red) and dummy (black) for all of the 
elastic kinematics. The efficiency is plotted as a function of run number. The run number is sorted by increasing $Q^2$ and 
for each $Q^2$ by $\varepsilon$ (low to high). Kinematics $a$ and $b$ have lower S$_1$ scintillator efficiency for their LH$_2$
and dummy runs compared to the rest of the kinematics because one of the six paddles in the S$_1$ scintillator plane was tilted 
at the time of the run forming a gap with the adjacent paddle. Because the gap was far from the elastic acceptance, we increased 
the efficiency of the S$_1$ scintillator by 0.15\% for all the runs in kinematics $a$ and $b$.}
\label{fig:left_s1s2_scint_effic}
\end{figure}

Figures \ref{fig:right_s1s2_scint_effic} and \ref{fig:left_s1s2_scint_effic} show the efficiency 
of the two scintillators planes, $\epsilon_{S_1}$ and $\epsilon_{S_2}$, for all the 
LH$_2$ and dummy data runs for both arms. For the both arms, we estimate the random and scale 
uncertainties to be 0.05\% and 0.10\%, respectively. There is not any significant $\varepsilon$ 
dependence observed in the results, therefore a 0.0\% slope uncertainty is assigned.

\section{Particle Identification Efficiency} \label{pid_cuts}

Particle identification (PID) cuts are needed in order to obtain a clean
proton sample and the efficiency of these cuts must be accurately determined as well
as any misidentification of other particles as protons. 
Two spectrometers were used, the low $Q^2$ spectrometer (right arm) and the 
high $Q^2$ spectrometer (left arm), and PID cuts and their efficiencies are needed 
for both.

\subsection{The efficiency of the right arm $\beta$ cut} \label{beta_eff}

\begin{figure}[!htbp]
\begin{center}
\epsfig{file=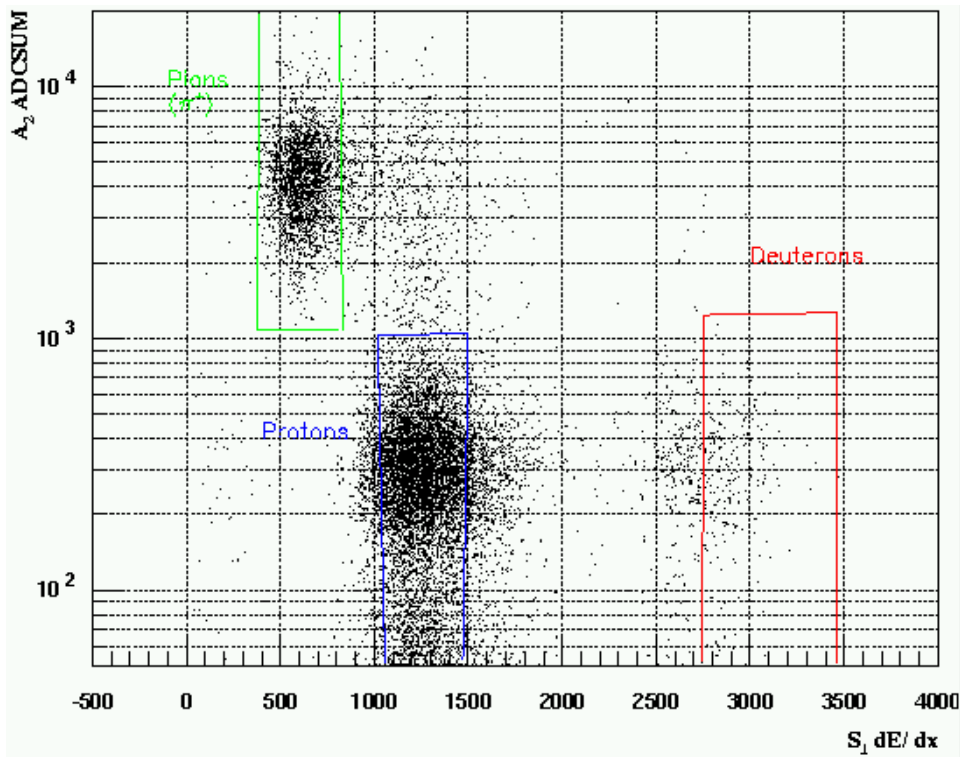,width=5in}
\end{center}
\caption[Two-dimensional plot of $A_{2}$ ADCSUM vs $S_{1}~dE/dx$.]
{Two-dimensional plot of $A_{2}$ ADCSUM vs $S_{1}~dE/dx$ for run 1730 from kinematics $n$.
Deuterons (red boundary), protons (blue boundary), and pions ($\pi^+$) (green boundary) have
been identified and separated from each other using the tight two-dimensional cuts shown.
The cut for deuterons is offset to reduce proton contamination since the efficiency of this cut
is not important.}
\label{fig:a2_s1dedx}
\end{figure}

\begin{figure}[!htbp]
\begin{center}
\epsfig{file=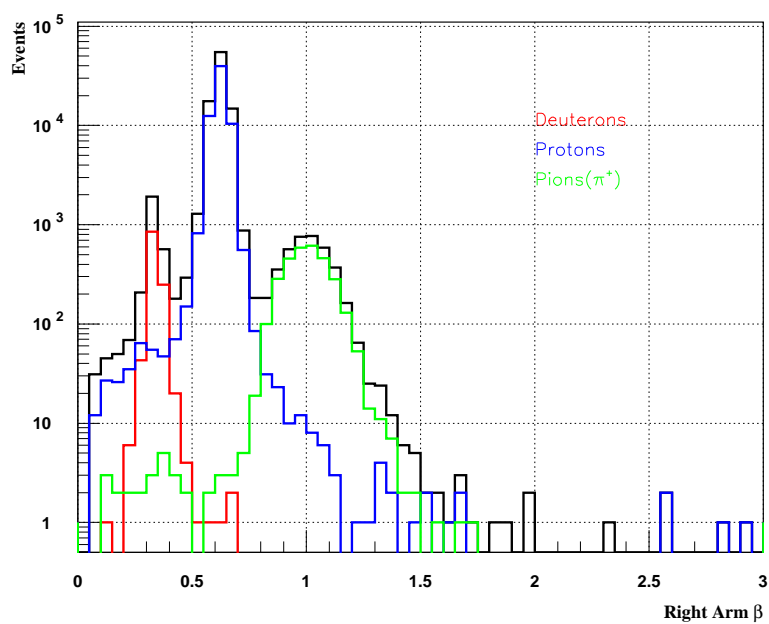,width=4in}
\end{center}
\caption[Right arm $\beta$ spectrum.]
{The contribution of pions (green), deuterons (red), and protons (blue) to the full right arm 
$\beta$ spectrum (solid black line).}
\label{fig:right_beta}
\end{figure}

In order to determine the efficiency of the $\beta$ cut, $\epsilon_{\beta}$, that we are 
applying, a two-dimensional plot of the sum of the ADC signals for the $A_{2}$ aerogel, 
$A_{2}$ ADCSUM, and the energy deposited in the first right arm scintillator, $S_{1}~dE/dx$, 
was generated. Figure \ref{fig:a2_s1dedx} shows such a plot.
In Figure \ref{fig:a2_s1dedx}, deuterons, protons, and pions have been identified and separated 
from each other by applying two-dimensional cuts on $A_{2}$ ADCSUM vs $S_{1} dE/dx$. 
A plot of the right arm $\beta$ for the three particle types was then generated using 
two-dimensional cuts shown in Figure \ref{fig:a2_s1dedx}. Figure \ref{fig:right_beta}
shows the contribution of each particle to the full $\beta$ spectrum. Pions (which usually 
measure 0.85$<\beta<$1.20) and deuterons (usually 0.02$<\beta<$0.45) leak into the protons area with 
0.45$<\beta<$0.85. Although the tails of the distributions (tails of the pions, deuterons, and protons distributions) 
with $\beta<$ zero and $\beta>$ 1.0 are unphysical and are due to finite timing resolution of the PMTs, 
we account for the events in these tails when we calculate the pion and deuteron contaminations as well as the proton loss.

Based on Figure \ref{fig:right_beta}, the deuteron and pion contamination and proton loss 
were determined to be:

\begin{itemize}
\item {Total number of deuterons = 1185} 
\item {Number of deuterons into protons area = 9}
\item {Number of deuterons into pions area = 0}
\item {Total number of pions = 3051}
\item {Number of pions into protons area = 132}
\item {Number of pions into deuterons area = 14}
\item {Total number of protons = 64587}
\item {Number of protons into pions area = 62} 
\item {Number of protons into deuterons area = 271}
\item {Number of protons into protons area = 64089}
\item {Deuteron Contamination =
(Number of deuterons into protons area/Number of protons into protons area)$\times$100\%
  = (9/64089)$\times$100\% = 0.014\%}
\item {Pion Contamination =
(Number of pions into protons area/Number of protons into protons area)$\times$100\%
  = (132/64089)$\times$100\% = 0.206\%}
\item {Proton Efficiency $(\epsilon_{\beta})$ = 
(Number of protons into protons area/Total number of protons)
  = (64089/64587) = 0.99243}
\end{itemize} 
\begin{table}[!htbp]
\begin{center}
\begin{tabular}{||c|c|c|c|c|c||}
\hline \hline
Run   & Incident Energy &  Maximum Deuteron & Maximum Pion  & Proton   \\
Number&  (MeV)          &  Contamination    & Contamination & Efficiency \\ 
      &                 &  (\%)             & (\%)          & ($\epsilon_{\beta}$) \\
\hline \hline
1252  & 2260.00         & 0.0100            & 0.1886        & 0.9943      \\
1653  & 2844.71         & 0.0086            & 0.1827        & 0.9932       \\
1730  & 4702.52         & 0.0140            & 0.2060        & 0.9924        \\ 
1772  & 1912.94         & 0.0000            & 0.2290        & 0.9934         \\ 
1823  & 3772.80         & 0.0120            & 0.1937        & 0.9925          \\
\hline \hline
\end{tabular}
\caption[The incident energy, deuteron contamination, pion contamination, and proton
efficiency for the $\beta$ cut.]     
{The incident energy, deuteron contamination, pion contamination, and proton
efficiency for the $\beta$ cut.}
\label{beta_effic}
\end{center}
\end{table}
\begin{figure}[!htbp]
\begin{center}
\epsfig{file=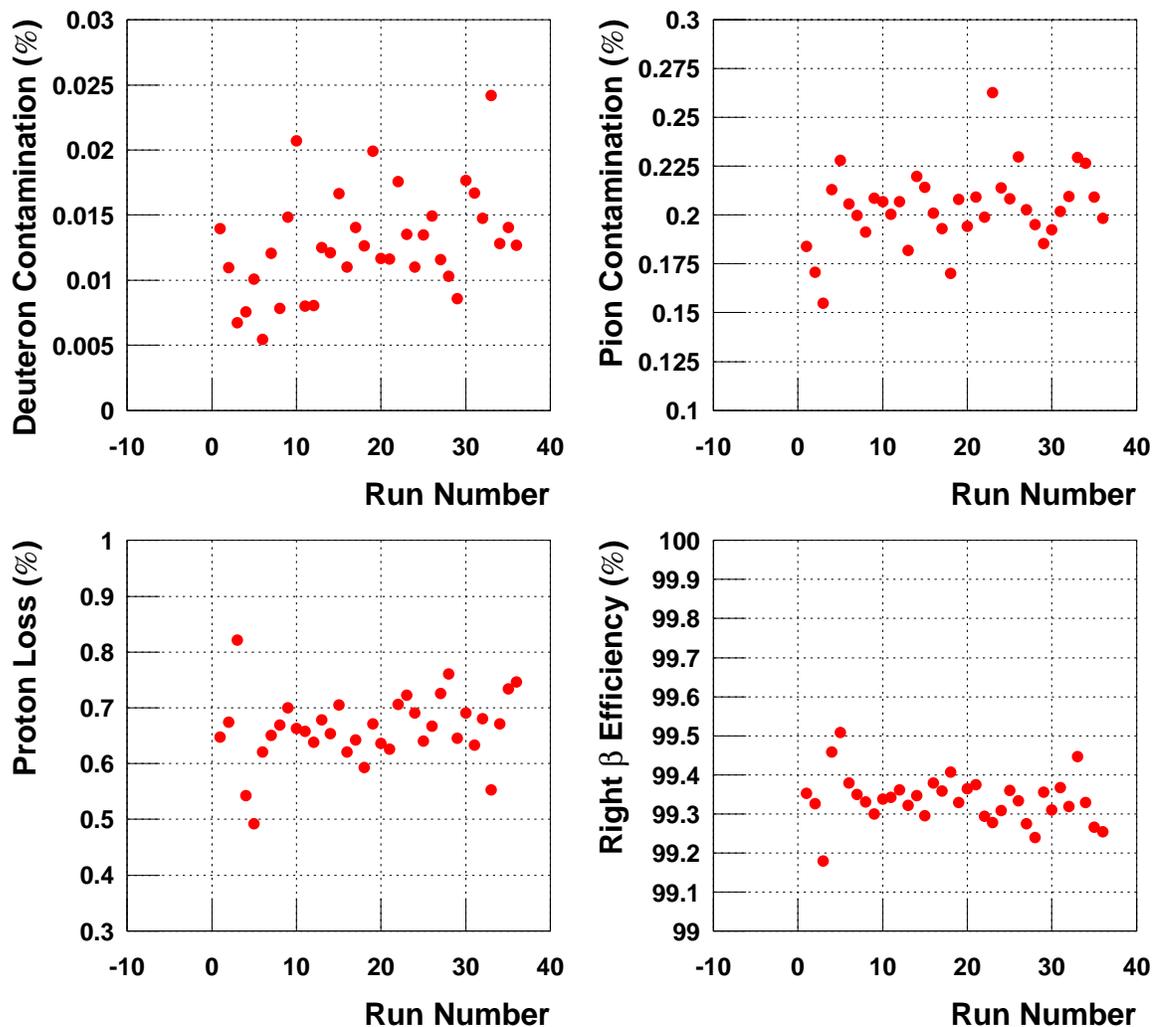,width=6in}
\end{center}
\caption[The right arm deuteron contamination, pion contamination, proton loss, and 
$\epsilon_{\beta}$ for sample runs from all kinematics.] 
{The right arm deuteron contamination (top left), pion contamination (top right), proton loss 
(bottom left), and efficiency of the $\beta$ cut, $\epsilon_{\beta}$, (bottom right) for 
sample runs from all kinematics. The points are sorted according to the kinematics of the left arm,
with three runs shown for each kinematics setting. Runs 1-15 correspond to $Q^2$ = 2.64 GeV$^2$, 
while 16-27 (28-36) correspond to 3.20 (4.10) GeV$^2$. For each $Q^2$ value, the points are 
sorted by $\varepsilon$ (low to high). See Table \ref{kinematics} for details.}
\label{fig:right_arm_beta_effic}
\end{figure}

This analysis was done at all 5 incident energies and for multiple runs. Table \ref{beta_effic} 
summarizes the results. The deuteron contamination is clearly negligible while the pion contamination 
is small and is negligible after applying the aerogel $A_2$ ADCSUM cut. 

Figure \ref{fig:right_arm_beta_effic} shows the deuteron contamination, pion contamination, proton loss, 
and the efficiency of the $\beta$ cut, $\epsilon_{\beta}$, for a multiple runs from all the
kinematics. The results are plotted in order of increasing $\varepsilon$. The efficiency of the
$\beta$ cut is typically $\geq$ 99.2\% with an average random fluctuation of 0.05\% and
without any noticeable $\varepsilon$ dependence. 

Having determined the proton efficiency from the $\beta$ cut, $\epsilon_{\beta}$, we would like to determine the 
uncertainty in $\epsilon_{\beta}$. Note that $\epsilon_{\beta}$ was determined assuming a pure proton sample. However, 
this is not the case since any pion and deuteron contamination will change the value of $\epsilon_{\beta}$.
In order to determine the size of possible pions/deuterons contamination, the $\beta$ spectrum is used 
and the number of deuterons and pions which can be misidentified as protons
determined. This was done for multiple runs at all 5 incident energies. First, the maximum number of
deuterons (maximum bin content) in the deuterons area is determined. Then the maximum number of protons 
which leaked into the deuterons area is determined. This is an upper limit estimate and based on the 
content of the bin directly below the bin that contained the maximum number of deuterons. This procedure
is illustrated below for run number 1250 as in Figure \ref{fig:right_beta} and for deuterons:

\begin{itemize}
\item {Total number of deuterons in all $\beta$ spectrum = 1326.} 
\item {Maximum deuterons bin content in deuterons area = 506.} 
\item {Maximum protons into deuterons area (upper limit estimate)= 65.}
\item {Ratio of protons to deuterons in deuterons area = $\alpha$ = 12.85\%.}
\item {Number of deuterons misidentified as protons = $N_{Deuterons}$ = $\alpha$(total number of deuterons) 
$\approx$ 170.} 
\item {Total number of protons in all $\beta$ spectrum = $N_{proton}$= 161019.}
\end{itemize} 

The same procedure used to determine the uncertainty in $\epsilon_{\beta}$ is used to determine the uncertainty
in $\epsilon_{A_2}$. Similarly the number of pions (in the pions area) which are misidentified as
protons are determined. The ratio of protons to pions is estimated to be $\gamma$ = 0.69\%, giving the number of pions
misidentified as protons = $N_{\pi^+}$ = $\gamma$(total number of pions) $\approx$ 47. The scale uncertainty
is determined as ($N_{Deuterons}$ + $N_{\pi^+}$)$\times$100\%/$N_{proton}$ = 0.135\%. The scale uncertainty 
in the efficiency of the $\beta$ cut as determined at all the incident energies was fairly constant and 
an overall scale uncertainty of 0.15\% was assigned.  

\subsection{The efficiency of the right arm $A_{2}$ Aerogel ($A_{2}$ ADCSUM) cut} \label{A2_eff}

\begin{figure}[!htbp]
\begin{center}
\epsfig{file=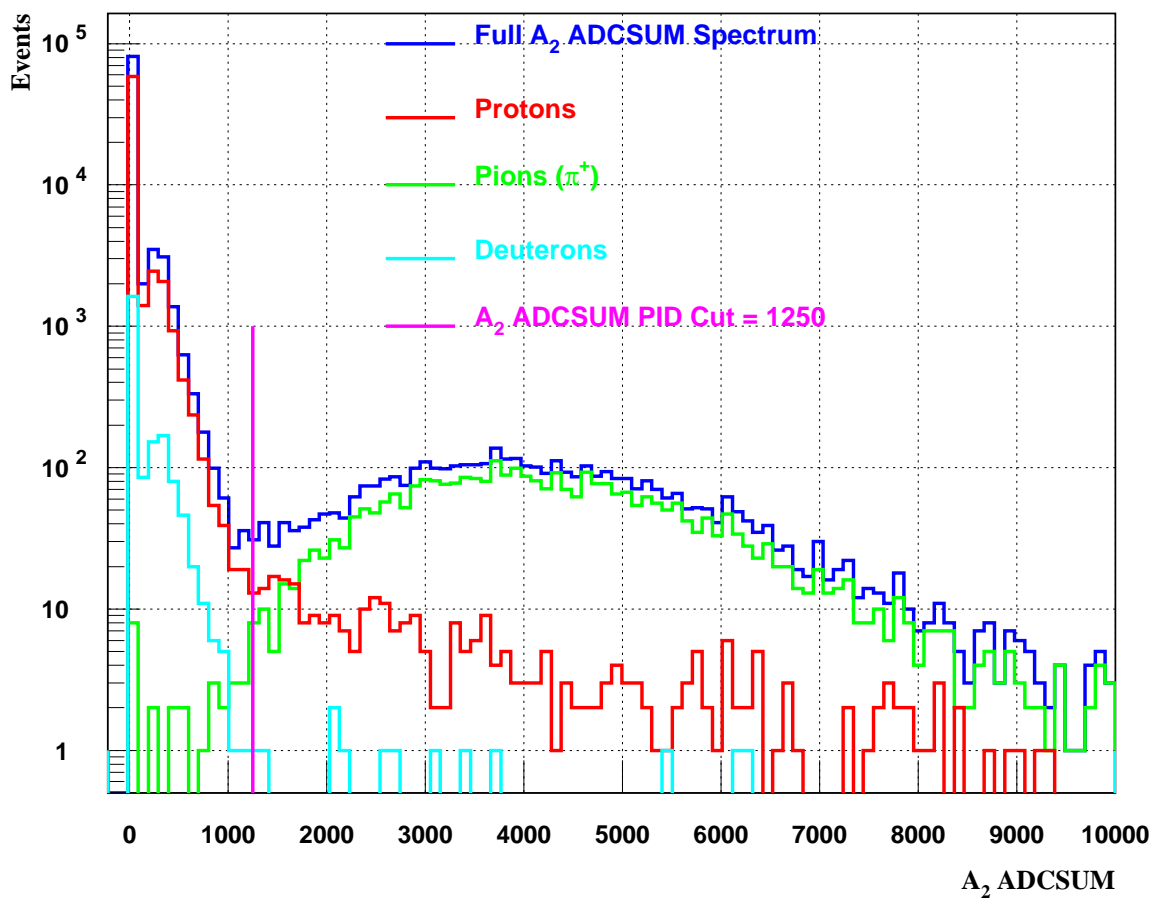,width=6in}
\end{center}
\caption[The full $A_{2}$ ADCSUM spectrum (blue). Also shown the deuterons (cyan), protons (red), and pions($\pi^+$) 
(green) contribution to the full $A_{2}$ ADCs sum signal.]
{The full $A_{2}$ ADCSUM spectrum (blue). Also shown the deuterons (cyan),
protons (red), and pions ($\pi^+$) (green) contribution to the full $A_{2}$ ADCSUM signal for run
1730 from kinematics $n$. The magenta line at $A_{2}$ ADCSUM = 1250 is the PID cut used.}
\label{fig:ra2_adcsum}
\end{figure}

We looked at the $A_2$ ADCSUM for each of the five runs listed in Table \ref{beta_effic}.
For each run, we overlayed the contribution from each of the three particles to 
the $A_{2}$ ADCSUM spectrum using tight cuts on $\beta$ and $S_{1}~dE/dx$ as shown
in Table \ref{beta_s1dedx}. Figure \ref{fig:ra2_adcsum} shows the full $A_2$ ADCSUM spectrum 
and its constituents. In addition, the PID boundary cut used in the analysis is also shown. 
\begin{table}[!htbp]
\begin{center}
\begin{tabular}{||c|c|c||}
\hline \hline
Particle    & $\beta$ range   &  $S_{1}~dE/dx$ range\\
\hline \hline
Deuterons  &0.20$<\beta<$0.40  & 2500$<S_{1}~dE/dx<$3500\\
Protons    &0.60$<\beta<$0.70  & 1000$<S_{1}~dE/dx<$1800 \\ 
Pions      &0.80$<\beta<$1.20  &  400$<S_{1}~dE/dx<$900   \\
\hline \hline
\end{tabular}
\caption[The tight $\beta$ and $S_{1}~dE/dx$ cuts range.] 
{The tight $\beta$ and $S_{1}~dE/dx$ cuts range.} 
\label{beta_s1dedx}
\end{center}
\end{table}

It can be seen that some of the protons leak into the pion area, $A_2$ ADCSUM $>$ 1250. 
Similarly, some pions make it into the proton region, $A_2$ ADCSUM $<$ 1250. 
Only events with $A_2$ ADCSUM $<$ 1250 are kept knowing that these good protons are definitely 
contaminated with some pions and deuterons while events with $A_2$ ADCSUM $>$ 1250 will be rejected
with some good protons lost. 

The contamination of pions and the protons efficiency were determined
for the data shown in Figure \ref{fig:ra2_adcsum} as the following:
\begin{itemize}
\item{Total Protons = 66503}
\item{Protons greater than 1250 = 322}
\item{Protons less than 1250 = 66181}
\item{Total Pions = 3208}
\item{Pions greater than 1250 = 3176}
\item{Pions less than 1250 = 32}
\item{Pion Contamination =
(Number of pions less than 1250/Number of protons less than 1250)$\times$100\%
                    = (32/66181)$\times$100\% = 0.05\%}
\item{Proton Efficiency $(\epsilon_{A_2})$ = 
(Number of protons less than 1250/Total number of protons)
                    = (66181/66503) = 0.99516}
\end{itemize}

Similar analyses were done on runs taken at the other four incident energies.
Table \ref{ra2_effic} summarizes the results. The final proton efficiency for each of the 5 
$\varepsilon$ points is defined as the product of the proton efficiency as determined from the 
$\beta$ cut and the proton efficiency as determined from the $A_{2}$ ADCSUM cut or 
$\epsilon_{\beta} \epsilon_{A_2}$. As can be seen from Table \ref{final_effic},
the final proton efficiency as determined for all of the 5 $\varepsilon$ points is 
fairly constant as expected since all data are taken at a fixed proton momentum. 
The particle identification efficiency, $\epsilon_{PID}$, for the right arm is taken as 
the product of the average value of $\epsilon_{\beta}$ and $\epsilon_{A_2}$ or 
$\epsilon_{PID} = <\epsilon_{\beta}><\epsilon_{A_2}>$ $=$ 0.9886. 
Table \ref{final_effic} summarizes the results. It can be seen from Table \ref{ra2_effic} 
that the pion contamination is $<$ 0.1\% and can be reduced by applying the $\beta$ cut. 
Such contamination can be reduced further by kinematics cuts and endcaps subtraction and it is taken 
to be negligible in the analysis. 
\begin{table}[!htbp]
\begin{center}
\begin{tabular}{||c|c|c|c|c|c||}
\hline \hline
Run   & Incident Energy& Pion          & Proton   \\
Number& (MeV)          & Contamination & Efficiency \\ 
      &                & (\%)          & ($\epsilon_{A_2}$) \\
\hline \hline
1252  & 2260.00        & 0.0410        & 0.9960     \\
1653  & 2844.71        & 0.0353        & 0.9950      \\
1730  & 4702.52        & 0.0500        & 0.9952       \\ 
1772  & 1912.94        & 0.0404        & 0.9960        \\
1823  & 3772.80        & 0.0397        & 0.9950         \\ 
\hline \hline
\end{tabular}
\caption[The pion contamination and proton efficiency as determined using the $A_{2}$ ADCSUM cut.]     
{The pion contamination and proton efficiency as determined using the $A_{2}$ ADCSUM cut.}
\label{ra2_effic}
\end{center}
\end{table}

\begin{figure}[!htbp]
\begin{center}
\epsfig{file=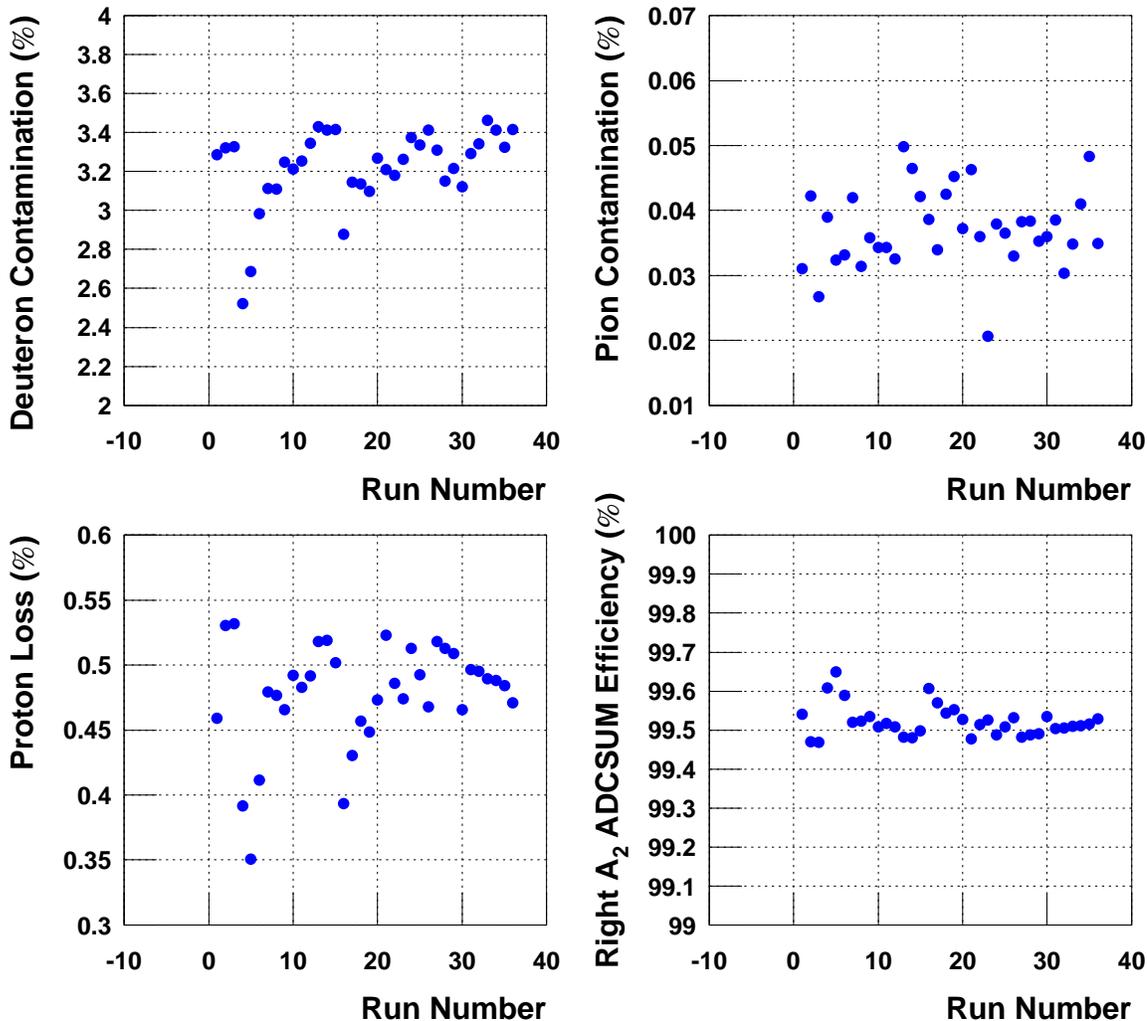,width=6in}
\end{center}
\caption[The right arm deuteron contamination, pion contamination, proton loss, and 
$\epsilon_{A_2}$ for sample runs from all kinematics.]
{The right arm deuteron contamination (top left), pion contamination (top right), 
proton loss (bottom left), and efficiency of the $A_2$ ADCSUM cut, $\epsilon_{A_2}$, 
(bottom right) for sample runs from all kinematics. The points are sorted according to the kinematics 
of the left arm, with three runs shown for each kinematics setting. Runs 1-15 correspond to 
$Q^2$ = 2.64 GeV$^2$, while 16-27 (28-36) correspond to 3.20 (4.10) GeV$^2$. For each $Q^2$ value, 
the points are sorted by $\varepsilon$ (low to high). See Table \ref{kinematics} for details.}
\label{fig:right_arm_a2sum_effic}
\end{figure}

Figure \ref{fig:right_arm_a2sum_effic} shows the deuteron contamination, pion contamination, 
proton loss, and the efficiency of the A$_2$ ADCSUM cut, $\epsilon_{A_2}$, for multiple runs 
from all the kinematics. Again, the results are plotted in order of increasing $\varepsilon$ and the 
efficiency of the $A_{2}$ ADCSUM cut is typically $\geq$ 99.4\% with an average random 
fluctuation of 0.05\% and without any significant $\varepsilon$ dependence. 
Similarly, and based on the same procedure used earlier with the $\beta$ cut efficiency, the scale 
offset uncertainty was determined by looking at the A$_2$ ADCSUM spectrum and determining the 
number of deuterons and pions which can be misidentified as protons. The scale uncertainty in the 
efficiency of the A$_2$ ADCSUM cut as determined at all the incident energies was fairly close
and an overall scale uncertainty of 0.25\% was assigned. Therefore, a scale uncertainty of
$\sim$ 0.30\% (adding in quadrature the scale uncertainty from $\epsilon_{\beta}$ and 
$\epsilon_{A_2}$) on the right arm particle identification efficiency $\epsilon_{PID}$ is assigned. 
\begin{table}[!htbp]
\begin{center}
\begin{tabular}{||c|c|c|c|c||}
\hline \hline
Run   & Energy          & Proton Efficiency  & Proton Efficiency & Final Proton\\
Number& (MeV)           & $\epsilon_{\beta}$ & $\epsilon_{A_2}$  &Efficiency $\epsilon_{\beta} \epsilon_{A_2}$\\
\hline \hline                                                                     
1252  & 2260.00          &0.9943              &      0.9960         &     0.9903          \\   
1653  & 2844.71          &0.9932              &      0.9950         &     0.9882           \\ 
1730  & 4702.52          &0.9924              &      0.9952         &     0.9876            \\
1772  & 1912.94          &0.9934              &      0.9960         &     0.9894             \\
1823  & 3772.80          &0.9925              &      0.9950         &     0.9875              \\
\hline \hline
      &                  &$<\epsilon_{\beta}>$ $=$ 0.9932 &$<\epsilon_{A_2}>$ $=$ 0.9954  & $\epsilon_{PID}$ $=$ 0.9886\\
\hline \hline
\end{tabular}
\caption[The final proton efficiency as determined from the efficiency of the $\beta$ and
$A_{2}$ ADCSUM cuts.]
{The final proton efficiency as determined from the efficiency of the $\beta$ and $A_{2}$ 
ADCSUM cuts.}
\label{final_effic}
\end{center}
\end{table}

\newpage
\subsection{The efficiency of the left arm $A_{1}$ Aerogel ($A_{1}$ ADCSUM) cut} \label{A1_eff}

Understanding the PID cuts and efficiencies for the left arm is not as straightforward 
as it is for the right arm.  In the right arm, we had a good
separation of particles using $\beta$, $dE/dx$, and the $A_{2}$ ADCSUM cuts. 
For the left arm, using the $A_{1}$ aerogel to separate the protons from pions works well, 
however, the $\beta$ cut cannot be used because at such high $Q^2$ protons and pions 
have almost the same value for $\beta$.

\begin{figure}[!htbp]
\begin{center}
\epsfig{file=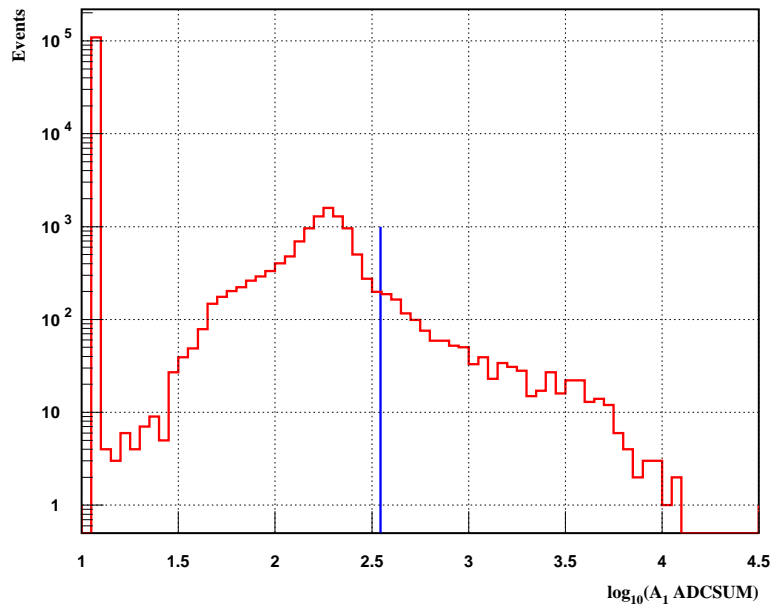,width=4in}
\end{center}
\caption[The $A_{1}$ ADCSUM spectrum for a pure proton sample from coincidence kinematics coin1.]
{The $A_{1}$ ADCSUM spectrum for a pure proton sample from coincidence kinematics coin1. 
The blue line is the PID cut used.} 
\label{fig:la1sum_protoneffic}
\end{figure}

The coincidence runs with protons in the left spectrometer and electrons in the right 
spectrometer were used to generate a pure proton spectrum.
First, all the runs in each of the three coincidence kinematics were added. A plot of  
log$_{10}(A_{1})$ for each coincidence kinematics was generated. For simplicity,
I will refer to log$_{10}(A_{1})$ by $A_{1}$ ADCSUM throughout this section.
The total number of events under the $A_{1}$ ADCSUM spectrum as
well as the number below and above $A_{1}$ ADCSUM = 2.544, which is the PID cut used 
for $A_{1}$ ADCSUM and is represented by the blue line in Figure \ref{fig:la1sum_protoneffic}, 
was determined for pure proton sample. A pure proton sample was generated by applying cuts on 
the elastic peak $\Delta P$ and/or y-coordinate of the extended target length $y_{tg}$. The results
were insensitive to the exact cuts applied. This allowed for determination of the fraction of 
protons (proton inefficiency) that leaked into $A_{1}$ ADCSUM $>$ 2.544 which was fairly constant 
regardless of the various cuts applied. The two $Q^2$ = 2.64 GeV$^2$ spectra yield consistent proton 
inefficiencies, (1.037$\pm$0.029)\% and (1.030$\pm$0.013)\%, while the $Q^2$ = 4.1 GeV$^2$ has 
a higher proton inefficiency of (1.940$\pm$0.054)\%. Figure \ref{fig:la1sum_protoneffic} shows the 
$A_{1}$ ADCSUM spectrum for kinematics coin1 (see Table \ref{kinematics}). 
We do not have any coincidence kinematics with $Q^2$ = 3.2 GeV$^2$ to extract clean proton sample. 
Instead, we take the $A_{1}$ ADCSUM spectrum for each of the kinematics at this $Q^2$ value, and get clean 
proton sample by applying kinematics cuts on the elastic peak $\Delta P$, -15.5$<\Delta P<$30.5 MeV,
and $y_{tg}$, -0.0044$<y_{tg}<$-0.001 m, to eliminate the endcaps contribution. 
The same procedure applied above was used and gave a value of (1.550$\pm$0.049)\% for the 
proton inefficiency. 

\begin{figure}[!htbp]
\begin{center}
\epsfig{file=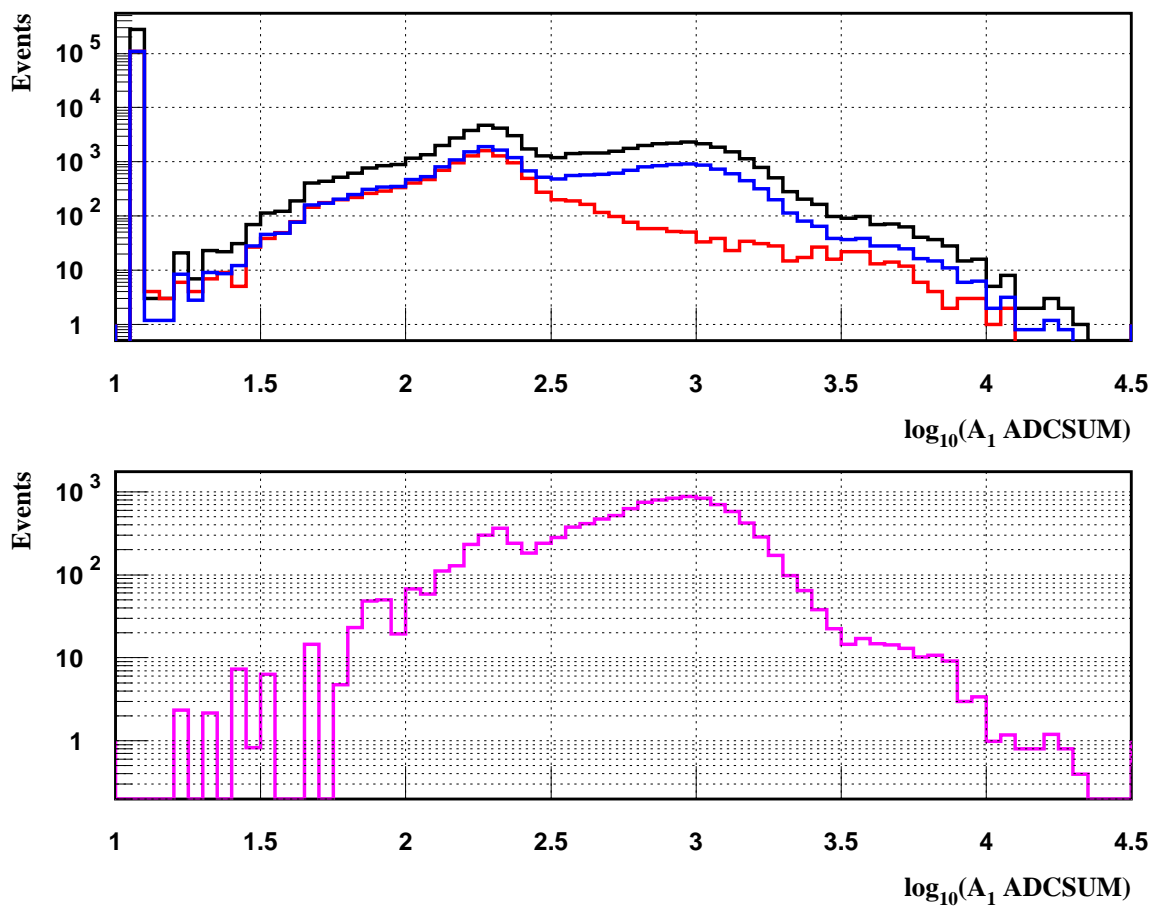,width=6in}
\end{center}
\caption[The $A_{1}$ ADCSUM spectra from singles kinematics $n$ (top black), coincidence kinematics 
coin1 (top red), scaled kinematics $n$ (top blue). The difference between the $A_{1}$ ADCSUM spectrum 
from the scaled kinematics $n$ and the coin1 spectrum (blue minus red) is the pions spectrum 
(bottom magenta).]
{The $A_{1}$ ADCSUM spectra from singles kinematics $n$ (top black), coincidence kinematics coin1 
(top red), scaled kinematics $n$ (top blue). The difference between the $A_{1}$ ADCSUM spectrum from
the scaled kinematics $n$ and the coin1 spectrum (blue minus red) is the pions spectrum (bottom magenta).}
\label{fig:pions_shape}
\end{figure}

We also need to determine the $\pi^+$ contamination. To do that, we need the number of $\pi^+$
and fraction of $\pi^+$ that are identified as protons ($\pi^+$ efficiency).
We compared the $A_{1}$ ADCSUM spectrum from trigger type $T_{3}$ events 
(LH$_2$ singles events, which include both protons and $\pi^+$) with that of the trigger type $T_{5}$ 
events (LH$_2$ coincidence events, which have only protons). The difference in the spectra should 
represent the $A_{1}$ ADCSUM spectrum for the background $\pi^+$. When we compare the two 
$A_{1}$ ADCSUM spectra, we scale the $A_{1}$ ADCSUM spectrum for the singles events so that the two peaks 
match. Figure \ref{fig:pions_shape} illustrates this procedure where the $A_{1}$ ADCSUM spectrum with trigger 
type $T_{3}$ events is compared with that of trigger type $T_{5}$ events.
The peak of the $A_{1}$ ADCSUM spectrum from singles was scaled to match that of coincidence. 
The difference between the $A_{1}$ ADCSUM spectrum from the scaled singles and the coincidence spectrum 
should represent the $A_{1}$ ADCSUM spectrum for pions. The total number of pions under the $A_{1}$ 
ADCSUM spectrum and the number of pions below and above $A_{1}$ ADCSUM = 2.544 are determined. 
The pion efficiency is defined as (number of pions below 
$A_{1}$ ADCSUM = 2.544)/(total number of pions). 

Having determined the pion efficiency and proton inefficiency, the pion contamination is then 
calculated for each kinematics based on the $A_{1}$ ADCSUM spectrum and without applying any
kinematics cuts or subtracting off the yield from the aluminum endcaps using the following procedure:
\begin{itemize}
\item {Total number of protons $\approx$ number of events  $<$ 2.544.}
\item {Total number of pions $\approx$ number of events $>$ 2.544 - (total number of protons)$\times$ 
(proton inefficiency).}
\item {Number of pions $<$ 2.544 = (pion efficiency)$\times$(total number of pions).}
\item{Pion contamination = (number of pions $<$ 2.544)/(total number of protons).}
\end{itemize}

\begin{figure}[!htbp]
\begin{center}
\epsfig{file=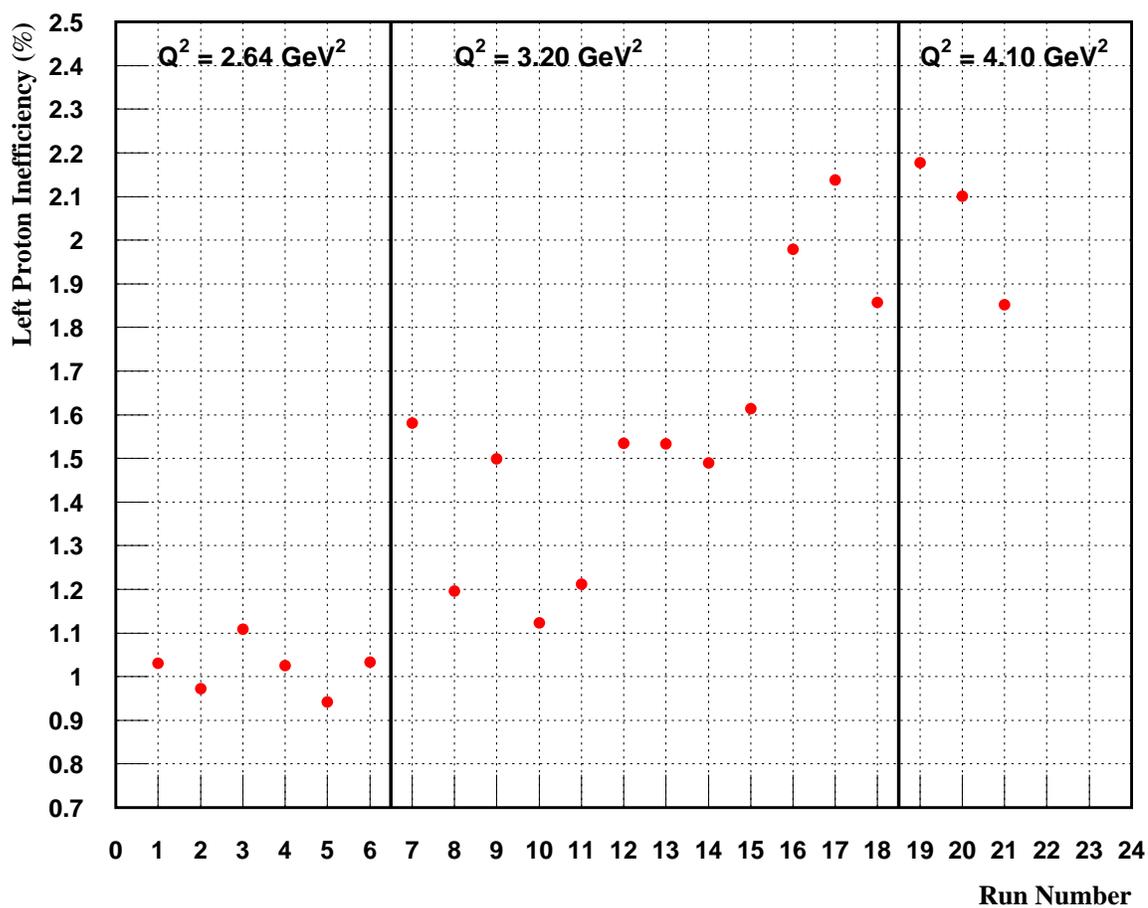,width=6in}
\end{center}
\caption[The left arm proton inefficiency at $Q^2$ = 2.64, 3.20, and 4.10 GeV$^2$ plotted
as a function of run number sorted in order of increasing $\varepsilon$.]
{The left arm proton inefficiency for sample runs. 
Runs 1-6 correspond to the two coincidence kinematics with $Q^2$ = 2.64 GeV$^2$,
7-18 correspond to singles kinematics with $Q^2$ = 3.20 GeV$^2$, and 19-21 correspond to coincidence 
kinematics with $Q^2$ = 4.10 GeV$^2$. For each $Q^2$ value, the points are sorted by $\varepsilon$ 
(low to high). See Table \ref{kinematics} for details. Note that for $Q^2$ = 3.20 GeV$^2$, we do not have a direct measurements, 
and so the scatter of the results shown in the plot is due to uncertainty in the procedure used rather than an actual variation 
in the proton inefficiency.}
\label{fig:left_a1sum_inefficiency}
\end{figure}

For the proton inefficiency we considered the average weight of (1.031$\pm$0.012)\%
for $Q^2$ = 2.64 GeV$^2$, while for $Q^2$ = 4.10 GeV$^2$, a value of 1.88\% was used. 
For $Q^2$ = 3.20 GeV$^2$, the 
average value of the proton inefficiency for $Q^2$ = 2.64 and 4.10 GeV$^2$ or 1.45\% was finally used. 
Figure \ref{fig:left_a1sum_inefficiency} shows the results for the proton inefficiency
as determined by applying cuts on the elastic peak $\Delta P$ and $y_{tg}$ for multiple runs selected at 
the three $Q^2$ values. For $Q^2$ = 2.64 GeV$^2$, the proton inefficiency seems to have, on the average,
a random fluctuation and scale offset uncertainties of $\leq$ 0.1\% each, and no significant 
$\varepsilon$ dependence. For $Q^2$ = 3.20 GeV$^2$, the proton inefficiency shows, on the average,
a random fluctuation and scale offset uncertainties of $\leq$ 0.1\% and (0.10-0.20)\% respectively, and 
does not show any significant $\varepsilon$ dependence based on the data centered around the 1.5\% proton 
inefficiency. Finally, for $Q^2$ = 4.10 GeV$^2$, the proton inefficiency shows a random 
fluctuation and scale offset uncertainties of $\leq$ 0.1\% and 0.10\% respectively, and does not
show any significant $\varepsilon$ dependence. Therefore, and mainly based on the results of the $Q^2$ = 3.20 GeV$^2$ point, 
a scale, random, and slope uncertainties of 0.20\%, 0.10\%, and 0.0\%, respectively, will be assigned for the $A_1$ PID 
efficiency at all three $Q^2$ points. 

To estimate the pion contamination, we also need the pion efficiency. We take the worst case estimate to give a conservative
limit on the pion contamination. The pion efficiency for $Q^2$ = 2.64 and 4.10 GeV$^2$ was found to be (4.230$\pm$1.32)\% or 
5.5\% as a worst case (defined as the maximum value) and (8.935$\pm$1.93)\% or 11\% as a worst case, respectively. 
For $Q^2$ = 3.20 GeV$^2$, pion efficiency value of 11\% was used as a worst case.
Table \ref{proton_ineffic} summarizes the results for the pion efficiency, proton inefficiency,
pion contamination at the three $Q^2$ values. The proton efficiency $\epsilon_{A_1}$, 
$\epsilon_{A_1}$ = 1.0 $-$ proton inefficiency, or the efficiency of the $A_{1}$ ADCSUM cut 
is also listed. The pion contamination is always $\leq$ 1.0\%. Applying kinematics cuts and subtracting 
the target endcaps reduces the pions contaminations as most of the pions are generated by 
bremsstrahlung scattering in the endcaps. Thus, pion contamination in the final results, after the dummy subtraction,
is always $\leq$ 0.1\%.
\begin{table}[!htbp]
\begin{center}
\begin{tabular}{||c|c|c|c|c|c||}
\hline \hline
Kinematics&$Q^2$          &Pion          &Proton         &Pion            &Proton \\
          &(GeV$^2$)      &Efficiency    &Inefficiency   &Contamination   &Efficiency \\ 
          &(\%)           &(\%)          &(\%)           &(\%)            &($\epsilon_{A_1}$)\\
\hline \hline
o         &2.64           &5.5           &1.02           &1.012E-02       &0.9898      \\  
a         &2.64           &5.5           &1.02           &1.590E-02       &0.9898       \\
i         &2.64           &5.5           &1.02           &1.585E-01       &0.9898        \\
q         &2.64           &5.5           &1.02           &2.230E-01       &0.9898         \\
l         &2.64           &5.5           &1.02           &2.155E-01       &0.9898          \\
\hline
b         &3.20           &11.0          &1.45           &9.153E-02       &0.9855            \\
j         &3.20           &11.0          &1.45           &3.100E-01       &0.9855             \\
p         &3.20           &11.0          &1.45           &6.216E-01       &0.9855              \\
m         &3.20           &11.0          &1.45           &6.098E-01       &0.9855               \\
\hline
k         &4.10           &11.0          &1.88           &7.630E-02       &0.9812                \\
r         &4.10           &11.0          &1.88           &7.944E-01       &0.9812                 \\
n         &4.10           &11.0          &1.88           &8.065E-01       &0.9812                  \\ 
\hline \hline
\end{tabular}
\caption[The pion efficiency, proton inefficiency, pion contamination, and proton efficiency 
as determined using the $A_{1}$ ADCSUM cut.]
{The pion efficiency as determined using the $A_{1}$ ADCSUM cut. In addition, the proton inefficiency, 
pion contamination, and proton efficiency are also listed.}
\label{proton_ineffic}
\end{center}
\end{table}

\newpage
\section{Proton Absorption} \label{proton_absorp}

Once the protons are struck, they have to travel through material
in the target, spectrometer, and the detector stack before they can 
cause a trigger in the scintillators. While traveling, some protons undergo nuclear 
interaction in these materials and hence get absorbed or scattered. 
Such loss due to the nuclear interaction is called nuclear absorption. 
In order to determine the proton absorption for each arm, we have to account for all the 
target materials the protons have to pass through on their way to the scintillators. 
Also the thickness \cite{alcorn04,schultethesis,lingyanzhuthesis,jones_priv,arrington_priv} and 
density of each absorber \cite{hagiwara02,eidelman04,schultethesis,lingyanzhuthesis}, and the mean 
free path between nuclear collision $\bar{\lambda}$ (effective absorption length) are needed 
in order to calculate the absorption. 

To determine $\bar{\lambda}$, the mean free path between nuclear collision (total interaction
length), $\lambda_{T}$, and the mean free path between inelastic interactions 
(inelastic interaction length), $\lambda_{I}$, must be known \cite{hagiwara02,eidelman04} . 
In the analysis of the E01-001 experiment, $\bar{\lambda}$ was determined using two different definitions. 
The first definition, $\bar{\lambda}$ is estimated as $2\lambda_{T}\lambda_{I}/(\lambda_{T}+\lambda_{I})$ 
assuming that half of the elastic and inelastic scattering contribute to the absorption. 
The second definition, $\bar{\lambda}$ is estimated as the average of two lengths or 
($\lambda_{T}+\lambda_{I}$)/2 since we cannot determine exactly the full contribution from the elastic 
scattering to the absorption. I will refer to $\bar{\lambda}$ estimated by the first(second) definition 
as $\lambda_{act}$($\lambda_{avg}$) where $act$ and $avg$ stand for actual and average. 

The ratio of $X/\bar{\lambda}$ is calculated for each absorber first using $\lambda_{act}$
and then $\lambda_{avg}$. Here $X$ is the product of the absorber's density and thickness. The ratios 
are then added, i.e, $\sum_{i=1}^n (X_{i}/\bar{\lambda_{i}})$ where $i$ runs over all absorbers $n$.  
The proton absorption is given by:
\begin{equation}
\mbox{proton absorption} = 1.0 - e^{-\sum_{i=1}^n (X_{i}/\bar{\lambda_{i}})}~, 
\end{equation}
where $e^{-\sum_{i=1}^n (X_{i}/\bar{\lambda_{i}})}$ defines the proton transmission. 
The final proton absorption used is the average value of the calculated proton absorption from 
the two definitions or 5.19\% for the right arm and 4.91\% for the left arm giving a proton 
absorption correction of $C_{Absorption}$ = 0.948 and 0.951 for the right arm and left arm, 
respectively. Tables \ref{r_proton_absorp} and \ref{l_proton_absorp} list the absorbers used and their 
properties for both arms. Uncertainty in the final proton absorption used rise due to our lack of precise 
knowledge of the actual thickness of each absorber used and the use of different cross sections in determining 
the mean free path lengths discussed above. Therefore, a scale offset of 1\% is assigned. 
Due to the fact that the scattered proton will have to travel different path length inside the LH$_2$ 
target at each kinematics, calculations of the proton absorption inside the LH$_2$ target at different 
scattering angles for both arms show an $\varepsilon$ dependence uncertainty of 0.03\% and 0.10\% for the
left and right arm, respectively. Random uncertainty of 0.0\% is assigned. 
\begin{table}[!htbp]
\begin{center}
\begin{tabular}{||c|c|c|c||}
\hline \hline
Target&Density&Thickness&X\\
      &(gm/cm$^3$)&(cm)&(gm/cm$^2$)\\
\hline \hline
LH$_2$ Target&0.708E-01&1.69934&0.1203\\
Al Target&2.70&0.164484E-01&0.444E-01\\
Al Chamber&2.70&0.33E-01&0.891E-01\\
Kapton&1.42&0.254E-01&0.361E-01\\
Titanium&4.54&0.10E-01&0.454E-01\\
Air (Spectrometer exist to VDC1)&0.121E-02&80.00&0.968E-01\\
Mylar (Wire Chamber)&1.39&0.12E-01&0.167E-01\\
Wire VDC (effective)&19.30&0.40E-04&0.772E-03\\
Ar/Ethan&0.107E-02&20.00&0.214E-01\\
S$_1$ Scintillator&1.032&0.507&0.523\\
A$_2$ Aerogel&0.22&5.00&1.100\\
AM Aerogel&0.10&9.00&0.900\\
Short Gas Cerenkov(CO$_2$)&0.1977E-02&100.00&0.197\\
S$_2$ Scintillator&1.032&0.52&0.536\\
Air (VDC1 to S$_2$ Scintillator)&0.121E-02&220.20&0.266\\
\hline \hline
$\lambda_T$&$\lambda_I$&X/$\lambda_{avg}$&X/$\lambda_{act}$ \\
(gm/cm$^2$)&(gm/cm$^2$)& &                                         \\
\hline \hline
43.30  &50.80&    0.2557E-02&0.2573E-02\\
70.60  &106.40&   0.5018E-03&0.5232E-03\\
70.60  &106.40&   0.1006E-02&0.1049E-02\\
60.30  &85.80&    0.4937E-03&0.5092E-03\\
79.90  &124.90&   0.4433E-03&0.4658E-03\\
62.00  &90.00&    0.1273E-02&0.1318E-02\\
62.50  &85.70&    0.2251E-03&0.2307E-03\\
110.30 &185.00&   0.5228E-05&0.5586E-05\\
68.572 &101.43&   0.2517E-03&0.2615E-03\\
58.50  &81.90&    0.7453E-02&0.7666E-02\\
66.30  &96.90&    0.1348E-01&0.1397E-01\\
66.30  &96.90&    0.1102E-01&0.1143E-01\\
62.40  &89.70&    0.2599E-02&0.2686E-02\\
58.50  &81.90&    0.7644E-02&0.7862E-02\\
62.00  &90.00&    0.3505E-02&0.3628E-02\\
\hline \hline 
$\sum_{i=1}^n$(X/$\bar{\lambda}$)& - &0.0525 & 0.0542 \\  
\hline 
Transmission  & - & 0.9488& 0.9472\\       
\hline \hline 
Absorption (\%)      &   -    & 5.114  & 5.275  \\ 
\hline \hline
\end{tabular}
\caption[The absorbers used and their properties for the right arm spectrometer.]
{The absorbers used and their properties for the right arm spectrometer.}
\label{r_proton_absorp}
\end{center}
\end{table}
\begin{table}[!htbp]
\begin{center}
\begin{tabular}{||c|c|c|c||}
\hline \hline
Target&Density&Thickness&X\\
      &(gm/cm$^3$)&(cm)&(gm/cm$^2$)\\
\hline \hline
LH$_2$ Target&0.708E-01&1.94808&0.1379              \\
Al Target&2.70&0.147682E-01&0.3987E-01               \\   
Al Chamber&2.70&0.33E-01&0.891E-01                     \\
Kapton&1.42&0.254E-01&0.360E-01                        \\
Titanium&4.54&0.10E-01&0.454E-01                          \\
Air (Spectrometer exist to VDC1)&0.121E-02&80.00&0.968E-01 \\
Mylar (Wire Chamber)&1.39&0.12E-01&0.166E-01               \\
Wire VDC (effective)&19.30&0.40E-04&0.772E-03                \\ 
Ar/Ethan&0.107E-02&20.00&0.214E-01                           \\
S$_0$ Scintillator&1.032&1.00&1.032                             \\
S$_1$ Scintillator&1.032&0.507&0.523                                \\
A$_1$ Aerogel&0.60E-01&9.00&0.540                                        \\ 
Short Gas Cerenkov (CO$_2$)&0.1977E-02&100.00&0.197                     \\
S$_2$ Scintillator&1.032&0.52&0.536                                     \\
Air (VDC1 to S$_2$ Scintillator)&0.121E-02&220.20&0.266                 \\
\hline \hline
$\lambda_T$&$\lambda_I$&X/$\lambda_{avg}$&X/$\lambda_{act}$ \\
(gm/cm$^2$)&(gm/cm$^2$)&  &                                           \\
\hline \hline
43.30 &  50.80 &   0.2931E-02&0.2950E-02                                 \\
70.60 &  106.40&   0.4505E-03&0.4697E-03                                 \\
70.60 &  106.40&   0.1006E-02&0.1049E-02                                  \\
60.30 &  85.80 &   0.4937E-03&0.5092E-03                                    \\
79.90 &  124.90&   0.4433E-03&0.465E-03                                   \\
62.00 &  90.00 &   0.1273E-02&0.1318E-02                                      \\
62.50 &  85.70 &   0.2251E-03&0.2307E-03                                       \\
110.30&  185.00&   0.5228E-05&0.5586E-05                                     \\
68.572&  101.43&   0.2517E-03&0.2615E-03                                      \\ 
58.50 &  81.90 &   0.1470E-01&0.1512E-01                                          \\
58.50 &  81.90 &   0.7453E-02&0.7666E-02                                          \\
66.30 &  96.90 &   0.6617E-02&0.6858E-02                                         \\
62.40 &  89.70 &   0.2599E-02&0.2686E-02                                            \\
58.50 &  81.90 &   0.7644E-02&0.7862E-02                                            \\
62.00 &  90.00 &   0.3505E-02&0.3628E-02                                           \\
\hline \hline 
$\sum_{i=1}^n$(X/$\bar{\lambda}$)& - & 0.0496& 0.0510\\  
\hline 
Transmission  & - & 0.9516& 0.9502\\       
\hline \hline 
Absorption (\%)      &   -    & 4.839& 4.980\\ 
\hline \hline
\end{tabular}
\caption[The absorbers used and their properties for the left arm spectrometer.]
{The absorbers used and their properties for the left arm spectrometer.}
\label{l_proton_absorp}
\end{center}
\end{table}

\newpage
\section{Target Length Correction} \label{target_leng_corr}

\begin{figure}[!htbp]
\begin{center}
\epsfig{file=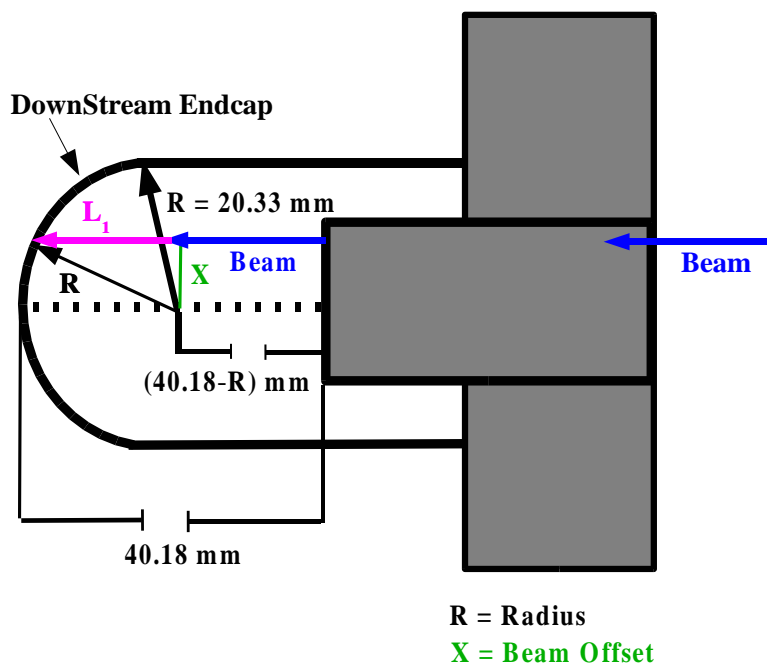,width=4in}
\end{center}
\caption[Geometry of the 4 cm LH$_2$ cell.]
{Geometry of the 4 cm LH$_2$ cell. The electron beam (blue arrow) is going from
the right to the left side with an offset X (green line) from the central axis of the cell 
(black dashed line).}
\label{fig:target_offset}
\end{figure}

Figure \ref{fig:target_offset} shows the geometry of the 4 cm LH$_2$ cell used
in the experiment. The cell wall is made of 0.14 mm Al. 
The cell has two endcaps. The upstream endcap (not shown) is the beam entrance window 
and it is made of Al 7075 T6 with 0.142 mm thickness, while the downstream endcap  
window is uniformly machined Al in the shape of a hemisphere and has a thickness of 
0.15 mm and radius R = 20.33 mm. The length of the central axis of the cell is 40.18 mm 
(black dashed line) \cite{meekins_priv,dmeekins,xzheng_priv,arrington_priv}. 

The 4 cm dummy target used is made of Al 6061 T6 with density 2.85 g/cm$^3$. The thickness 
is 0.2052 g/cm$^2$ for the upstream foil and 0.2062 g/cm$^2$ for the downstream one. The distance between
the two foils is (40$\pm$0.13) mm \cite{meekins_priv,dmeekins,xzheng_priv,arrington_priv}.     

If the beam (the blue arrow going from left to right) is displaced a distance X (green line) from the 
central axis, then the length that the beam will travel in the target is
L = L$_{1}$ + (40.18 - R) mm where L$_{1}<$ R and is shown as a magenta arrow. 
This length is effectively the new target length.
A survey showed that the target was displaced down (2.0$\pm$0.5) mm relative to the center of 
the beam. Furthermore, the beam was rastered to produce an $\sim$ 2 mm $\times$ 2 mm spot.
Therefore, the average beam path length in the target (new target length) is then L = $<$L$_{1}$$>$ + (40.18 - R) where $<$L$_{1}$$>$ is given by:
\begin{equation}
<L_{1}>~= \frac{\int_{X_{min}}^{X_{max}} \sqrt{R^{2} - X^{2}}~dX}{\int_{X_{min}}^{X_{max}}dX}~,
\end{equation}
and X$_{min}$ and X$_{max}$ are the minimum and maximum beam displacement. 

The target length correction needed is defined as $\delta$L = (40.18-L)$\times$100\%/40.18.
Table \ref{target_beam_offset} lists the results of the calculations. The nominal value of
$\delta$L, $\delta$L = 0.266\%, gives a target length correction to $Q_{eff}$ of $C_{TL}$ = 0.9973,
which changes by $\sim$ 0.12\% (average difference from the nominal value) for a 0.50 mm beam shift.
Therefore, a scale uncertainty in the target length of 0.12\% is assigned for both arms.

The beam position measurements are discussed in detail in section \ref{beam_position}. 
Figure \ref{fig:beam_position} shows the x and y coordinates of the beam at the target. 
The beam was well focused on the target with an average beam drift of 0.30 mm. This drift in the 
beam position translates into an uncertainty in the effective target length and introduces a correction to 
Q$_{eff}$. Table \ref{target_beam_drift} lists the results. The target length correction $\delta$L 
changes by $\sim$ 0.074\% (average difference from the nominal value) for a 0.30 mm beam drift. 
Therefore, a random uncertainty in the target length of 0.074\% is assigned for both arms. 
\begin{table}[!htbp]
\begin{center}
\begin{tabular}{||c|c|c|c|c||}
\hline \hline
X$_{min}$ & X$_{max}$ & $<$L$_{1}$$>$ & L       & $\delta$L   \\
(mm)      & (mm)      & (mm)          & (mm)    & (\%)         \\ 
\hline \hline
0.5       & 2.5       & 20.2663       & 40.1163 & 0.158540       \\  
1.0       & 3.0       & 20.2231       & 40.0731 & 0.266053        \\  
1.5       & 3.5       & 20.1673       & 40.0173 & 0.404930         \\  
\hline \hline
Average   & -         & -             &  -      & 0.27651           \\
\hline \hline
\end{tabular}
\caption[The target length correction $\delta$L as determined for several beam offset ranges.]
{The target length correction $\delta$L as determined for several beam offset ranges.} 
\label{target_beam_offset}
\end{center}
\end{table}
\begin{table}[!htbp]
\begin{center}
\begin{tabular}{||c|c|c|c|c||}
\hline \hline
X$_{min}$ & X$_{max}$ & $<$L$_{1}$$>$ & L       & $\delta$L   \\
(mm)      & (mm)      & (mm)          & (mm)    & (\%)         \\ 
\hline \hline
0.7       & 2.7       & 20.2505       & 40.1005 & 0.197860       \\  
1.0       & 3.0       & 20.2231       & 40.0731 & 0.266053        \\  
1.3       & 3.3       & 20.1911       & 40.0411 & 0.345694         \\  
\hline \hline
Average   & -         & -             &  -      & 0.26987          \\
\hline \hline
\end{tabular}
\caption[The target length correction $\delta$L as determined for several beam drift ranges.]
{The target length correction $\delta$L as determined for several beam drift ranges.} 
\label{target_beam_drift}
\end{center}
\end{table}

\newpage
\section{Computer and Electronics Deadtime} \label{comp_elec_deadtime}

The computer deadtime is the portion of the time that the DAQ system is unable to
record an event because it is busy recording another event. Scalers are used to count 
the total number of triggers generated for each event or trigger type. By knowing the 
number of events recored by the DAQ system, $N_{DAQ}^i$, and the total number of 
events fed to the DAQ, $N_{Total}^i$, as recorded by scalers for a particular event 
type ($T_{i}$ where $i$ = 1, $\cdots$, 5), one can account for the missing events 
and determine the computer livetime $CLT = ps_{i}N_{DAQ}/N_{Total}$. 
The computer deadtime is then defined as $\eta_{CDT}$ = 1.0 - $CLT$ where $ps_{i}$ 
is the prescale factor for that event type. 

The electronics deadtime is the portion of the time during which the triggers are missed 
because the hardware is busy when a second event comes in. When an event causes a trigger, 
the logic gate in the trigger is activated and the TDC signal stays high
for a fixed time $\tau$, which is the gate time width of the logic signal and 
has a value of (100$\pm$20) nsec. If a second event tries to trigger and activate the gate 
within that time window $\tau$, the second event will be ignored. If the event rate 
is $R$, then the probability of having $n$ counts in a time $t$ is given 
by the Poisson distribution $P(n) = (Rt)^{n}e^{-Rt}/n!$, and the probability 
distribution for the time between the events is $P(t) = Re^{-Rt}$. For small electronics 
deadtime, the fraction of measured events or the electronics livetime $ELT$ 
is equal to the probability that the time between events will be greater than $\tau$ or:
\begin{equation}
ELT~= \frac{N_{Measured}}{N_{Total}}~= \int_{\tau}^{\infty} Re^{-Rt}~dt~ = e^{-R\tau}~,
\end{equation}
and the electronics deadtime is then defined as $\eta_{EDT}$ = 1.0 - $ELT$.

The raw event rates were $\leq$ 1KHz for the left arm and $\leq$ 30KHz for the right arm, respectively,
but the right arm was heavily prescaled. 
The computer deadtime is in the range of (9$-$20)\% for the left arm and (2$-$16)\% for the right arm 
with an average of $\sim$ 10\% correction on both arms. The computer deadtime is well measured and known to within 1.0\%. 
Therefore a 1.0\% uncertainty in a 10\% correction yields a scale uncertainty of 0.10\% for both arms. Similarly, we assign 
a slope uncertainty of 0.10\% for the left arm, and $\sim$ 1.0\% (0.10\%/0.07) for the right arm when we consider the 
$\Delta \varepsilon$ range of 0.07. A random uncertainty in the computer deadtime of 0.0\% is assigned for both arms.

The electronic deadtime is in the range of (0.001$-$0.008)\% for the left arm with 
a random uncertainty, determined based on the 20\% uncertainty in $\tau$, of 0.002\% (set to 0.0\%),
and (0.03$-$0.30)\% for the right arm with a maximum random uncertainty of 0.06\%. If we take the 
average random uncertainty of 0.04\% for the right arm and consider the $\Delta \varepsilon$ range of 0.07,
we estimate the slope uncertainty to be 0.04\%/0.07 $\approx$ 0.50\%. The slope uncertainty is 0.0\% for the left arm.  
A 0.0\% scale uncertainty in the electronic deadtime is assigned for both arms.  

\section{Beam Current Monitor Correction} \label{BCM_correct}

The beam current measurement was described in detail in section \ref{beam_current}. 
The basic technique of calibrating the beam current monitors is well documented 
\cite{alcorn04,xzhengbcm,mjonesbcm}. 
The idea is to normalize the BCMs cavities to the OLO2 cavity (Unser monitor) at the injector. 
During a calibration run, the beam was interrupted using a Faraday cup inserted after
the OLO2 cavity so that the zeros of the cavities and the Unser could be determined.
The beam current was first set to 80 $\micro$A for one minute and then the Faraday cup 
was inserted for another minute. The beam current was then stepped down to 70 $\micro$A and the 
Faraday cup was inserted again for another minute. This procedure was done down to a beam current of 
10 $\micro$A and then reversed up to beam current of 80 $\micro$A. 
     
While the beam was stepped down through various currents, the EPICS information
such as the values of the upstream and downstream BCM voltage, Unser current, injector 
Faraday cup current, and the OLO2 current were saved to a file. The EPICS BCM constants 
were determined from these values as:
\begin{equation}
\mbox{constant} = \frac{\mbox{$<I_{OLO2}>$}}{\mbox{$<V_{cavity}>$ - zero offset}}~,
\end{equation}
where $<I_{OLO2}>$ and $<V_{cavity}>$ are the average value for the OLO2 current and cavity voltage, 
respectively. Table \ref{epics_Bbcm_const} summarizes the results. The zero offsets for the cavities 
and Unser were determined from the beam off periods when the Faraday cup was inserted and the 
calibration constants and the zero offsets for converting the voltage-to-frequency (V-to-F) scalers to 
charge were also determined to give an average current of:
\begin{equation}
\mbox{Average current}  = \frac{\frac{\mbox{Scaler}}{\mbox{time}} - \mbox{zero offset}}{\mbox{constant}}~.
\end{equation}
The values for V-to-F zero offsets and constants are listed in Table \ref{vtof_const_offset}. 
Based on the analysis of all of the calibration runs, the gain and offset drifts were stable to within
0.1\% and 0.01\%, respectively. The effect of the offset drift is negligible, so a random uncertainty 
in the charge measurement of 0.1\% is assigned for both arms. It is reported that \cite{alcorn04} the BCMs 
with the Unser are calibrated at the same time as the charge monitors every 2-3 months, and the results are 
stable to within $\pm$0.5\%. Therefore, a scale uncertainty in the charge measurement of 0.5\% is assigned 
for both arms. 
\begin{table}[!htbp]
\begin{center}
\begin{tabular}{||c|c||}
\hline \hline
Constant Type& Constant Value \\
\hline \hline
EPICS Upstream cavity~~~ & 77.44 $\pm$ 0.04 \\ 
EPICS Downstream cavity& 79.08 $\pm$ 0.04 \\
\hline \hline
\end{tabular}
\caption[The EPICS BCM calibration constants.]
{The EPICS BCM calibration constants.}
\label{epics_Bbcm_const}
\end{center}
\end{table}
\begin{table}[!htbp]
\begin{center}
\begin{tabular}{||c|c|c||}
\hline \hline
Constant Type & Constant Value & Constant Zero Offsets \\
\hline \hline
V-to-F U1x  & 1331.2 $\pm$ 0.7   & 152.78   \\
V-to-F U3x  & 4094.8 $\pm$ 2.1   & 163.88   \\
V-to-F U10x & ~12451.7 $\pm$ 9.6~~ & 360.37    \\
V-to-F D1x  & 1353.4 $\pm$ 0.7   &~~34.63   \\
V-to-F D3x  & 4190.0 $\pm$ 2.2   & 110.01     \\
V-to-F D10x & ~~~13187.7 $\pm$ 10.2~~ & 307.91     \\
\hline \hline
\end{tabular}
\caption[The V-to-F calibrations constants and offsets.]
{The V-to-F calibrations constants and offsets.}
\label{vtof_const_offset}
\end{center}
\end{table}

\section{Target Boiling Correction} \label{target_boiling}

The LH$_{2}$ target density $\rho_{0}$ can decrease due to boiling caused by 
the electron beam current. In order to study the boiling effect in the LH$_{2}$,
runs were taken at several beam current values using LH$_{2}$ and carbon targets. 
We define the normalized yield $Y$ as the total number of events $N$ detected in a chosen acceptance 
region by applying kinematics and acceptance cuts and then normalized to the effective beam charge 
$Q_{eff}$ or:
\begin{equation} \label{yield_boiling1}
Y = \frac{N}{Q_{eff}}~.
\end{equation}

The normalized yield as a function of current showed a linearly decreasing
relationship. The deviation of the normalized yield from 1.0 over the current 
range represents the instability of the target density due to target boiling effect. 
The target density $\rho$ was parameterized as a function of current $I$ as:
\begin{equation} \label{yield_boiling2}
\rho(I) = \rho_{0}(1.0 - B I)~,
\end{equation}
where $\rho_{0}$ = $\rho(I = 0)$ and $B$ is the slope giving a target boiling correction 
$C_{TB} = \rho(I)/\rho_{0}$ as listed in Table \ref{target_boiling_corr} for the different kinematics.

\begin{table}[!htbp]
\begin{center}
\begin{tabular}{||c|c|c|c||}
\hline \hline
Kinematics & $I(\micro$A) &$C_{TB}$ $\pm$ $\delta C_{TB}$&$\delta C_{TB}(relative)$ \%\\
\hline \hline
a and run$\leq$ 1269 & 30  &  0.9925 $\pm$ 0.0045 &$\pm$ 0.60\\
a and run$>$  1269   & 50  &  0.9875 $\pm$ 0.0075 &$\pm$ 0.30\\
b                    & 50  &  0.9875 $\pm$ 0.0075 &$\pm$ 0.30\\
i-r                  & 70  &  0.9825 $\pm$ 0.0105 &$\pm$ 0.00\\
\hline \hline
\end{tabular}
\caption[The target boiling correction $C_{TB}$ as determined for different kinematics.]  
{The target boiling correction $C_{TB}$ and its uncertainty $\delta C_{TB}$ as determined for different 
kinematics. In determining $\delta C_{TB} = I\delta B$, a $\delta B$ value of 1.5\%/100$\micro$A is used. 
See text for details. Note that $\delta C_{TB}(relative)$ is the uncertainty relative to data at 
70$\micro$A.}
\label{target_boiling_corr}
\end{center}
\end{table}

\begin{figure}[!htbp]
\begin{center}
\epsfig{file=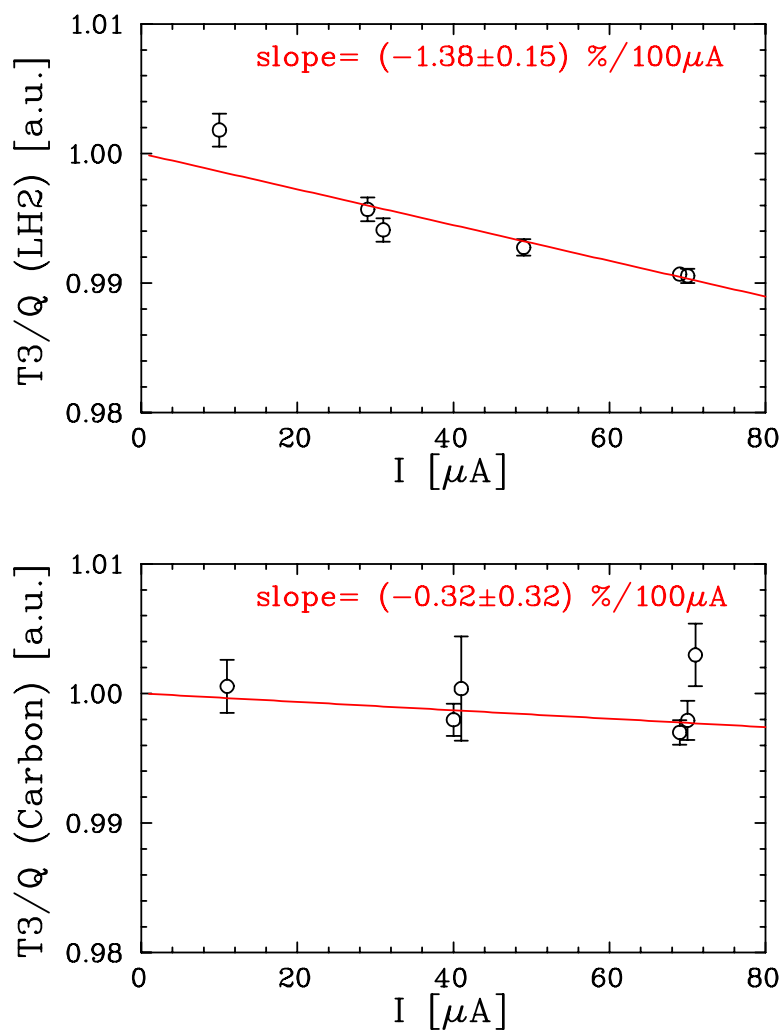,width=4in}
\end{center}
\caption[Target boiling studies. The left arm normalized yield for LH$_2$ and carbon targets 
as a function of current.]
{Target boiling studies. The left arm normalized yield for LH$_2$ target (top) and carbon 
target (bottom) as a function of current.}
\label{fig:left_boiling}
\end{figure}

In the boiling studies, we analyze the data in two different methods. First, we plot the events/charge
from scalers corrected for current-independent rate (rate of cosmic ray triggers) as a function 
of current. Figure \ref{fig:left_boiling} shows the results. For carbon, the slope is consistent with zero 
(0.32 $\pm$ 0.32)\%/100$\micro$A, as it should be.
For LH$_2$, the slope is (1.38 $\pm$ 0.15)\%/100$\micro$A. However, there is a slight 
nonlinearity at lower current, possibly due to nonlinearity in the BCM calibration at very low current
or uncertainty in the correction for cosmic trigger ray, so this may be a slight overestimate. 
This may also be a slight underestimate, since the thickness of the endcaps does not depend on 
current. 

In the second method, we plot the yield as defined by equation (\ref{yield_boiling1}) as a function
of current where $N$ is the number of the T$_3$ type tracked events. For carbon, the check is not of 
good quality due to poor statistics. For LH$_2$, the results yield a slope of 
(1.12 $\pm$ 0.65)\%/100$\micro$A, but the statistics are low and so the uncertainty is large. 
In addition, the results are fairly cut-dependent, yielding slopes up to 3\%, but with large
uncertainties.

Clearly, neither test is ideal and both give lower corrections than observed by previous experiments. 
The scalers analysis is sensitive to the endcaps
contribution and the current-independent rate. The tracked analysis is cleaner, but has low statistics.  
So we take a correction of 2.5\%/100$\micro$A, but apply a large uncertainty (2.5$\pm$1.5)\%/100$\micro$A
or (1.75$\pm$1.05)\%/70$\micro$A. Therefore a scale uncertainty of 1.05\% in the target boiling correction is assigned. 
One thing to mention is that this effect cancels out if we take the ratio of the left to the right arm cross sections, 
and even for the left arm alone, it is mostly a normalization uncertainty since most of the kinematics were run at
the same current and at a fixed $Q^2$ the proton corrections do not vary much with $\varepsilon$.

\begin{figure}[!htbp]
\begin{center}
\epsfig{file=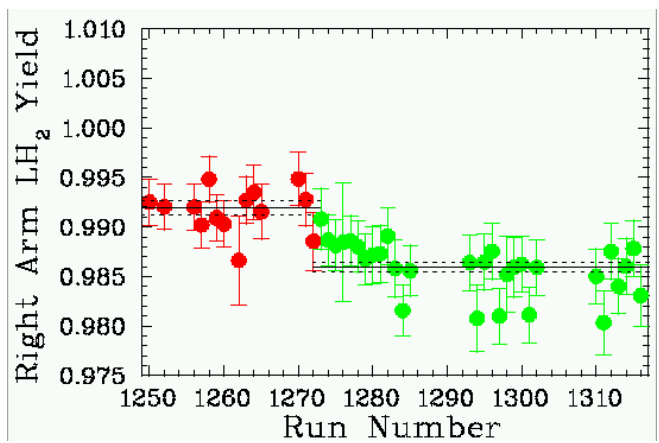,width=3.4in}
\end{center}
\caption[Target boiling studies. The right arm LH$_2$ normalized yield for kinematics $a$ and $b$
as a function of run number before applying the target boiling correction.]
{The right arm LH$_2$ yield before applying the target boiling correction for all the runs from kinematics $a$ (1250-1285) 
and $b$ (1293-1316) normalized to the yield of run 1250. The yield is shown in red(green) for runs taken at $I$ = 30(50)$\micro$A. 
Runs 1272-1285 from kinematics $a$ are taken at $I$ = 50$\micro$A and their yield is shown in green as well. The solid black line is 
the average normalized yield for that current setting. The dashed black lines are the average uncertainty in the normalized yield.}
\label{fig:rightkinab_noboil}
\end{figure}

\begin{figure}[!htbp]
\begin{center}
\epsfig{file=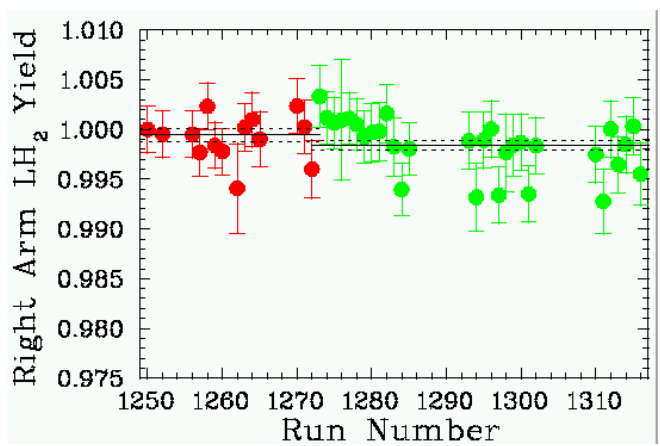,width=3.4in}
\end{center}
\caption[Target boiling studies. The right arm LH$_2$ normalized yield for kinematics $a$ and $b$
as a function of run number after applying the target boiling correction.]
{The right arm LH$_2$ yield after applying the target boiling correction for all the runs from kinematics $a$ (1250-1285) 
and $b$ (1293-1316) normalized to the yield of run 1250. The yield is shown in red(green) for runs taken at $I$ = 30(50)$\micro$A. 
Runs 1272-1285 from kinematics $a$ are taken at $I$ = 50$\micro$A and their yield is shown in green as well. The solid black line is 
the average normalized yield for that current setting. The dashed black lines are the average uncertainty in the normalized yield.}
\label{fig:rightkinab_boil}
\end{figure}

We check this correction by looking at kinematics a and b, where part of the data is taken at 30$\micro$A 
and part at 50$\micro$A. The left arm has a relatively low rate (making it more sensitive to cosmic ray rates) 
and has a larger contribution from the aluminum endcaps, and so we focus on the right arm to check the correction. 
Also, since we mainly want this correction for the left arm, it is good to have an independent check of the boiling.
For the right arm data, we see a clear difference between the yield at 30 and 50$\micro$A if we do not apply the boiling 
correction as can be seen in Figure \ref{fig:rightkinab_noboil}. The yield differs by roughly (0.6 $\pm$ 0.1)\% for a 20$\micro$A 
difference, yielding a boiling correction of (3.0 $\pm$ 0.5)\%/100$\micro$A. This is consistent with the (2.5$\pm$1.5)\%/100$\micro$A,
so we stick with the more conservative number and use the consistency at 30 and 50$\micro$A as a test of the correction. 
Figure \ref{fig:rightkinab_boil} shows the yield after applying the boiling correction.

Since most of the data are taken at 70$\micro$A, we want the uncertainty relative to the data at 70$\micro$A. 
So we set the random uncertainty for all kinematics to 0.0\% except that of kinematics a and b. We estimate the random 
uncertainty for kinematics a and b to be 0.45\% or 1.5\%(70$\micro$A - 40$\micro$A)/100$\micro$A assuming an average current 
of 40$\micro$A and 0.30\% or 1.5\%(20$\micro$A)/100$\micro$A, respectively. 
Furthermore, the beam was rastered to produce an $\sim$ 2 mm $\times$ 2 mm spot with uncertainty in the 
raster size of $< \pm$0.2 mm which translates into a $\sim$ 0.1\% random uncertainty in the target boiling. 


\section{Summary of Systematic Uncertainties}
Tables \ref{r_systematic_summary} and \ref{l_systematic_summary} summarize the systematic 
uncertainties for the right and left arm, respectively.
The systematic uncertainties are discussed in the sections listed in these tables.
The systematic uncertainty in each listed source is broken down into three types: 
scale, random, and slope. The contribution of each uncertainty type from all sources is 
then added in quadrature to form the total uncertainty in $\sigma_R$ for that uncertainty type. 
Only the range on total $\delta_{Random}$ is given in these tables, while the actual value of 
total $\delta_{Random}$ in $\sigma_R$ at each $\varepsilon$ point is listed in Tables \ref{rsigma_R_data} 
and \ref{lsigma_R_data}. It must be mentioned that the scale uncertainty in the pion subtraction for the left arm is 0.20\% 
for all kinematics except those of $Q^2$ = 4.10 GeV$^2$ which is 0.40\%. This results in a $\delta_{Slope} = $ 0.539\% for all 
kinematics with $Q^2$ = 2.64 and 3.20 GeV$^2$, and $\delta_{Slope} = $ 0.641\% for all kinematics with $Q^2$ = 4.10 GeV$^2$.
In addition, $\delta_{Slope}$ for $Q^2$ = 4.10 GeV$^2$ is then scaled up by 25\% to account for the $\Delta \varepsilon$ range
difference among the three $Q^2$ points which results in a $\delta_{Slope} = $ 0.801\%. See Table \ref{l_systematic_summary} for details.
\begin{table}[!htbp]
\begin{center}
\begin{tabular}{||c|c|c|c|c||}
\hline \hline
Source                                   &Section                 &Right         &Right          &Right\\  
                                         &                        &Scale         &Random         &Slope\\
                                         &                        &(\%)          &(\%)           &(\%)   \\
\hline \hline
BCM Calibration($^{\star}$)              &\ref{BCM_correct}       &0.50          &\textbf{0.10}  &0.00\\
Target Boiling at 70$\micro$A($^{\star}$)&\ref{target_boiling}    &\textbf{1.05} &0.30-0.45($\dagger a$)  &0.00\\
Raster Size($^{\star}$)                  &\ref{target_boiling}    &0.00          &\textbf{0.10}  &0.00\\
Target Length($^{\star}$)                &\ref{target_leng_corr}  &0.12          &0.07           &0.00 \\
\hline \hline 
Electronic Deadtime                      &\ref{comp_elec_deadtime}&0.00          &0.04           &\textbf{0.50}\\
Computer Deadtime                        &\ref{comp_elec_deadtime}&0.10          &0.00           &\textbf{1.00}\\
VDC Zero-Track Inefficiency              &\ref{vdcs_track_eff}    &0.00          &0.00           &0.00\\
VDC Multiple-Track Inefficiency          &\ref{vdcs_track_eff}    &0.10          &0.00           &\textbf{0.70}\\
VDC Hardware Cuts                        &\ref{vdcs_hardware_eff} &0.50          &\textbf{0.10}  &0.00\\
Scintillator Efficiency                  &\ref{scint_effic}       &0.10          &0.05           &0.00\\
PID Efficiency ($\beta$)                 &\ref{beta_eff}          &0.15          &0.05           &0.00\\ 
PID Efficiency ($A_2$)                   &\ref{A2_eff}            &0.25          &0.05           &0.00\\ 
PID Efficiency ($A_1$)                   &\ref{A1_eff}            &0.00          &0.00           &0.00\\  
Pion Contamination                       &\ref{pid_cuts}          &0.00          &0.00           &0.00\\
Proton Absorption                        &\ref{proton_absorp}     &\textbf{1.00} &0.00           &0.10\\ 
Solid Angle Cut                          &\ref{e01001_optics}     &\textbf{2.06} &0.00           &0.00\\
Pion Subtraction                         &\ref{extract_sigma}     &0.00          &0.05           &0.00\\
Dummy Subtraction                        &\ref{extract_sigma}     &0.50          &0.00           &0.40\\
$\Delta P$ Cut-Dependence                &\ref{extract_sigma}     &0.50          &\textbf{0.14}  &\textbf{2.00}\\
$\delta$ Cut-Dependence                  &\ref{extract_sigma}     &0.30          &\textbf{0.30}  &\textbf{4.30}\\
0.18 mrad Angle Offset                   &\ref{spect_mispoint}    &0.20          &0.02           &\textbf{0.67}\\
0.10 mrad Angle Fluctuations             &\ref{spect_mispoint}    &0.00          &0.097-0.125($\dagger b$) &0.00\\
0.03\% Beam Energy                       &\ref{beam_energy_measurements}&0.03    &0.01           &0.29\\
0.02\% Beam Energy Fluctuations          &\ref{beam_energy_measurements}&0.00    &0.015-0.029($\dagger c$) &0.00\\
Radiative Corrections                    &\ref{rad_corrections}         &\textbf{1.00} &\textbf{0.20}  &\textbf{2.00}\\
\hline \hline 
 & &$\delta_{Scale}$&$\delta_{Random}$&$\delta_{Slope}$\\
\hline
Total (\%)                    & &2.93  &0.454-0.640 &5.38    \\
\hline \hline 
\end{tabular}
\caption[Summay of the systematic uncertainties for the right arm.]
{Summay of the systematic uncertainties for the right arm. Numbers in bold show the biggest 
contribution to that uncertainty type. Sources marked with ($^{\star}$) contribute to the uncertainty 
in the luminosity monitor. For ($\dagger a$), the value is 0\% for all kinematics except $a$ and $b$ which is
0.45\% and 0.30\%, respectively. For ($\dagger b$), the value is applied on kinematics-by-kinematics basis as shown in Figure 
\ref{fig:right_proton_sigma0.10mrad_epsilon} and it ranges from 0.097\%-0.125\%. For ($\dagger c$), the value is applied on kinematics-by-kinematics basis as shown in Figure \ref{fig:right_proton_sigma0.02e0_epsilon} and it ranges from 0.015\%-0.029\%.
Since $\Delta \varepsilon$ $\sim$ 0.07 for the right arm, the slope uncertainty is much larger for the right arm.}
\label{r_systematic_summary}
\end{center}
\end{table}
\begin{table}[!htbp]
\begin{center}
\begin{tabular}{||c|c|c|c|c||}
\hline \hline
Source                                   &Section                 &Left          &Left           &Left\\  
                                         &                        &Scale         &Random         &Slope\\
                                         &                        &(\%)          &(\%)           &(\%)   \\
\hline \hline
BCM Calibration($^{\star}$)              &\ref{BCM_correct}       &0.50          &\textbf{0.10}  &0.00\\
Target Boiling at 70$\micro$A($^{\star}$)&\ref{target_boiling}    &\textbf{1.05} &0.30-0.45($\dagger a$)  &0.00\\
Raster Size($^{\star}$)                  &\ref{target_boiling}    &0.00          &\textbf{0.10}  &0.00\\
Target Length($^{\star}$)                &\ref{target_leng_corr}  &0.12          &0.07           &0.00 \\
\hline \hline 
Electronic Deadtime                      &\ref{comp_elec_deadtime}&0.00          &0.00           &0.00\\
Computer Deadtime                        &\ref{comp_elec_deadtime}&0.10          &0.00           &\textbf{0.10}\\
VDC Zero-Track Inefficiency              &\ref{vdcs_track_eff}    &0.10          &0.00           &0.00\\
VDC Multiple-Track Inefficiency          &\ref{vdcs_track_eff}    &0.10          &0.02           &0.00\\
VDC Hardware Cuts                        &\ref{vdcs_hardware_eff} &0.50          &\textbf{0.10}  &\textbf{0.25}\\
Scintillator Efficiency                  &\ref{scint_effic}       &0.10          &0.05           &0.00\\
PID Efficiency ($\beta$)                 &\ref{beta_eff}          &0.00          &0.00           &0.00\\ 
PID Efficiency ($A_2$)                   &\ref{A2_eff}            &0.00          &0.00           &0.00\\ 
PID Efficiency ($A_1$)                   &\ref{A1_eff}            &0.20          &\textbf{0.10}  &0.00\\  
Pion Contamination                       &\ref{pid_cuts}          &0.00          &\textbf{0.10}  &0.00\\
Proton Absorption                        &\ref{proton_absorp}     &\textbf{1.00} &0.00           &0.03\\ 
Solid Angle Cut                          &\ref{e01001_optics}     &\textbf{2.06} &0.00           &0.00\\
Pion Subtraction                         &\ref{extract_sigma}     &0.00          &0.15-0.30($\dagger b$) &0.20-0.40\\
Dummy Subtraction                        &\ref{extract_sigma}     &0.50          &0.00           &\textbf{0.10}\\
$\Delta P$ Cut-Dependence                &\ref{extract_sigma}     &0.50          &\textbf{0.14}  &\textbf{0.20}\\
$\delta$ Cut-Dependence                  &\ref{extract_sigma}     &0.00          &0.00           &0.00\\
0.18 mrad Angle Offset                   &\ref{spect_mispoint}    &0.13          &0.01           &\textbf{0.18}\\
0.10 mrad Angle Fluctuations             &\ref{spect_mispoint}    &0.00          &0.03-\textbf{0.10}&0.00\\
0.03\% Beam Energy                       &\ref{beam_energy_measurements}&0.13    &0.02           &0.07\\
0.02\% Beam Energy Fluctuations          &\ref{beam_energy_measurements}&0.00    &0.043-0.081    &0.00\\
Radiative Corrections                    &\ref{rad_corrections}   &\textbf{1.00} &\textbf{0.20}  &\textbf{0.30}\\
\hline \hline 
 & &$\delta_{Scale}$&$\delta_{Random}$&$\delta_{Slope}$\\
\hline
Total (\%)                    & &2.91  &0.384-0.593  &0.539-0.801 \\
\hline \hline 
\end{tabular}
\caption[Summay of the systematic uncertainties for left arm.]
{Summay of the systematic uncertainties for the left arm. Numbers in bold show the biggest 
contribution to that uncertainty type. Sources marked with ($^{\star}$) contribute to the uncertainty 
in the luminosity monitor. For ($\dagger a$), the value is 0\% for all kinematics except $a$ and $b$ which is
0.45\% and 0.30\%, respectively. For ($\dagger b$), the value is 0.15\% for all kinematics except those of
$Q^2$ = 4.10 GeV$^2$ which is 0.30\%. The random uncertainty in the 0.10 mrad angle fluctuations and 0.02\% beam energy fluctuations 
is applied on kinematics-by-kinematics basis as shown in Figure \ref{fig:left_proton_sigma0.10mrad_epsilon} and 
Figure \ref{fig:left_proton_sigma0.02e0_epsilon}, respectively.}
\label{l_systematic_summary}
\end{center}
\end{table}

\chapter{Data Analysis II: Reduced Cross Sections Extraction}\label{chap_cross_sections}
\pagestyle{plain}
\section{Overview}
\pagestyle{plain}
\begin{figure}[!htbp]
\begin{center}
\epsfig{file=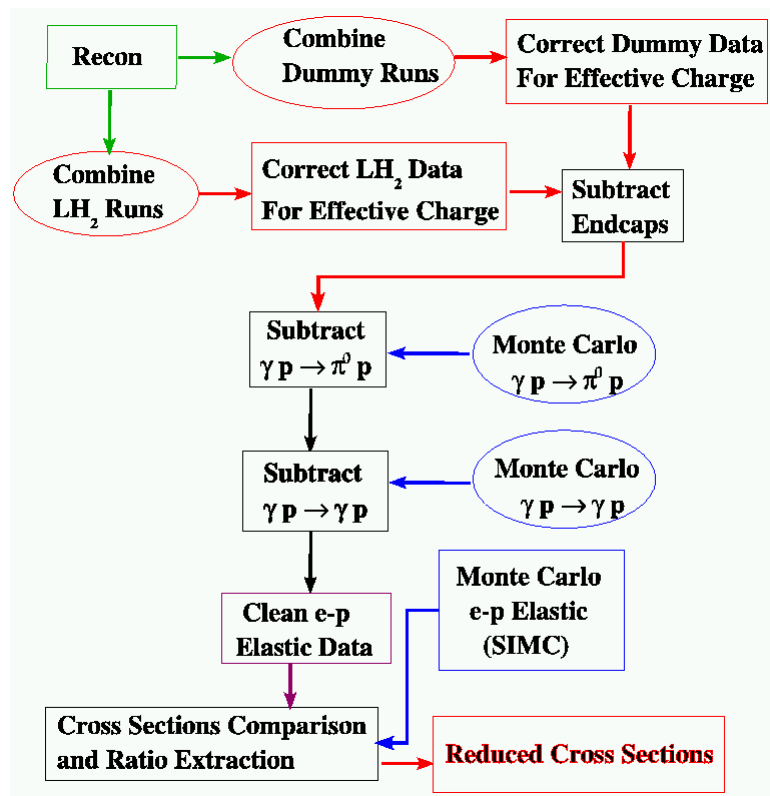,width=4in}
\end{center}
\caption[E01-001 experiment analysis flow chart.]
{E01-001 experiment analysis flow chart.}
\label{fig:e01001_flow_chart}
\end{figure}

In this chapter I will present the procedure used to extract the
e-p reduced cross section $\sigma_{R}$. Figure \ref{fig:e01001_flow_chart}
shows a flow chart of the procedure used. First, ESPACE generates HBOOKs
for a list of LH$_2$ and dummy target runs in a particular kinematics which are read by Recon
(section \ref{recon}). Recon applies a set of cuts and corrections on an event-by-event basis and 
generates a new HBOOK file for each run on the list. In addition, Recon generates the $\Delta P$ spectrum
for each LH$_2$ and dummy run on the list. For each kinematics the $\Delta P$ spectra from all of the 
LH$_2$ and dummy runs are added separately after applying a new set of cuts to form the final $\Delta P$ 
spectrum for both LH$_2$ and dummy (section \ref{add_histo}). The final $\Delta P$ spectrum for both the 
LH$_2$ and the dummy is then normalized to its total $Q_{eff}$ (section \ref{qeff}). The final corrected dummy 
$\Delta P$ spectrum is subtracted from the final corrected LH$_2$ $\Delta P$ spectrum. 
Monte Carlo simulations of the $\gamma p \to \pi^{0} p$ and $\gamma p \to \gamma p$ backgrounds are 
performed (section \ref{pi0}), and then normalized and subtracted from the final corrected LH$_2$ $\Delta P$ 
spectrum to produce the residual elastic e-p spectrum. The net number of the elastic e-p events is then compared 
to the number of elastic events in the e-p peak as simulated using the Monte Carlo simulation program SIMC 
(section \ref{ep_simc}) in a window in the $\Delta P$ spectrum. The reduced cross section $\sigma_{R}$ is taken 
as the value of the input e-p cross section used in the simulation normalized by the ratio of number of elastic 
events in the data to that in the simulation (section \ref{sigma_R}). Finally, a linear fit of 
$\sigma_{R}$ to $\varepsilon$ gives $\tau G_{Mp}^2(Q^2)$ as the intercept and $G_{Ep}^2(Q^2)$ as the 
slope.

\section{Event Reconstruction by Recon} \label{recon}
\pagestyle{myheadings}

The HBOOK file for each physics run generated by ESPACE (see section \ref{event_reconstruction}) 
was read using the in-house event analyzer Recon. 
Recon reads in the HBOOK file for a given run as Column-Wise Ntuples, 
applies a set of cuts on an event-by-event basis on all the physical variables saved in 
these Ntuples, and calculates the necessary physics quantities. A new HBOOK file for each run is then 
generated for the final data analysis. Table \ref{recon_cuts} lists the cuts applied on these physical 
variables. Events that satisfy these cuts are accepted for the final data analysis.
  
First, a single-track event type T$_1$(T$_3$) for the right(left) arm is selected.
The multiplicity of a cluster of hit wires in each plane is defined as the number of hit wires
in the cluster. A single-track event making an angle of 45$^o$ with the VDC surface has typically
a multiplicity of 5. We require the multiplicity of each of the four planes for both VDCs to be 
greater than two. The $Q_3$ cut is applied to reject events that scrape off the exit pipe of the $Q_3$ 
quadrupole. The $Q_3$ cut is defined as:
\begin{equation}
Q_{3} = \sqrt{(xfp-2.64xpfp)^{2} + (yfp-2.64ypfp)^{2}} < 0.29~m~, 
\end{equation}
\begin{figure}[!htbp]
\begin{center}
\epsfig{file=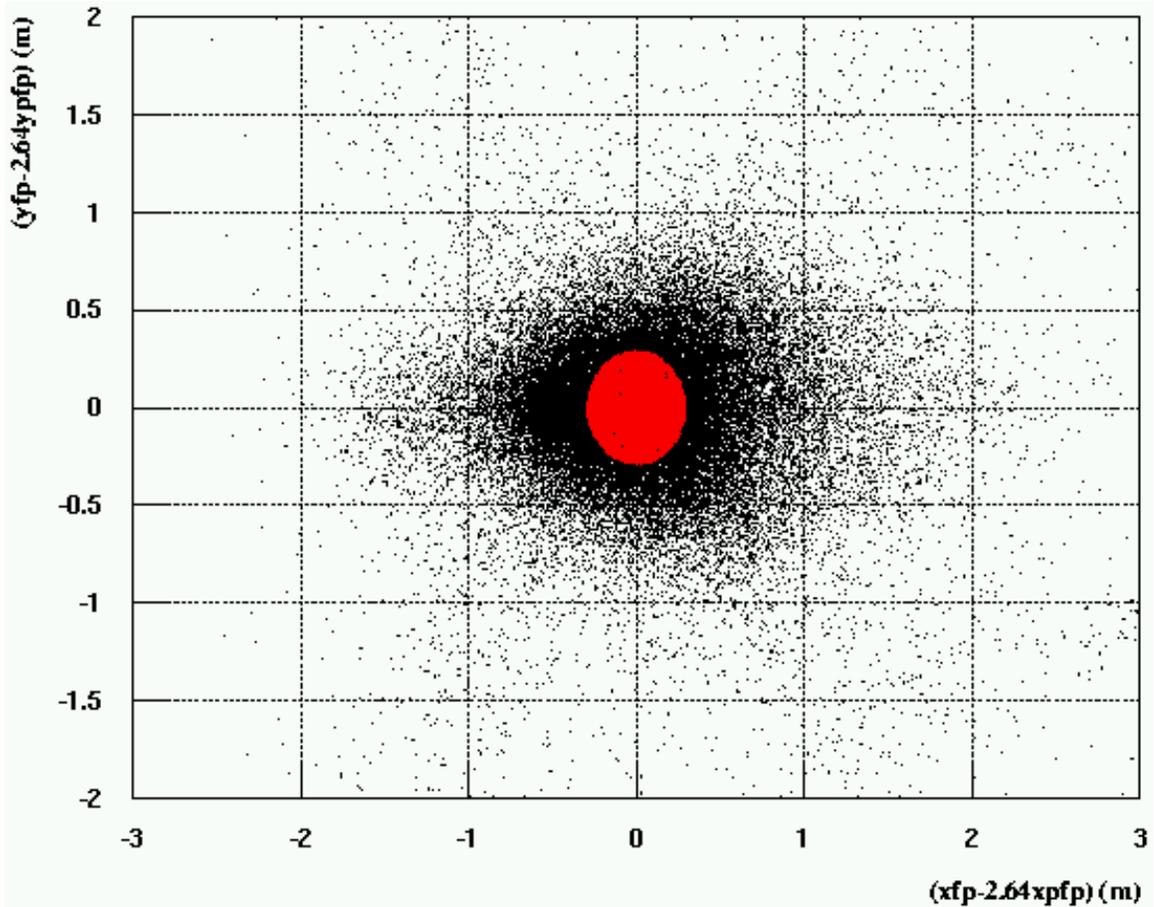,width=6in}
\end{center}
\caption[The $Q_3$ cut applied to the left arm events.]
{The $Q_3$ cut, solid red circle with radius of 0.29 m, applied to the left arm events (black) for 
a single run from kinematics k to reject events that were scraping off the exit pipe of the $Q_3$ 
quadrupole. Only events within the area of the solid red circle are accepted.} 
\label{fig:q3_cut}
\end{figure}
where $xfp$, $xpfp$, $yfp$, and $ypfp$ are the focal plane positions ($xfp$, $yfp$)
and angles ($xpfp$, $ypfp$) variables. The $Q_3$ was applied to the left arm events only since
there is no evidence that events were scraping off the exit pipe of the $Q_3$ 
quadrupole in the right arm spectrometer. It must be mentioned that the elastic events are
far from the $Q_3$ edges and are not affected by the $Q_3$ cut. Figure \ref{fig:q3_cut} shows how 
the $Q_3$ cut was applied. 

\begin{figure}[!htbp]
\begin{center}
\epsfig{file=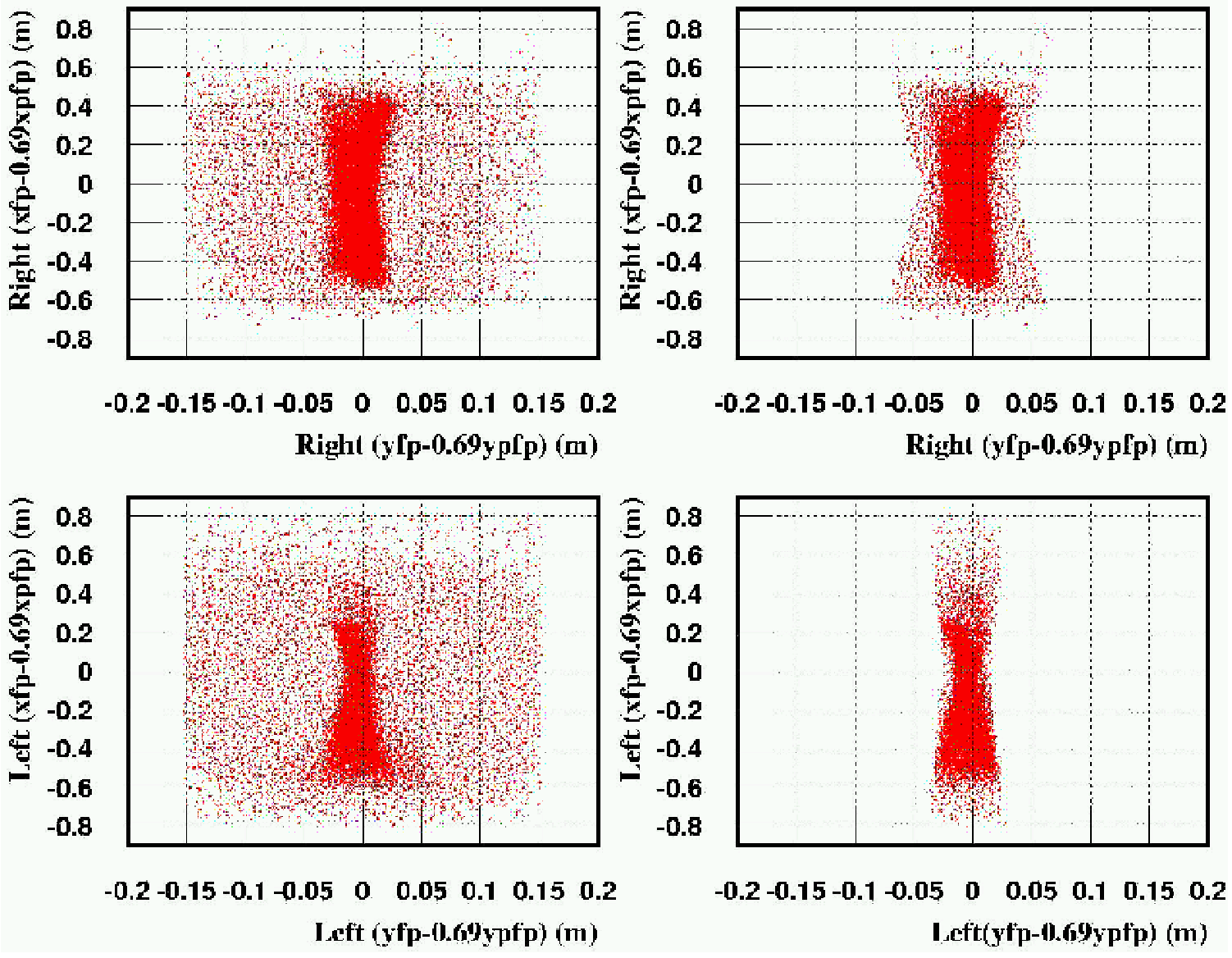,width=6in}
\end{center}
\caption[The hourglass cut applied to the right and left arms events.]
{The hourglass cut applied to the right and left arms events from run 1282. 
The right arm events before the cut (top left) and after the cut (top right). 
The left arm events before the cut (bottom left) and after the cut (bottom right).}
\label{fig:hourglass_cut}
\end{figure}

The hourglass cut (all good trajectories that make it 
through the HRS form an hourglass pattern $\sim$ 69 cm before the first VDC) is then applied to 
remove events which come from multiple scattering inside the spectrometer and located outside the 
acceptance. The hourglass cut is defined as:
\begin{equation}
\Big|yfp-0.69ypfp+0.005 \Big| - C_{1}\Big|xfp-0.69xpfp\Big|~, 
\end{equation}
where $C_{1}$ has a value of 0.045(0.017) for the right(left) arm. Figure \ref{fig:hourglass_cut}
shows how the hourglass cut is applied.

\begin{figure}[!htbp]
\begin{center}
\epsfig{file=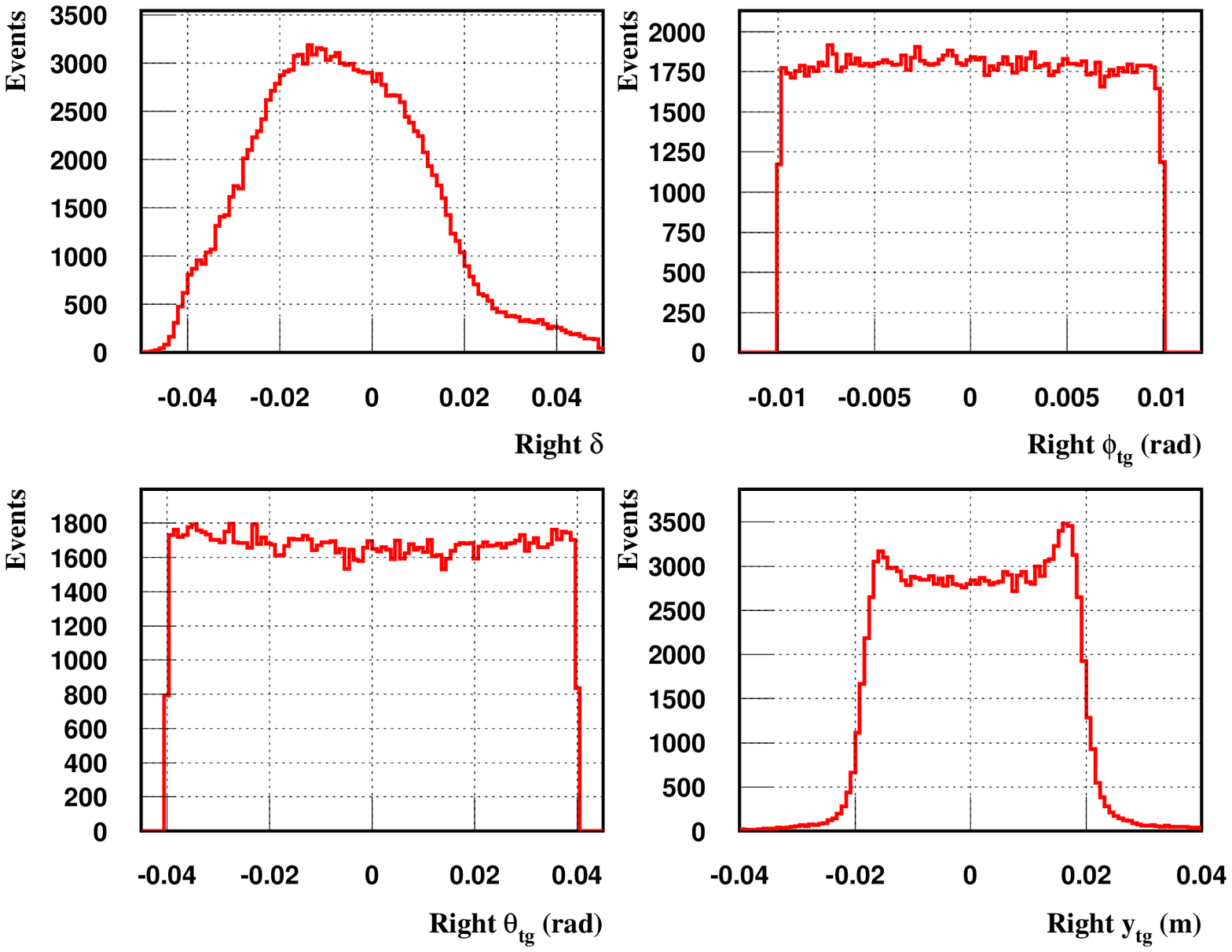,width=6in}
\end{center}
\caption[The right arm target reconstructed variables with cuts.]
{The right arm target reconstructed variables with cuts from run 1282. The right arm momentum 
acceptance $\delta$ (top left), in-plane angle $\phi_{tg}$ (top right), out-of-plane angle 
$\theta_{tg}$ (bottom left), and y-coordinate of the extended target length 
$y_{tg}$ (bottom right).}
\label{fig:rtarget_cuts}
\end{figure}

\begin{figure}[!htbp]
\begin{center}
\epsfig{file=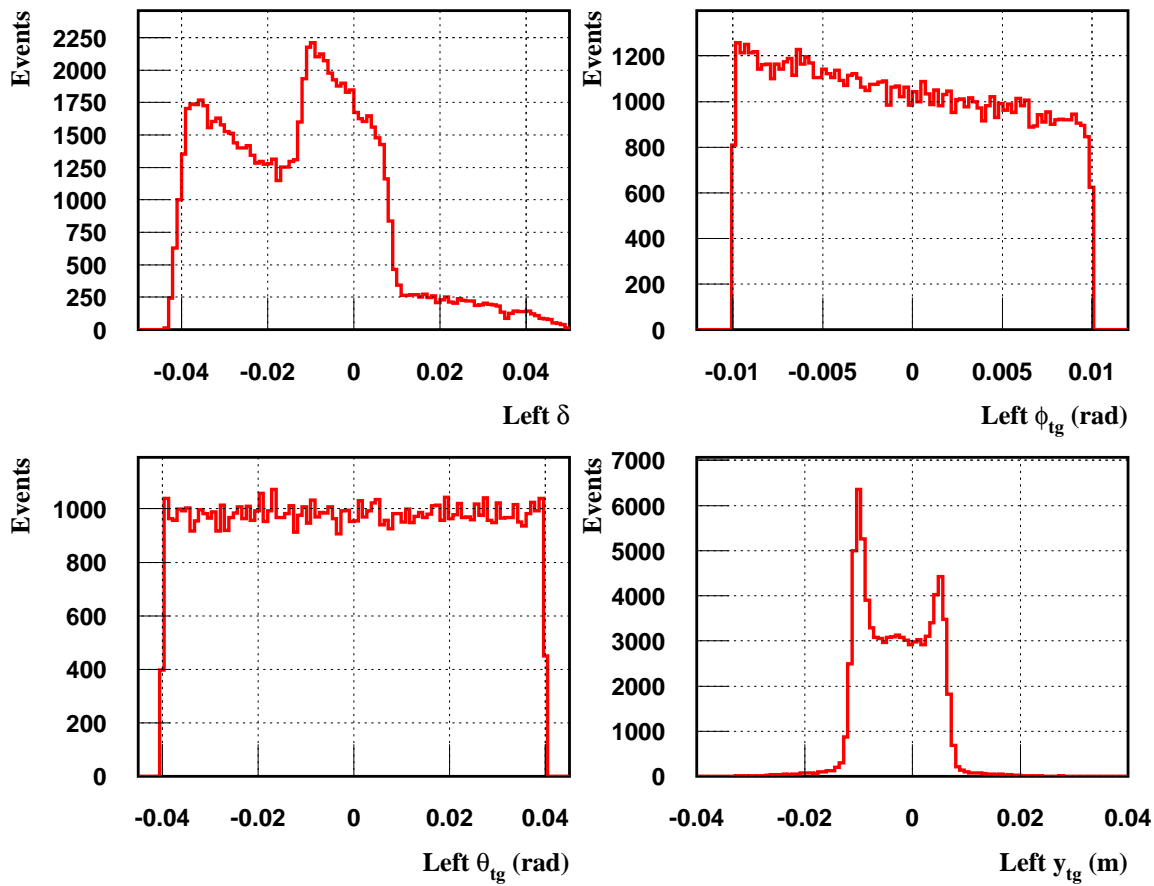,width=6in}
\end{center}
\caption[The left arm target reconstructed variables with cuts.]
{The left arm target reconstructed variables with cuts from run 1282. The left arm momentum 
acceptance $\delta$ (top left), in-plane angle $\phi_{tg}$ (top right), out-of-plane angle 
$\theta_{tg}$ (bottom left), and y-coordinate of the extended target length 
$y_{tg}$ (bottom right).} 
\label{fig:ltarget_cuts}
\end{figure}

Finally cuts on the target reconstructed variables such as the momentum acceptance
$\delta$, the y-coordinate of the extended target length $y_{tg}$, the out-of-plane
angle $\theta_{tg}$, and the in-plane angle $\phi_{tg}$ are applied. Figures \ref{fig:rtarget_cuts} 
and \ref{fig:ltarget_cuts} show the target reconstructed variables after the cuts have been applied.

\begin{figure}[!htbp]
\begin{center}
\epsfig{file=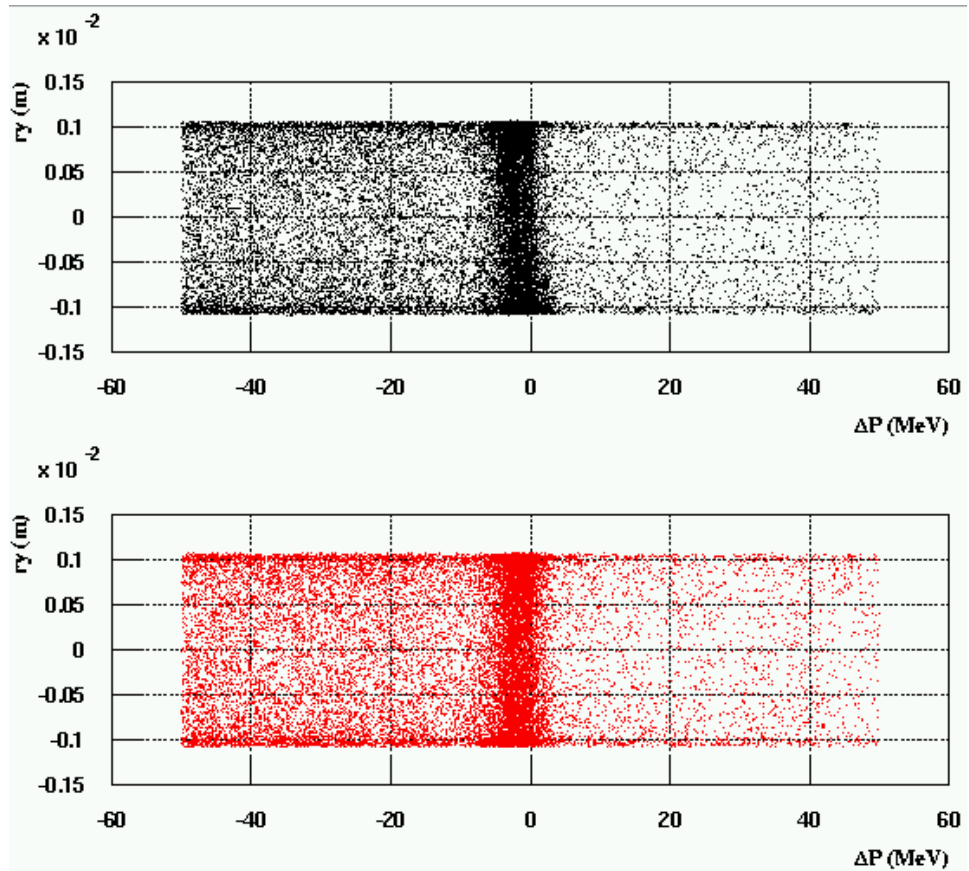,width=5in}
\end{center}
\caption[Two dimensional histogram of $\Delta P$ vs raster ry.]
{Two dimensional histogram of $\Delta P$ (x-axis) vs raster ry (y-axis). Top black: the $\Delta P$ 
spectrum before the correction shows a raster ry dependence as the events centered around 
$\Delta P$ = 0 MeV are slightly tilted counter clockwise. Bottom red: the $\Delta P$ spectrum 
after the correction where the events are tilted clockwise (see equation (\ref{eq:dp_vs_raster})) 
and become mainly centered around $\Delta P$ = 0 MeV.}
\label{fig:ldp_raster_correct}
\end{figure}

It should be mentioned that the momentum acceptance, $\delta$, showed an out-of-plane angle dependence on 
both the right and left arms, and a raster, ry, dependence on the left arm only. The momentum acceptance 
$\delta$ had first to be adjusted to correct for the $\theta_{tg}$ dependence for the left 
arm as:
\begin{equation}
\delta_{1} = \delta_{0} + 0.025\theta_{tg} + 0.250\theta_{tg}^2~,
\end{equation}
while for the right arm:
\begin{equation}
\delta_{1} = \delta_{0} + 0.02742\theta_{tg}~,
\end{equation}
where $\delta_{0}$ and $\delta_{1}$ are the momentum acceptance before and after the adjustment, 
respectively.
The raster correction was then applied as: 
\begin{equation} \label{eq:dp_vs_raster}
\delta_{2} = \delta_{1}  + \mbox{constant} \times \mbox{ry}~,
\end{equation}
where ry is the raster's ADC signal (raster number 2 or rast2adc):
\begin{equation} 
\mbox{ry} = 0.002\frac{(\mbox{rast2adc}-1560)}{\mbox{raster width}}~,
\end{equation}
constant is the correction applied, and the raster width is the width of the ADC signal 
whose average width is 1560 and is energy dependent. Again, $\delta_{1}$ and $\delta_{2}$ are the 
momentum acceptance before and after the raster correction, respectively. 
Figure \ref{fig:ldp_raster_correct} shows the $\Delta P$ spectrum before and after the raster correction.
Table \ref{fry_correct} lists the values for the constant and raster width. 
\begin{table}[!htbp]
\begin{center}
\begin{tabular}{||c|c|c||} \hline
\hline
Tiefenback Energy & constant & raster width \\ 
(MeV)           &          &               \\
\hline \hline
1912.94           & 0.42 & -1400              \\
2260.00           & 0.38 & +1100               \\
2844.71           & 0.00 & -1000                \\
3772.80           & 0.00 & +1200                 \\
4702.52           & 0.30 & -600                   \\ 
\hline \hline 
\end{tabular}
\caption[The Tiefenback energy, constant (correction applied) and raster width needed to
correct the momentum acceptance $\delta$ for the fast raster (fry).]
{The Tiefenback energy, constant(correction applied) and raster width needed to
correct the momentum acceptance $\delta$ for the raster dependence.}
\label{fry_correct}
\end{center}
\end{table}

In order to reconstruct $\Delta P$, equation (\ref{eq:delta_p}), the following procedure is used.
The final momentum of the scattered protons $P_{measured}$ (measured using the high resolution
spectrometer) is reconstructed as:
\begin{equation}
P_{measured} = P_{L(R)}(1 + \delta_{2})~, 
\end{equation}
where $P_{L(R)}$ is the central momentum setting for the left(right) arm spectrometer as
listed by Table \ref{kinematics}. 
The scattering angle of the proton is then reconstructed:
\begin{equation} \label{eq:scatter_angle}
\cos\theta_{p} = \frac{\cos\theta_{R(L)} \pm \phi_{tg} \sin\theta_{R(L)}}{\sqrt{1 + \phi_{tg}^2 + \theta_{tg}^2}}~,
\end{equation}
where the plus(minus) sign is used with the right(left) arm respectively, and $\theta_{R(L)}$
is the right(left) arm spectrometer scattering angle as listed by Table \ref{angles_used}. 
Finally, the momentum of the scattered protons $P_{calculated}(\theta_{p})$, equation (\ref{eq:pcal}),
is then obtained 

\begin{table}[!htbp]
\begin{center}
\begin{tabular}{||c|c|c||} \hline
\hline
Cut Type & Left Arm  & Right Arm                                                \\
Applied  & Cut Value & Cut Value                                                 \\
\hline \hline
Event Type                       & 3                              & 1              \\
Number of Tracks                 & 1                              & 1               \\
VDC Multiplicity                 & Multiplicity$>$2               & Multiplicity$>$2 \\
$Q_3$ (m)                        & $<$0.29                        & -                 \\
Hourglass (m)    &$<$$0.02\sin\theta_{L}$+0.01& $<$$0.02\sin\theta_{R}$+0.02\\
$\delta$ Momentum (\%)           &-5.0$<$$\delta$$<$5.0               &-5.0$<$$\delta$$<$5.0 \\
\hline \hline
PID Aerogel ADCSUM               &A$_1$ADCSUM$<$350           & A$_2$ADCSUM$<$1250   \\ 
PID $\beta$                      & -                              & 0.45$<$$\beta$$<$0.85    \\
Number of Clusters/VDC Plane     & 1                              & 1                         \\
Minimum Number of Hits/Cluster   & 3                              & 3                         \\
Maximum Number of Hits/Cluster   & 6                              & 6                         \\
$y_{tg}$ (m)                     & -0.05$<$$y_{tg}$$<$0.05          & -0.05$<$$y_{tg}$$<$0.05      \\
Out-of-Plane Angle ($\theta_{tg}$) (mrad)& -40.0$<$$\theta_{tg}$$<$40.0 & -40.0$<$$\theta_{tg}$$<$40.0 \\
In-Plane Angle ($\phi_{tg}$) (mrad)& -10.0$<$$\phi_{tg}$$<$10.0   & -10.0$<$$\phi_{tg}$$<$10.0    \\
\hline \hline
\end{tabular}
\caption[Cuts applied in the analysis of the E01-001 data.]
{Cuts applied in the analysis of the E01-001 data. Cuts in the upper
half of table are applied in Recon and cuts in the bottom half are applied later as discussed
in section \ref{add_histo}. See text for more details.}
\label{recon_cuts}
\end{center}
\end{table}
using the measured scattering angle of the proton as given by equation 
(\ref{eq:scatter_angle}).

\section{Adding Histograms} \label{add_histo}

For the data analysis, the HBOOK file generated by Recon for each physics run in 
a given kinematics is read by a kumac file (code) written using PAW the Physics 
Analysis Workstation software. The PAW kumac starts by reading the run 
number and then loading its HBOOK file. A new set of cuts such as the PID cuts
A$_2$ ADCSUM and $\beta$ cuts for the right arm and A$_1$ ADCSUM cut for the left arm 
(to select good protons) and tracking cuts (number of clusters for each VDC's plane = 1, 
minimum number of hits per cluster in each VDC's plane = 3, and maximum number of hits per cluster 
in each VDC's plane = 6) are applied to the LH$_2$(dummy) $\Delta P$ ntuple of that run. 
See section \ref{pid_cuts} for the range of PID cuts used. The $\Delta P$ ntuple for LH$_2$ or dummy 
is then projected onto a histogram where a distribution of the number of counts per one MeV bin 
can be plotted. This procedure was done for all the LH$_2$ and dummy runs in that kinematics. 
The LH$_2$ and dummy $\Delta P$ histograms from all the runs are added to form the final $Q_{eff}$ 
uncorrected LH$_2$ and dummy $\Delta P$ histogram. 

The LH$_2$ $\Delta P$ spectrum is made of several contributions. The main contribution  
is from elastic $ep \to ep$ scattering which has a peak and a radiative tail. 
In addition there are backgrounds due to quasi-elastic and inelastic scattering in the aluminum 
target windows and walls and protons generated the photoreactions $\gamma p \to \pi^0 p$ and 
$\gamma p \to \gamma p$. 
\begin{figure}[!htbp]
\begin{center}
\epsfig{file=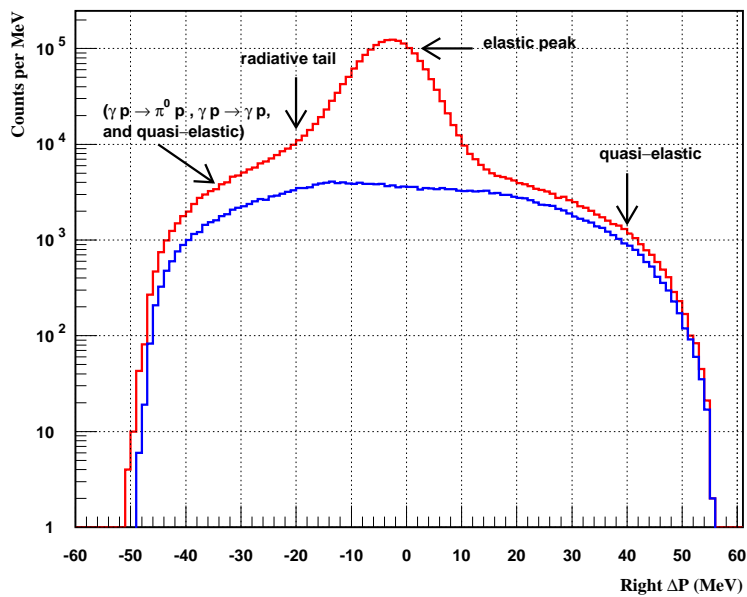,width=3.88in}
\end{center}
\caption[$Q_{eff}$ uncorrected right arm $\Delta P$ spectrum for LH$_2$ and dummy targets 
from kinematics $b$.]
{$Q_{eff}$ uncorrected right arm $\Delta P$ spectrum for LH$_2$ (red) and dummy (blue) targets 
from kinematics $b$.}
\label{fig:uncorr_rdelta_p}
\end{figure}
\begin{figure}[!htbp]
\begin{center}
\epsfig{file=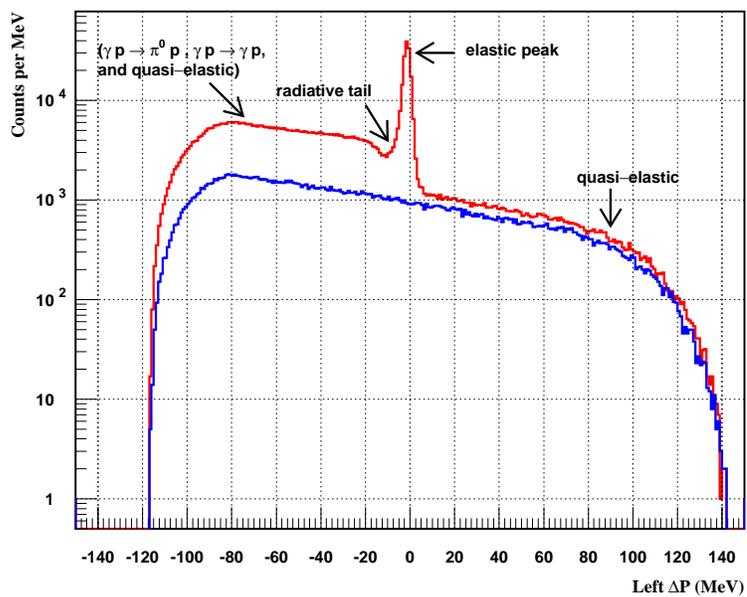,width=3.88in}
\end{center}
\caption[$Q_{eff}$ uncorrected left arm $\Delta P$ spectrum for LH$_2$ and dummy targets 
from kinematics $b$.]
{$Q_{eff}$ uncorrected left arm $\Delta P$ spectrum for LH$_2$ (red) and dummy (blue) targets 
from kinematics $b$.}
\label{fig:uncorr_ldelta_p}
\end{figure}

To account for the quasi-elastic contribution to the LH$_2$ $\Delta P$ spectrum, events scattered from 
the aluminum target windows were measured using an empty dummy target.
The windows of the dummy target were thicker than the windows of the LH$_2$ target and the determination
of the effective thickness of the dummy target, referred to as ``dummy thickness'' will be discussed
is section \ref{sigma_R}. Figures \ref{fig:uncorr_rdelta_p} and \ref{fig:uncorr_ldelta_p} show 
a final $Q_{eff}$ uncorrected $\Delta P$ spectra for the LH$_2$ and dummy targets for both the 
right and left arm spectrometers. The resolution of the elastic peak is dominated by the angular 
resolution. The right arm elastic peak has a broader width than the left arm elastic peak. 
This is due to the fact that at low momentum, $\Delta P$ is more sensitive to the angular resolution of
the spectrometer, causing a broadening of the $\Delta P$ peak.

\section{The Effective Charge Normalization} \label{qeff}

After adding all the LH$_2$ and all of the dummy histograms from the runs in a given kinematics,
the final $Q_{eff}$ uncorrected $\Delta P$ histogram must be divided by $Q_{eff}$. 
The effective charge as defined in section \ref{qeff_intro} was calculated for each LH$_2$ run 
using equation (\ref{eq:qeff}).

The dummy target is an empty target with thicker aluminum windows than the actual
entrance and exit windows of the LH$_2$ target. In determining $Q_{eff}$, equation (\ref{eq:qeff}),
we must correct for the effective thickness of the dummy target. 
The reason we use an effective thickness for the dummy target is the bremsstrahlung initiating reactions, 
primarily in the downstream endcap. The bremsstrahlung are different for LH$_2$ and dummy 
because of the different thicknesses and also the presence of the hydrogen. The dummy target spectrum is 
used to determine the shape of the endcaps contribution and the procedure of determining the effective
dummy thickness is discussed in section \ref{sigma_R}.  
The final $Q_{eff}$ is the sum of the effective charge from 
all the LH$_2$(dummy) runs, i.e., $\sum_{i}^{N_{run}} Q^{i}_{eff}$ where $Q^{i}_{eff}$ is the effective 
charge for run number $i$ and $N_{run}$ is the total number of LH$_2$(dummy) runs.

\begin{figure}[!htbp]
\begin{center}
\epsfig{file=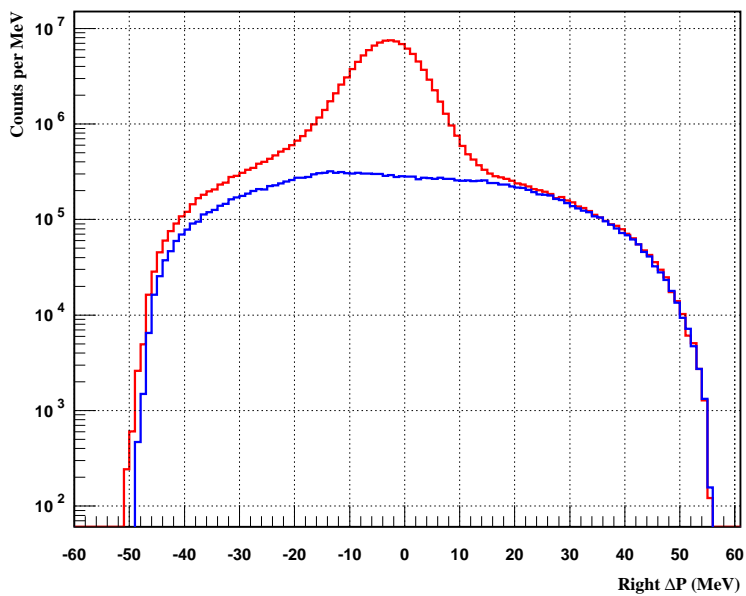,width=3.88in}
\end{center}
\caption[$Q_{eff}$ corrected right arm $\Delta P$ spectrum for LH$_2$ and dummy targets.]
{$Q_{eff}$ corrected right arm $\Delta P$ spectrum for LH$_2$ (red) and dummy (blue) targets
from kinematics $b$.}
\label{fig:corr_rdelta_p}
\end{figure}

\begin{figure}[!htbp]
\begin{center}
\epsfig{file=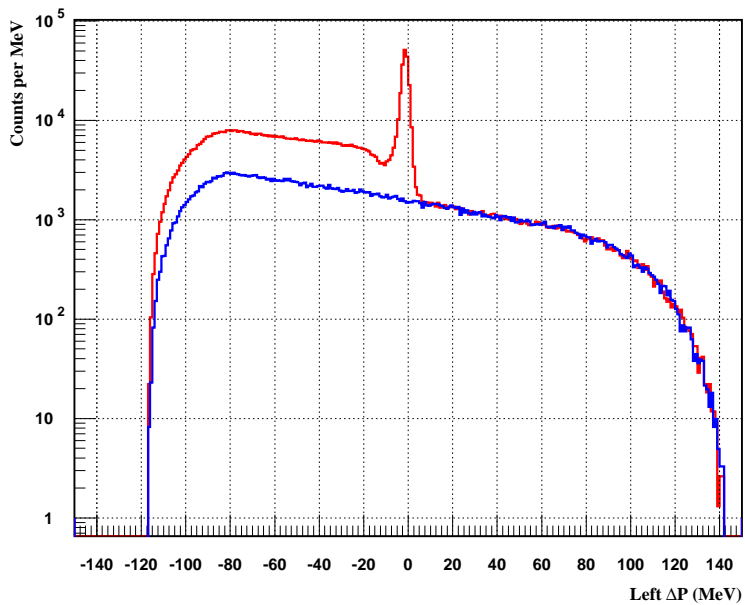,width=3.88in}
\end{center}
\caption[$Q_{eff}$ corrected left arm $\Delta P$ spectrum for LH$_2$ and dummy targets.]
{$Q_{eff}$ corrected left arm $\Delta P$ spectrum for LH$_2$ (red) and dummy (blue) targets 
from kinematics $b$.}
\label{fig:corr_ldelta_p}
\end{figure}

The default and initial value of 4.11 for the effective dummy thickness was used to calculate 
$Q_{eff}$ for the dummy data. The procedure of extracting the final effective dummy thickness 
used to extract the reduced cross section $\sigma_{R}$ for each kinematics is discussed in section 
\ref{dummy_extraction}. This initial value seems to do a good job producing the high $\Delta P$ 
side of the LH$_2$ spectrum which is mainly dominated by the quasi-elastic contribution from the endcaps. 
Figures \ref{fig:corr_rdelta_p} and \ref{fig:corr_ldelta_p} show normalized 
LH$_2$ and dummy $\Delta P$ spectra for the right and left arms, respectively.

\section{Monte-Carlo Simulations}

\subsection{Elastic e-p Simulations (SIMC)} \label{ep_simc}
The Monte Carlo simulation program SIMC is used to simulate elastic scattering for all kinematics and for both arms. 
The elastic e-p simulations are a crucial component of the analysis as they are used to extract the reduced cross section. 
SIMC was adapted from the (e,e'p) simulation program named SIMULATE that was written for SLAC experiment NE18 
\cite{naomithesis,tomthesis}, and was converted to simulate the Jefferson Lab Hall C and A spectrometers. 
The two main components of SIMC are the event generator, which includes the cross-section weighting and radiative corrections,
and the spectrometer models. First, SIMC randomly generates the energy and position of the incident electron at the target to match 
the energy and spatial spread of the actual beam. In doing so, SIMC takes into account the target length and the beam raster.
The beam energy is then corrected for ionization losses in the target. Moreover, SIMC randomly generates the momenta and angles of the 
scattered electron and proton vectors with a flat distribution over limits exceeding the actual experimental spectrometer acceptance. 
In our case, we generate electrons and protons in coincidence and use the Bosted cross section to weight each generated event. 
Having generated a basic event at the scattering vertex, SIMC allows for any or all of these events 
to emit real or virtual photons where the corresponding event vectors are adjusted and radiatively corrected \cite{ent01}.
The scattered electron and proton vectors are transported through the target where ionization energy losses and multiple 
scattering in the target material, cells, and chamber are applied. Finally, only the scattered protons are transported through the 
spectrometers. 

Transporting the protons through the spectrometer was done using the spectrometer optics models built in the 
Monte Carlo simulation program COSY \cite{cosy95}. COSY generates both the forward and backward sets of matrix elements to simulate
the optical resolution of the magnetic systems in the spectrometer. Note that all the material the protons have to travel through in 
the spectrometer and the detector stack are included, and the multiple scattering due to all the material is calculated. However,
we do not account for proton absorption in the simulations since we correct the data for that.
The forward matrix elements transport the particle vectors from the entrance window of the spectrometer to its focal plane going through 
every major aperture in the spectrometer. SIMC assures that these particles have gone through each of these apertures by checking the 
acceptance of each aperture using a set of apertures cuts. These apertures include the front, middle, and back aperture of each magnetic 
element ($QQDQ$ configuration), the aperture of the vacuum can after the $Q_3$ quadrupole, the aperture of the rectangular collimator, 
and the aperture of each detector that was used as a fiducial cut. In addition, VDCs smearing is applied to the particle positions at 
the two VDCs to match those of the actual VDCs. The backward matrix elements then reconstruct or transport back the particle vectors to 
the target vertex where they are corrected for energy loss and multiple scattering and the elastic $\Delta P$ spectrum for a given 
kinematics is then reconstructed. 

\begin{figure}[!htbp]
\begin{center}
\epsfig{file=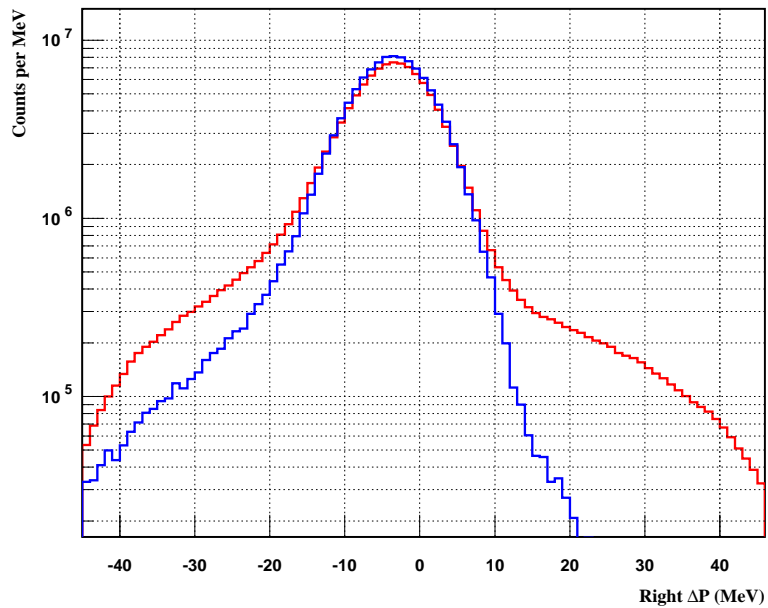,width=4in} 
\end{center}
\caption[$Q_{eff}$ corrected right arm LH$_2$ $\Delta P$ spectrum and elastic e-p simulation.]
{$Q_{eff}$ corrected right arm LH$_2$ $\Delta P$ spectrum (red) and elastic e-p simulation (blue)
for kinematics $b$.}
\label{fig:r_lh2_ep}
\end{figure}

\begin{figure}[!htbp]
\begin{center}
\epsfig{file=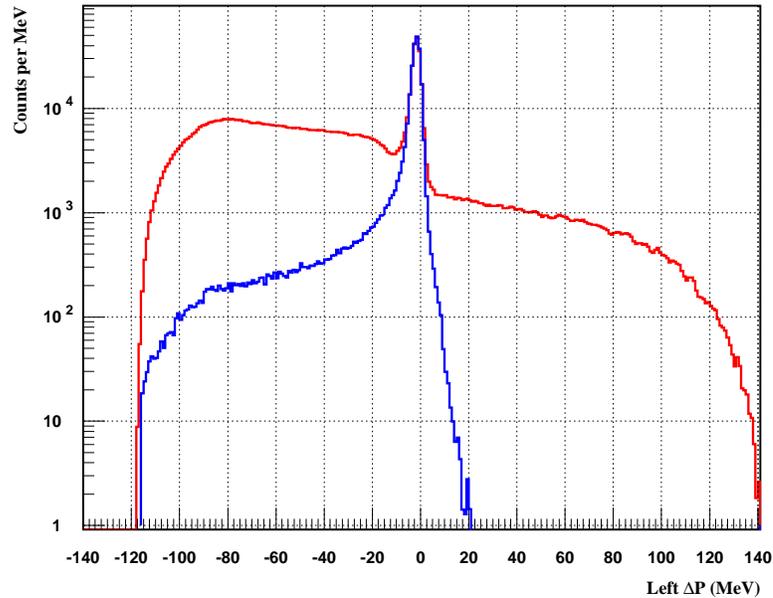,width=4in}
\end{center}
\caption[$Q_{eff}$ corrected left arm LH$_2$ $\Delta P$ spectrum and elastic e-p simulation.]
{$Q_{eff}$ corrected left arm LH$_2$ $\Delta P$ spectrum (red) and elastic e-p simulation (blue)
for kinematics $b$.}
\label{fig:l_lh2_ep}
\end{figure}

When comparing the shape of the elastic peak from data to that of simulations, it was obvious that the width of the elastic peak from data 
was broader than that of simulations. In addition, there was a clear mismatch between the shape of elastic tail from data and that of 
simulations. To resolve these issues, we compare the $\Delta P$ spectrum from coincidence data to that of simulations since the coincidence 
data represent a pure protons sample without any background. Since the peak width is limited by the angular resolution of the spectrometer, 
we smear the angular resolution of the spectrometer in the simulations using a symmetric gaussian function to better fit the observed peak 
widths. We also apply an additional non-gaussian smearing to better match the tails. 
Figures \ref{fig:r_lh2_ep} and \ref{fig:l_lh2_ep} show the normalized LH$_2$ $\Delta P$ spectrum along with the elastic e-p simulations 
for the right and left arms, respectively.


\subsection{$\gamma p \to \pi^0 p$ and $\gamma p \to \gamma p$ Simulations} \label{pi0}
When the electron beam passes through the target (protons in our case), electrons lose energy by radiating 
real photons which is known by bremsstrahlung. These real photons which impinge upon the target have a maximum 
energy at the beam energy. To model the $\Delta P$ spectrum for these high energy $\pi^0$ protons, we first 
calculate the bremsstrahlung cross section and reconstruct the $E_{\gamma}$ spectrum for these photons \cite{schultethesis}, and
then randomly generate photons according to the $E_{\gamma}$ spectrum. The next step is to uniformly and randomly generate protons 
over the acceptance and this is done according to the event generation procedure discussed before.   
The generated event is then weighted by an $s^{-7}$ cross section dependence, as predicted by the high energy 
approximation and based on the constituent counting rules or $s^{-n}$, forming the shape of the $\Delta P$ spectrum used for the 
$\gamma p \to \pi^0 p$ contribution. Note that the generated proton is transported through the spectrometer using the
spectrometer optics models mentioned before. Finally, the shape of the $\gamma p \to \pi^0 p$ $\Delta P$ spectrum from simulations is 
normalized to that of data as discussed in section \ref{normalized_pi0_simul}.
  
\begin{figure}[!htbp]
\begin{center}
\epsfig{file=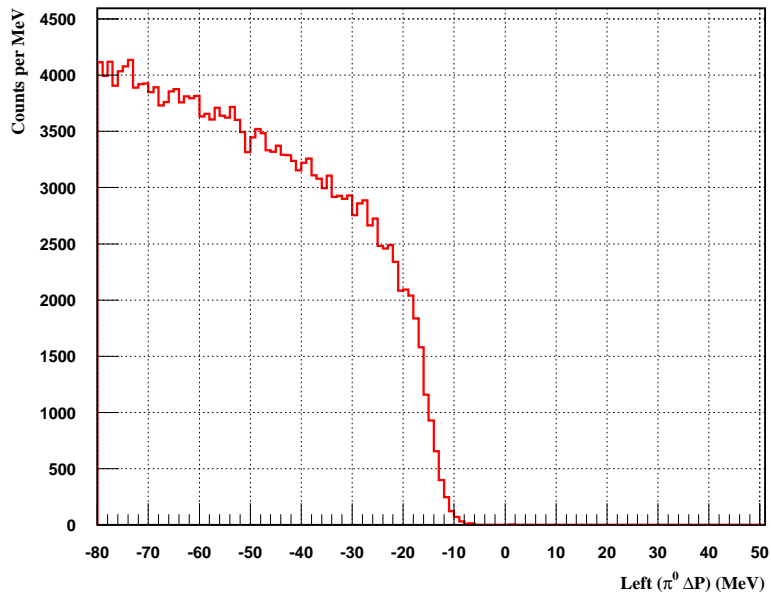 ,width=4in}
\end{center}
\caption[Left arm $\gamma p \to \pi^{0} p$ $\Delta P$ simulation.]
{Left arm $\Delta P$ simulation for $\gamma p \to \pi^{0} p$ background ($\pi^{0} \Delta P$) 
for kinematics $b$.}
\label{fig:l_pi0}
\end{figure}

\begin{figure}[!htbp]
\begin{center}
\epsfig{file=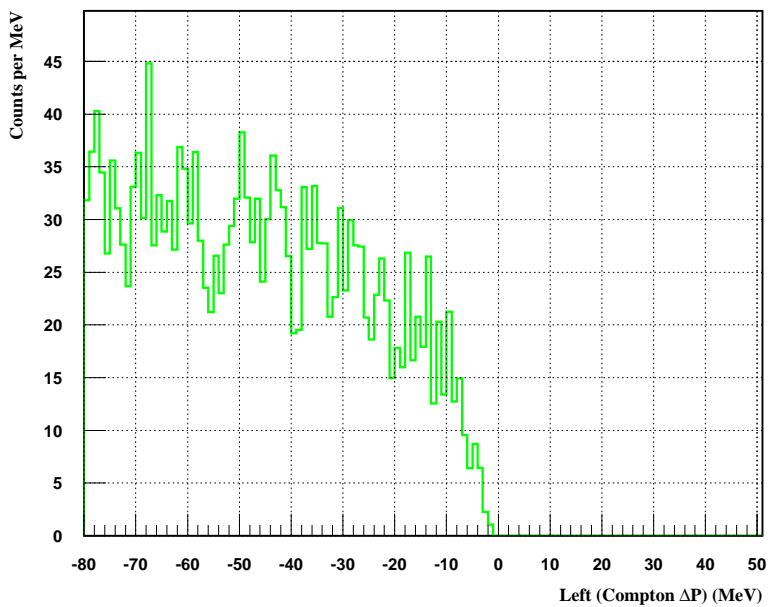,width=4in}
\end{center}
\caption[Left arm $\gamma p \to \gamma p$ $\Delta P$ simulation.]
{Left arm $\Delta P$ simulation for $\gamma p \to \gamma p$ background (Compton $\Delta P$) 
for kinematics $b$.}
\label{fig:l_compton}
\end{figure}

Previous experiments \cite{shupe79} indicated that the ratio of the $\gamma p \to \gamma p$ 
to the $\gamma p \to \pi^0 p$ cross sections (relative cross section) was 1-5\%.
We took the ratio from the original reference \cite{shupe79} and plotted it as a function of beam energy.
Note that the beam energy range quoted was very similar to that of the E01-001. There was a clear dependence on energy, 
with the ratio well fitted by $(0.92E - 1.2)$\%. Therefore, we rescaled the $\gamma p \to \pi^0 p$ $\Delta P$ spectrum by
$(0.92E - 1.2)$\% to generate the Compton $\Delta P$ background. Figures \ref{fig:l_pi0} and \ref{fig:l_compton} show the 
simulated $\Delta P$ spectrum for the $\gamma p \to \pi^0 p$ and $\gamma p \to \gamma p$ backgrounds for the left arm, 
respectively. 


\section{Radiative Corrections} \label{rad_corrections}

In this section I will start by describing the radiative corrections procedures for inclusive
p(e,e')p scattering, and then discuss details of implementation for E01-001. 
The Lorentz invariant transition scattering amplitude 
equation (\ref{eq:lorentz_trans_matrix}) and hence the elastic e-p 
scattering differential cross section equation (\ref{eq:ep_diff_cross_section}) 
were derived to lowest order in $\alpha$ including only the amplitude due to
the exchange of a single virtual photon between the incident electron and struck proton 
as shown in Figure \ref{fig:Feynman}. 
This is known as the Born approximation or single-photon exchange. However, in reality 
higher order processes in $\alpha$ and beyond the Born approximation also affect the cross 
section such as radiation of additional real and virtual photons. 
These processes clearly must be included. 
The incident electrons radiate in the presence of a nuclear field due to changes in their 
velocities caused by Coulomb interactions. We refer to the radiation resulting from such 
deceleration of the electron by bremsstrahlung or ``Braking Radiation''. 
Corrections for these higher order processes are referred to as radiative corrections. 
Radiative corrections are classified into internal and external radiative corrections.
\begin{figure}[!htbp]
\begin{center}
\epsfig{file=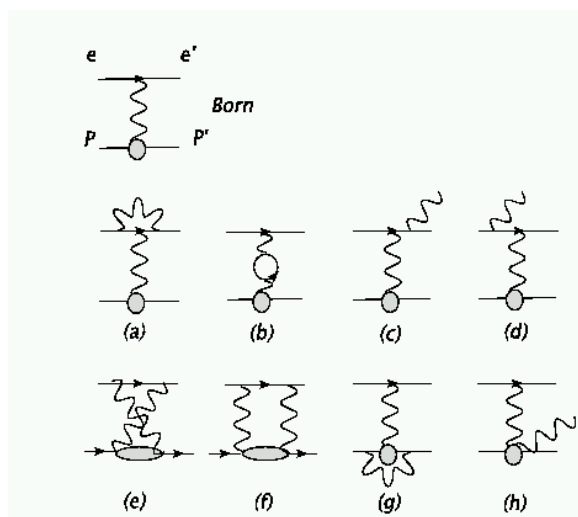,width=3in}
\end{center}
\caption[Feynman diagrams for the elastic e-p scattering including both the first-order (Born) 
and higher-order QCD radiative corrections.]
{Feynman diagrams for the elastic e-p scattering including both the first-order (Born) 
and higher-order QCD radiative corrections (a)-(h).}
\label{fig:RCdiagrams}
\end{figure}

\begin{itemize}
\item {\textbf{Internal Radiative Corrections}:
When the incoming and outgoing electrons interact with the Coulomb field
of the nucleus involved primary in the scattering process, this results in
emission and reabsorption of virtual photons and emission of real soft photons. 
Figure \ref{fig:RCdiagrams} shows the Born term and higher-order Feynman diagrams used in the calculation
of the internal radiative corrections. 
Note that these corrections are corrections to the principle scattering vertex itself.
These internal radiative corrections are also classified into elastic radiative corrections 
such as vertex corrections (for both the electron and proton) Figure \ref{fig:RCdiagrams}(a) and (g), 
vacuum polarization or loop diagrams Figure \ref{fig:RCdiagrams}(b), and two-photon-exchange (TPE) Figure 
\ref{fig:RCdiagrams}(e) and (f). The inelastic radiative corrections involve the emission of real photons 
such as internal bremsstrahlung (for either electron or proton) Figure \ref{fig:RCdiagrams}(c), (d), and (h). 
Higher-order processes due to three or more photon exchange (Coulomb corrections) and emission of
multiple real photons are also possible. 

Corrections that involve only lepton-photon or lepton-lepton vertices (electron vertex 
and electron bremsstrahlung) are exactly calculable in QED. Corrections that involve 
photon-proton vertices (proton vertex, TPE, proton bremsstrahlung, and Coulomb corrections) are not 
exactly calculable in QED and must be either estimated or measured.}    
\item {\textbf{External Radiative Corrections}:
When the incoming and outgoing electrons interact with the Coulomb field of a nucleus other than the 
one involved in the primary scattering vertex, this also results in emission of real photons. Such radiative 
processes are caused by bremsstrahlung in material that the electron passes through before and after the scattering.}
\end{itemize}

As we see, the radiative processes are real physical processes which depend on the target material, geometry, 
and kinematics of the scattering. They affect the data by modifying the cross section of the process and the 
kinematics (momentum, energy, and angle) of the incident electron or final state of the electron and proton. 
Emission of these real photons causes the detected particle's momentum to be different from the actual 
momentum at the scattering vertex and therefore 
distorts the experimental spectra. On the other hand, amplitudes involving the emission of additional virtual
particles such as TPE processes affect only the magnitude of the measured cross section. The procedure for doing 
such radiative corrections was first derived by Schwinger \cite{schwinger49} and later modified by Mo and Tsai 
\cite{mo69}.

The $\delta$-function that multiplies equation (\ref{eqnarray:diffrentional1})
is used to assure elastic scattering in the case of the electron, that is, at a given energy $E$ 
and angle $\theta_{e}$, the elastic differential cross section is a $\delta$-function in $E'$ at 
$E'= E_{elastic} = E - \frac{Q^2}{2M_{p}}$
where $E'$ is given by equation (\ref{eq:recoil_effect}). See section \ref{elastic_scattering}
for more details. Bremsstrahlung radiative processes cause the elastic cross section to 
change from the simple $\delta$-function (modified by finite resolution, small angle scattering, 
and energy spread of the beam) to an asymmetric peak with an extended tail at lower energies
as shown in Figures \ref{fig:uncorr_rdelta_p} and \ref{fig:uncorr_ldelta_p} in the case of the 
proton. Note that this radiative tail extends down to values of $\Delta P$ where other inelastic processes 
such as pion production yields high energy electron, or for the case of proton detection, high energy
protons generated from photoreaction $\gamma p \to \pi^{0} p$ and $\gamma p \to \gamma p$ also occur. 
The procedure of extracting the elastic e-p scattering reduced cross section $\sigma_{R}$ will 
be discussed in section \ref{sigma_R} and sub-sections therein.

To convert the measured cross section to the single-photon-exchange cross section, the Born cross section of $O(\alpha)$, 
we use the radiative corrections procedure of Mo and Tsai \cite{mo69} as modified by Walker \cite{walker94,walkerphd} and 
Ent \cite{ent01,naomithesis} and implemented in the elastic e-p simulation code SIMC \cite{naomithesis,tomthesis}. 
Therefore, we write:
\begin{equation}
\frac{d\sigma}{d\Omega}|_{Measured} = (1 + \delta_{corr})\frac{d\sigma}{d\Omega}\Big|_{Born}~,
\end{equation}
where $\delta_{corr} = (\delta_{int} + \acute{\delta_{int}} + \delta_{ext})$ and $\delta_{int}$ and 
$\delta_{ext}$ represent the internal and external radiative corrections, respectively. The additional
term $\acute{\delta_{int}}$ results from the improvements made to the internal radiative corrections by
Walker. For corrections higher than $O(\alpha^3)$, we exponentiate $\delta_{corr}$ and we write:
\begin{equation} \label{higher_order_RADC}
\frac{d\sigma}{d\Omega}|_{Measured} = e^{\delta_{corr}} \frac{d\sigma}{d\Omega}\Big|_{Born}~.
\end{equation}

\subsection{Internal Radiative Corrections $\delta_{int}$}
The internal radiative corrections $\delta_{int}$ are determined based on the modification made
by Walker \cite{walker94,walkerphd} and Ent \cite{ent01} to the work of Mo and Tsai \cite{mo69}. 
They include the processes of: vacuum polarization, TPE, electron and proton vertex corrections, and internal 
bremsstrahlung. Note that the approximation used in equation (\ref{higher_order_RADC}) for higher order corrections is 
mainly valid only for the infrared divergent terms and the error caused by neglecting the nondivergent terms is estimated 
to be $<$ 1\% \cite{drell57,drell59,werthamer61,greenhut69,lewis56}. The contributions from vacuum polarization and electron vertex corrections are calculated exactly.
However, the contributions from TPE are limited only to the infrared divergent 
contributions (nondivergent terms have been neglected). Therefore, the contributions to $\delta_{int}$ can be written as:
\begin{equation} \label{RADC_delta_int}
\delta_{int} = \frac{-\alpha}{\pi}\Big(\frac{28}{9} - \frac{13}{6}ln(\frac{Q^2}{m^2_{e}}) + \delta_{int.bremss.}\Big)~,
\end{equation}
where $\delta_{int.bremss.}$ is the internal bremsstrahlung contribution and will be discussed next.

For the internal bremsstrahlung, the calculations are done under the assumption used by Tsai \cite{tsai61} 
that $\Delta E (1+\frac{2E}{M_{p}}) \ll E'$ where $M_{p}$ is the mass of the proton, $E$ is the incident energy of 
the electron, $E'$ is the final energy of the electron, and $\Delta E$ is the $E'$ cutoff of the elastic peak or 
$\Delta E = E_{elast} - E_{cutoff}$. The resulting expression is rather lengthy and complicated. 
In general it has a complicated dependence on the incident energy of the electron, final energy of the 
electron, final energy of the proton, and $\Delta E$. To achieve uncertainty better than 1\% in the radiative corrections,
corrections to this approximation were made by Walker \cite{walker94,walkerphd} as part of the improvements he made to the 
internal radiative corrections represented by $\acute{\delta_{int}}$. 

\subsection{Walker's Improved Internal Radiative Corrections $\acute{\delta_{int}}$}

For a full description of the improvements made to the internal radiative corrections, the reader
is referred to Walker \cite{walker94,walkerphd} where he discussed these corrections in detail. However,
I will just summarize briefly these improvements. The term $\acute{\delta_{int}}$ is just an improvement on
the precision to the original internal radiative correction. These improvements include correction for
the $\Delta E (1+\frac{2E}{M_{p}}) \ll E'$ approximation made by Tsai to the internal bremsstrahlung,
inclusion of the $q \bar{q}$ and $\mu^{+}\mu^{-}$ contributions to the vacuum polarization or loop diagrams which
were neglected previously, and correction for a sign error in Tsai's paper \cite{tsai61} for the Schwinger's correction 
\cite{schwinger49} to the noninfrared divergent part of the soft photon emission cross section. 

\subsection{External Radiative Corrections $\delta_{ext}$}
The external radiative corrections $\delta_{ext}$ are determined based on the modification made
by Walker \cite{walker94,walkerphd} and Ent \cite{ent01} to the work of Mo and Tsai \cite{mo69}. 
The external radiative corrections are corrections applied to the cross section due to bremsstrahlung 
in the target material (major contribution), spectrometer material, and the small effects of the Landau tail 
of the ionization energy loss spectrum \cite{landau44}.
When the incident electron interacts with the Coulomb field of a nucleus other than the one involved primary 
scattering vertex, it will emit bremsstrahlung and hence lose energy.
The cross section of an electron of initial energy $E$ that scatters elastically from a proton
and has a final energy $E' = E_{elast} - \Delta E'$, when the electron is emitting bremsstrahlung photons
with $t_{i}$ and $t_{f}$ radiation lengths of material before and after the scattering, is:
\begin{eqnarray} \label{eq:external_RADC_sigma}
\frac{d^2 \sigma(E,E')}{d\Omega dE'} = \Big(\frac{R \Delta E'}{E}\Big)^{b_{i}t_{i}} 
\Big(\frac{R \Delta E'}{E_{elast}}\Big)^{b_{f}t_{f}}\frac{1}{\Gamma (1 + b_{i}t_{i})} 
\frac{1}{\Gamma (1 + b_{f}t_{f})} {}
                                    \nonumber\\
\times \Big[\frac{d \sigma(E)}{d\Omega} \frac{b_{f}t_{f}}{\Delta E'} \phi\Big(\frac{\Delta E'}{E_{elast}}\Big) + 
\frac{d \sigma(E - R \Delta E')}{d\Omega} \frac{b_{I}t_{I}}{\Delta E'} \phi\Big(\frac{R \Delta E'}{E}\Big) \Big] {}~,
\end{eqnarray}
where the parameter $b \sim$ 3/4 \cite{tsai74}, the parameter $R$ represents the recoil of the proton and is 
$\approx (\frac{E}{E'})^2$, and the function $\phi\Big(\frac{R \Delta E'}{E}\Big) = \phi\Big(\frac{\omega}{E}\Big)$
gives the shape of the bremsstrahlung spectrum which is normalized to one at $\frac{\omega}{E}$ = zero where
$\omega$ is the lost energy of the electron after passing a thickness of $t$ radiation lengths. The external radiative
corrections $\delta_{ext}$ can be then expressed in terms of the electron cross section equation 
(\ref{eq:external_RADC_sigma}) as:
\begin{equation}
\delta_{ext} = ln\Bigg(\frac{1}{\frac{d \sigma(E)}{d\Omega}} \Big [LT_{corr}
\int_{E_{elast}-\Delta E'}^{E_{elast}} \frac{d^2 \sigma(E,E')}{d \Omega dE'} dE'\Big]\Bigg)~,
\end{equation}
where $LT_{corr}$ is the correction for the Landau tail as calculated from the Landau distribution \cite{landau44}. 
Note that the integrand diverges as $\Delta E' \to$ zero due to the soft photon bremsstrahlung spectrum. 
To handle this problem, the integral is usually calculated for $\Delta E'< \delta_{\gamma}$ where $\delta_{\gamma}$ is 
the soft-photon cut. When calculating $\delta_{ext}$, one needs an accurate geometrical model of the target along with 
precise knowledge of the materials of which the target is made. This information is built in the elastic e-p simulation 
code SIMC.

\subsection{E01-001 Specific Radiative Corrections}
As mentioned before, we use the radiative corrections procedure of Mo and Tsai as modified by Walker 
and Ent and implemented in the elastic e-p simulation code SIMC. Full coincidence (e,e'p) simulations 
are performed using the prescription of Ent \cite{ent01} taking into account bremsstrahlung from all three tails 
(bremsstrahlung from the incident electron, scattered electron, and scattered proton). 
The bremsstrahlung from these particles is calculated in the extended peaking approximation \cite{ent01,naomithesis}. 
Furthermore, only the contribution from the infrared divergent part of TPE is included. 

Based on equation (\ref{eq:reduced}) and because of the factor $\tau$ that multiplies $G^2_{Mp}$,
the cross section is dominated by $G^2_{Mp}$ at large $Q^2$. That is, $\sigma_{R}$ in the Born approximation has 
only a small $\varepsilon$ dependence at large $Q^2$. Therefore, understanding the $\varepsilon$ dependence of the radiative 
correction $\delta_{corr}$ at large $Q^2$ becomes absolutely important as a few percent change in the
$\varepsilon$ slope of $\sigma_{R}$ would have a sizable effect on $\frac{\mu_{p}G_{Ep}}{G_{Mp}}$.

The main $\varepsilon$ dependence to the radiative corrections comes from the internal and external bremsstrahlung corrections 
(and possibly the neglected TPE terms). Bremsstrahlung corrections are usually accounted for in the conventional radiative 
corrections procedure used and they enter differently depending on 
whether the electron or proton are detected in the final state. On the other hand, several calculations in 
the 1950s and 1960s estimated the size of TPE to be extremely small ($\leq$ 1\%) 
\cite{drell57,drell59,werthamer61,greenhut69,lewis56} and therefore was neglected in the conventional radiative corrections 
procedure used. An important point is that the infrared divergent part of the TPE contribution cancels fully the 
corresponding infrared divergent contribution from the interference between the electron 
and proton bremsstrahlung. Furthermore, the finite part of TPE has always been neglected. Recently it has been found that 
the finite part of TPE has a significant $\varepsilon$ dependence and it should not be neglected 
\cite{blunden03,blunden05,kondratyuk05}.  
Figure \ref{fig:rad_corr_epsilon} shows the radiative correction factor calculated for internal contributions only 
as a function of $\varepsilon$ for Q$^2$ = 2.64 GeV$^2$. While the magnitude of the corrections is similar and both show 
an approximately linear dependence on $\varepsilon$, the dependence is much smaller for protons, $\sim -$8\%, than for the 
electron $\sim$ 17\%. Note that when detecting the electron, the $\varepsilon$ dependence of the bremsstrahlung 
correction exceeds the $\varepsilon$ dependence coming from $G_{Ep}$ which is not the case in the E01-001 
experiment. 

Previous Rosenbluth measurements \cite{walker94,andivahis94,christy04} estimated scale and random uncertainties in the 
radiative corrections of $\sim$ 1.0\% and $\sim$ 0.60\%, respectively. In these measurements, electrons rather than
protons were detected. In our case, we assign the same scale uncertainty of 1.0\% for both arms and take a 50\% 
of the random uncertainty as quoted by the previous measurements or $\sim$ 0.30\%. 
This is due to the fact that detecting protons will yield a smaller $\varepsilon$ dependence bremsstrahlung correction
as discussed above.
We split the 0.30\% uncertainty into a slope and random uncertainties with more emphasize on the slope contribution of 
0.30\% and random contribution of 0.20\%. For the right arm, the same random uncertainty of 0.20\% 
is assigned, while the slope uncertainty is expected to be small due to the small $\Delta \varepsilon$ range of 0.07.
Note that for the left arm, the average $\Delta \varepsilon$ range of 0.70 is accounted for when the slope uncertainty 
is estimated. Therefore, we estimate the slope uncertainty for the right arm to be 3.0\% ($0.30\%\times0.70/0.07$) but use
a smaller and more realistic figure of 2.0\% since the radiative corrections are expected to be small in a such limited 
$\Delta \varepsilon$ range.

Finally and based on Figure \ref{fig:rad_corr_epsilon}, a relative $\sim$ 30\% radiative correction 
is needed to bring the cross sections measured by detecting electrons and those measured by 
detecting protons into agreement. That is, before the radiative corrections are applied the electron and proton
cross sections have an $\varepsilon$ dependence that differs by as much as $\sim$ 30\%.
The extracted slope from the two measurements after the correction is applied agree to within uncertainty of 1-2\%. 
If we assume an average uncertainty in the slope of 1.5\%, 
this yields a fractional uncertainty in a 30\% correction of 1.5\%/30 = 5\%. For the proton, a 5\% uncertainty in a 
10\% $\varepsilon$ dependence bremsstrahlung correction yields a $\sim$ 0.50\% uncertainty which is comparable to the 
combined 0.36\% uncertainty estimated for the E01-001 measurements. It is important to mention that this is the first
time a comparison between the electrons and protons measurements has been made.  
The excellent agreement in the extracted slope or $G^2_{Ep}$ from the two measurements provides a strong test of the 
validity of the standard radiative corrections used.

\section{The e-p Reduced Cross Section $\sigma_{R}$} \label{sigma_R}

\subsection{Overview}

In this section the procedure for extracting $\sigma_{R}$ is discussed. Figure \ref{fig:e01001_flow_chart} 
shows a flow chart of the procedure used. From Figures \ref{fig:uncorr_rdelta_p} and \ref{fig:uncorr_ldelta_p}, we see 
that the LH$_2$ spectrum is dominated by the elastic e-p peak. In addition, there are backgrounds due to 
quasi-elastic scattering from the aluminum target windows and protons generated from photoreactions 
$\gamma p \to \pi^0 p$ and $\gamma p \to \gamma p $ that contribute to the $\Delta P$ spectrum.
There is no region where only one component, the elastic peak or inelastic backgrounds, is significant. 
Therefore, the spectra has to be unraveled and this is done with aid of the simulated spectra but first the 
simulated and measured spectra have to be put on the same energy scale as will be discussed below.

\subsection{The Effective Dummy Thickness and Endcaps Subtraction} \label{dummy_extraction}

\begin{figure}[!htbp]
\begin{center}
\epsfig{file=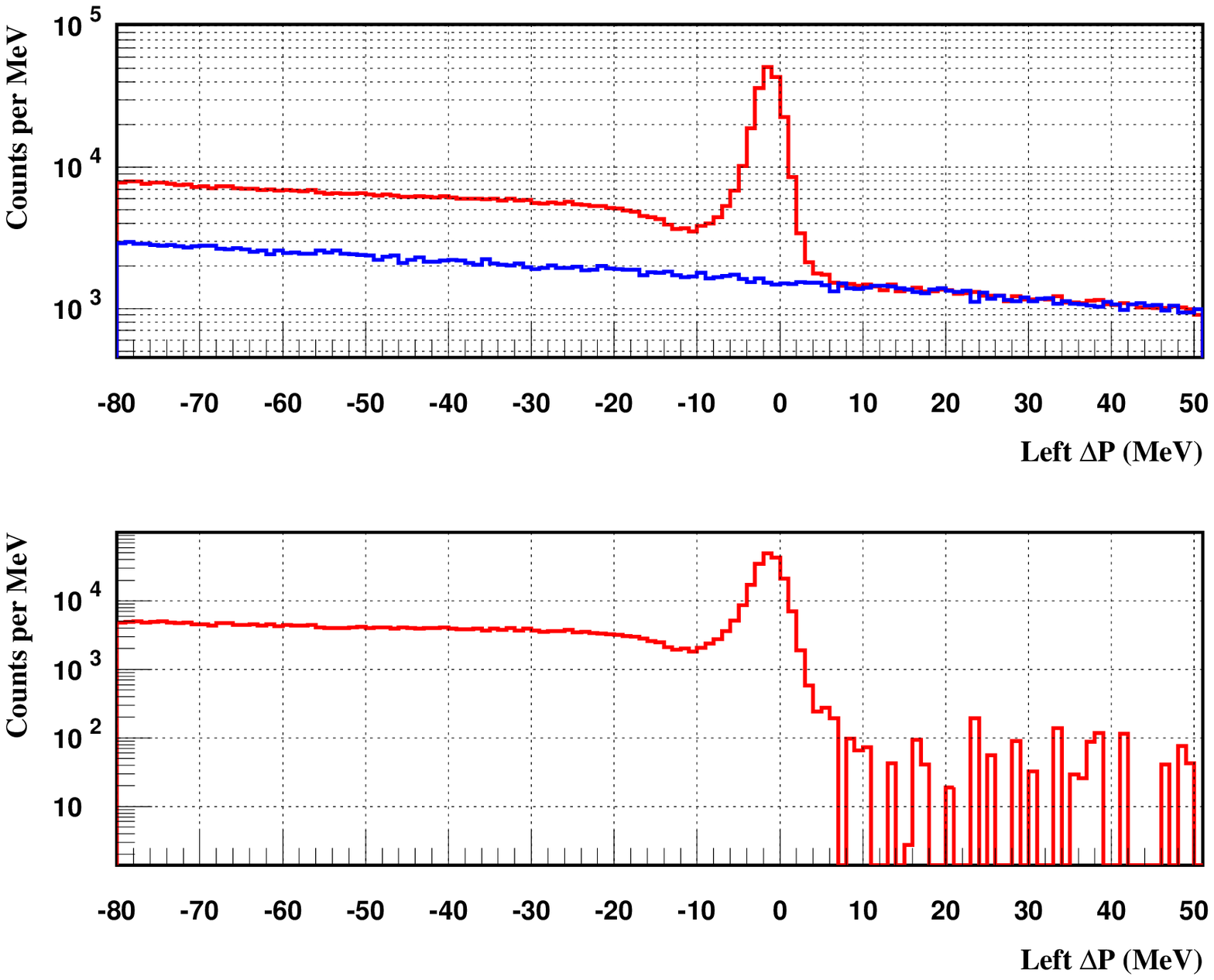,width=6in} 
\end{center}
\caption[The initial left arm dummy subtraction for kinematics b.]
{The initial left arm dummy subtraction for kinematics b (low $\varepsilon$). 
Top: the left arm LH$_2$ $\Delta P$ spectrum (red) and dummy $\Delta P$ spectrum (blue).
Bottom: the dummy subtracted LH$_2$ $\Delta P$ spectrum. An initial dummy thickness factor of 4.11 was used.}
\label{fig:kinb_lunshift_lh2_dummy_diff}
\end{figure}

\begin{figure}[!htbp]
\begin{center}
\epsfig{file=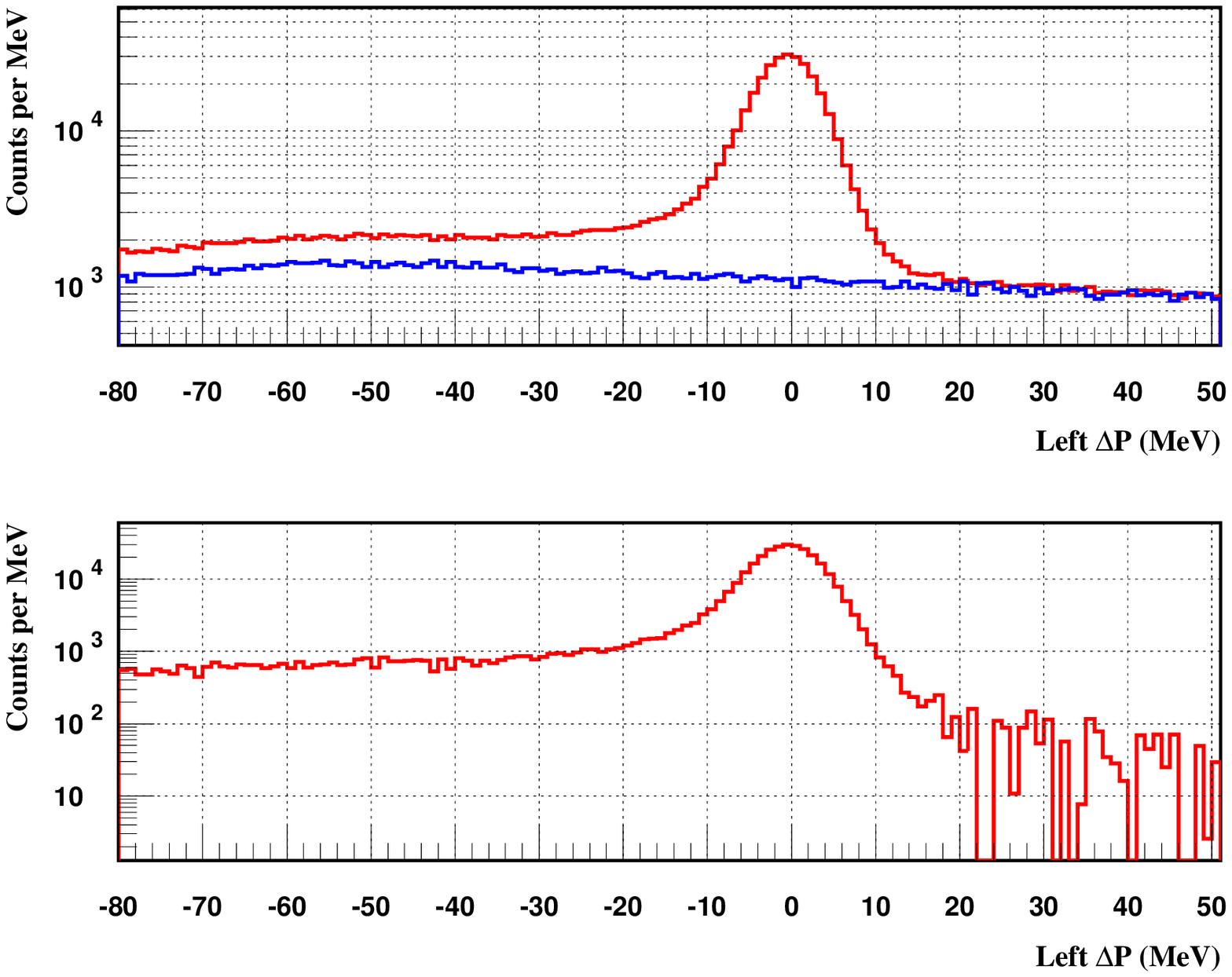,width=6in} 
\end{center}
\caption[The initial left arm dummy subtraction for kinematics m.]
{The initial left arm dummy subtraction for kinematics m (high $\varepsilon$). 
Top: the left arm LH$_2$ $\Delta P$ spectrum (red) and dummy $\Delta P$ spectrum (blue).
Bottom: the dummy subtracted LH$_2$ $\Delta P$ spectrum. An initial dummy thickness factor of 4.11 was used.}
\label{fig:kinm_lunshift_lh2_dummy_diff}
\end{figure}

\begin{figure}[!htbp]
\begin{center}
\epsfig{file=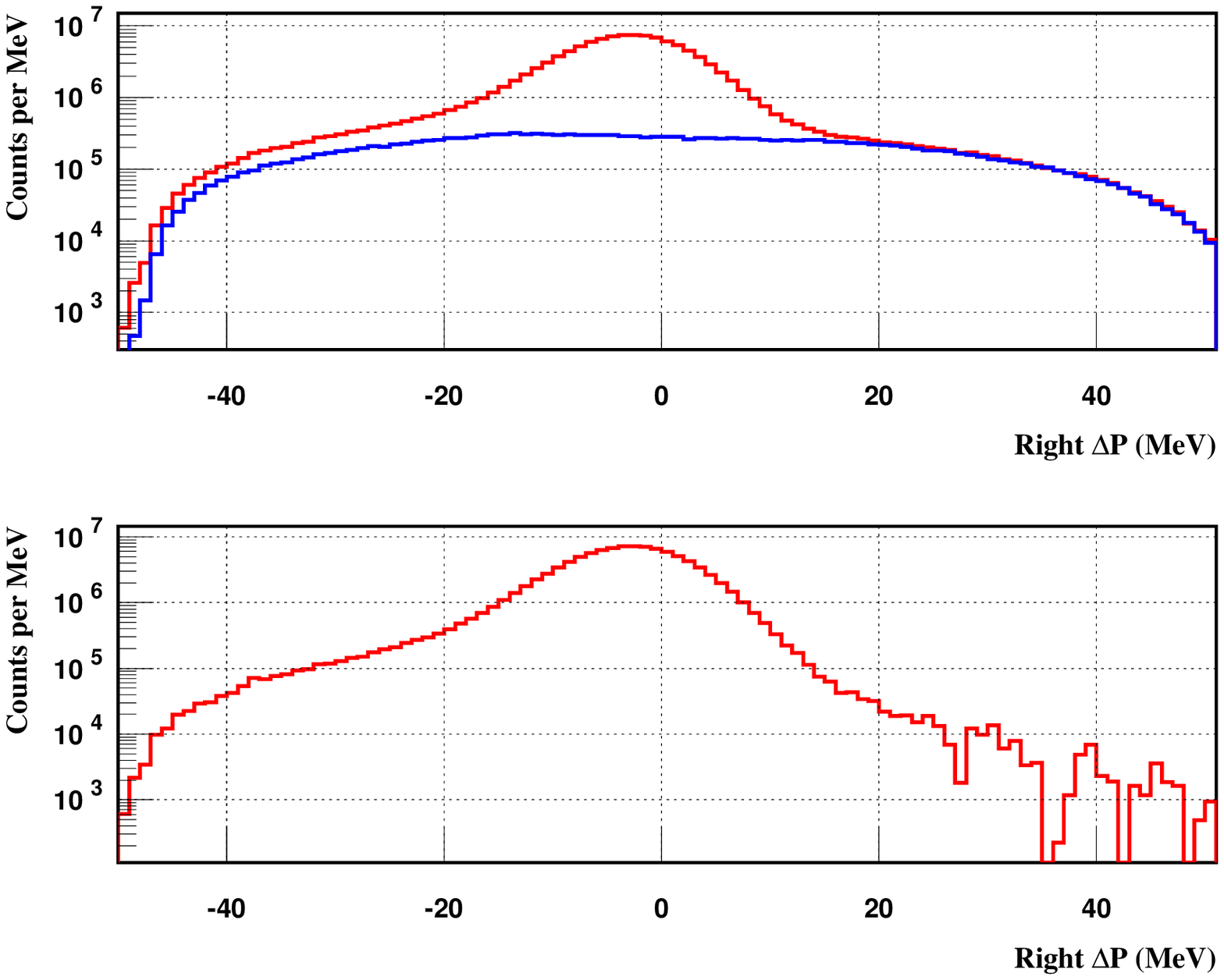,width=6in} 
\end{center}
\caption[The initial right arm dummy subtraction for kinematics b.]
{The initial right arm dummy subtraction for kinematics b. Top: the right arm LH$_2$ 
$\Delta P$ spectrum (red) and dummy $\Delta P$ spectrum (blue).
Bottom: the dummy subtracted LH$_2$ $\Delta P$ spectrum. An initial dummy thickness factor of 4.11 was used.}
\label{fig:kinb_runshift_lh2_dummy_diff}
\end{figure}

In this section the procedure of extracting the effective dummy thickness will be discussed. We start by 
subtracting the $Q_{eff}$ corrected dummy spectrum using an initial dummy thickness factor  
($\frac{\mbox{dummy}}{\mbox{endcap}}$ ratio)
of 4.11 from the $Q_{eff}$ corrected LH$_2$ spectrum. 
Figures \ref{fig:kinb_lunshift_lh2_dummy_diff}, \ref{fig:kinm_lunshift_lh2_dummy_diff}, 
and \ref{fig:kinb_runshift_lh2_dummy_diff} show this procedure for a low $\varepsilon$ left arm
spectrum, a high $\varepsilon$ left arm spectrum, and a right arm spectrum, respectively.

\begin{figure}[!htbp]
\begin{center}
\epsfig{file=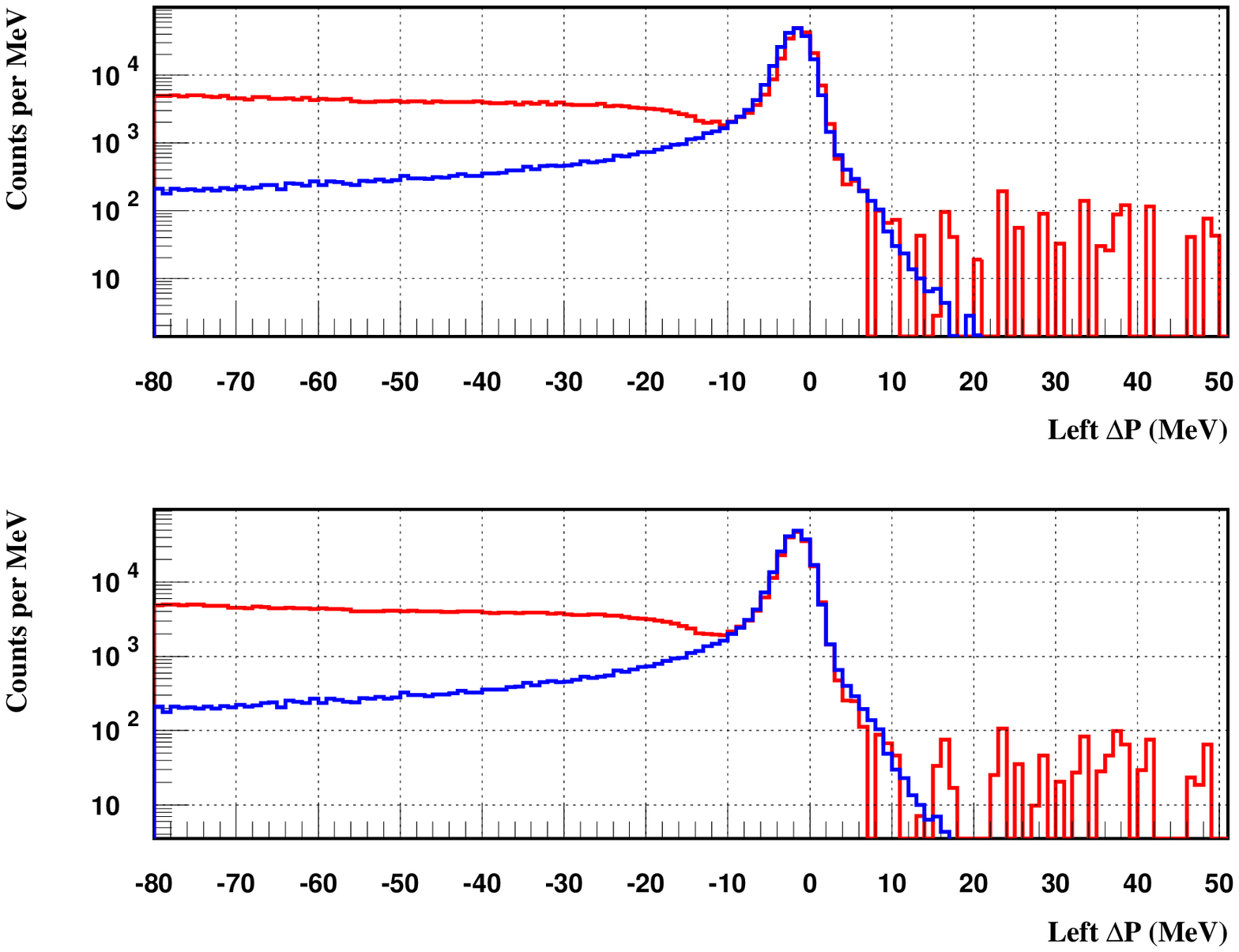,width=6in} 
\end{center}
\caption[The left arm dummy subtracted LH$_2$ $\Delta P$ spectrum and elastic e-p simulation for kinematics b.]
{The left arm dummy subtracted LH$_2$ $\Delta P$ spectrum and elastic e-p simulation for kinematics b (low $\varepsilon$).
Top: the left arm unshifted dummy subtracted LH$_2$ $\Delta P$ spectrum (red) and elastic e-p simulation 
(blue). 
Bottom: the left arm shifted dummy subtracted LH$_2$ $\Delta P$ spectrum (red) and elastic e-p simulation (blue). 
This data are from kinematics $b$, using the nominal incident energy and scattering angle. The final
version after applying the angular shift to the scattering angle (see section \ref{spect_mispoint}) yields much smaller 
shift.}
\label{fig:kinb_lunshift_shift_lh2_ep}
\end{figure}

\begin{figure}[!htbp]
\begin{center}
\epsfig{file=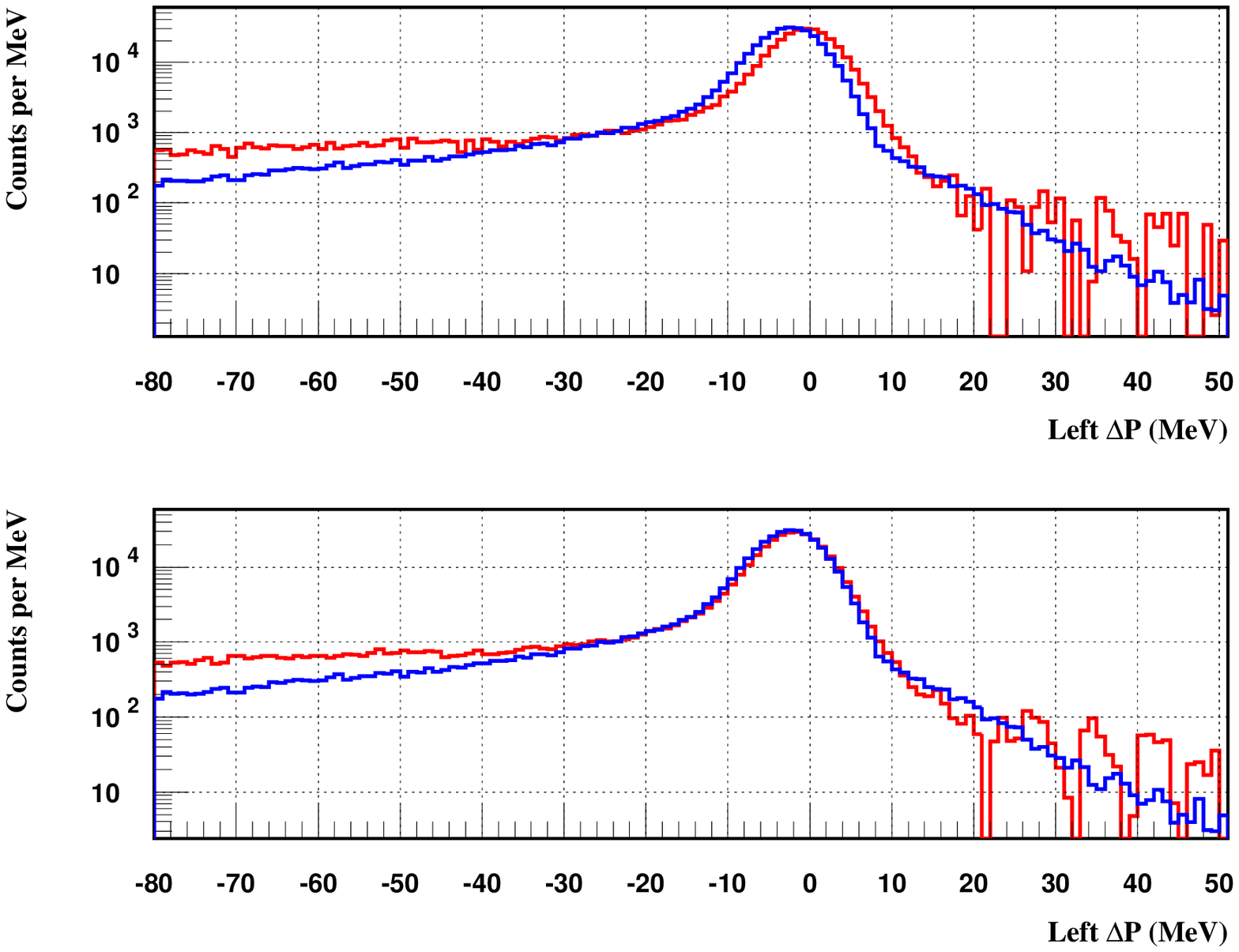,width=6in} 
\end{center}
\caption[The left arm dummy subtracted LH$_2$ $\Delta P$ spectrum and elastic e-p simulation for kinematics m.]
{The left arm dummy subtracted LH$_2$ $\Delta P$ spectrum and elastic e-p simulation for kinematics m (high $\varepsilon$).
Top: the left arm unshifted dummy subtracted LH$_2$ $\Delta P$ spectrum (red) and elastic e-p 
simulation (blue).
Bottom: the left arm shifted dummy subtracted LH$_2$ $\Delta P$ spectrum (red) and elastic e-p simulation (blue). 
This data are from kinematics $m$, using the nominal incident energy and scattering angle. The final 
version after applying the angular shift to the scattering angle (see section \ref{spect_mispoint}) yields much smaller 
shift.}
\label{fig:kinm_lunshift_shift_lh2_ep}
\end{figure}

\begin{figure}[!htbp]
\begin{center}
\epsfig{file=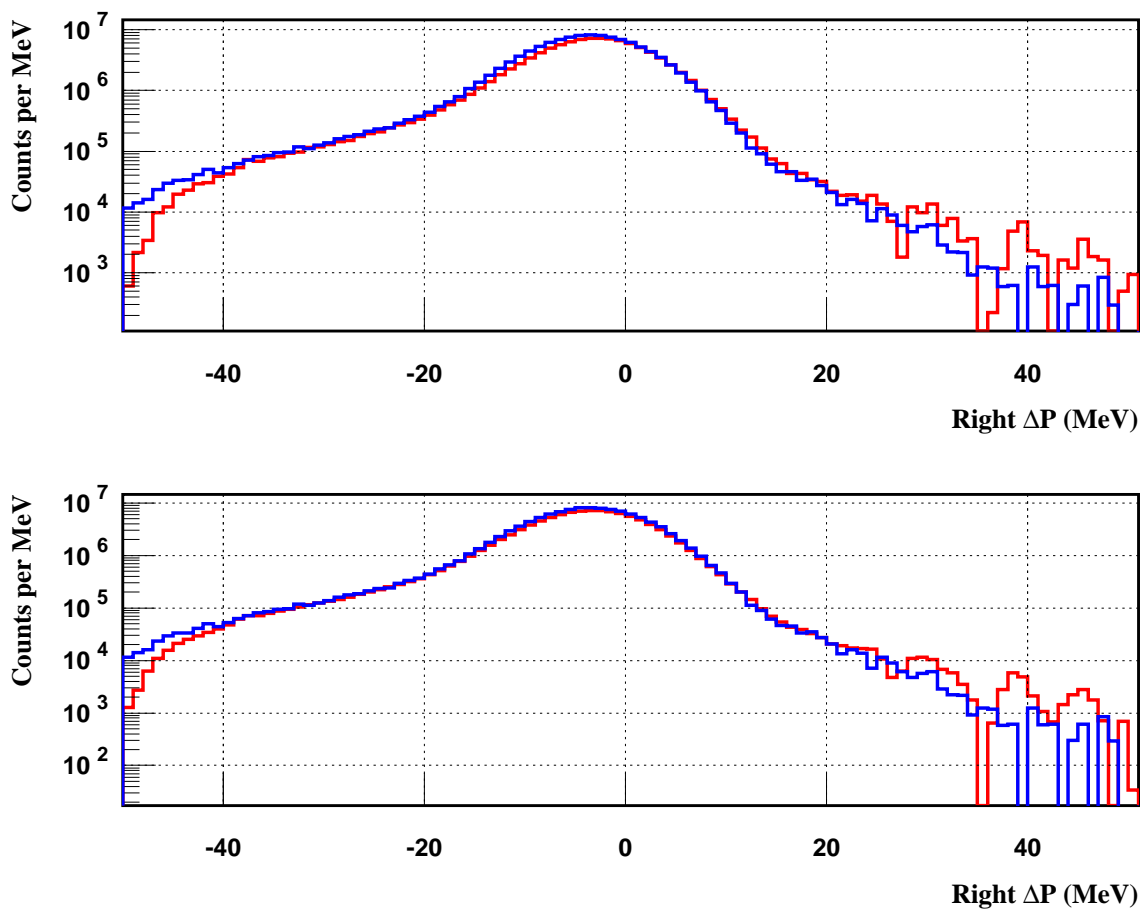,width=6in} 
\end{center}
\caption[The right arm dummy subtracted LH$_2$ $\Delta P$ spectrum and elastic e-p simulation for kinematics b.]
{The right arm dummy subtracted LH$_2$ $\Delta P$ spectrum and elastic e-p simulation for kinematics b.
Top: the right arm unshifted dummy subtracted LH$_2$ $\Delta P$ spectrum (red) and elastic e-p simulation 
(blue). Bottom: the right arm shifted dummy subtracted LH$_2$ $\Delta P$ spectrum (red) and elastic e-p  
simulation (blue). This data are from kinematics $b$.}
\label{fig:kinb_runshift_shift_lh2_ep}
\end{figure}

The resultant LH$_2$ $\Delta P$ spectrum is then compared to that of elastic e-p simulation 
as generated by SIMC. The experimental e-p peak is shifted relative to that of the simulation. 
Therefore, we shift the position of the e-p peak from data to match that of simulation. 
Figures \ref{fig:kinb_lunshift_shift_lh2_ep}, \ref{fig:kinm_lunshift_shift_lh2_ep},
and \ref{fig:kinb_runshift_shift_lh2_ep} illustrate this procedure.  
The contribution of the $\gamma p \to \pi^{0} p$ and $\gamma p \to \gamma p$ backgrounds to the  
LH$_2$ $\Delta P$ spectrum is the difference between the dummy subtracted LH$_2$ $\Delta P$ spectrum and elastic e-p  
simulation. From Figure \ref{fig:kinb_runshift_shift_lh2_ep} we see that the contribution of these backgrounds 
to the right arm LH$_2$ $\Delta P$ spectrum is almost negligible. 
The momentum acceptance of the spectrometers is $\pm$5\%. For the higher momentum protons detected in the left arm, the
momentum acceptance is great enough to be flat over the region of interest. However, for the low momentum protons detected
in the right arm, the falling acceptance at the lowest $\Delta P$ results in a distortion of the spectrum at the low momentum
end. This region was not used in extracting the cross sections. However, because the shape of the right arm spectrum is almost
the same at all the kinematics, this distortion would only introduce a scale uncertainty in the reduced cross sections.
For the left arm, Figures \ref{fig:kinb_lunshift_shift_lh2_ep} and 
\ref{fig:kinm_lunshift_shift_lh2_ep} show that the contribution of these backgrounds to the LH$_2$ $\Delta P$ spectrum is 
significant. As the cross sections of the $\pi^{0}$ protons are forward peaked, the contribution to the LH$_2$ 
$\Delta P$ spectrum is larger for low $\varepsilon$ than for large $\varepsilon$. 

\begin{figure}[!htbp]
\begin{center}
\epsfig{file=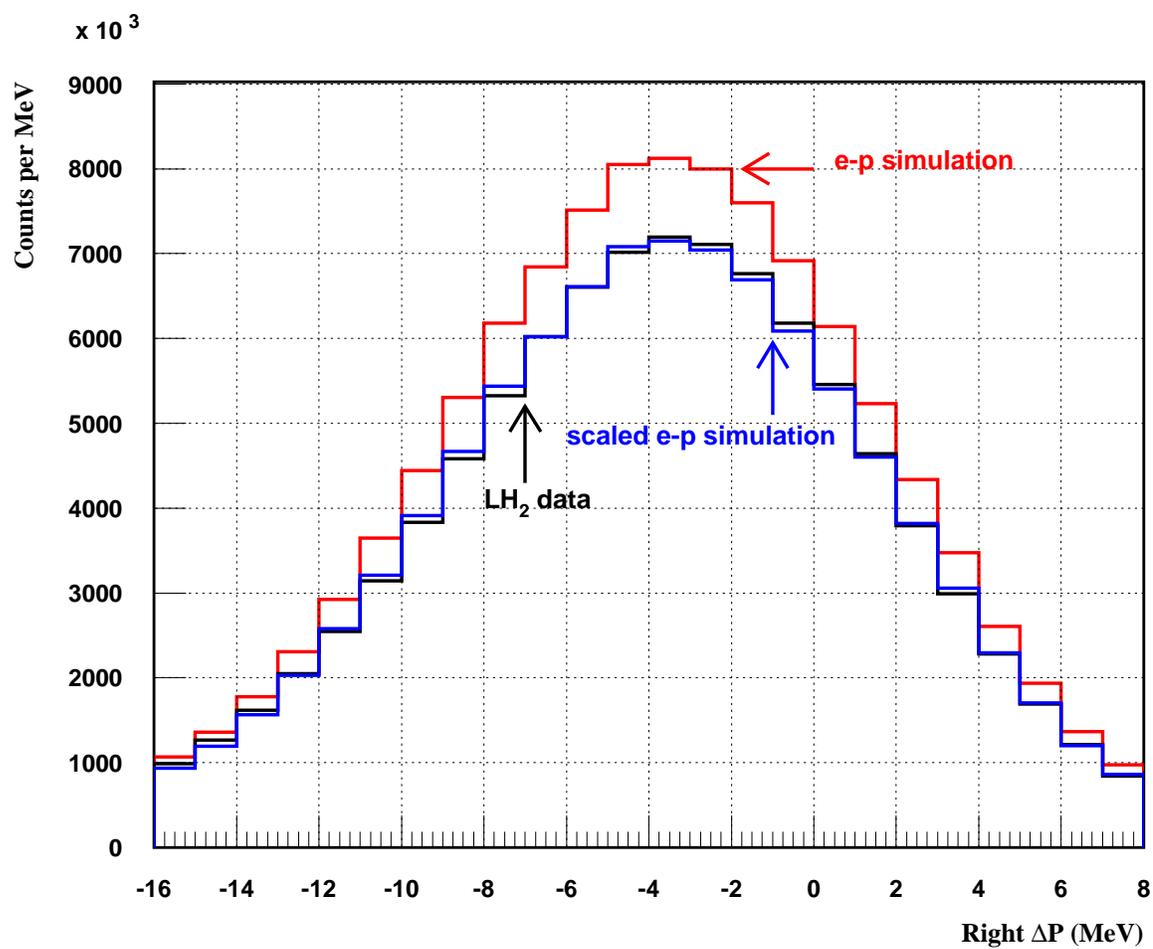,width=6in} 
\end{center}
\caption[Right arm $\Delta P$ spectrum for LH$_2$ data, elastic e-p simulation, and scaled
elastic e-p simulation for kinematics $b$.]
{Right arm $\Delta P$ spectrum for LH$_2$ data (black), elastic e-p simulation (red), and scaled
elastic e-p simulation (blue) for kinematics $b$.}
\label{fig:rlh2_ep_scaled}
\end{figure}

\begin{figure}[!htbp]
\begin{center}
\epsfig{file=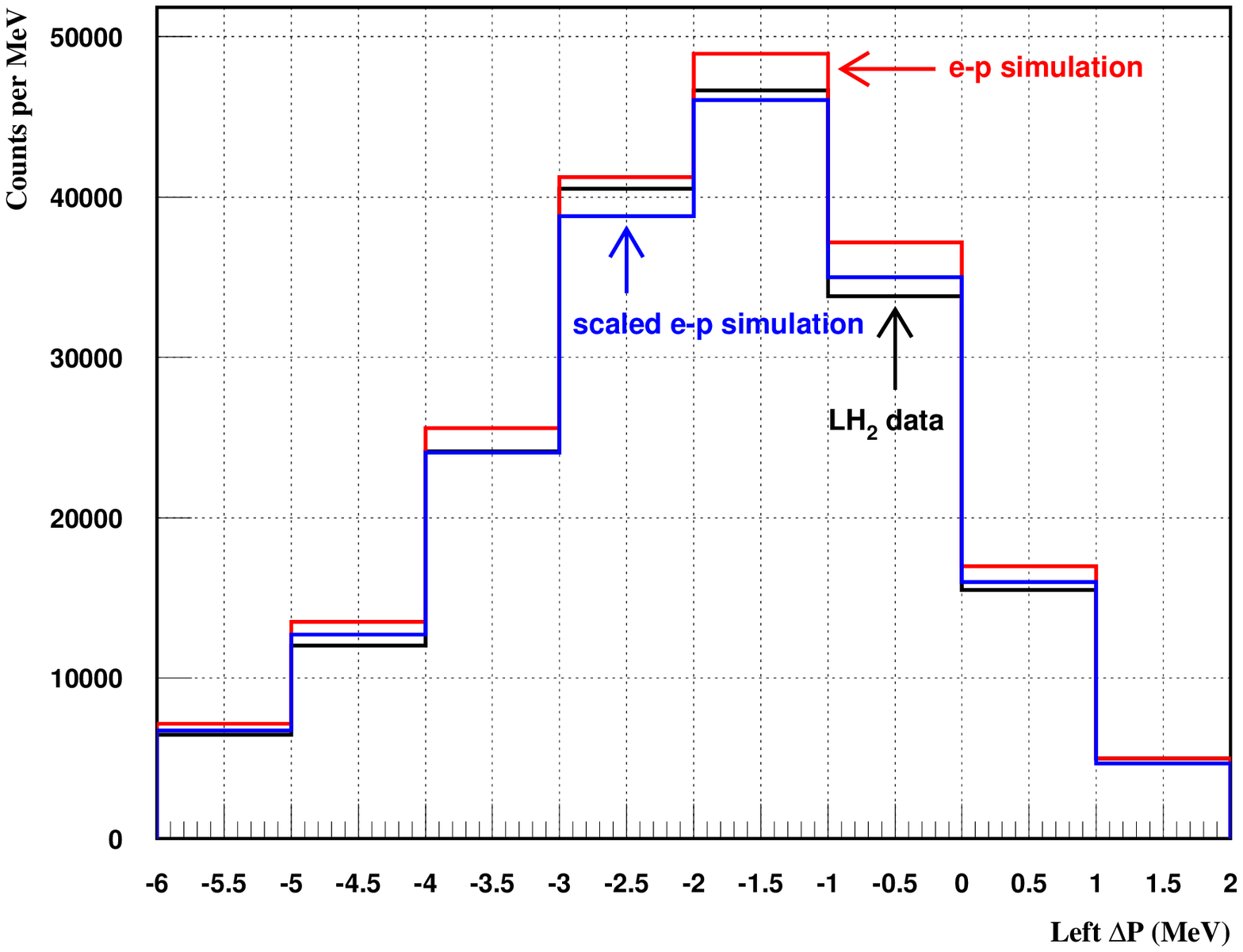,width=6in} 
\end{center}
\caption[Left arm $\Delta P$ spectrum for LH$_2$ data, elastic e-p simulation, and scaled
elastic e-p simulation for kinematics $b$.]
{Left arm $\Delta P$ spectrum for LH$_2$ data (black), elastic e-p simulation (red), and scaled
elastic e-p simulation (blue) for kinematics $b$.}
\label{fig:llh2_ep_scaled}
\end{figure}

After we shift the position of the e-p peak from data to match that of the simulation, we compare the two $\Delta P$ spectra over 
a region where the elastic e-p simulation $\Delta P$ bin content is larger than 10\% of that of the peak. 
The number of LH$_2$ events, $N_{LH_{2}}$, and number of e-p elastic, $N_{e-p}$, for that window is used to determine the scaling 
factor $\gamma_{e-p} = (N_{LH_{2}}/N_{e-p})$ needed to scale the elastic e-p simulation to 
match that of LH$_2$. Figures \ref{fig:rlh2_ep_scaled} and \ref{fig:llh2_ep_scaled} illustrate
this procedure.

\begin{figure}[!htbp]
\begin{center}
\epsfig{file=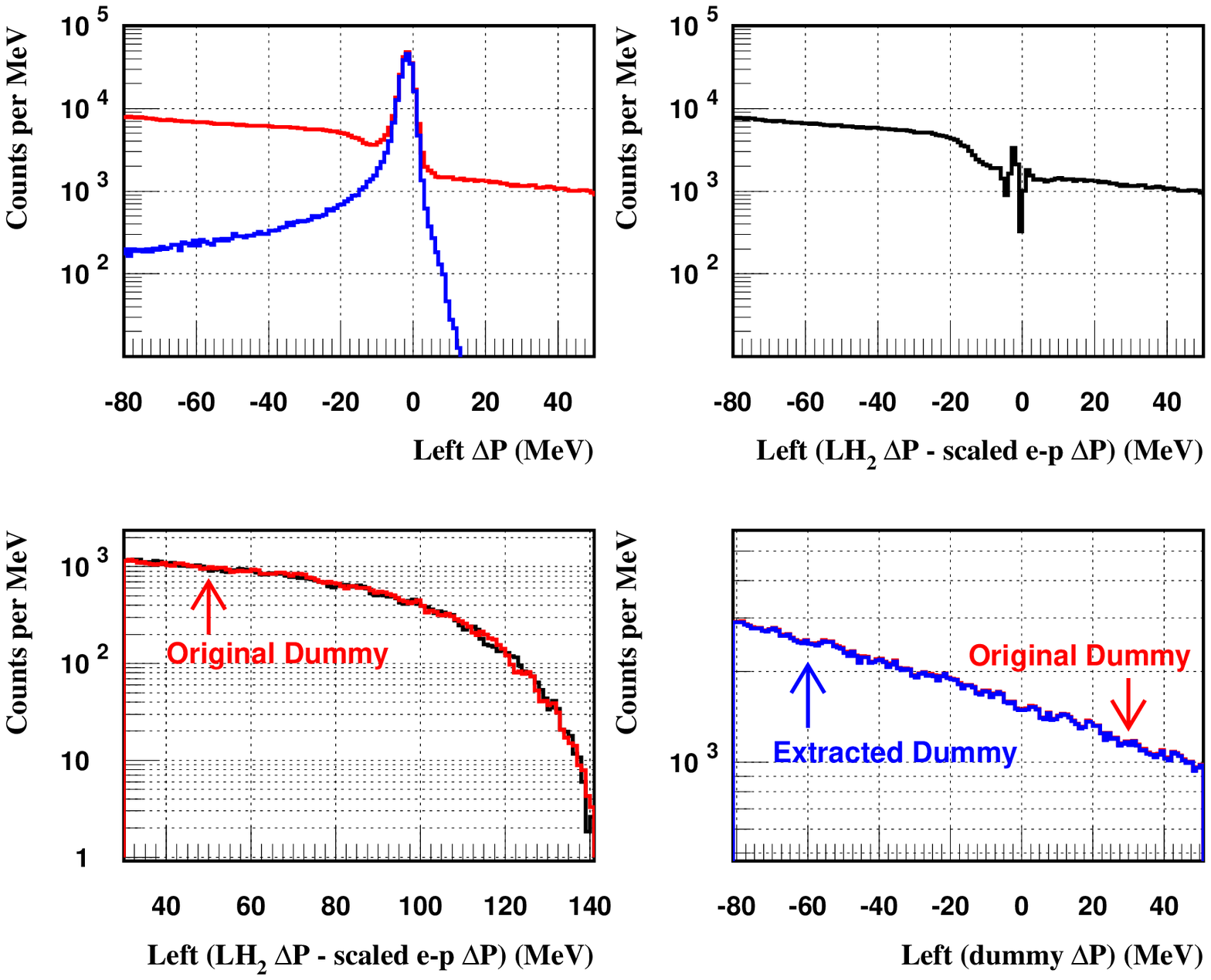,width=6in} 
\end{center}
\caption[The left arm effective dummy target thickness extraction for kinematics $b$.]
{The left arm effective dummy target thickness extraction for kinematics $b$.
Top left: The left arm dummy subtracted LH$_2$ $\Delta P$ spectrum (red) and scaled elastic e-p  
simulation (blue). Top right: the left arm dummy subtracted LH$_2$ $\Delta P$ spectrum minus scaled 
e-p $\Delta P$ spectrum. 
Bottom left: The difference between the left arm dummy subtracted LH$_2$ $\Delta P$ spectrum and 
scaled e-p $\Delta P$ spectrum (black) is then normalized to the left arm original dummy $\Delta P$ 
spectrum (red) in the range of 30-140 MeV. Bottom right: The left arm original dummy spectrum (red) and 
the new extracted dummy spectrum (blue).} 
\label{fig:kinb_get_ldummy}
\end{figure}

\begin{figure}[!htbp]
\begin{center}
\epsfig{file=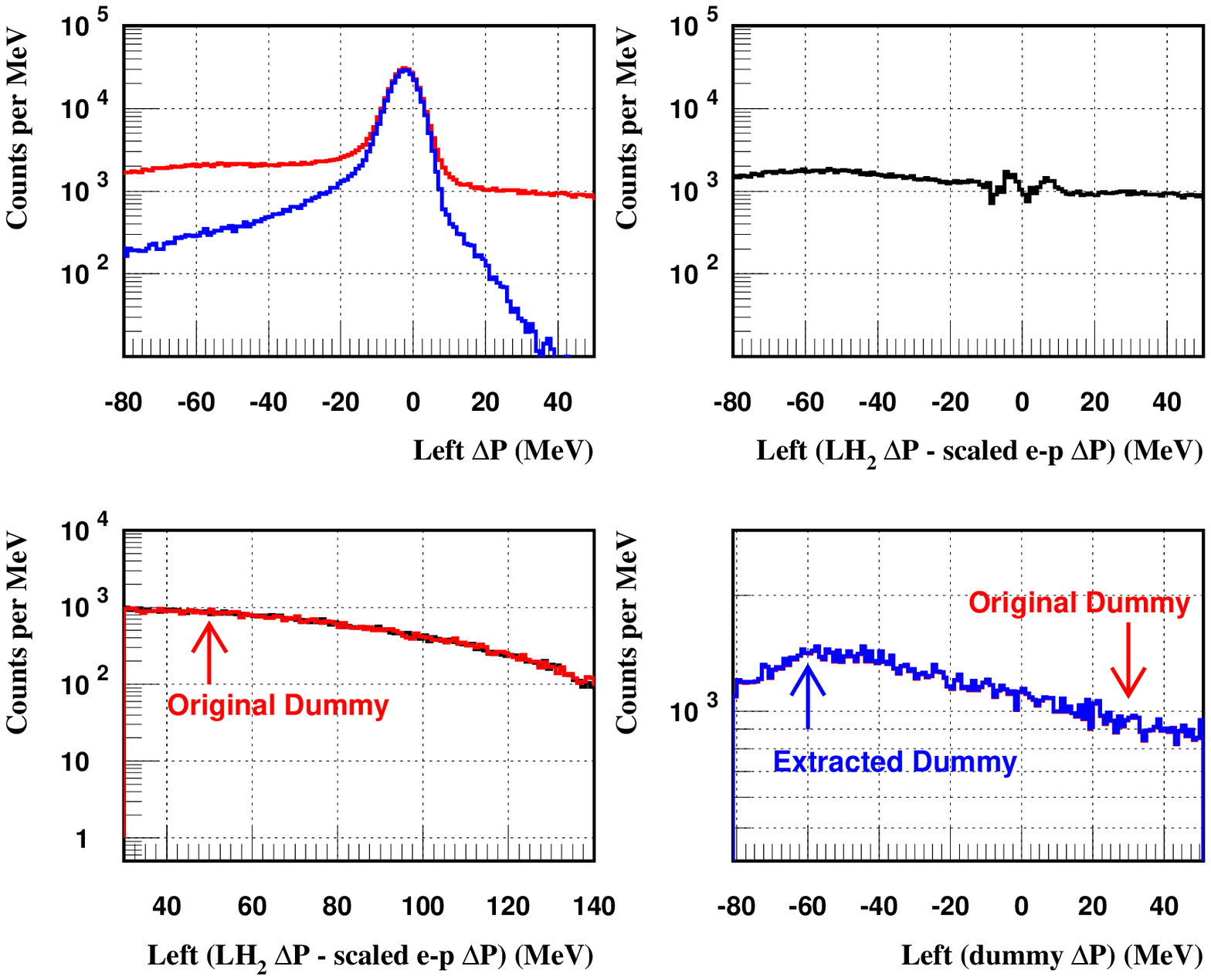,width=6in} 
\end{center}
\caption[The left arm effective dummy target thickness extraction for kinematics $m$.]
{The left arm effective dummy target thickness extraction for kinematics $m$.
Top left: The left arm dummy subtracted LH$_2$ $\Delta P$ spectrum (red) and scaled elastic e-p  
simulation (blue). Top right: the left arm dummy subtracted LH$_2$ $\Delta P$ spectrum minus scaled 
e-p $\Delta P$ spectrum. 
Bottom left: The difference between the left arm dummy subtracted LH$_2$ $\Delta P$ spectrum and 
scaled e-p $\Delta P$ spectrum (black) is then normalized to the left arm original dummy $\Delta P$ 
spectrum (red) in the range of 30-140 MeV. Bottom right: The left arm original dummy spectrum (red) and 
the new extracted dummy spectrum (blue).} 
\label{fig:kinm_get_ldummy}
\end{figure}

\begin{figure}[!htbp]
\begin{center}
\epsfig{file=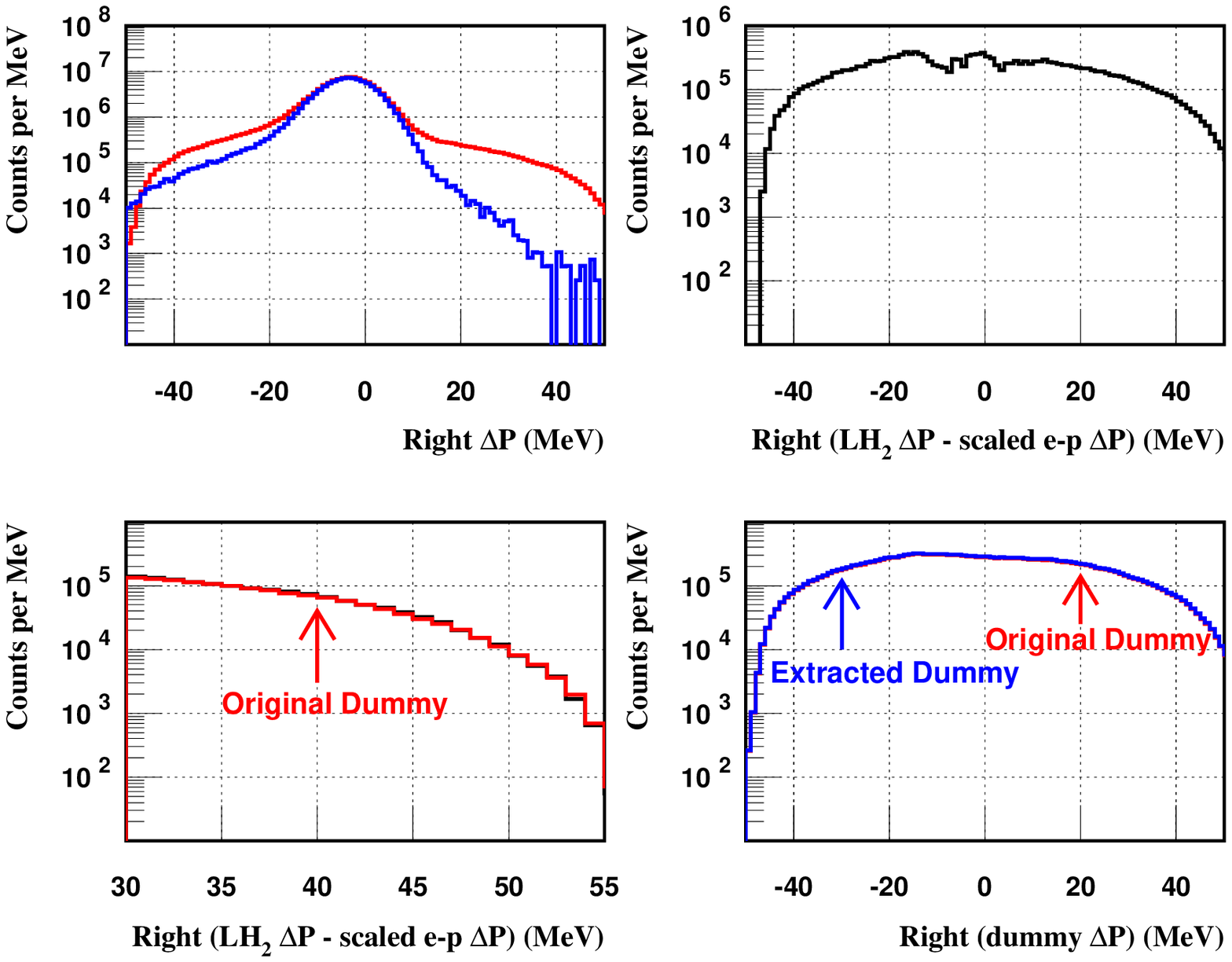,width=6in} 
\end{center}
\caption[The right arm effective dummy target thickness extraction for kinematics $b$.] 
{The right arm effective dummy target thickness extraction for kinematics $b$.
Top left: The right arm dummy subtracted LH$_2$ $\Delta P$ spectrum (red) and scaled elastic e-p  
simulation (blue). Top right: the right arm dummy subtracted LH$_2$ $\Delta P$ spectrum minus scaled 
e-p $\Delta P$ spectrum. 
Bottom left: The difference between the right arm dummy subtracted LH$_2$ $\Delta P$ spectrum and 
scaled e-p $\Delta P$ spectrum (black) is then normalized to the right arm original dummy $\Delta P$ 
spectrum (red) in the range of 30-55 MeV. Bottom right: The right arm original dummy spectrum (red) and 
the new extracted dummy spectrum (blue).} 
\label{fig:kinb_get_rdummy}
\end{figure}

Knowing that the contribution from the elastic LH$_2$ events to the LH$_2$ $\Delta P$ spectrum at large 
$\Delta P$ is small, the shape of the endcaps spectrum can be reliably determined by normalizing the 
difference between the LH$_2$ $\Delta P$ and that of the scaled elastic e-p simulation 
``extracted dummy'' to that of the original $Q_{eff}$ corrected dummy spectrum in the range of $\Delta P$$>$30 
MeV. Figures \ref{fig:kinb_get_ldummy}, \ref{fig:kinm_get_ldummy}, and \ref{fig:kinb_get_rdummy}
illustrate this procedure. 

\begin{figure}[!htbp]
\begin{center}
\epsfig{file=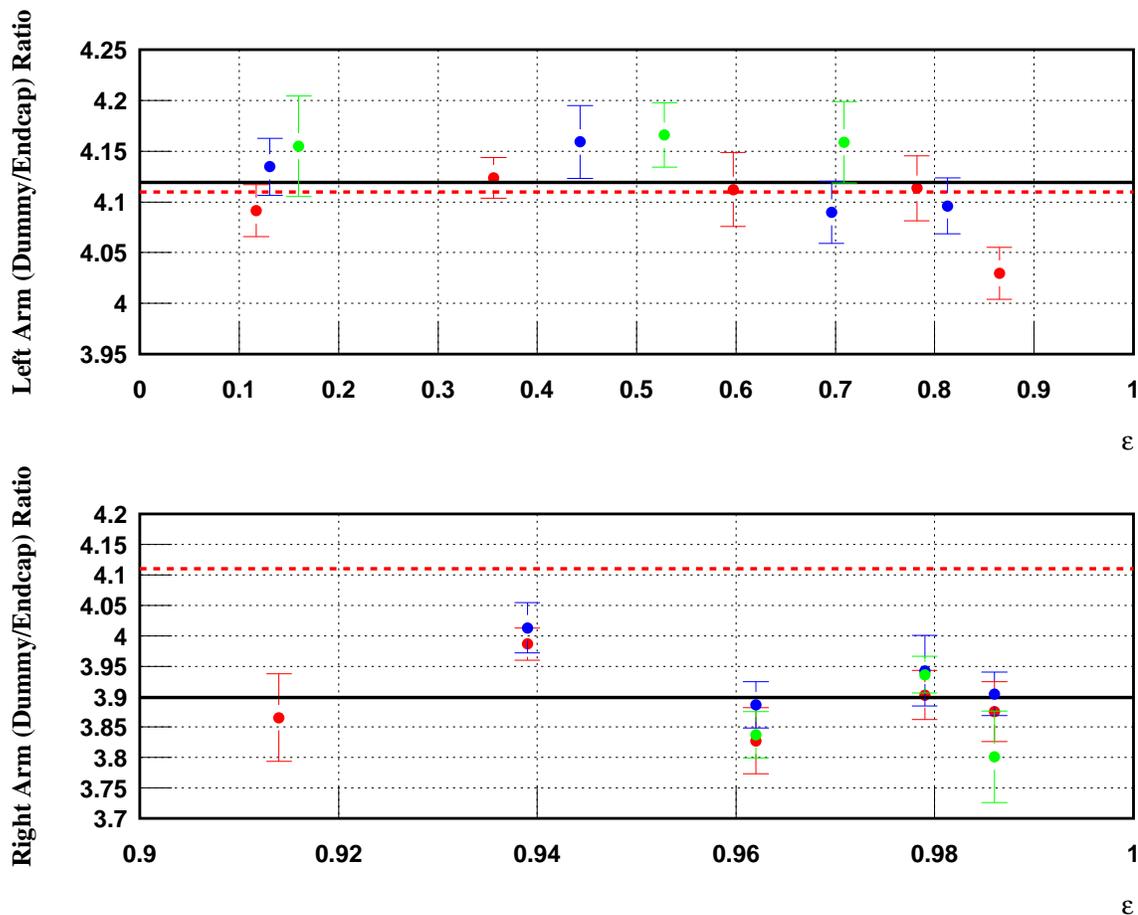,width=6in} 
\end{center}
\caption[The left and right arms effective dummy thickness.]
{The left and right arms effective dummy thickness.
Top: The left arm dummy/endcap ratio plotted as a function of increasing $\varepsilon$
for the Q$^2$ = 2.64 GeV$^2$ kinematics (solid red circles), Q$^2$ = 3.20 GeV$^2$ kinematics 
(solid blue circles), and  Q$^2$ = 4.10 GeV$^2$ kinematics (solid green circles). The solid
black line is the unweighted average dummy/endcap ratio and the red dashed line is the initial default 
value used of 4.11. 
Bottom: The right arm dummy/endcap ratio plotted as a function of increasing $\varepsilon$
for all the Q$^2$ = 0.5 GeV$^2$ kinematics. The same color was assigned on data so the kinematics
on the right arm can be associated with their left arm counterpartner. Again, the solid black 
line is the average dummy/endcap ratio and the red dashed line is the initial default value used 
of 4.11. The uncertainty in the data is statistical only.}
\label{fig:effective_dummy}
\end{figure}

The number of events in the extracted dummy spectrum, $N_{extracted}$, and the number of events in the 
original dummy spectrum, $N_{original}$, for that window is used to determine the correction factor 
$\gamma_{dummy} = (N_{original}/N_{extracted})$ needed to correct the initial default value of
the dummy thickness and hence define the new effective dummy thickness:
\begin{equation}
\mbox{Effective Dummy Thickness}~= \frac{\mbox{Default Dummy Thickness}}{\gamma_{dummy}}= \frac{4.11}{\gamma_{dummy}},
\end{equation}
which is used to correct $Q_{eff}$ for the dummy and hence scale the dummy target spectrum
for the final subtraction. The original and extracted dummy spectra are shown for both arms in the 
bottom right hand side plots of Figures \ref{fig:kinb_get_ldummy}, \ref{fig:kinm_get_ldummy}, 
and \ref{fig:kinb_get_rdummy}. Figure \ref{fig:effective_dummy} shows
the effective dummy thickness for all kinematics and for both arms.
For the left arm, the effective dummy thicknesses are consistent with random 
fluctuation of (1-2)\% from the average. The average effective dummy thickness is 0.22\% thicker 
than the default dummy thickness used initially. For the right arm, the effective dummy thicknesses 
are consistent with random fluctuation of (2-3)\% from the average, but the average effective dummy 
thickness is 5.15\% less than the default dummy thickness value used initially.
\subsection{Subtracting the $\gamma p \to \pi^{0} p$ and $\gamma p \to \gamma p$ Backgrounds} \label{normalized_pi0_simul}

\begin{figure}[!htbp]
\begin{center}
\epsfig{file=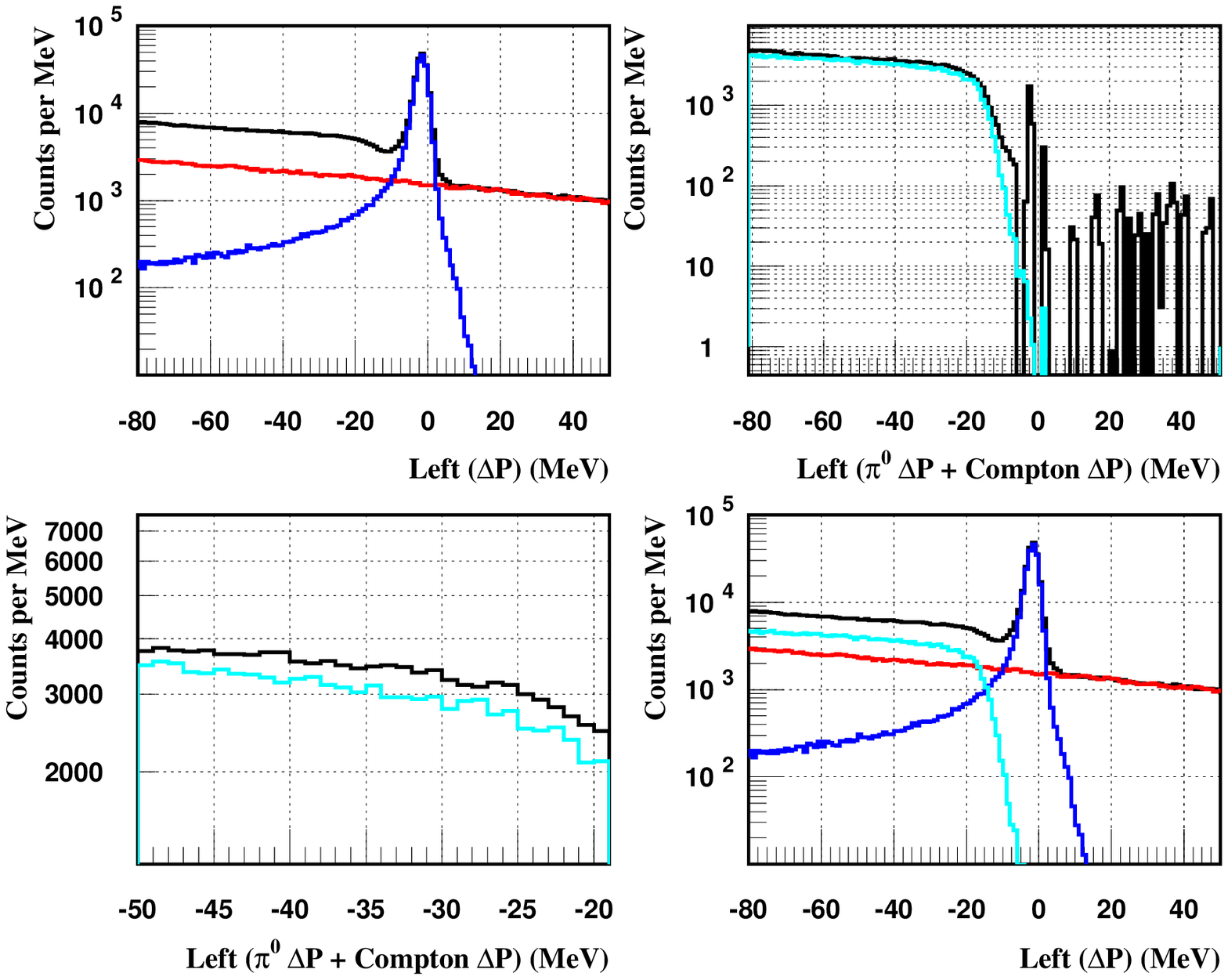,width=6in} 
\end{center}
\caption[The left arm $\gamma p \to \pi^{0} p$ and $\gamma p \to \gamma p$ backgrounds subtraction
for kinematics $b$.]
{The left arm $\gamma p \to \pi^{0} p$ and $\gamma p \to \gamma p$ backgrounds subtraction for 
kinematics $b$ (low $\varepsilon$).
Top left: the LH$_2$ $\Delta P$ (black), extracted dummy (red), and scaled elastic e-p simulation (blue). 
Top right: the sum of the $\gamma p \to \pi^{0} p$ and 
$\gamma p \to \gamma p$ data backgrounds remaining after the subtraction of the extracted dummy and 
scaled elastic e-p simulation spectra from the LH$_2$ $\Delta P$ (black), 
and the sum of the $\gamma p \to \pi^{0} p$ and $\gamma p \to \gamma p$ from simulations (cyan). 
Bottom left: the sum of the $\gamma p \to \pi^{0} p$ and $\gamma p \to \gamma p$ 
as in top right above (black) and simulations (cyan) from -50 to -20 MeV. This window is used for 
normalization. Bottom right: the LH$_2$ $\Delta P$ (black), extracted dummy (red), scaled elastic e-p 
simulation (blue), and normalized sum of the $\gamma p \to \pi^{0} p$ and 
$\gamma p \to \gamma p$ from simulations (cyan).}
\label{fig:kinb_get_lpions_compton}
\end{figure}

\begin{figure}[!htbp]
\begin{center}
\epsfig{file=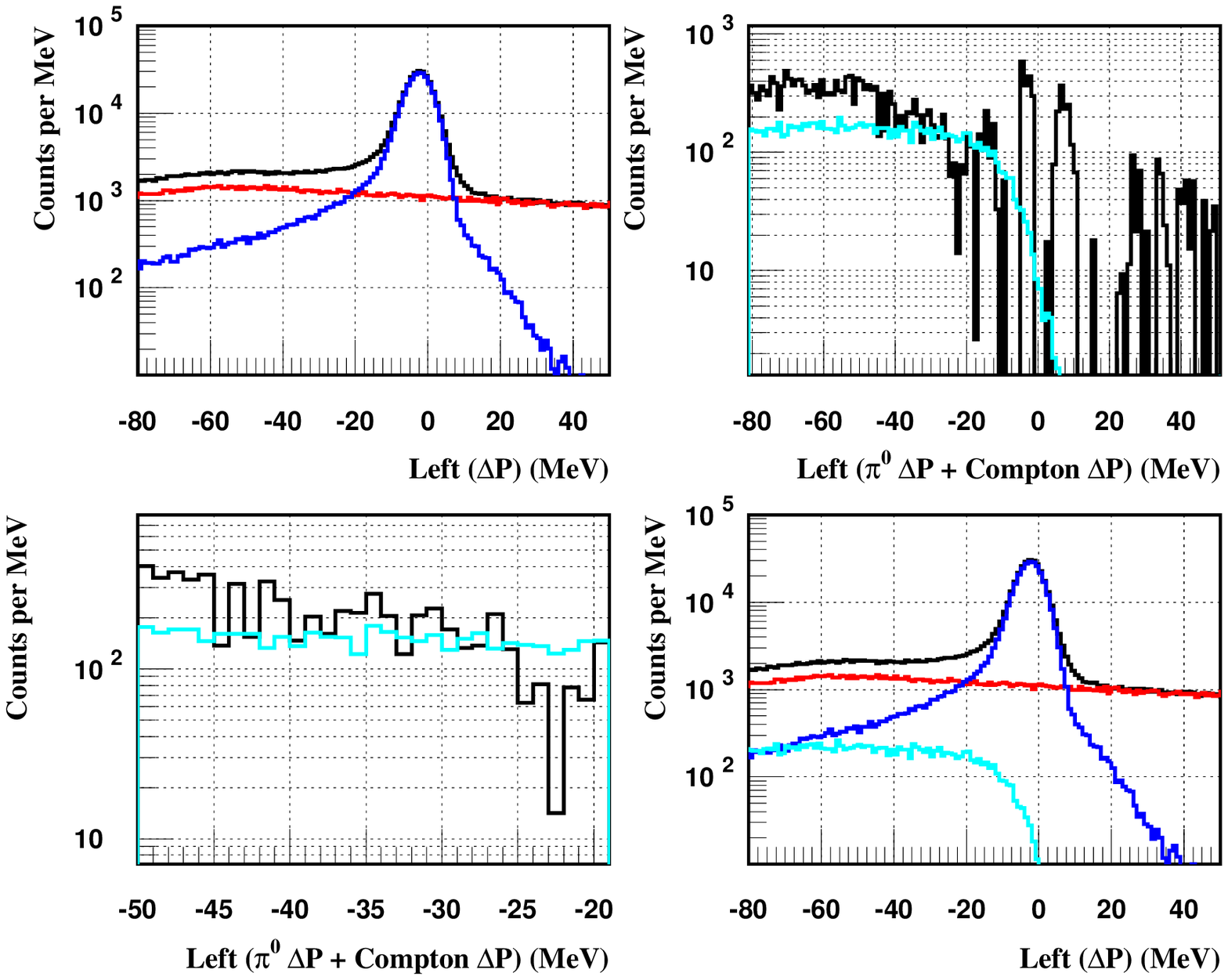,width=6in} 
\end{center}
\caption[The left arm $\gamma p \to \pi^{0} p$ and $\gamma p \to \gamma p$ backgrounds subtraction 
for kinematics $m$.]
{The left arm $\gamma p \to \pi^{0} p$ and $\gamma p \to \gamma p$ backgrounds subtraction 
for kinematics $m$ (high $\varepsilon$).
Top left: the LH$_2$ $\Delta P$ (black), extracted dummy (red), and scaled e-p elastic simulation (blue). 
Top right: the sum of the $\gamma p \to \pi^{0} p$ and $\gamma p \to \gamma p$ data backgrounds 
remaining after the subtraction of the extracted dummy and scaled e-p elastic simulation spectra from 
the LH$_2$ $\Delta P$ (black), and the sum of the $\gamma p \to \pi^{0} p$ and $\gamma p \to \gamma p$ 
from simulations (cyan). Bottom left: the sum of the $\gamma p \to \pi^{0} p$ and $\gamma p \to \gamma p$ 
as a result of the subtraction of the extracted dummy and scaled e-p elastic simulation spectra from the 
LH$_2$ $\Delta P$ (black) and from simulations (cyan) from -50 to -20 MeV. 
This window is used for normalization.
Bottom right: the LH$_2$ $\Delta P$ (black), extracted dummy (red), 
scaled e-p elastic simulation (blue), and normalized sum of the $\gamma p \to \pi^{0} p$ and 
$\gamma p \to \gamma p$ from simulations (cyan).}
\label{fig:kinm_get_lpions_compton}
\end{figure}

\begin{figure}[!htbp]
\begin{center}
\epsfig{file=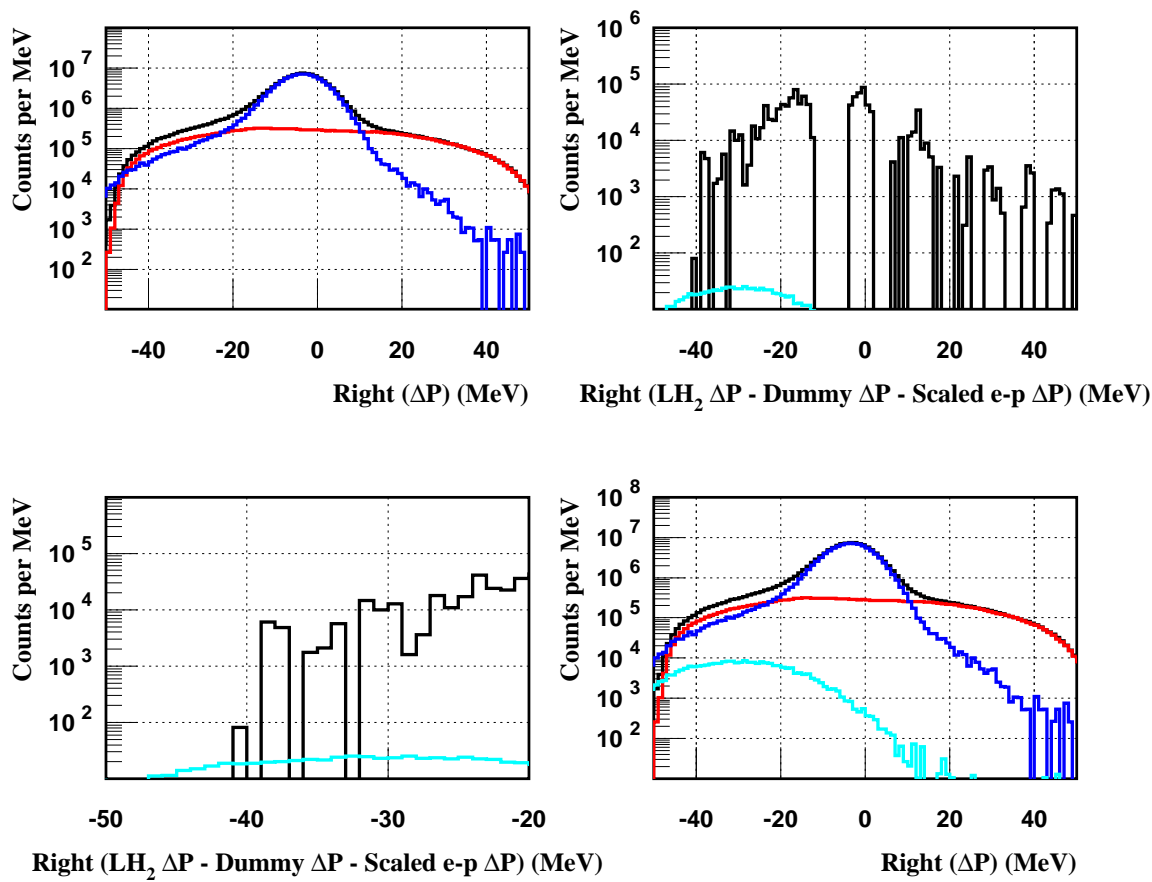,width=6in} 
\end{center}
\caption[The right arm $\gamma p \to \pi^{0} p$ and $\gamma p \to \gamma p$ backgrounds subtraction
for kinematics $b$]
{The right arm $\gamma p \to \pi^{0} p$ and $\gamma p \to \gamma p$ backgrounds subtraction 
for kinematics $b$.
Top left: the LH$_2$ $\Delta P$ (black), extracted dummy (red), and scaled e-p elastic simulation (blue). 
Top right: the sum of the $\gamma p \to \pi^{0} p$ and 
$\gamma p \to \gamma p$ data backgrounds remaining after the subtraction of the extracted dummy and 
scaled e-p elastic simulation spectra from the LH$_2$ $\Delta P$ (black), 
and the sum of the $\gamma p \to \pi^{0} p$ and $\gamma p \to \gamma p$ from simulations (cyan). 
Bottom left: the sum of the $\gamma p \to \pi^{0} p$ and $\gamma p \to \gamma p$ as a result of 
the subtraction of the extracted dummy and scaled e-p elastic simulation spectra from the 
LH$_2$ $\Delta P$ (black) and from simulations (cyan) from -50 to -20 MeV. 
This window is used for normalization.
Bottom right: the LH$_2$ $\Delta P$ (black), extracted dummy (red), scaled e-p elastic simulation (blue), 
and normalized sum of the $\gamma p \to \pi^{0} p$ and $\gamma p \to \gamma p$ from simulations (cyan).}
\label{fig:kinb_get_rpions_compton}
\end{figure}

The next step is to subtract the protons generated from photoreactions 
$\gamma p \to \pi^{0} p$ and $\gamma p \to \gamma p$ from the LH$_2$ $\Delta P$ spectrum. When the 
endcaps background and the scaled elastic e-p simulation events are 
subtracted from the LH$_2$ $\Delta P$ spectrum, the resultant spectrum 
represents the sum of the $\gamma p \to \pi^{0} p$ and $\gamma p \to \gamma p$ backgrounds. 
For the left arm, the sum of $\gamma p \to \pi^{0} p$ and $\gamma p \to \gamma p$ spectra from simulations 
is normalized to the (LH$_2$ - endcaps - scaled e-p simulation) spectrum
in a window of -50$<\Delta P<$-20 MeV. The normalized simulation is then subtracted.
Figures \ref{fig:kinb_get_lpions_compton}, \ref{fig:kinm_get_lpions_compton}, and
\ref{fig:kinb_get_rpions_compton} show this procedure for three representative spectra.
While the uncertainty in extracting the backgrounds is larger for large $\varepsilon$, where the
backgrounds contribution is small, even a large uncertainty in this small background yields a small uncertainty
in the final extracted cross sections.
The contribution of the $\gamma p \to \pi^{0} p$ and $\gamma p \to \gamma p$ backgrounds 
to the right arm LH$_2$ spectrum is almost negligible although this contribution has 
a threshold that is close to the elastic peak as can be seen clearly from the bottom right
hand plot of Figure \ref{fig:kinb_get_rpions_compton}.
On the other hand, the contribution of the $\gamma p \to \pi^{0} p$ and $\gamma p \to \gamma p$ 
backgrounds to the left arm LH$_2$ spectrum is much larger but with a lower threshold for the low
$\varepsilon$ case as shown in the bottom right hand plots of Figures \ref{fig:kinb_get_lpions_compton} and 
\ref{fig:kinm_get_lpions_compton}. It is important to mention that for the right arm the same procedure for
extracting the $\gamma p \to \pi^{0} p$ and $\gamma p \to \gamma p$ backgrounds was used, and the background
was found to be consistent with zero, with upper limit contribution of $<<$0.10\%. Thus in the analysis, we do not
make any correction for these processes and assign a 0.05\% random uncertainty.

The procedure for unpealing the spectra can be iterated if necessary. However,
the residual spectra after backgrounds and normalized simulated e-p are subtracted off show
that this is not necessary.

\subsection{Extracting the Reduced Cross Section $\sigma_{R}$} \label{extract_sigma}

\begin{figure}[!htbp]
\begin{center}
\epsfig{file=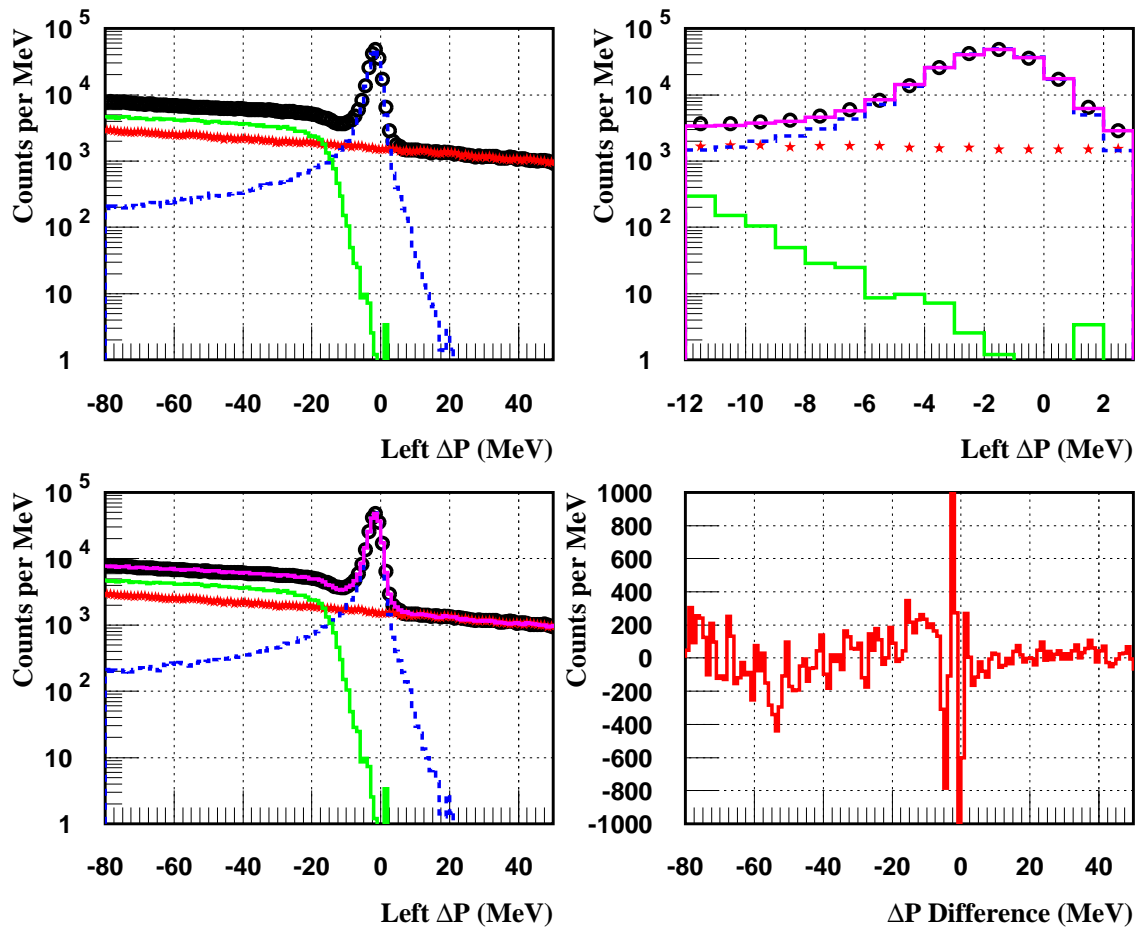,width=6in}
\end{center}
\caption[All contributions to the left arm LH$_2$ $\Delta P$ spectrum for kinematics $b$.]
{All contributions to the left arm LH$_2$ $\Delta P$ spectrum for kinematics $b$ (low $\varepsilon$). 
Top left: the LH$_2$ $\Delta P$ data (black circles), unscaled elastic e-p simulation (blue dashed line),
extracted dummy (red stars), normalized sum of the $\gamma p \to \pi^{0} p$ and $\gamma p \to \gamma p$ 
backgrounds (solid green). Top right: the same as in Top left but for a smaller $\Delta P$ window cut to 
show the final $\Delta P$ window used for the final ratio extraction. The magenta line through the data 
is the sum of all contributions. See text for more details. Bottom left: the same as in Top right but for 
a larger $\Delta P$ window. Bottom right: the difference between the LH$_2$ $\Delta P$ data and the sum of 
all contributions.}
\label{fig:kinb_all_ldp}
\end{figure}

\begin{figure}[!htbp]
\begin{center}
\epsfig{file=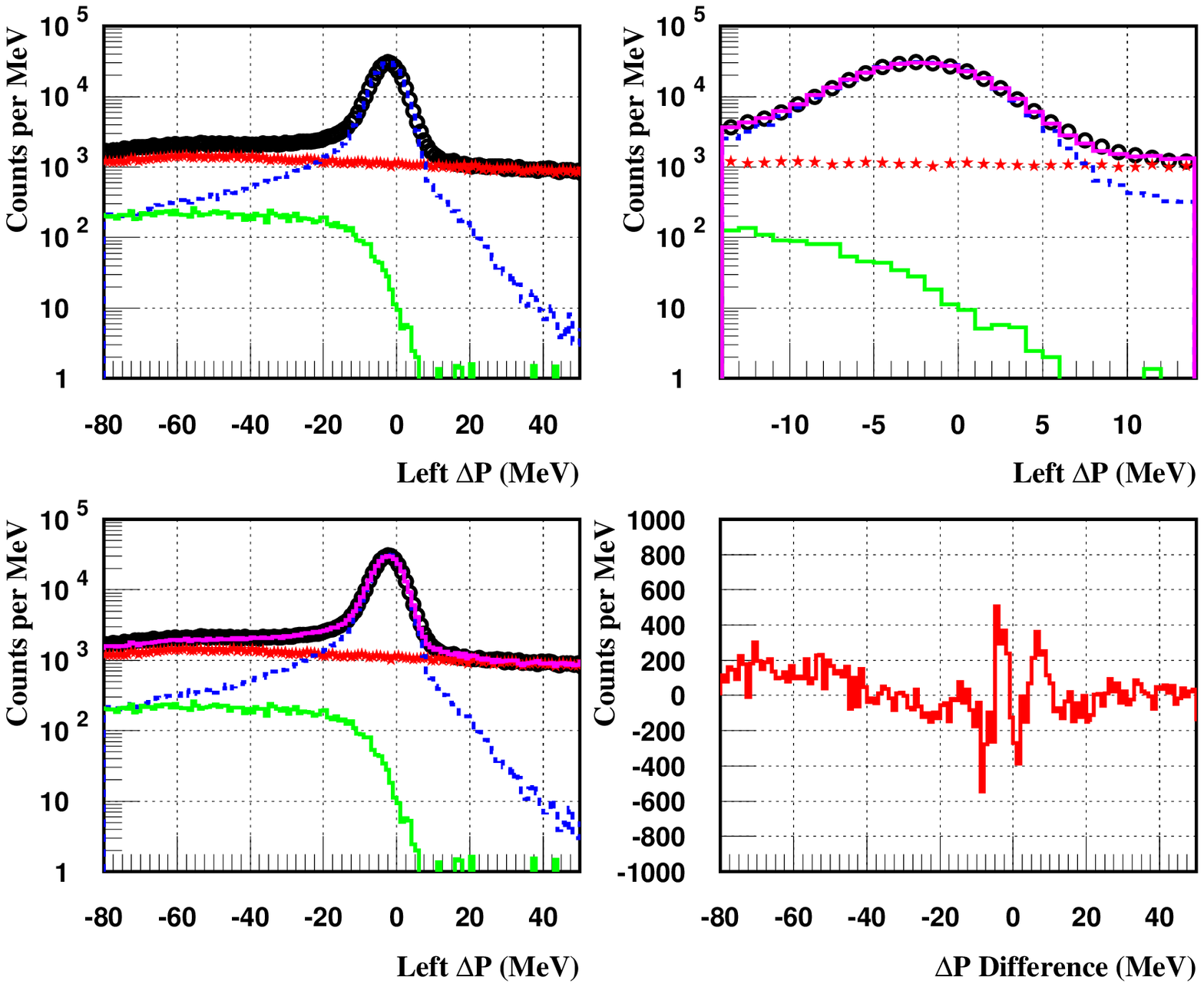,width=6in}
\end{center}
\caption[All contributions to the left arm LH$_2$ $\Delta P$ spectrum for kinematics $m$.]
{All contributions to the left arm LH$_2$ $\Delta P$ spectrum for kinematics $m$ (high $\varepsilon$).
Top left: the LH$_2$ $\Delta P$ data (black circles), unscaled elastic e-p simulation (blue dashed line),
extracted dummy (red stars), normalized sum of the $\gamma p \to \pi^{0} p$ and $\gamma p \to \gamma p$ 
backgrounds (solid green). Top right: the same as in Top left but for a smaller $\Delta P$ window cut to 
show the final $\Delta P$ window used for the final ratio extraction. The magenta line through the data 
is the sum of all contributions. See text for more details. Bottom left: the same as in Top right but for 
a larger $\Delta P$ window. Bottom right: the difference between the LH$_2$ $\Delta P$ data and the sum of 
all contributions.}
\label{fig:kinm_all_ldp}
\end{figure}

\begin{figure}[!htbp]
\begin{center}
\epsfig{file=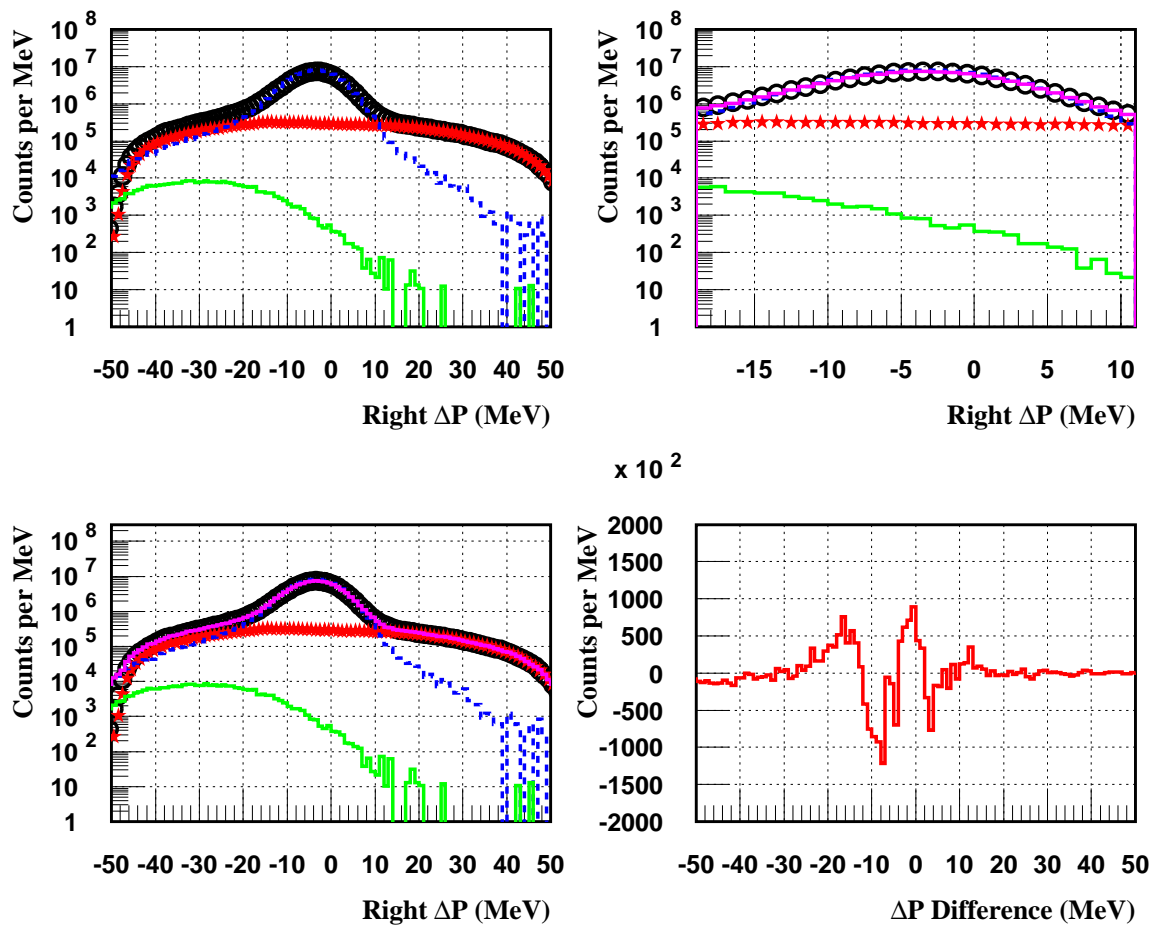,width=6in} 
\end{center}
\caption[All contributions to the right arm LH$_2$ $\Delta P$ spectrum for kinematics $b$.]
{All contributions to the right arm LH$_2$ $\Delta P$ spectrum for kinematics $b$. 
Top left: (black circles) LH$_2$ $\Delta P$ data, (blue dashed line) unscaled elastic e-p simulation, 
(red stars) the modified target walls background, (solid green) the modified sum of the 
$\gamma p \to \pi^{0} p$ and $\gamma p \to \gamma p$ backgrounds. 
Top right: the same as in Top left but for a smaller $\Delta P$ window cut to show the final 
$\Delta P$ window needed for the final ratio extraction. The magenta line through the data is the sum of 
all contributions. See text for more details. 
Bottom left: the same as in Top right but for a larger $\Delta P$ window cut. 
Bottom right: the difference between the LH$_2$ $\Delta P$ data and the sum of all contributions.}
\label{fig:kinb_all_rdp}
\end{figure}

The top left plots of Figures \ref{fig:kinb_all_ldp}, \ref{fig:kinm_all_ldp},
and \ref{fig:kinb_all_rdp} show the individual contributions to the $\Delta P$ spectrum 
for the three representative spectra. In order to extract $\sigma_{R}$, the ratio of elastic e-p 
data events to unscaled elastic e-p simulation events $R = (N_{\mbox{elastic-data}}/N_{e-p}$)
is determined in a $\Delta P$ window around the elastic peak. In determining the range of the
$\Delta P$ window used for each kinematics, the $\Delta P$ window cut was varied to simultaneously
minimize the $\varepsilon$-dependence of the dummy and pions subtraction, as well as the elastic acceptance
correction. 
The top right plots of Figures \ref{fig:kinb_all_ldp}, \ref{fig:kinm_all_ldp},
and \ref{fig:kinb_all_rdp} show the individual contributions in this $\Delta P$ window cut 
while the magenta solid line represents the sum of all contributions. The number of elastic events $N_{\mbox{elastic-data}}$ 
in that $\Delta P$ window is determined as:
\begin{equation} \label{eq:No_elastic_events}
N_{\mbox{elastic-data}} = N_{LH_{2}} - N_{dummy} - N_{\gamma p \to \pi^{0} p} - N_{\gamma p \to \gamma p}~.
\end{equation}
\begin{table}[!htbp]
\begin{center}
\begin{tabular}{||c|c|c|c||} 
\hline \hline
$Q^2$&Kinematics&Right $\Delta P$ Window&Right Ratio       \\
(GeV$^2$)&      & From : To             &$R \pm \delta_R$\\    
         &      & (MeV)                 &                         \\ 
\hline 
0.50 &$o$         &-19.0 : +11.0    & 0.9970 $\pm$ 0.0024 \\          
     &$a$         &-19.0 : +11.0    & 1.0021 $\pm$ 0.0020  \\
     &$i$         &-18.0 : +12.0    & 0.9881 $\pm$ 0.0022   \\
     &$q$         &-18.0 : +12.0    & 0.9919 $\pm$ 0.0021    \\
     &$l$         &-18.0 : +12.0    & 0.9896 $\pm$ 0.0028     \\
\hline 
0.50 &$b$         &-19.0 : +11.0    & 1.0010 $\pm$ 0.0021      \\                
     &$j$         &-18.0 : +12.0    & 0.9883 $\pm$ 0.0019       \\
     &$p$         &-18.0 : +12.0    & 0.9921 $\pm$ 0.0022        \\
     &$m$         &-18.0 : +12.0    & 0.9884 $\pm$ 0.0026         \\
\hline 
0.50 &$k$         &-18.0 : +12.0    & 0.9865 $\pm$ 0.0020          \\              
     &$r$         &-18.0 : +12.0    & 0.9923 $\pm$ 0.0020           \\
     &$n$         &-18.0 : +12.0    & 0.9838 $\pm$ 0.0029            \\
\hline 
\end{tabular}
\caption[The $\Delta P$ window range used and the extracted ratio $R$ for the right arm 
kinematics.] 
{The $\Delta P$ window range used and the extracted ratio $R$ for the right arm kinematics. 
The listed uncertainties in the ratios are statistical only.}
\label{R_Ratio_data}
\end{center}
\end{table}
\begin{table}[!htbp]
\begin{center}
\begin{tabular}{||c|c|c|c||} 
\hline 
$Q^2$&Kinematics  &Left $\Delta P$ Window&Left Ratio  \\
(GeV$^2$)&        &From : To            &$R \pm \delta_R$\\    
         &        & (MeV)               &                                    \\ 
\hline 
2.64  &$o$        &-12.0 : +2.0~             &1.0156 $\pm$ 0.0030 \\          
      &$a$        &-12.0 : +6.0~             &1.0084 $\pm$ 0.0026 \\
      &$i$        &-12.0 : +9.0~             &1.0054 $\pm$ 0.0033 \\
      &$q$        &-13.0 : +11.0             &1.0038 $\pm$ 0.0029  \\
      &$l$        &-14.0 : +12.0             &1.0022 $\pm$ 0.0026  \\
\hline 
3.20 &$b$         &-12.0 : +3.0~             &1.0236 $\pm$ 0.0036  \\                
     &$j$         &-12.0 : +8.0~             &1.0241 $\pm$ 0.0033  \\
     &$p$         &-12.0 : +12.0             &1.0152 $\pm$ 0.0033  \\
     &$m$         &-14.0 : +14.0             &1.0234 $\pm$ 0.0029  \\
\hline 
4.10 &$k$         &-12.0 : +5.0~             &1.0296 $\pm$ 0.0049  \\             
     &$r$         &-15.0 : +10.0             &1.0368 $\pm$ 0.0037  \\
     &$n$         &-17.0 : +13.0             &1.0454 $\pm$ 0.0040  \\
\hline 
\end{tabular}
\caption[The $\Delta P$ window range used and the extracted ratio $R$ for the left arm 
kinematics.] 
{The $\Delta P$ window range used and the extracted ratio $R$ for the left arm kinematics. 
The listed uncertainties in the ratios are statistical only.}
\label{L_Ratio_data}
\end{center}
\end{table}

Tables \ref{R_Ratio_data} and \ref{L_Ratio_data} list the range of the $\Delta P$ window used and the extracted ratio $R$ for 
each kinematics for both arms. Having determined the ratio $R$, the unscaled elastic e-p simulation is scaled as
$N^{new}_{e-p} = R~N_{e-p}$ and then added to the remaining contributions:
\begin{equation} 
N_{\mbox{total}}~= N^{new}_{e-p} + N_{dummy} + N_{\gamma p \to \pi^{0} p} + N_{\gamma p \to \gamma}~,
\end{equation}
to generate the sum spectrum shown which reproduces the LH$_2$ $\Delta P$ spectrum very well. 
The reduced cross section $\sigma_{R}$ is
taken to be the value of the reduced cross section from the Bosted fit, $\sigma_{R~\mbox{(Bosted)}}$, 
used in the simulation scaled by the ratio $R$ or:
\begin{equation}
\sigma_{R}~= R~\sigma_{R~\mbox{(Bosted)}}~= R\Big[\tau G^{2}_{Mp}(Q^2) + \varepsilon G^{2}_{Ep}(Q^2) \Big]~,
\end{equation}
where the form factors $G_{Ep}(Q^2)$ and $G_{Mp}(Q^2)$ at a given $Q^2$ point are determined
from the Bosted parameterization equations (\ref{eq:bosted_gep}) and (\ref{eq:bosted_gmp}).
Tables \ref{rsigma_R_data} and \ref{lsigma_R_data} list the reduced cross sections and their
uncertainties for each kinematics for both arms.  
Note that the statistical uncertainty in $\sigma_{R}$ includes the statistical uncertainty in all of the yields in
equation \ref{eq:No_elastic_events}, as well as statistical uncertainty in the extracted dummy normalization factor.

\begin{figure}[!htbp]
\begin{center}
\epsfig{file=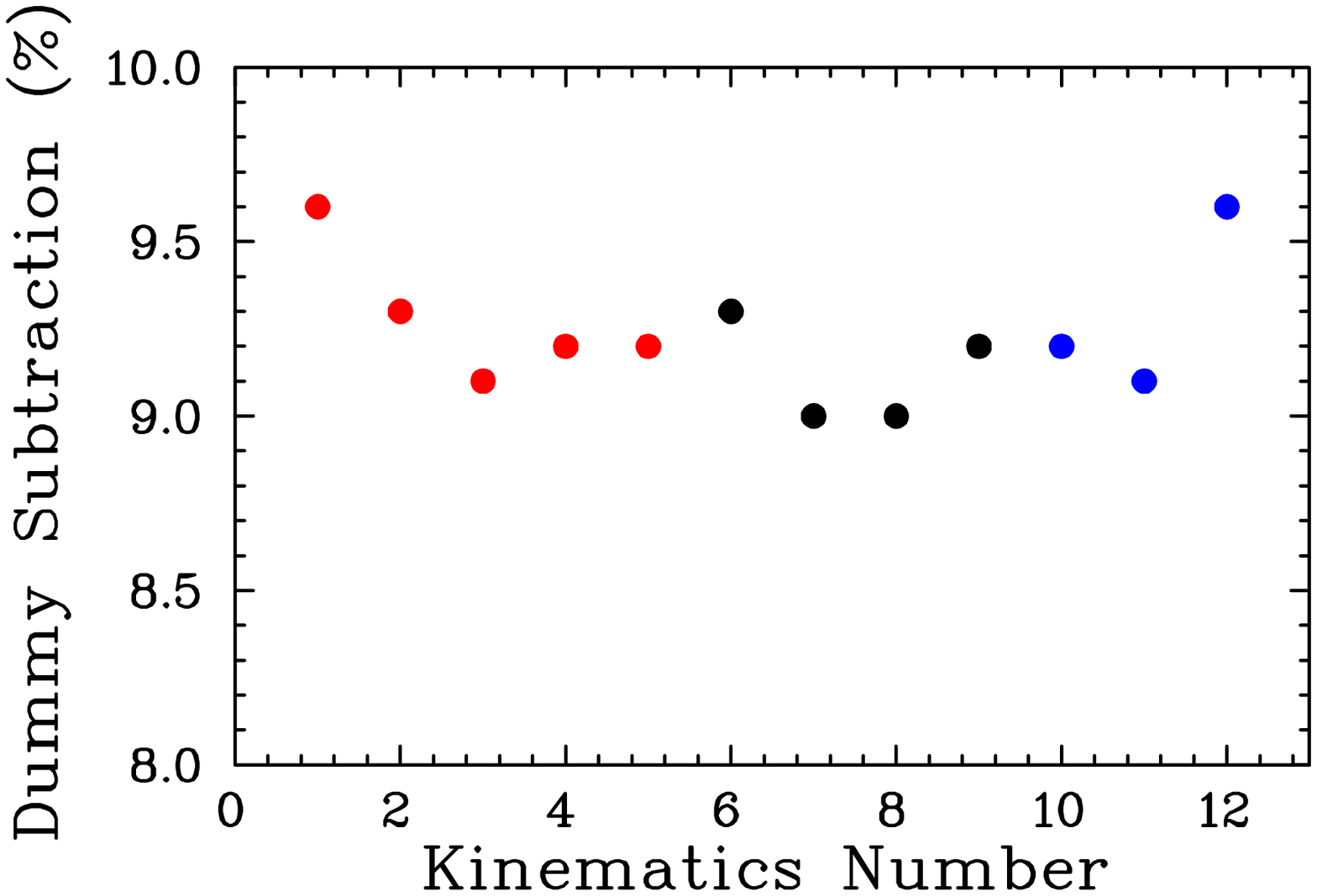,width=3.4in}
\end{center}
\caption[The ratio of $N_{dummy}$ to $N_{LH_2}$ events or ``Dummy Subtraction (\%)'' for the right arm in the $\Delta P$ window cut 
used as a function of kinematics number.] 
{The ratio of $N_{dummy}$ to $N_{LH_2}$ events or ``Dummy Subtraction (\%)'' for the right arm in the $\Delta P$ window cut used 
as a function of kinematics number. The points are sorted according to the kinematics of the left arm. 
Kinematics 1-5 correspond to $Q^2$ = 2.64 GeV$^2$ and shown in red, while kinematics 6-9(10-12) correspond to 3.20(4.10) GeV$^2$ 
and shown in black(blue). For each $Q^2$ value, the points are sorted by $\varepsilon$ (low to high).}
\label{fig:show_dummysub_right}
\end{figure}

\begin{figure}[!htbp]
\begin{center}
\epsfig{file=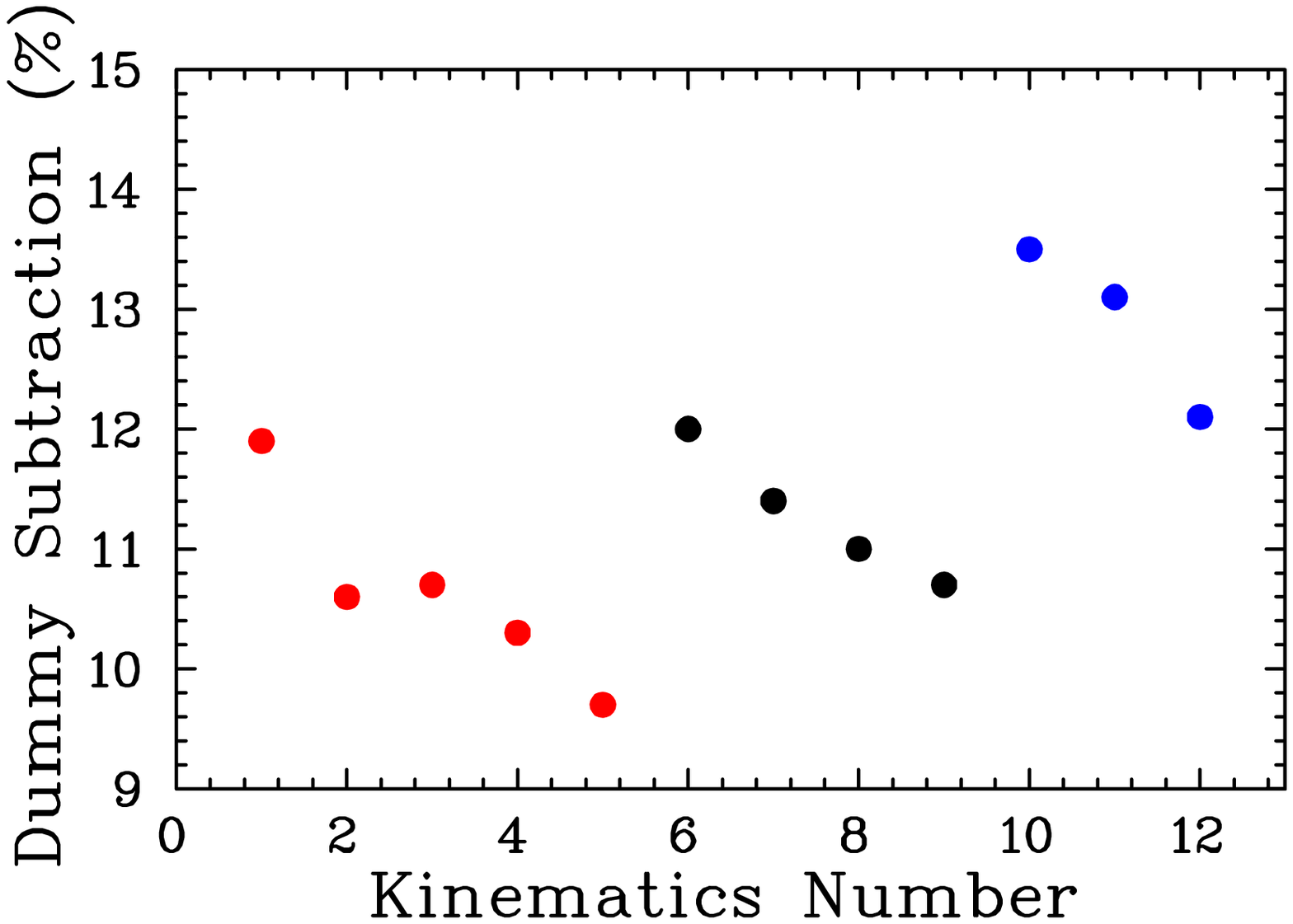,width=3.4in}
\end{center}
\caption[The ratio of $N_{dummy}$ to $N_{LH_2}$ events or ``Dummy Subtraction (\%)'' for the left arm in the $\Delta P$ window cut 
used as a function of kinematics number.] 
{The ratio of $N_{dummy}$ to $N_{LH_2}$ events or ``Dummy Subtraction (\%)'' for the left arm in the $\Delta P$ window cut used 
as a function of kinematics number. Kinematics 1-5 correspond to $Q^2$ = 2.64 GeV$^2$ and shown in red, while kinematics 
6-9(10-12) correspond to 3.20(4.10) GeV$^2$ and shown in black(blue). For each $Q^2$ value, the points are sorted by $\varepsilon$ 
(low to high).}
\label{fig:show_dummysub_left}
\end{figure}

To estimate the size of the uncertainty in the dummy subtraction, we plot the ratio of $N_{dummy}$ to $N_{LH_2}$ events 
or ``Dummy Subtraction'' in the $\Delta P$ window cut used as a function of kinematics number and for both arms. 
Figures \ref{fig:show_dummysub_right} and \ref{fig:show_dummysub_left} show the results for the right and left arm, respectively. 
The dummy subtraction is a $\sim$ 10\% correction for both arms and with a $\sim$ 2\% slope dependence for the left arm and $\sim$ 
0.5\% for the right arm. We assign a 5\% systematic uncertainty in the dummy subtraction correction which yields a slope uncertainty 
of 0.10\% for the left arm and 0.36\% (0.025\%/0.07) for the right arm when we consider the $\Delta \varepsilon$ range of 0.07. 
Furthermore, a 5\% uncertainty in a 10\% correction yields a scale uncertainty of 0.5\% for the both arms.
The random uncertainty is set to 0.0\% to avoid double counting the random uncertainty in the dummy subtraction since the statistical 
uncertainty in the effective dummy thickness as shown in Figure \ref{fig:effective_dummy} is taken into account when calculating 
the statistical uncertainty $\delta_{R}$ in the ratio $R$ for both arms. See Tables \ref{R_Ratio_data} and \ref{L_Ratio_data}.
\begin{table}[!htbp]
\begin{center}
\begin{tabular}{||c|c|c|c|c|c|c||} 
\hline \hline
$Q^2$&Kinematics &$\varepsilon$&$\delta_{Stat}$&$\delta_{Random}$&$\delta_{\sigma_{R}}$&$\sigma_{R}$\\
(GeV$^2$)&       &               &                 &            &      &             \\           
\hline                                                                          
0.50&$o$        &0.914 &0.054E-02 &0.104E-02 &0.117E-02 &0.22856 \\
    &$a$        &0.939 &0.046E-02 &0.149E-02 &0.156E-02 &0.23253 \\ 
    &$i$        &0.962 &0.051E-02 &0.106E-02 &0.118E-02 &0.23187 \\
    &$q$        &0.979 &0.051E-02 &0.107E-02 &0.119E-02 &0.23468 \\
    &$l$        &0.986 &0.066E-02 &0.108E-02 &0.127E-02 &0.23491 \\
\hline    
0.50&$b$        &0.939 &0.048E-02 &0.148E-02 &0.156E-02 &0.23231 \\
    &$j$        &0.962 &0.045E-02 &0.106E-02 &0.115E-02 &0.23192 \\
    &$p$        &0.979 &0.052E-02 &0.107E-02 &0.119E-02 &0.23492 \\
    &$m$        &0.986 &0.063E-02 &0.108E-02 &0.125E-02 &0.23462 \\
\hline 
0.50&$k$        &0.962 &0.047E-02 &0.105E-02 &0.115E-02 &0.23150 \\
    &$r$        &0.979 &0.047E-02 &0.107E-02 &0.117E-02 &0.23477 \\
    &$n$        &0.986 &0.070E-02 &0.107E-02 &0.127E-02 &0.23353 \\
\hline 
\end{tabular}
\caption[The elastic e-p reduced cross section $\sigma_{R}$ for the right arm kinematics.] 
{The elastic e-p reduced cross section $\sigma_{R}$ for the right arm kinematics. In addition, the
statistical uncertainty $\delta_{Stat}$, random systematic uncertainty $\delta_{Random}$, and total uncertainty
$\delta_{\sigma_{R}}$ (combined $\delta_{Stat}$ and $\delta_{Random}$) in $\sigma_{R}$ are also given.} 
\label{rsigma_R_data}
\end{center}
\end{table}
\begin{table}[!htbp]
\begin{center}
\begin{tabular}{||c|c|c|c|c|c|c||} 
\hline \hline
$Q^2$&Kinematics &$\varepsilon$&$\delta_{Stat}$&$\delta_{Random}$&$\delta_{\sigma_{R}}$&$\sigma_{R}$\\
(GeV$^2$)&       &      &             &             &             &           \\ 
\hline 
2.64&$o$         &0.117 &0.40E-04 &0.51E-04 &0.65E-04 &0.13402E-01\\
    &$a$         &0.356 &0.35E-04 &0.82E-04 &0.89E-04 &0.13823E-01\\
    &$i$         &0.597 &0.47E-04 &0.55E-04 &0.72E-04 &0.14301E-01\\
    &$q$         &0.782 &0.44E-04 &0.57E-04 &0.72E-04 &0.14677E-01\\
    &$l$         &0.865 &0.38E-04 &0.58E-04 &0.69E-04 &0.14832E-01\\
\hline 
3.20&$b$         &0.131 &0.30E-04 &0.42E-04 &0.52E-04 &0.08686E-01\\
    &$j$         &0.443 &0.29E-04 &0.35E-04 &0.45E-04 &0.09075E-01\\ 
    &$p$         &0.696 &0.30E-04 &0.36E-04 &0.47E-04 &0.09305E-01\\ 
    &$m$         &0.813 &0.27E-04 &0.37E-04 &0.46E-04 &0.09525E-01\\ 
\hline 
4.10&$k$         &0.160 &0.23E-04 &0.22E-04 &0.31E-04 &0.04726E-01\\
    &$r$         &0.528 &0.17E-04 &0.23E-04 &0.29E-04 &0.04973E-01\\ 
    &$n$         &0.709 &0.19E-04 &0.24E-04 &0.31E-04 &0.05121E-01\\ 
\hline 
\end{tabular}
\caption[The elastic e-p reduced cross section $\sigma_{R}$ for the left arm kinematics.] 
{The elastic e-p reduced cross section $\sigma_{R}$ for the left arm kinematics. In addition, the
statistical uncertainty $\delta_{Stat}$, random uncertainty $\delta_{Random}$, and total uncertainty
$\delta_{\sigma_{R}}$ (combined $\delta_{Stat}$ and $\delta_{Random}$) in $\sigma_{R}$ are also given.} 
\label{lsigma_R_data}
\end{center}
\end{table}

\begin{figure}[!htbp]
\begin{center}
\epsfig{file=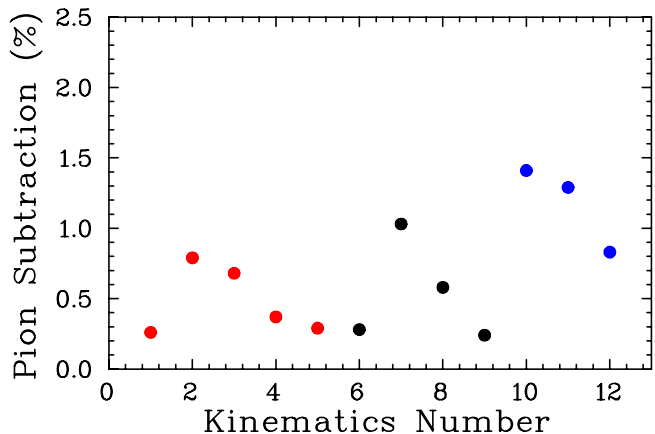,width=3.4in}
\end{center}
\caption[The ratio of ($N_{\gamma p \to \pi^{0} p} + N_{\gamma p \to \gamma p}$) to $N_{LH_2}$ events or ``Pion Subtraction (\%)'' 
for the left arm in the $\Delta P$ window cut used as a function of kinematics number.] 
{The ratio of ($N_{\gamma p \to \pi^{0} p} + N_{\gamma p \to \gamma p}$) to $N_{LH_2}$ events or ``Pion Subtraction (\%)'' 
for the left arm in the $\Delta P$ window cut used as a function of kinematics number. Kinematics 1-5 correspond 
to $Q^2$ = 2.64 GeV$^2$ and shown in red, while kinematics 6-9(10-12) correspond to 3.20(4.10) GeV$^2$ and shown in black(blue). 
For each $Q^2$ value, the points are sorted by $\varepsilon$ (low to high).}
\label{fig:show_pionsub_left}
\end{figure}

Similarly, to estimate the size of the uncertainty in the pion subtraction, we plot the ratio of 
($N_{\gamma p \to \pi^{0} p} + N_{\gamma p \to \gamma p}$) to $N_{LH_2}$ events ``Pion Subtraction''
in the $\Delta P$ window cut used as a function of kinematics number and for both arms. 
Figure \ref{fig:show_pionsub_left} shows the results for the left arm. 
The pion subtraction is typically a $\sim$ 0.5\% correction for the left arm with a large scatter and $\varepsilon$ dependence. 
We assign a $\sim$ 50\% uncertainty in the pion subtraction correction or 0.25\% with almost equal contribution from the slope 
(0.20\%) and random (0.15\%) uncertainties. The slope and random uncertainties are doubled for $Q^2$ = 4.10 GeV$^2$ to account for 
the larger observed $\varepsilon$ dependence. For the right arm, the pion subtraction correction is set to zero and a random 
uncertainty of 0.05\% is assigned. It is crucial to mention that when determining the ratio $R$ for the right arm, the number of 
elastic events $N_{\mbox{elastic-data}}$ as given by equation (\ref{eq:No_elastic_events}) is now given by 
$N_{\mbox{elastic-data}} = N_{LH_{2}} - N_{dummy}$. This is equivalent to set the normalization factor or the $\pi^{0}$-factor needed 
to normalize the sum of $\gamma p \to \pi^{0} p$ and $\gamma p \to \gamma p$ from simulations to the resultant
(LH$_2$ - endcaps - scaled e-p simulation) spectrum to zero. 
The statistical uncertainty in the $\pi^{0}$-factor is taken into account when calculating the statistical uncertainty $\delta_{R}$ 
in the ratio $R$ for the left arm. See section \ref{normalized_pi0_simul} for details.

\begin{figure}[!htbp]
\begin{center}
\epsfig{file=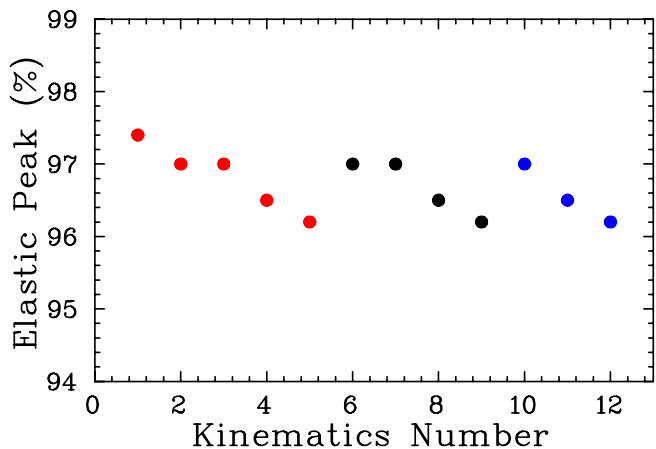,width=3.4in}
\end{center}
\caption[The percentage of the $N_{e-p}$ events or ``Elastic Peak (\%)'' for the right arm in the $\Delta P$ window cut used 
as a function of kinematics number.] 
{The percentage of the events or ``Elastic Peak (\%)'' for the right arm in the $\Delta P$ window cut used 
as a function of kinematics number. 
The points are sorted according to the kinematics of the left arm. 
Kinematics 1-5 correspond to $Q^2$ = 2.64 GeV$^2$ and shown in red, 
while kinematics 6-9(10-12) correspond to 3.20(4.10) GeV$^2$ and shown in black(blue). For each $Q^2$ value, the points are 
sorted by $\varepsilon$ (low to high).}
\label{fig:show_elasaccep_right}
\end{figure}

\begin{figure}[!htbp]
\begin{center}
\epsfig{file=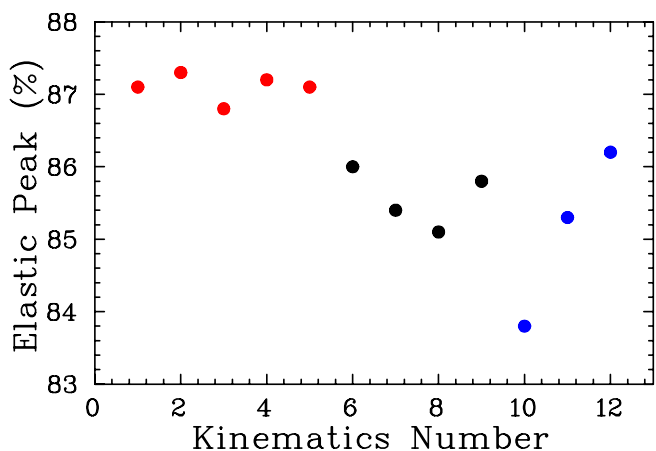,width=3.4in}
\end{center}
\caption[The percentage of the $N_{e-p}$ events or ``Elastic Peak \%'' for the left arm in the $\Delta P$ window cut used 
as a function of kinematics number.] 
{The percentage of the $N_{e-p}$ events or ``Elastic Peak \%'' for the left arm in the $\Delta P$ window cut used 
as a function of kinematics number. Kinematics 1-5 correspond to $Q^2$ = 2.64 GeV$^2$ and shown in red, 
while kinematics 6-9(10-12) correspond to 3.20(4.10) GeV$^2$ and shown in black(blue). For each $Q^2$ value, the points are 
sorted by $\varepsilon$ (low to high).}
\label{fig:show_elasaccep_left}
\end{figure}

To estimate the uncertainty due to the $\Delta P$ cut used, we plot the percentage of the $N_{e-p}$ events or 
``Elastic Peak Acceptance'' within the $\Delta P$ window cut used as a function of kinematics number and for both arms. 
Figures \ref{fig:show_elasaccep_right} and \ref{fig:show_elasaccep_left} show the results for the right and left arm, respectively. 
The elastic acceptance is $\sim$ 85\% and $\sim$ 95\% for the left and right arm, respectively. Of the 15\% events lost on the 
left arm, $\sim$ 14\% are events lost in the radiative tail and $\sim$ 1\% are events lost in the smeared non-gaussian tails of 
the elastic peak. 
The uncertainty in the $\Delta P$ cut used due to the correction that results from the events lost in 
the radiative tail is accounted for in the radiative corrections uncertainties. See section \ref{rad_corrections} for details.
The smeared non-gaussian tails are known to within $\sim$ 50\%, and therefore, a 1\% correction 
due to events lost in the non-gaussian tails yields a 0.50\% scale uncertainty for the left arm. 
We vary the $\Delta P$ window cut used on the left arm by $\pm$ 2 MeV which changes the ratio $R$ by $\sim$ 0.20\%. 
The 0.20\% uncertainty in $R$ is then broken down equally as 0.14\% random uncertainty and 0.14\% slope uncertainty. 
However, if we consider the average $\Delta \varepsilon$ range of 0.70, that will modify the slope uncertainty to 0.20\% 
(0.14\%/0.70). 

The right arm is more complicated than the left arm since the HRS momentum acceptance cuts off the $\Delta P$ peak. 
The loss due to a non-gaussian tails is $\sim$ 1\%, and we take the same error estimates as for the left arm but account for the 
$\Delta \varepsilon$ range of 0.07 which modifies the slope uncertainty to 2.0\% (0.14\%/0.07).


It is crucial to realize that although the elastic peak is broadened with increasing $\varepsilon$ for all the kinematics on 
the left arm, it is well within the momentum acceptance $\delta$ of the HRS and thus no elastic events are lost 
after applying the momentum acceptance $\delta$ cut. On the other hand, this is not the case for the right arm since the elastic 
peak is broader than that of the left arm and extends a bit beyond the edges of the momentum acceptance. Therefore, 
applying the $\delta$ cut will get rid of some elastic events. The elastic acceptance for the right arm shows a $\sim$ 1.5\% 
$\varepsilon$ dependence as can be seen from Figure \ref{fig:show_elasaccep_right}. Varying the $\delta$ cut changes the correction 
by $\sim$ 30\%. Therefore, a 30\% uncertainty in a 1.5\% correction yields 
a $\sim$ 0.5\% uncertainty. However, the 30\% uncertainty is not simply a scale, random, or slope uncertainty and so 
we break the 0.5\% uncertainty equally among the scale, random, and slope uncertainties and assign a $\sim$ 0.30\% uncertainty for 
each. Finally, if we consider the $\Delta \varepsilon$ range of 0.07 for the right arm, that will modify the slope uncertainty to 4.28\%.

\chapter{Results and Two-Photon Exchange} \label{chap_results}
\pagestyle{plain}
\section{Overview} \label{results_overview}

In this chapter I will discuss the procedure used to extract the electromagnetic form factors of 
the proton, $G_{Ep}$ and $G_{Mp}$, and their ratio $\mu_{p}G_{Ep} \over G_{Mp}$. 
Rosenbluth extractions of the proton form factors and their ratio using the reduced cross sections of each arm separately 
will be discussed. Rosenbluth extractions using the left arm reduced cross sections were recently published \cite{qattan05}. 
Initially, we planned to use the Super-Rosenbluth ratio, $R_{Super}$, 
defined as the left arm reduced cross section normalized to the right arm ratio at each kinematics point or:
\begin{equation} \label{eq:super_ratio}
R_{Super} = \frac{\mbox{Left}~\sigma_{R}}{\mbox{Right}~R}~,
\end{equation}
to extract the form factor ratios. Note that by using $R_{Super}$, we take advantage of the right arm spectrometer 
which served as a luminosity monitor to remove any uncertainties due to beam charge, current, target length, raster 
size, and target density fluctuations. See chapter \ref{chap_corrections} for a detailed discussion of the 
corrections and efficiencies applied to the measured beam charge, and Tables \ref{r_systematic_summary} 
and \ref{l_systematic_summary} for a summary of the systematic uncertainties in these corrections and efficiencies. 
However, we decided not to use $R_{Super}$ to extract the form factors and their ratio since it is estimated that by
dividing by the luminosity monitor, that will introduce more uncertainty in $R_{Super}$ than it will eliminate.
By looking at Table \ref{rsigma_R_data}, we see that the right arm cross sections have $\sim$ 0.50\% random
uncertainty, while the luminosity-related random uncertainty is $<$ 0.20\% as can be seen from Table
\ref{r_systematic_summary}. Instead we use the right arm results as a consistency check of the assumed uncertainties.

The form factors and their ratio extracted from this work are discussed and compared to previous 
Rosenbluth extractions and recent polarization transfer results. 
See section \ref{sect_previous_data} and subsections therein for a full discussion of previous Rosenbluth 
extractions and recent polarization transfer results. An introduction to the two-photon-exchange (TPE) and 
Coulomb distortion corrections as possible sources for the discrepancy between the Rosenbluth 
extractions and polarization transfer results is given. The linearity in the Born 
approximation of the Rosenbluth separation plots (L-T plots) as a function of $\varepsilon$ is 
examined for this work because any deviation from linearity would provide a clear signature of 
the effects beyond traditional radiative corrections, and would provide information on the nonlinear 
component of the TPE. Finally, as the results of this work raised the interest in the physics of the 
TPE and Coulomb distortion corrections and laid down the foundations for new experiments aimed at 
measuring the size of the TPE corrections, I conclude with a brief discussion of the TPE calculations and presentations
of these future experiments.

\section{Form Factors Extraction} \label{form_factors_extraction}
\pagestyle{myheadings}

In this section I will describe the general procedure used to extract $G_{Ep}$, $G_{Mp}$, and 
$\mu_{p}G_{Ep} \over G_{Mp}$ by using the reduced cross sections $\sigma_{R}$, as listed in Tables \ref{rsigma_R_data} 
and \ref{lsigma_R_data}. 

\subsection{Form Factors Extraction Using $\sigma_{R}$ (Single Arm)} \label{extractions_sigmaR}
Equation (\ref{eq:reduced}) defines the elastic e-p reduced cross section $\sigma_{R}$. 
A linear fit of $\sigma_{R}$ to $\varepsilon$ at a fixed $Q^2$ gives $\tau G_{Mp}^2(Q^2)$ 
as the intercept and $G_{Ep}^2(Q^2)$ as the slope. 
If the goal is to extract $G_{Mp}$ and $G_{Ep}$, then a linear fit of $\sigma_{R}$ to $\varepsilon$ is performed 
using the following form:
\begin{equation} \label{eq:fit_ge_gm1}
\sigma_{R} = \tau A^{2} + B^{2} \varepsilon~,
\end{equation}
where $A$ and $B$ are the parameters of the fit and represent $G_{Mp}$ and $G_{Ep}$, and the 
uncertainties in $A$ and $B$ represent $\delta G_{Mp}$ and $\delta G_{Ep}$, respectively.
On the other hand, if the goal is to extract $\frac{G_{Ep}}{G_{Mp}}$, then a linear fit of 
$\sigma_{R}$ to $\varepsilon$ is performed using:
\begin{equation} \label{eq:fit_ge_gm2}
\sigma_{R} = \tau A^{2} \Bigg(1.0 + B^{2} \frac{\varepsilon}{\tau} \Bigg)~,
\end{equation}
where $A$ and $B$ represent $G_{Mp}$ and $\frac{G_{Ep}}{G_{Mp}}$, and the uncertainties in $A$ and $B$ 
represent $\delta G_{Mp}$ and $\delta \frac{G_{Ep}} {G_{Mp}}$, respectively. Note that the ratio $\frac{G_{Ep}} {G_{Mp}}$
and its uncertainty can be determined first by taking the ratio of the individual form factors as extracted using equation 
(\ref{eq:fit_ge_gm1}) and then propagating the (correlated) errors on $G_{Ep}$ and $G_{Mp}$,  
however, equation (\ref{eq:fit_ge_gm1}) is used since it directly provides the ratio $\frac{G_{Ep}} {G_{Mp}}$ and its 
uncertainty.

\subsection*{(a) $G_{Ep}$ and $G_{Mp}$ Extraction Using Equation \ref{eq:fit_ge_gm1}}
The classification of the systematic uncertainties into scale, random, and slope uncertainties and their
effects on the extracted form factors has been discussed in section \ref{qeff_intro}.
To determine the individual form factors, we start by adding in quadrature the total random 
uncertainty in $\sigma_{R}$, $\delta_{Random}$, to the statistical uncertainty in each $\sigma_{R}$, $\delta_{Stat}$, as listed in 
Tables \ref{rsigma_R_data} and \ref{lsigma_R_data} or:
\begin{equation} \label{eq:sigma_R_stat_random}
\delta_{\sigma_{R}} = \sqrt{\delta_{Stat}^2 + \delta_{Random}^2}~.
\end{equation}

\begin{figure}[!htbp]
\begin{center}
\epsfig{file=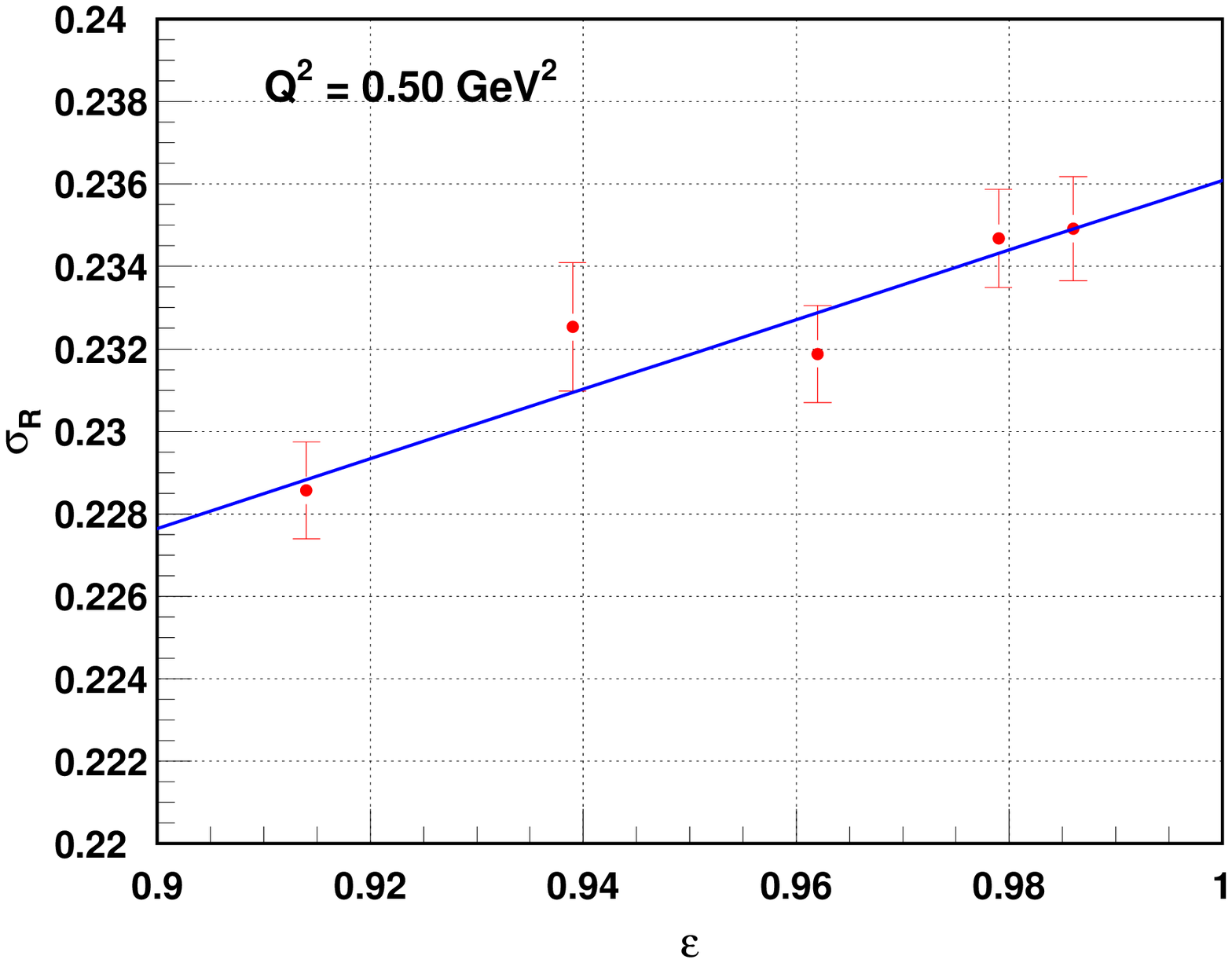,width=3.55in}
\end{center}
\caption[A linear fit of the right arm $\sigma_{R}$ to $\varepsilon$ at $Q^2$ = 0.50 GeV$^2$ using 
equation (\ref{eq:fit_ge_gm1}).]
{A linear fit of the right arm $\sigma_{R}$ to $\varepsilon$ at $Q^2$ = 0.50 GeV$^2$ using 
equation (\ref{eq:fit_ge_gm1}). The shown uncertainties are the combined statistical and random from
equation (\ref{eq:sigma_R_stat_random}).}
\label{fig:0.50_lt_sigma_r}
\end{figure}

\begin{figure}[!htbp]
\begin{center}
\epsfig{file=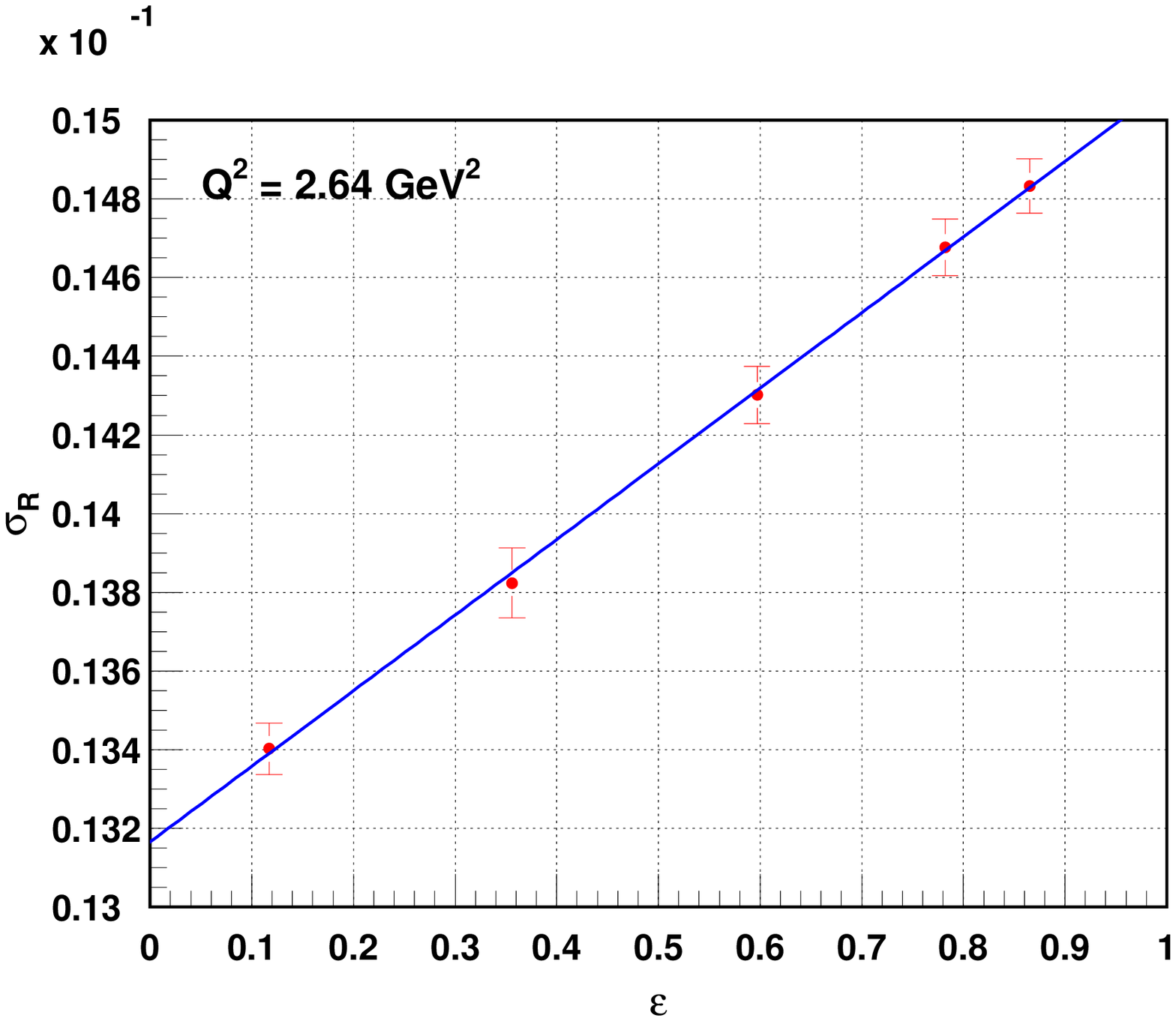,width=3.55in}
\end{center}
\caption[A linear fit of the left arm $\sigma_{R}$ to $\varepsilon$ at $Q^2$ = 2.64 GeV$^2$ using 
equation (\ref{eq:fit_ge_gm1}).]
{A linear fit of the left arm $\sigma_{R}$ to $\varepsilon$ at $Q^2$ = 2.64 GeV$^2$ using 
equation (\ref{eq:fit_ge_gm1}). The shown uncertainties are the combined statistical and random
from equation (\ref{eq:sigma_R_stat_random}).}
\label{fig:2.64_lt_sigma_r}
\end{figure}

\begin{figure}[!htbp]
\begin{center}
\epsfig{file=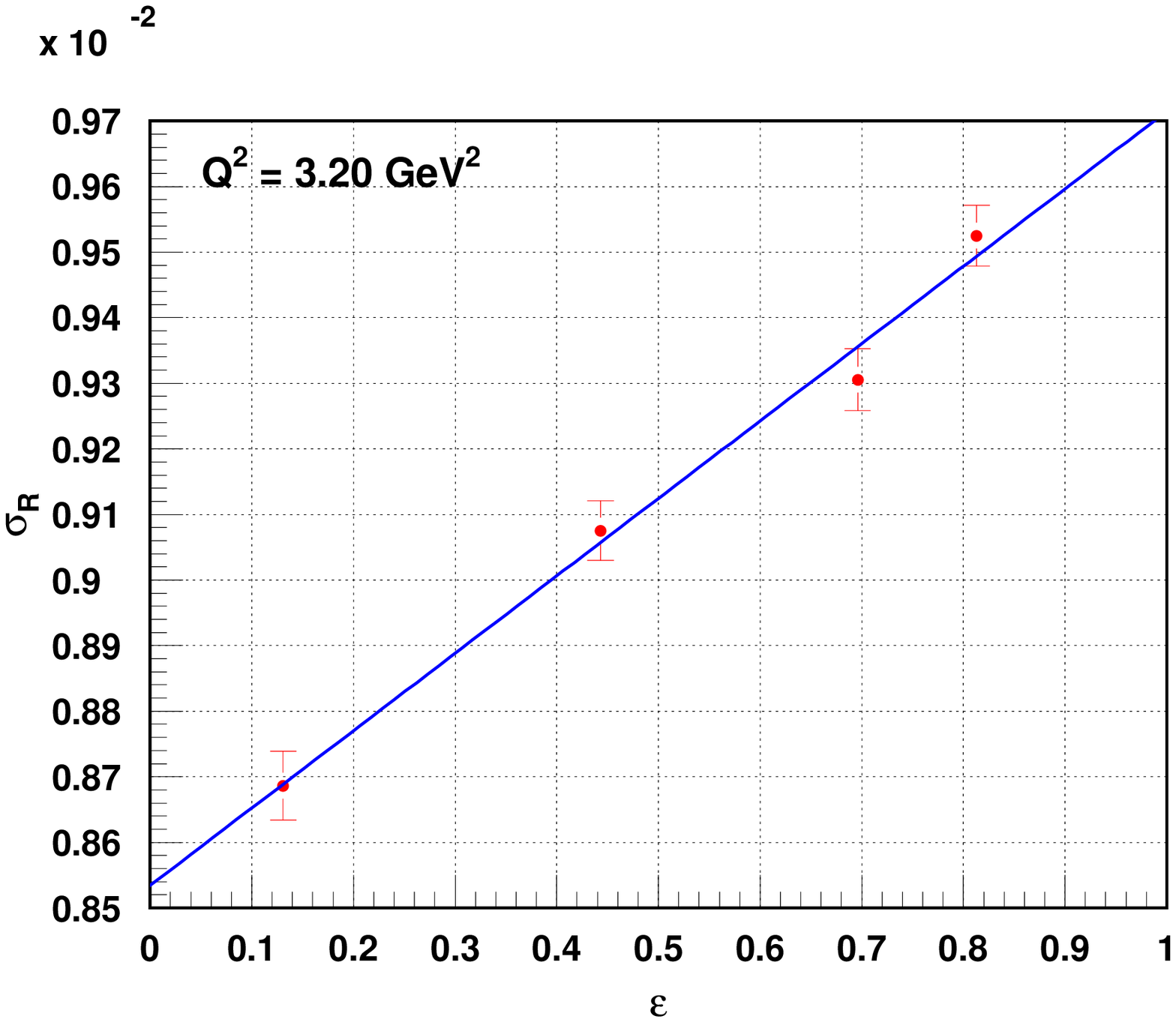,width=3.55in}
\end{center}
\caption[A linear fit of the left arm $\sigma_{R}$ to $\varepsilon$ at $Q^2$ = 3.20 GeV$^2$ using 
equation (\ref{eq:fit_ge_gm1}).]
{A linear fit of the left arm $\sigma_{R}$ to $\varepsilon$ at $Q^2$ = 3.20 GeV$^2$ using 
equation (\ref{eq:fit_ge_gm1}). The shown uncertainties are the combined statistical and random from
equation (\ref{eq:sigma_R_stat_random}).}
\label{fig:3.20_lt_sigma_r}
\end{figure}

\begin{figure}[!htbp]
\begin{center}
\epsfig{file=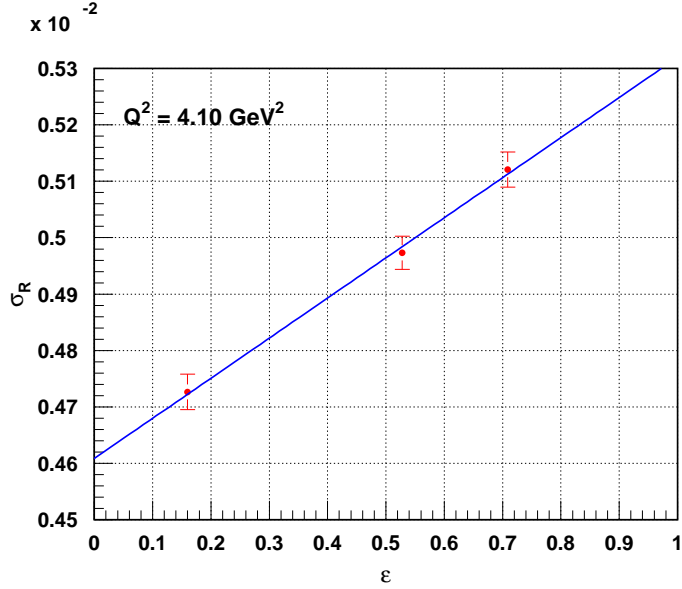,width=3.55in}
\end{center}
\caption[A linear fit of the left arm $\sigma_{R}$ to $\varepsilon$ at $Q^2$ = 4.10 GeV$^2$ using 
equation (\ref{eq:fit_ge_gm1}).]
{A linear fit of the left arm $\sigma_{R}$ to $\varepsilon$ at $Q^2$ = 4.10 GeV$^2$ using 
equation (\ref{eq:fit_ge_gm1}). The shown uncertainties are the combined statistical and random from
equation (\ref{eq:sigma_R_stat_random}).}
\label{fig:4.10_lt_sigma_r}
\end{figure}

A linear fit of $\sigma_{R}$ to $\varepsilon$ at a fixed $Q^2$ using equation 
(\ref{eq:fit_ge_gm1}) yields a value of $G_{Mp} = G_{Mp}^{(1)} \pm \delta G_{Mp}^{(1)}$ and  
$G_{Ep} = G_{Ep}^{(1)} \pm \delta G_{Ep}^{(1)}$ where the contribution from the random uncertainty 
to the total uncertainty in $G_{Mp}$ and $G_{Ep}$ is given by
$\delta G_{Mp}^{Random} = \delta G_{Mp}^{(1)}$ and $\delta G_{Ep}^{Random} = \delta G_{Ep}^{(1)}$. 
Figures \ref{fig:0.50_lt_sigma_r}, \ref{fig:2.64_lt_sigma_r}, \ref{fig:3.20_lt_sigma_r}, and 
\ref{fig:4.10_lt_sigma_r} show such fits done at $Q^2$ = 0.50, 2.64, 3.20, and 4.10 GeV$^2$, respectively. 

To determine the contribution of the slope uncertainty in $\sigma_{R}$ to the total uncertainty 
in $G_{Mp}$ and $G_{Ep}$, we vary $\sigma_{R}$ by the
uncertainties that vary linearly with $\varepsilon$ or the total slope uncertainty, 
$\delta_{Slope}$, as given at the bottom of Tables \ref{r_systematic_summary} and \ref{l_systematic_summary}:
\begin{equation} \label{eq:sigma_R_stat_random_slope_correc}
\sigma_{R}^{Slope} = \sigma_{R}\Bigg(1.0 + \varepsilon \delta_{Slope}\Bigg)~,
\end{equation}
then, a linear fit of $\sigma_{R}^{Slope}$ to $\varepsilon$ using equation 
(\ref{eq:fit_ge_gm1}), with uncertainties in $\sigma_{R}^{Slope}$ given by 
$\delta_{\sigma_{R}^{Slope}} = \delta_{\sigma_{R}}(1.0 +\varepsilon \delta_{Slope})$, 
yields a value of $G_{Mp} = G_{Mp}^{(2)} \pm \delta G_{Mp}^{(2)}$
and $G_{Ep} = G_{Ep}^{(2)} \pm \delta G_{Ep}^{(2)}$. 
Therefore, the contribution from the slope uncertainty to the total uncertainty in $G_{Mp}$ and 
$G_{Ep}$ is given by $\delta G_{Mp}^{Slope} = \Big|G_{Mp}^{(1)}-G_{Mp}^{(2)}\Big|$ and 
$\delta G_{Ep}^{Slope} = \Big|G_{Ep}^{(1)}-G_{Ep}^{(2)}\Big|$. Note that for the right arm we used
$\delta_{\sigma_{R}^{Slope}} = \delta_{\sigma_{R}}\Big(1.0 +(\varepsilon-1.0)\delta_{Slope}\Big)$ since all the 
measurements were taken near $\varepsilon$=1.0. 

Similarly, we determine the contribution of the scale uncertainty in $\sigma_{R}$ to the total 
uncertainty in $G_{Mp}$ and $G_{Ep}$ by varying $\sigma_{R}$ by the total scale uncertainty, 
$\delta_{Scale}$, as given at the bottom of Tables \ref{r_systematic_summary} and \ref{l_systematic_summary}:
\begin{equation} \label{eq:sigma_R_stat_random_scale_correc}
\sigma_{R}^{Scale} = \sigma_{R}\Bigg(1.0 + \delta_{Scale}\Bigg)~,
\end{equation}
and a linear fit of $\sigma_{R}^{Scale}$ to $\varepsilon$ using equation 
(\ref{eq:fit_ge_gm1}), with uncertainties in $\sigma_{R}^{Scale}$ given by 
$\delta_{\sigma_{R}^{Scale}} = \delta_{\sigma_{R}}(1.0 + \delta_{Scale})$, 
yields a value of $G_{Mp} = G_{Mp}^{(3)} \pm \delta G_{Mp}^{(3)}$
and $G_{Ep} = G_{Ep}^{(3)} \pm \delta G_{Ep}^{(3)}$. The contribution from the 
scale uncertainty to the total uncertainty in $G_{Mp}$ and $G_{Ep}$ is therefore given by 
$\delta G_{Mp}^{Scale} = \Big|G_{Mp}^{(1)}-G_{Mp}^{(3)}\Big|$ and 
$\delta G_{Ep}^{Scale} = \Big|G_{Ep}^{(1)}-G_{Ep}^{(3)}\Big|$.

Finally, $G_{Mp}$ and $G_{Ep}$ are given by $G_{Mp} = G_{Mp}^{(1)} \pm \delta G_{Mp}^{Total}$
and $G_{Ep} = G_{Ep}^{(1)} \pm \delta G_{Ep}^{Total}$ where $\delta G_{Mp}^{Total}$ and 
$\delta G_{Ep}^{Total}$ are given by:
\begin{equation} \label{eq:total_uncertainty_gmp}
\delta G_{Mp}^{Total} = \sqrt{ (\delta G_{Mp}^{Random})^2 + (\delta G_{Mp}^{Slope})^2 
+ (\delta G_{Mp}^{Scale})^2}~,
\end{equation}
and
\begin{equation} \label{eq:total_uncertainty_gep}
\delta G_{Ep}^{Total} = \sqrt{ (\delta G_{Ep}^{Random})^2 + (\delta G_{Ep}^{Slope})^2 
+ (\delta G_{Ep}^{Scale})^2}~.
\end{equation}

The electric and magnetic form factors for the proton with their uncertainties were determined at 
$Q^2$ = 0.50, 2.64, 3.20, and 4.10 GeV$^2$. Tables \ref{gmp_value_alluncertainties1} and 
\ref{gep_value_alluncertainties1} list the results.

\begin{table}[!htbp]
\begin{center}
\begin{tabular}{|c|c|c|c|c|c|c|} 
\hline \hline
$Q^2$&$G_{Mp}$&$\delta G_{Mp}^{Random}$&$\delta G_{Mp}^{Slope}$&$\delta G_{Mp}^{Scale}$&$\delta G_{Mp}^{Total}$&$\frac{G_{Mp}}{\mu_{p}G_{D}}$\\
(GeV$^2$)& & & & & & $\pm$\\
         & & & & & & $\delta (\frac{G_{Mp}}{\mu_{p}G_{D}})$\\
\hline \hline 
0.50 &1.03340 &0.06726   &0.01247   &0.01702   &0.07049 & 1.07467 \\
     &        &          &          &          &        & $\pm$   \\
     &        &          &          &          &        & 0.07330 \\
\hline 
2.64 &0.1325  &0.0003466 &6.988E-06 &0.001913  &0.001944& 1.05632 \\
     &        &          &          &          &        & $\pm$   \\
     &        &          &          &          &        & 0.01549  \\
\hline     
3.20 &0.09691 &0.0003204 &7.398E-06 &0.001398  &0.001435& 1.05233\\
     &        &          &          &          &        & $\pm$   \\
     &        &          &          &          &        & 0.01558  \\
\hline 
4.10 &0.06291 &0.0002822 &4.701E-06 &0.0009097 &0.0009525 & 1.03390 \\
     &        &          &          &          &          & $\pm$   \\
     &        &          &          &          &          & 0.01565\\
\hline \hline
\end{tabular}
\caption[The magnetic form factor for the proton $G_{Mp}$ at $Q^2$ = 0.50, 2.64, 3.20, and 4.10 GeV$^2$ as 
determined by fitting $\sigma_{R}$ to $\varepsilon$ using equation (\ref{eq:fit_ge_gm1}).]
{The magnetic form factor for the proton $G_{Mp}$ at $Q^2$ = 0.50, 2.64, 3.20, and 4.10 GeV$^2$ as 
determined by fitting $\sigma_{R}$ to $\varepsilon$ using equation (\ref{eq:fit_ge_gm1}). In addition, the 
contributions from the random, slope, and scale uncertainties to the total uncertainty in $G_{Mp}$ is also 
given.}
\label{gmp_value_alluncertainties1}
\end{center}
\end{table}
\newpage
\begin{table}[!htbp]
\begin{center}
\begin{tabular}{|c|c|c|c|c|c|c|} 
\hline \hline
$Q^2$&$G_{Ep}$&$\delta G_{Ep}^{Random}$&$\delta G_{Ep}^{Slope}$&$\delta G_{Ep}^{Scale}$&$\delta G_{Ep}^{Total}$&$\frac{G_{Ep}}{G_{D}}$\\
(GeV$^2$)& & & & & &$\pm$\\
         & & & & & &$\delta (\frac{G_{Ep}}{G_{D}})$\\
\hline \hline 
0.50 &0.2906 &0.03589  &0.02682  &0.003182 &0.04492 &0.8440\\ 
     &       &         &         &         &        &$\pm$\\
     &       &         &         &         &        &0.1304\\
\hline \
2.64 &0.04385&0.001276 &0.0009129&0.0006389&0.001694&0.9761\\  
     &       &         &         &         &        &$\pm$\\ 
     &       &         &         &         &        &0.03772\\
\hline 
3.20 &0.03436&0.001377&0.0007532&0.0004959&0.001646&1.04198\\
     &       &         &         &         &        &$\pm$\\
     &       &         &         &         &        &0.04994\\
\hline 
4.10 &0.02666&0.001489 &0.0007684&0.0003775&0.001718&1.2238\\  
     &       &         &         &         &        &$\pm$\\ 
     &       &         &         &         &        &0.07886\\
\hline \hline 
\end{tabular}
\caption[The electric form factor for the proton $G_{Ep}$ at $Q^2$ = 0.50, 2.64, 3.20, and 4.10 GeV$^2$ as 
determined by fitting $\sigma_{R}$ to $\varepsilon$ using equation (\ref{eq:fit_ge_gm1}).]
{The electric form factor for the proton $G_{Ep}$ at $Q^2$ = 0.50, 2.64, 3.20, and 4.10 GeV$^2$ as 
determined by fitting $\sigma_{R}$ to $\varepsilon$ using equation (\ref{eq:fit_ge_gm1}). In addition, the 
contributions from the random, slope, and scale uncertainties to the total uncertainty in $G_{Ep}$ is also 
given.}
\label{gep_value_alluncertainties1}
\end{center}
\end{table}

\subsection*{(b) $G_{Mp}$ and $G_{Ep} \over G_{Mp}$ Extraction Using Equation \ref{eq:fit_ge_gm2}}
A linear fit of $\sigma_{R}$ to $\varepsilon$ at a fixed $Q^2$ using equation (\ref{eq:fit_ge_gm2})
yields $G_{Mp}$ and $\frac{G_{Ep}}{G_{Mp}}$. The same procedure outlined in subsection 
(a) is used with the following changes:
\begin{itemize}
\item{$\frac{G_{Ep}}{G_{Mp}}$ replaces $G_{Ep}$ whenever it occurs or: $G_{Ep} \to \frac{G_{Ep}}{G_{Mp}}$.}
\end{itemize}

Therefore $G_{Mp} = G_{Mp}^{(1)} \pm \delta G_{Mp}^{Total}$
and $\frac{G_{Ep}}{G_{Mp}} = (\frac{G_{Ep}}{G_{Mp}})^{(1)} \pm \delta (\frac{G_{Ep}}{G_{Mp}})^{Total}$ where 
$\delta G_{Mp}^{Total}$ and $\delta (\frac{G_{Ep}}{G_{Mp}})^{Total}$ are given by:
\begin{equation} \label{eq:total_uncertainty_gmp2}
\delta G_{Mp}^{Total} = \sqrt{ (\delta G_{Mp}^{Random})^2 + (\delta G_{Mp}^{Slope})^2 
+ (\delta G_{Mp}^{Scale})^2}~,
\end{equation}
and
\begin{equation} \label{eq:total_uncertainty_gepgmp_sigma_R}
\delta (\frac{G_{Ep}}{G_{Mp}})^{Total} = \sqrt{\Big(\delta (\frac{G_{Ep}}{G_{Mp}})^{Random}\Big)^2 
  + \Big(\delta (\frac{G_{Ep}}{G_{Mp}})^{Slope}\Big)^2 + \Big(\delta (\frac{G_{Ep}}{G_{Mp}})^{Scale}\Big)^2}~.   
\end{equation}

The magnetic form factor and the ratio of electric to magnetic form factor for the proton with their 
uncertainties were determined at $Q^2$ = 0.50, 2.64, 3.20, and 4.10 GeV$^2$. The results obtained for $G_{Mp}$ 
are identical to those listed in Table \ref{gmp_value_alluncertainties1}.
Table \ref{gepgmp_value_alluncertainties1} lists the results obtained for the ratio $\frac{G_{Ep}}{G_{Mp}}$.

\begin{table}[!htbp]
\begin{center}
\begin{tabular}{|c|c|p{2.3cm}|p{1.9cm}|p{1.9cm}|p{1.9cm}|p{1.4cm}|} 
\hline \hline
$Q^2$&$\frac{G_{Ep}}{G_{Mp}}$&$\delta (\frac{G_{Ep}}{G_{Mp}})^{Random}$&$\delta (\frac{G_{Ep}}{G_{Mp}})^{Slope}$&$\delta (\frac{G_{Ep}}{G_{Mp}})^{Scale}$&$\delta (\frac{G_{Ep}}{G_{Mp}})^{Total}$&~~$\frac{\mu_{p} G_{Ep}}{G_{Mp}}$\\
(GeV$^2$)& & & & & &~~~~$\pm$\\
         & & & & & &$\delta (\frac{\mu_{p} G_{Ep}}{G_{Mp}})$\\
\hline \hline 
0.50 &0.2803 &~~0.05303  &~~0.03039  &~0.000655   &~~0.06112 &~0.7828\\ 
     &       &           &           &            &          &~~~~$\pm$\\ 
     &       &           &           &            &          &~0.1707\\
\hline 
2.64 &0.3308 &~~0.01040  &~~0.006930 &~6.202E-05  &~~0.01249 &~0.9240\\ 
     &       &           &           &            &          &~~~~$\pm$\\ 
     &       &           &           &            &          &~0.03490\\
\hline 
3.20 &0.3545 &~~0.01529  &~~0.007770 &~4.100E-05  &~~0.01715 &~0.9900\\ 
     &       &           &           &            &          &~~~~$\pm$\\ 
     &       &           &           &            &          &~0.04789\\
\hline
4.10 &0.4238 &~~0.02540  &~~0.01222  &~6.601E-05  &~~0.02819 &~1.1837\\
     &       &           &           &            &          &~~~~$\pm$\\
     &       &           &           &            &          &~0.07873\\
\hline \hline
\end{tabular}
\caption[The ratio of electric to magnetic form factor for the proton $\frac{G_{Ep}}{G_{Mp}}$ at 
$Q^2$ = 0.50, 2.64, 3.20, and 4.10 GeV$^2$ as determined by fitting $\sigma_{R}$ to $\varepsilon$ 
using equation (\ref{eq:fit_ge_gm2}).]
{The ratio of electric to magnetic form factor for the proton $\frac{G_{Ep}}{G_{Mp}}$ at 
$Q^2$ = 0.50, 2.64, 3.20, and 4.10 GeV$^2$ as determined by fitting $\sigma_{R}$ to $\varepsilon$ using 
equation (\ref{eq:fit_ge_gm2}). In addition, the contributions from the random, slope, and 
scale uncertainties to the total uncertainty in $\frac{G_{Ep}}{G_{Mp}}$ is also given.}
\label{gepgmp_value_alluncertainties1}
\end{center}
\end{table}

\section{Discussion of the Results} \label{results_conclusion}
\begin{figure}[!htbp]
\begin{center}
\epsfig{file=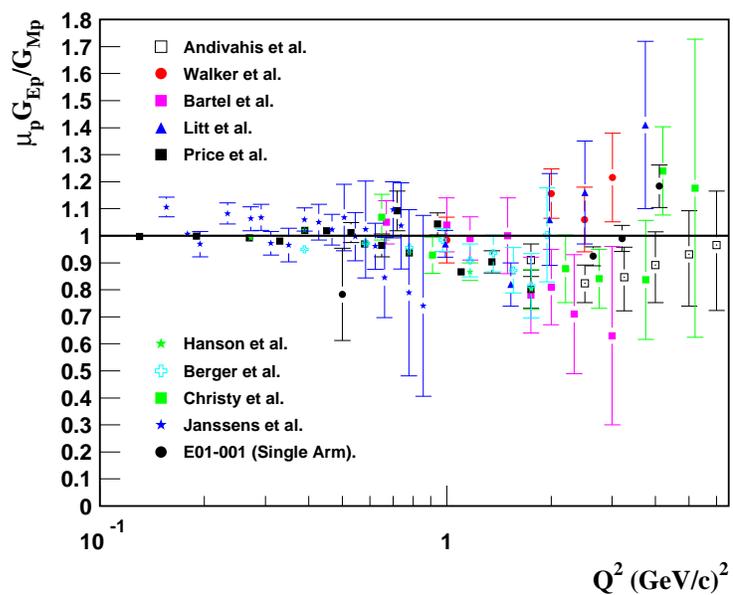,width=3.82in}
\end{center}
\caption[The world data of Rosenbluth separation determination of $\mu_{p}G_{Ep}/G_{Mp}$ ratio compared to that of 
E01-001 (single arm extraction).]
{The world data of Rosenbluth separation determination of $\mu_{p}G_{Ep}/G_{Mp}$ ratio compared to that of E01-001 
(single arm extraction) (Table \ref{gepgmp_value_alluncertainties1}).}
\label{fig:gep_gmp_normale01001_allrosenbluth}
\end{figure}

\begin{figure}[!htbp]
\begin{center}
\epsfig{file=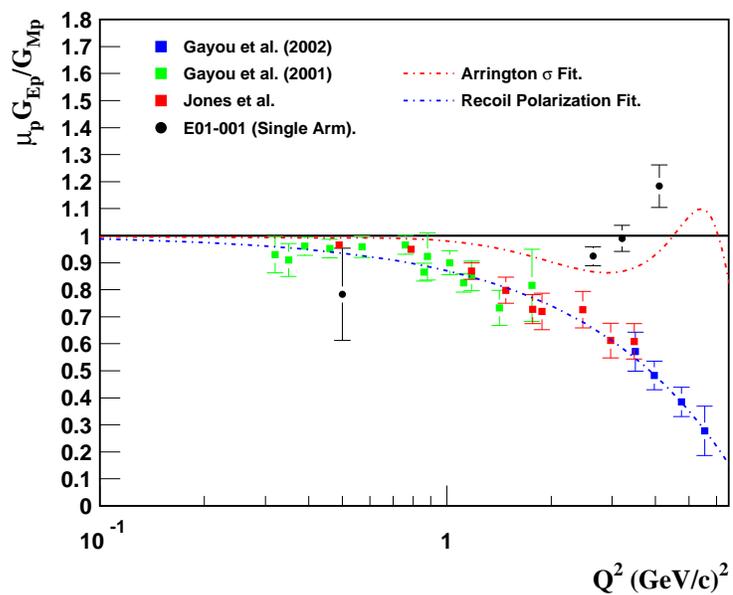,width=3.82in}
\end{center}
\caption[The values of $\mu_{p}G_{Ep}/G_{Mp}$ for the proton from E01-001 (single arm extraction) 
and recoil polarization.]
{The values of $\mu_{p}G_{Ep}/G_{Mp}$ for the proton from E01-001 (single arm extraction) 
(Table \ref{gepgmp_value_alluncertainties1}) and recoil polarization.}
\label{fig:e01001_singlearm_polarization}
\end{figure}

Figures \ref{fig:gep_gmp_normale01001_allrosenbluth} shows the world data on $\frac{\mu_{p}G_{Ep}}{G_{Mp}}$ 
ratio for the proton from previous Rosenbluth measurements along with the results of E01-001 single arm 
(Table \ref{gepgmp_value_alluncertainties1}). In addition, Figures \ref{fig:e01001_singlearm_polarization} 
shows the $\frac{\mu_{p}G_{Ep}}{G_{Mp}}$ ratio from the E01-001 single arm extractions along with the results of 
recoil polarization. The global analysis of previous cross section data by Arrington \cite{arrington03a} is also shown. 
By comparing the results of the E01-001 with those of previous Rosenbluth separations and 
recoil polarization measurements, we conclude that: 
\begin{enumerate}
{\it
\item \textbf{The values of $\frac{\mu_{p}G_{Ep}}{G_{Mp}}$ ratio from the single arm extractions 
are in good agreement with the previous Rosenbluth measurements but with much smaller statistical 
and systematic uncertainties than the previous best measurements \cite{andivahis94}. The total uncertainty 
is comparable to those of recoil polarization}.
}
{\it
\item \textbf{The high precision achieved in the E01-001 extractions:}
}
\begin{itemize}
{\it
\item \textbf{Leaves little room for doubting that the values of $\frac{\mu_{p}G_{Ep}}{G_{Mp}}$ ratio
extracted from Rosenbluth separations are inconsistent with those reported from the high-$Q^2$ recoil polarization measurements.}
}
{\it
\item \textbf{Makes it clear that the problem is not simply an experimental error in the Rosenbluth 
measurements or technique.}
}
{\it
\item \textbf{Confirms the reliability of the elastic e-p cross sections extracted from previous
Rosenbluth separations.}
}
\end{itemize}
{\it
\item \textbf{(This point will be discussed below):\\
Provides a significant test of the validity of the commonly used radiative corrections 
(bremsstrahlung corrections).} 
}
\end{enumerate}

In summary, we see that the results of the E01-001 are in agreement with the previous Rosenbluth data. 
The experiment provided systematic uncertainties much smaller than the best previous Rosenbluth measurements,
\cite{andivahis94}, and comparable to those of the recoil polarization clearly establishing the discrepancy
between the Rosenbluth separations and recoil polarization results. These results confirmed that the
discrepancy is not due to experimental error in the Rosenbluth measurements and provided a strong
test of the conventional radiative corrections used.

A possible source for the discrepancy is the effect of higher-order corrections to the reduced 
cross sections. Suggested sources for the discrepancy will be discussed in section
\ref{two_photon_exchange_coulomb}. The reduced cross sections as defined by equation (\ref{eq:reduced}) 
and listed in Tables \ref{rsigma_R_data} and \ref{lsigma_R_data} are derived in the Born approximation 
and have been corrected for many of the higher-order radiative corrections terms. 
The form factors and thus their ratio have been extracted using these reduced cross sections without including 
fully the effects of these higher-order corrections on the reduced cross sections for consistency with previous 
Rosenbluth measurements.

The standard radiative correction procedure used was discussed in section \ref{rad_corrections}. 
We correct the data for bremsstrahlung, vertex corrections, and loop diagrams, with almost all of
the $\varepsilon$ dependence coming from bremsstrahlung.
Figure \ref{fig:rad_corr_epsilon} shows the calculated radiative correction factor (internal corrections only) 
as a function of $\varepsilon$ for Q$^2$ = 2.64 GeV$^2$. Again, while the magnitude of the corrections is similar 
and both show an approximately linear dependence on $\varepsilon$, the dependence is much smaller for protons, 
$\sim -$8\%, than for the electron $\sim$ 17\%, and of opposite sign. 
Note that when detecting the electron, the $\varepsilon$ dependence of the bremsstrahlung correction exceeds the 
$\varepsilon$ dependence coming from $G_{Ep}$ which is not the case in the E01-001 experiment.

\section{Possible Sources for the Discrepancy } \label{two_photon_exchange_coulomb}

The results of the E01-001 experiment have clearly established and confirmed the discrepancy
between the Rosenbluth separations and recoil polarization results. The questions that need 
to be answered are: why do the two techniques disagree? and which form factors are the 
correct ones? 

As noted above, we have essentially ruled out the possibility that the discrepancy is due to an 
experimental error in the Rosenbluth separations measurements or technique. That confirms that 
previous elastic e-p cross sections are reliable which is crucial since these cross sections are used for 
normalization or as an input to the analysis of different experiments. In addition, 
reanalysis of the existing form factor data by Arrington \cite{arrington03a}, see section 
\ref{sect_discussion} for details, has confirmed that all the different data from previous Rosenbluth 
separations measurements are consistent. This reanalysis combined with the new results from E01-001 rule out any 
possibility for a bad data sets or incorrect normalization in the combined Rosenbluth analysis. 
\textbf{It should be noted that while the E01-001 experiment provided an independent check on previous Rosenbluth
measurements there has been, to the best of my knowledge, no independent check of the high-$Q^2$ recoil polarization results.}

As for which form factors are the correct ones to use, again, if the cross section measurements and 
hence the Rosenbluth extractions are incorrect, that still will not solve the problem since the recoil 
polarization technique provides the ratio of $G_{Ep}$ to $G_{Mp}$ and not the actual values for the 
individual form factors. While the form factors extracted by Rosenbluth separations may possibly not
represent the true form factors for the proton, they do indeed provide the correct elastic e-p cross sections
and presumably provide the best parameterization when elastic cross sections are used for normalization or as an input 
to the analysis of different experiments.    

\subsection{Two-Photon-Exchange (TPE) Correction}
The difference in the $\frac{\mu_{p}G_{Ep}}{G_{Mp}}$ ratio from Rosenbluth separations and recoil polarizations results 
can be explained by a common $\varepsilon$-dependent systematic error in the cross section measurements of 
(5-8)\% \cite{arrington04a,arrington03a,guichon03}. 
The establishment of the discrepancy has led people to suggest that the discrepancy is due to
a missing corrections in the cross section due to a higher-order processes such as 
two-photon-exchange (TPE). 
The standard radiative correction procedure used was discussed in 
section \ref{rad_corrections}. Figure \ref{fig:RCdiagrams} shows the full contribution of the elastic and 
inelastic radiative corrections to the elastic e-p cross section including the first-order QED (Born) radiative
corrections or single-photon-exchange. Figure \ref{fig:RCdiagrams} (e), crossed-box, and (f), box, are the contributions 
from TPE to the elastic e-p cross section which are not fully included in the standard radiative correction procedures.
See \cite{maximon02,blunden03,blunden05} for details.  
\textbf{It is believed that inclusion of TPE contributions may remove the discrepancy}. 

\begin{figure}[!htbp]
\begin{center}
\epsfig{file=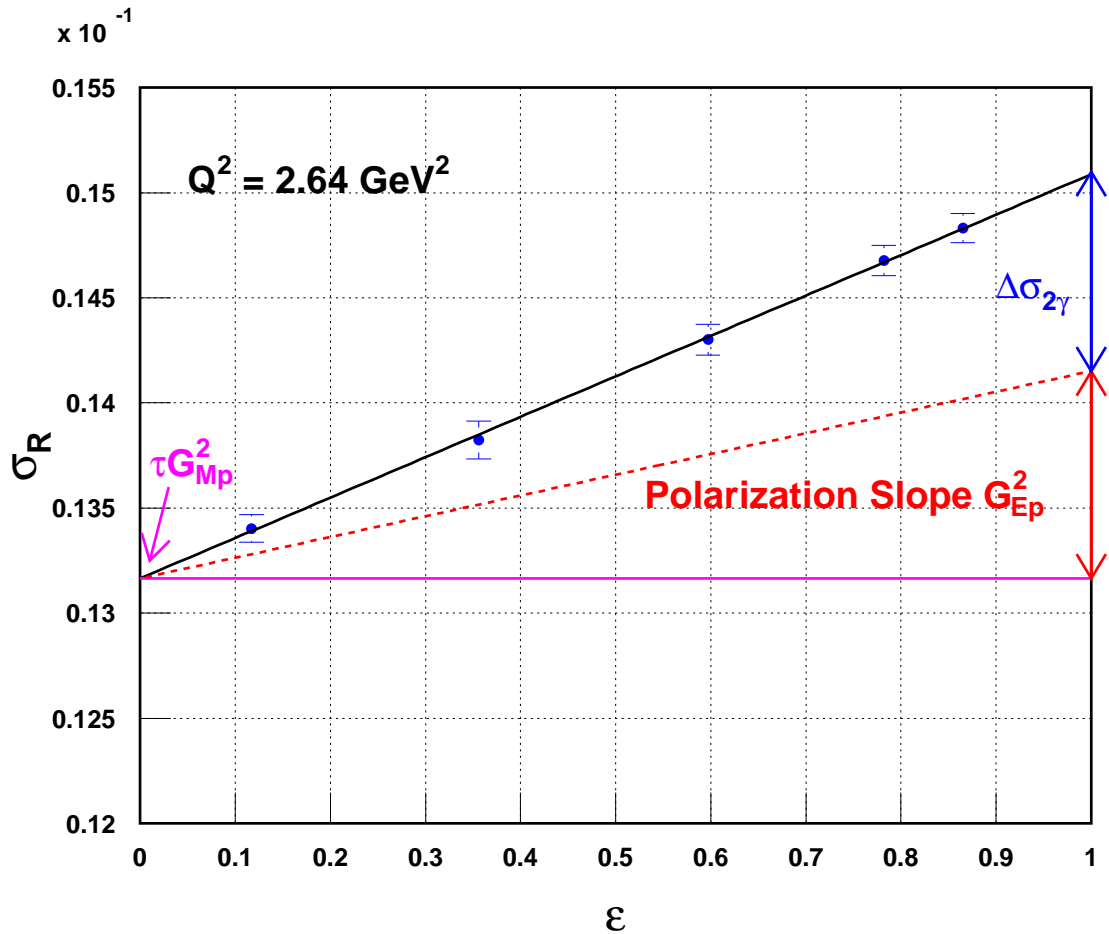,width=5.8in}
\end{center}
\caption[The $\varepsilon$ dependence of the reduced cross section at $Q^2$ = 2.64 GeV$^2$ as measured in 
the E01-001 experiment, and as predicted from the recoil polarization.]
{The $\varepsilon$ dependence of the reduced cross section at $Q^2$ = 2.64 GeV$^2$ as measured in 
the E01-001 experiment (blue solid circles), and as predicted from the recoil polarization (red dashed line). The
black solid line is a linear fit to the data. If the recoil polarization results represent the true form factors
(note the smaller slope or $G^2_{Ep}$ predicted by the recoil polarization and shown in red double arrow), 
then TPE or $\Delta \sigma_{2 \gamma}$ shown as blue double arrow will yield roughly about half of the $\varepsilon$ 
dependence at $Q^2$ = 2.64 GeV$^2$.}
\label{fig:2.64_sigma_lt_pol_2gamma}
\end{figure}

\begin{figure}[!htbp]
\begin{center}
\epsfig{file=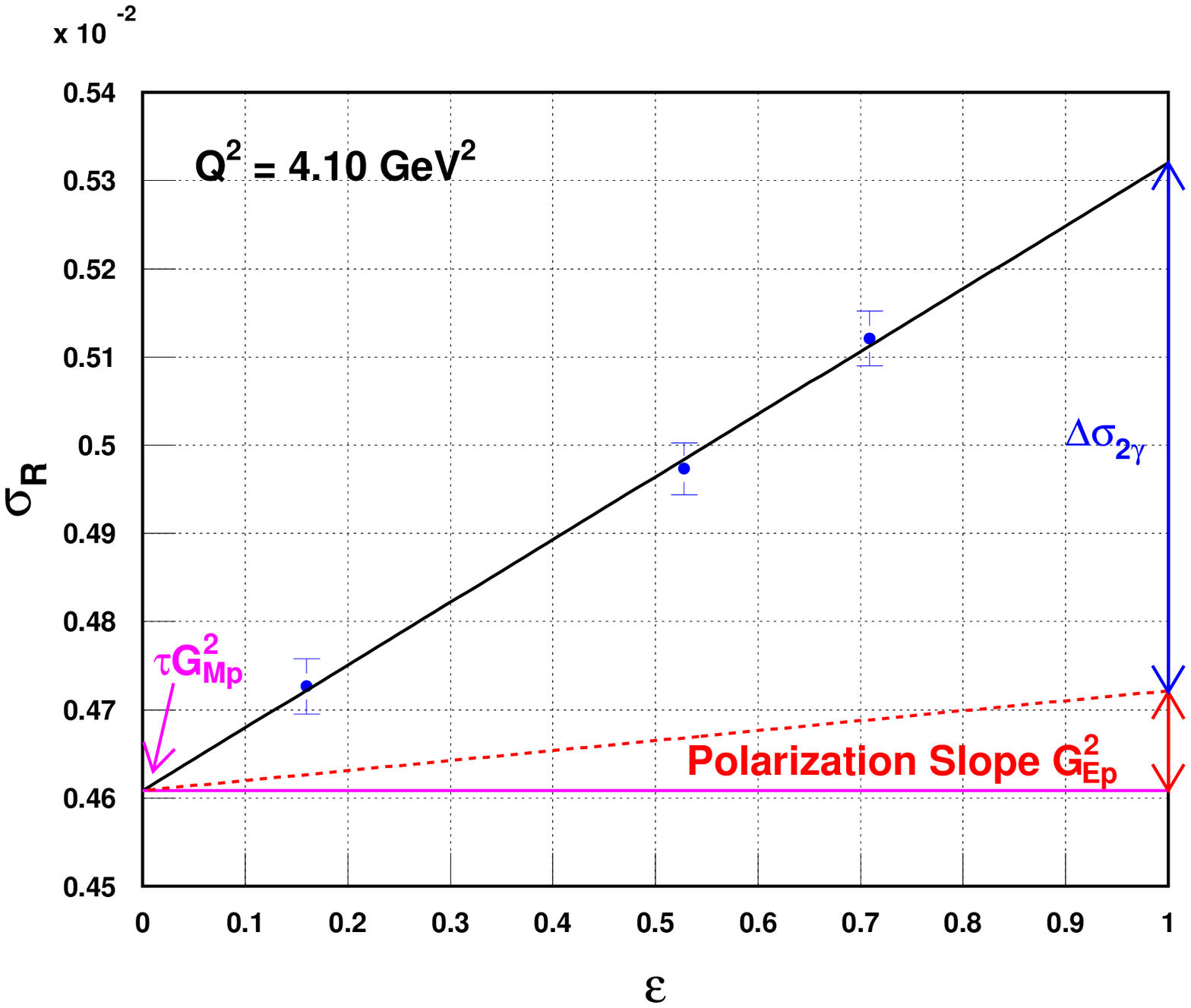,width=5.8in}
\end{center}
\caption[The $\varepsilon$ dependence of the reduced cross section at $Q^2$ = 4.10 GeV$^2$ as measured in 
the E01-001 experiment, and as predicted from the recoil polarization.]
{The $\varepsilon$ dependence of the reduced cross section at $Q^2$ = 4.10 GeV$^2$ as measured in 
the E01-001 experiment (blue solid circles), and as predicted from the recoil polarization (red dashed line). The
black solid line is a linear fit to the data. If the recoil polarization results represent the true form factors
(note the smaller slope or $G^2_{Ep}$ predicted by the recoil polarization and shown in red double arrow), 
then TPE or $\Delta \sigma_{2 \gamma}$ shown as blue double arrow will yield about 85\% of the $\varepsilon$ 
dependence at 4.10 GeV$^2$.}
\label{fig:4.10_sigma_lt_pol_2gamma}
\end{figure}

If we assume that the discrepancy is in fact due to TPE and the recoil polarization results are approximately
correct, then the $\frac{\mu_{p}G_{Ep}}{G_{Mp}}$ ratio becomes small enough as $Q^2$ gets large (note the
smaller slope predicated by recoil polarization) that the $\varepsilon$ dependence of the reduced cross section is 
dominated by TPE. Figures \ref{fig:2.64_sigma_lt_pol_2gamma} and \ref{fig:4.10_sigma_lt_pol_2gamma} show the contribution
of the TPE, $\Delta_{2\gamma}$, to $\varepsilon$ dependence of the L-T plots at $Q^2$ = 2.64 and 4.10 GeV$^2$, respectively. 
At large $Q^2$ where $G_{Ep}$ is suppressed and the reduced cross section in the Born approximation
has almost no $\varepsilon$ dependence, the discrepancy implies cross section difference of 5-10\%. 
In this region, almost all of the $\varepsilon$ dependence would come from TPE and the $\varepsilon$ dependence 
of the reduced cross section will represent the full contributions of TPE. 


\textbf{The high precision provided by the E01-001 experiment on $\sigma_{R}$ and hence on $\frac{\mu_{p}G_{Ep}}{G_{Mp}}$  
is of great importance as it constitutes a precise measurement of the discrepancy between the Rosenbluth separations
and recoil polarization results, thus allowing for a precise test as to whether TPE effects can explain the difference between 
Rosenbluth and recoil polarization measurements}. The reduced cross section in the Born approximation varies linearly with 
$\varepsilon$. Therefore, at low $Q^2$ where $G_{Ep}$ contributes most, any deviation from linearity would have to come from 
higher-order terms that are not included in the standard radiative correction procedures. Such deviation from linearity would 
provide a clean signature of TPE and would provide information about the nonlinear component of TPE. If the TPE correction is to 
explain the discrepancy, it would have to introduce (5-8)\% $\varepsilon$-dependent correction to the cross sections. Such correction 
would have to be roughly linear since large nonlinearities are not observed.    

An important point to emphasize is that \textbf{the standard radiative corrections have an $\varepsilon$ dependence that is 
comparable to the slope brought about by the form factors, (see Figure \ref{fig:rad_corr_epsilon})}.
In addition, if missing radiative correction terms are the reason for such discrepancy, then the form factors
extracted from Rosenbluth separations and to lesser extent those from recoil polarization do not represent 
the true form factors for the proton since we are not correctly isolating the single-photon-exchange process. 
Therefore, it is absolutely crucial to understand the radiative corrections and in particular higher-order 
processes such as TPE theoretically and experimentally before we can be confident in our knowledge of the 
proton form factors. 

Several calculations in the 1950s and 1960s tried to estimate the size of the TPE contributions to the 
unpolarized elastic e-p cross sections \cite{drell57,drell59,werthamer61,greenhut69,lewis56}. While some calculations 
used only the proton intermediate state \cite{lewis56}, other included the excited intermediate states of the proton as well 
\cite{drell57,drell59,werthamer61,greenhut69}. 
The TPE corrections estimated from these calculations were extremely small ($\leq$ 1\%) and were not included 
in the standard radiative correction procedures.  

Motivated in part by the results of the E01-001 experiment, the interest in the physics of TPE 
and Coulomb distortion corrections and their impact on electron scattering observables has recently risen both experimentally 
and theoretically. In addition, the results of the E01-001 experiment have laid down the foundation for new experiments 
aimed at measuring the size of the TPE corrections (to be discussed in section \ref{future_experiments}).
All of the new calculations have predicted a nonlinearities in the $\varepsilon$ dependence of TPE.
A brief description of these recent calculations will be given below. In addition, their effect on the
$\frac{\mu_{p}G_{Ep}}{G_{Mp}}$ ratio from the E01-001 extractions will be examined where possible.

\subsection{Recent Calculations of the TPE Corrections}

\begin{itemize}
{\it
\item \textbf{Calculation by Blunden, Melnitchouk, and Tjon et al. \cite{blunden03}:}
}
\end{itemize}

\begin{figure}[!htbp]
\begin{center}
\epsfig{file=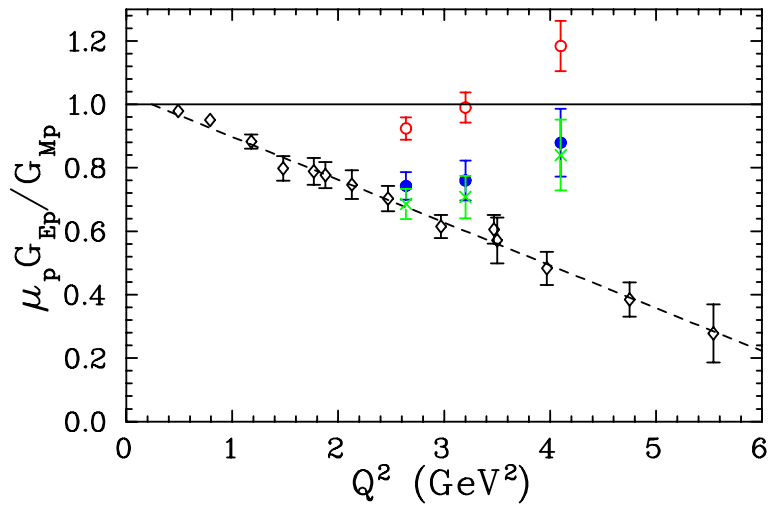,width=4in}
\end{center}
\caption[The $\mu_{p}G_{Ep}/G_{Mp}$ ratio for the proton from the E01-001 experiment corrected for TPE correction by Blunden 
et al \cite{blunden03} and Coulomb correction by Sick and Arrington \cite{arrington04c}.] 
{The $\mu_{p}G_{Ep}/G_{Mp}$ ratio for the proton from the E01-001 experiment (red open circles) and recoil polarization 
(black open diamonds). The E01-001 $\mu_{p}G_{Ep}/G_{Mp}$ corrected for TPE correction by Blunden et al \cite{blunden03} 
(blue solid circles). The E01-001 $\mu_{p}G_{Ep}/G_{Mp}$ ratio corrected for TPE correction by Blunden et al 
including the Coulomb correction by Sick and Arrington \cite{arrington04c} (green crosses).}
\label{fig:e01001_blunden_coul}
\end{figure}

This is a low energy hadronic model that takes into account the proton intermediate state and neglects excited
intermediate states. It includes only the elastic contributions to the TPE correction or the box and 
crossed-box diagrams shown in Figure \ref{fig:RCdiagrams} (f) and (e), respectively. 
Their initial calculations yielded an $\varepsilon$ dependence of $\sim$ 2\% with small nonlinearities at small $\varepsilon$ and 
insignificant $Q^2$ dependence. These corrections became larger with the inclusion of an improved form factors 
\cite{blunden05}. Figure \ref{fig:e01001_blunden_coul} shows the effect of such a calculation on the ratio 
$\frac{\mu_{p}G_{Ep}}{G_{Mp}}$ extracted from the E01-001 experiment. In addition, the effect of Coulomb correction by Sick and 
Arrington \cite{arrington04c} (see section \ref{coulomb_distortion} for details) is also shown. 
While the discrepancy is largely resolved for $Q^2$ = 2-3 GeV$^2$, this calculation
fails to resolve the discrepancy for $Q^2 >$ 3 GeV$^2$. Recently, the authors have included
the $\Delta$ resonance as an intermediate excited state to their elastic box and crossed-box calculations of 
TPE corrections \cite{kondratyuk05}. The $\Delta$ contribution is smaller in magnitude than the contribution when the proton is in 
the intermediate state with a small modification to the $\varepsilon$ dependence of the TPE corrections. 
\newpage

\begin{itemize}
{\it
\item \textbf{Calculation by Chen et al. \cite{chen04}:}
}
\end{itemize}

\begin{figure}[!htbp]
\begin{center}
\epsfig{file=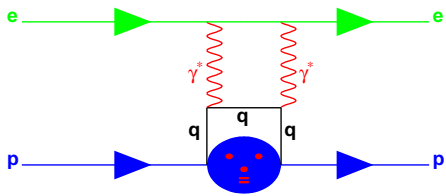,width=2.35in}
\end{center}
\caption[The handbag Feynman diagram used for the partonic calculation of TPE contribution to elastic e-p scattering by 
Chen et al \cite{chen04}.]
{Feynman diagram known as the handbag diagram used for the partonic calculation of TPE contribution to elastic e-p scattering by 
Chen et al \cite{chen04}. The electron (green) interacts with the proton via two virtual photons (red), 
while the proton (blue) shown as solid line and circle to represent the proton GPDs emits quark (black) which interacts with the 
two virtual photons. The quark gets re-absorbed again by the proton.}
\label{fig:2gamma_chen_hanbag}
\end{figure}

\begin{figure}[!htbp]
\begin{center}
\epsfig{file=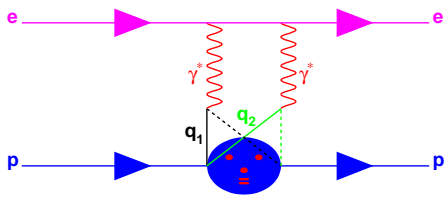,width=2.35in}
\end{center}
\caption[The cat's ears Feynman diagram used for the partonic calculation of TPE contribution to elastic e-p scattering by 
Chen et al \cite{chen04}.]
{Feynman diagram known as the cat's ears diagram used for the partonic calculation of TPE contribution to elastic e-p scattering by 
Chen et al \cite{chen04}. The electron (magenta) interacts with the proton via two virtual photons (red), 
while the proton (blue) shown as solid line and circle to represent the proton GPDs emits two quarks $q_1$ and $q_2$ (black and green solid
lines) which interact with the two virtual photons. The two quarks get re-absorbed again by the proton (black and green dashed lines).}
\label{fig:2gamma_chen_cat}
\end{figure}

\begin{figure}[!htbp]
\begin{center}
\epsfig{file=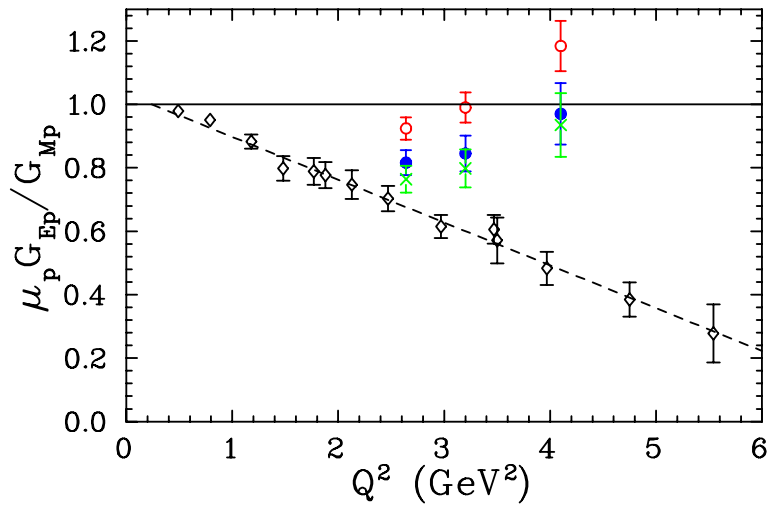,width=4in}
\end{center}
\caption[The $\mu_{p}G_{Ep}/G_{Mp}$ ratio for the proton from the E01-001 experiment corrected for TPE correction by 
Chen et al \cite{chen04} and Coulomb correction by Sick and Arrington \cite{arrington04c}.] 
{The $\mu_{p}G_{Ep}/G_{Mp}$ ratio for the proton from the E01-001 experiment (red open circles) and recoil 
polarization (black open diamonds). 
The E01-001 $\mu_{p}G_{Ep}/G_{Mp}$ ratio corrected for TPE correction by Chen et al \cite{chen04} (blue solid circles).
The E01-001 $\mu_{p}G_{Ep}/G_{Mp}$ ratio corrected for TPE correction by Chen et al including the 
Coulomb correction by Sick and Arrington (green crosses).}
\label{fig:e01001_chen_coul}
\end{figure}

This is a high energy model at the quark-parton level. It calculates the TPE correction using a generalized parton distribution 
(GPD) for the quark distribution to describe the emission and re-absorption of the partons by the nucleon as shown in Figures 
\ref{fig:2gamma_chen_hanbag} (the dominant contribution) and \ref{fig:2gamma_chen_cat}.
For large $Q^2$, the calculation shows a significant $\varepsilon$ dependence and nonlinearity to the correction with a weak $Q^2$ 
dependence. Since this is a high energy model, it is not expected to be valid at low $Q^2$ values. Figure \ref{fig:e01001_chen_coul} 
shows the effect of such a calculation on the $\frac{\mu_{p}G_{Ep}}{G_{Mp}}$ ratio from the E01-001 experiment. In addition, the effect 
of Coulomb correction by Sick and Arrington is also shown. Their correction predicts $\sim$ 50\% of what is needed to fully resolve the 
discrepancy. Their recent result \cite{afanasev05} also shows the model dependence due to the use of different GPDs. Using a different GPD
model can significantly enlarge their corrections. 


\newpage
%
\begin{itemize}
{\it
\item \textbf{Calculation by Afanasev et al. \cite{afanasev03}:}
}
\end{itemize}

This calculation is done at the quark-parton level in the double logarithm approximation. This calculation
predicts only the $\varepsilon$ and $Q^2$ dependence but not the overall magnitude, and therefore cannot be applied as 
a correction to the data without additional assumptions.
In addition, this calculation yields large nonlinearities at large $\varepsilon$ where the $\varepsilon$ dependence of the
TPE corrections behaves in the opposite direction to that of Blunden \cite{blunden03}.

\begin{itemize}
{\it
\item \textbf{Predictions by Rekalo and Tomasi-Gustafsson et al. \cite{rekalo04}:}
}
\end{itemize}

This work predicts the TPE correction based on symmetry arguments. The results show
large nonlinearities at large $\varepsilon$. They predict that TPE corrections should depend on 
$x = \sqrt{1+\varepsilon}/\sqrt{1-\varepsilon}$, but do not calculate the size or $Q^2$ dependence. 
Their calculation predicts a similar $\varepsilon$ dependence to that of Afanasev \cite{afanasev05}.
\newline
\newline

Clearly, if TPE explains the discrepancy it must be due to an $\varepsilon$ dependence modification
of the reduced cross sections and thus the overestimate of $G_{Ep}$ in the Rosenbluth separations. 
But this is not the whole story since TPE can have an effect on $G_{Mp}$ as well. 
Most of the previously described calculations predicated that the largest effect on the cross section 
would occur at small $\varepsilon$. This would reduce $\sigma_{R}$ at $\varepsilon$ = 0.0 ($\tau G^2_{Mp}$) by 3-5\% at 
large $Q^2$ with small $Q^2$ dependence. Therefore, it is crucial to know the exact $\varepsilon$ dependence
of the TPE correction and in particular the size of the nonlinearity as $\varepsilon \to $0 
in order to extract an accurate value of $G_{Mp}$.  
\subsection{Multiple Soft Photon Exchange (Coulomb Distortion)} \label{coulomb_distortion}

It has been reported that inclusion of the Coulomb distortion to the elastic e-p 
cross section can slightly reduce the discrepancy between the Rosenbluth separations and recoil 
polarization results \cite{higinbotham03,arrington04c}. Higinbotham \cite{higinbotham03} evaluated the effect
in the Effective Momentum Approximation (EMA) where the incident electron with energy $E$ approaches the nucleus 
is accelerated due to its Coulomb interaction with the nucleus. Consequently, the four momentum transfer 
$Q$ gets modified and the nucleus is probed at a slightly larger $Q$, $Q_{\mbox{Modified}}$:
\begin{equation}
Q_{\mbox{Modified}} = Q \Big(1+\frac{3}{2}\frac{Z \alpha \hbar c}{R_{eq}E}\Big)~,
\end{equation}
where $R_{eq}$ is the hard sphere equivalent radius of the proton. Therefore, $Q_{\mbox{Modified}}$ will
affect the reduced cross sections such as it will raise up the reduced cross sections more at low $\varepsilon$ values
than at high $\varepsilon$. This in turn will reduce the Rosenbluth extracted slope, $G^2_{Ep}$, and
hence lower the form factors ratio. The Coulomb distortion correction to the elastic e-p cross section exhibits a small 
$\varepsilon$ dependence of 0.5\% which is $\sim$ 10\% of the effect needed to bring the Rosenbluth and recoil polarization results 
into agreement. 

The Coulomb distortion and in particular the effect of the proton charge on the ingoing and 
outgoing electron waves has been examined using the second order Born approximation based on the 
approach used in electron-deuteron scattering \cite{arrington04c}. 
Such distortions have usually been ignored for $Z = 1$, but have a significant effect on the proton rms radius. 
An exponential charge distribution for the proton was assumed for the calculation using the fact that
$G_{Ep} \sim G_D$ where $G_D$ is the dipole form factor. The Coulomb distortion correction to the elastic e-p 
cross section exhibits an $\varepsilon$ dependence of (1.0-2.5)\% and behaves approximately like $\frac{1}{Q^2}$. The correction
is maximum near $Q^2$ = 1 GeV$^2$ and then begins to decrease by increasing $Q^2$. 
The Coulomb distortion explains a portion of the discrepancy but is a smaller effect than TPE.
At large $Q^2$, $G_{Ep}$ is suppressed and the effect of the Coulomb distortion on $G_{Ep}$ is larger than one might expect.

\subsection{Search For Nonlinearities}

\begin{figure}[!htbp]
\begin{center}
\epsfig{file=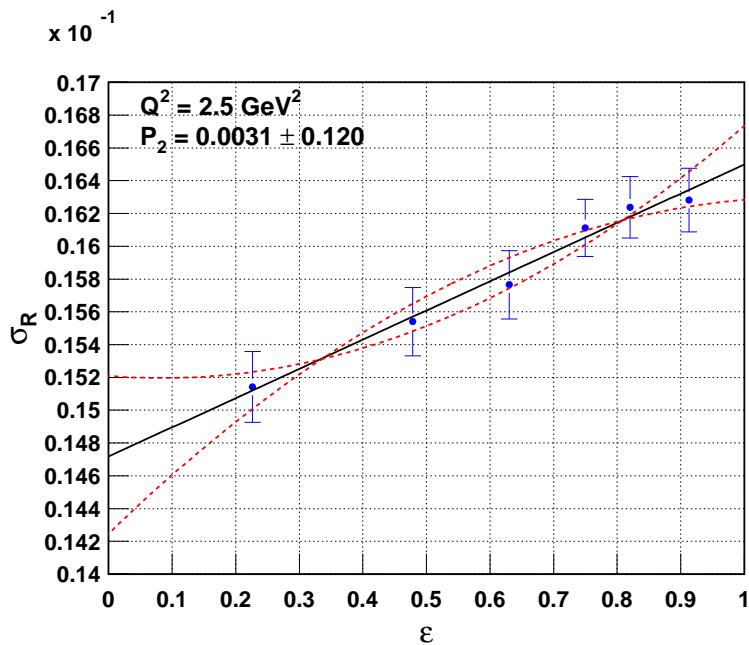,width=3.90in}
\end{center}
\caption[Search for nonlinearity in the SLAC NE11 $\sigma_{R}$ at $Q^2$ = 2.50 GeV$^2$.]
{Search for nonlinearity in the SLAC NE11 $\sigma_{R}$ at $Q^2$ = 2.50 GeV$^2$. The solid black
line is the linear fit, while the dashed red lines are quadratic fits with $P_{2} = \pm $ 0.120.}
\label{fig:2.50_lt_sigma_red_nonlinear}
\end{figure}

\begin{figure}[!htbp]
\begin{center}
\epsfig{file=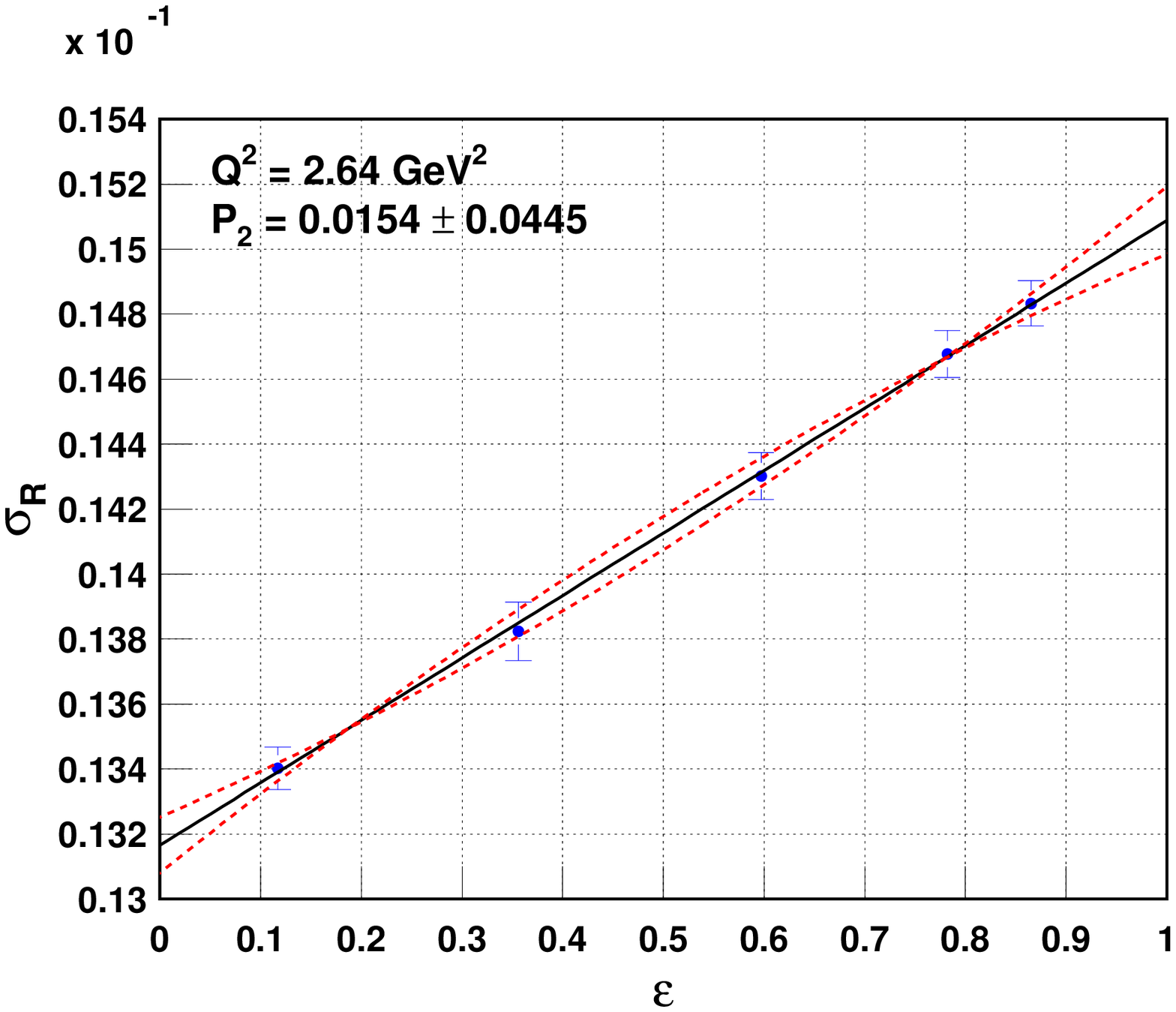,width=3.90in}
\end{center}
\caption[Search for nonlinearity in the E01-001 $\sigma_{R}$ at $Q^2$ = 2.64 GeV$^2$.]
{Search for nonlinearity in the E01-001 $\sigma_{R}$ at $Q^2$ = 2.64 GeV$^2$. The solid black
line is the linear fit, while the dashed red lines are quadratic fits with $P_{2} = \pm $ 0.0445.}
\label{fig:2.64_lt_sigma_red_nonlinear}
\end{figure}

Due to the linearity of the reduced cross section with $\varepsilon$ in the Born approximation, 
any deviation from linearity would have to come from higher-order terms that are not included
in the standard radiative correction procedures. At low $Q^2$, such a nonlinearity would
provide a clean signature of TPE and give information about the nonlinear component of TPE.
To search for a deviation from linearity in the reduced cross section, we fit the measured cross
sections to a second-order degree polynomial of the form:
\begin{equation} \label{linearity_test_equation}
\sigma_{R} = P_{0}\Big(1 + P_{1}\varepsilon + P_{2}\varepsilon^2 \Big)~,
\end{equation}  
where $P_{2}$ is the curvature parameter and provides a simple measure of the size of the 
nonlinear term relative to the cross section at $\varepsilon$ = 0.0. The uncertainty in $P_{2}$, $\delta P_{2}$, 
sets a limit on the $\varepsilon^2$ term.
Figures \ref{fig:2.50_lt_sigma_red_nonlinear} and \ref{fig:2.64_lt_sigma_red_nonlinear} show such fit done for the SLAC NE11 
\cite{andivahis94}, and E01-001 experiments at $Q^2$ = 2.50 and 2.64 GeV$^2$, respectively. In addition, the value of the extracted 
$P_{2}$ and $\delta P_{2}$ are also shown. We would like to compare the value of $\delta P_{2}$ from the two experiments at 
$Q^2$ = 2.64 GeV$^2$, however, since SLAC NE11 did not take any measurements at this $Q^2$, we had to compare to the closest 
$Q^2$ available which is $Q^2$ = 2.50 GeV$^2$. 
For the SLAC NE11, we get $P_{2}$ = 0.0031 with $\delta P_{2}$ = $\pm$0.12, while the E01-001 yields 
$P_{2}$ = 0.0154 with $\delta P_{2}$ = $\pm$0.0445 which is clearly a much better limit on $P_{2}$. 
Moreover, as $\varepsilon \to 0$, the variation of $P_{0}$ between the linear and quadratic fits can help estimate an upper 
limit on the TPE contribution to $\delta (\tau G^2_{Mp})$. 
\newpage
The E01-001 experiment provides lower uncertainty in $\tau G^2_{Mp}$ than 
SLAC NE11 which is the best previous Rosenbluth measurements. Similar analysis was done at other $Q^2$ 
values for the E01-001 experiment. Table \ref{nonlinearity_values} lists the results. 
\begin{table}[!htbp]
\begin{center}
\begin{tabular}{|c|p{2.63cm}|p{2.63cm}|p{2.63cm}|p{2.63cm}|}  
\hline \hline
Parameter              &~~~NE11        &~~~E01-001      &~~~E01-001     &~~~E01-001    \\
  &$Q^2$=2.50 GeV$^2$ &$Q^2$=2.64 GeV$^2$ &$Q^2$=3.20 GeV$^2$ &$Q^2$=4.10 GeV$^2$\\
\hline \hline
$P_2$                  &~~~~~0.003        &~~~~~~0.015        &~~~~~0.013       &~~~~~0.057          \\
$\delta P_{2}$         &~~~$\pm$0.120   &~~~$\pm$0.044   &~~~$\pm$0.056  &~~~$\pm$0.12       \\
$\delta (\tau G^2_{Mp})$ (\%) &~~~~~3.27  &~~~~~~0.87         &~~~~~1.10        &~~~~~2.19             \\ 
\hline \hline
\end{tabular}
\caption[The curvature parameter $P_{2}$, its uncertainty $\delta P_{2}$, and the uncertainty in 
$\tau G^2_{Mp}$ or $\delta (\tau G^2_{Mp})$ as extracted for the SLAC NE11 experiment at $Q^2$ = 2.50 GeV$^2$ 
and for the E01-001 experiment at $Q^2$ = 2.64, 3.20, and 4.10 GeV$^2$.]
{The curvature parameter $P_{2}$, its uncertainty $\delta P_{2}$, and the uncertainty in 
$\tau G^2_{Mp}$ or $\delta (\tau G^2_{Mp})$ as extracted for the SLAC NE11 experiment at $Q^2$ = 2.50 GeV$^2$ 
and for the E01-001 experiment at $Q^2$ = 2.64, 3.20, and 4.10 GeV$^2$ .}
\label{nonlinearity_values}
\end{center}
\end{table}

In order to be fully sensitive to nonlinearity that occurs at low $\varepsilon$, measurements of the 
reduced cross section must be made at very low $\varepsilon$. In fact, for an improved measurement of 
nonlinearities, it is crucial that one covers the maximum possible $\varepsilon$ range at both the
low and high $\varepsilon$ values, and to take enough $\varepsilon$ points in the linear region to
work as a baseline for deviations from linearity since most of TPE calculations predicted that
nonlinearities are likely to appear as $\varepsilon \to 0$ or $\varepsilon \to 1$.

\section{Future Experiments and TPE} \label{future_experiments}
\subsection{Introduction}
It has been pointed out previously that the results of the E01-001 experiment 
have clearly established and confirmed the discrepancy between the Rosenbluth separations and 
recoil polarization results. Consequently, the E01-001 results have laid down the foundation for new 
experiments aimed at measuring the size of the TPE corrections at Jefferson Lab. 

Based on the formalism of Guichon and Vanderhaeghen \cite{guichon03}, which provides the connection between
the cross section and polarization observables and goes beyond the Born approximation, 
the e-p cross section and the recoil polarization results can be expressed in terms of three 
generalized form factors or amplitudes, $\tilde{G}_{Ep}$, $\tilde{G}_{Mp}$, and $\tilde{F}_{3}$ which are complex functions  
that depend on both $Q^2$ and $\varepsilon$:
\begin{equation} \label{Vanderhaeghen_Pt}
P_{t} = -\sqrt{\frac{2\varepsilon(1-\varepsilon)}{\tau}} \frac{|\tilde{G}_{Mp}|^2}{\sigma_R}\Big(R+Y_{2\gamma}\Big)~,
\end{equation}
\begin{equation} \label{Vanderhaeghen_Pl}
P_{l} = \sqrt{(1+\varepsilon)(1-\varepsilon)} \frac{|\tilde{G}_{Mp}|^2}{\sigma_R}\Big(1+\frac{2\varepsilon}{1+\varepsilon}Y_{2\gamma}\Big)~,
\end{equation}
\begin{equation} \label{Vanderhaeghen_sigma_R}
\frac{\sigma_R}{|\tilde{G}_{Mp}|^2} = 1 + \frac{\varepsilon}{\tau}R^2 + 2\varepsilon\Big(1 + \frac{R}{\tau}\Big) Y_{2\gamma}~,
\end{equation}
where $\sigma_R$ is the reduced e-p cross section, $Y_{2\gamma}$ is a dimensionless variable replacing $\tilde{F}_{3}$ as it will be
discussed below, and $R$ is given by:
\begin{equation}
R = \frac{|\tilde{G}_{Ep}|}{|\tilde{G}_{Mp}|}~.
\end{equation}

Note that in the Born approximation, these amplitudes are real and depend only on $Q^2$ or:
$\tilde{G}_{Ep} \to G_{Ep}$, $\tilde{G}_{Mp} \to G_{Mp}$, and $\tilde{F}_{3} \to$ zero.
In order to separate the Born and higher order corrections, the generalized form factors are broken into
the Born and TPE contributions or:
\begin{equation}
\tilde{G}_{Ep}(\varepsilon,Q^2) = G_{Ep}(Q^2) + \Delta G_{Ep}(\varepsilon,Q^2)~,
\end{equation}
\begin{equation}
\tilde{G}_{Mp}(\varepsilon,Q^2) = G_{Mp}(Q^2) + \Delta G_{Mp}(\varepsilon,Q^2)~,
\end{equation}
\begin{equation}
\tilde{F}_{3}(\varepsilon,Q^2) = \tilde{F}_{3}(\varepsilon,Q^2)~.
\end{equation}

Therefore and following \cite{guichon03}, we can replace $\tilde{F}_{3}$ with a 
dimensionless variable $Y_{2\gamma}$:
\begin{equation}
Y_{2\gamma} = \mbox{Re} \Big(\frac{\nu \tilde{F}_{3}(\varepsilon,Q^2)}{M^2_{p} |\tilde{G}_{Mp}|}\Big)~,
\end{equation}
where $\nu = M^2_{p} \sqrt{(1+\varepsilon)/(1-\varepsilon)} \sqrt{\tau (1+\tau)}$. That way we have
the two real Born amplitudes or $G_{Ep}$ and $G_{Mp}$, and three two-photon amplitudes or 
$\Delta G_{Ep}$, $\Delta G_{Mp}$, and $Y_{2\gamma}$. Note that $\Delta G_{Ep}$ and $\Delta G_{Mp}$
are complex and expected to be small compared to the Born amplitudes. They interfere with the Born amplitudes, which are
real and much larger, so the imaginary part of two-photon amplitudes should be negligible. Therefore, the imaginary part 
of the two-photon amplitudes will be dropped and $\tilde{G}_{Ep}$ and $\tilde{G}_{Mp}$ will refer to the real part of the amplitudes.

In the generalized formalism, and if the TPE contributions are $\varepsilon$-independent, the measured ratio of the form factors 
can be written in terms of these generalized form factors for the recoil polarization as:
\begin{equation} \label{pol_2gamma}
R_{\mbox{Pol}} = \frac{\tilde{G}_{Ep}}{\tilde{G}_{Mp}} + \Big(1 - \frac{2\varepsilon}{1+\varepsilon} \frac{\tilde{G}_{Ep}}{\tilde{G}_{Mp}} \Big) Y_{2\gamma}~,
\end{equation}
and for the Rosenbluth separations (L-T) as:
\begin{equation} \label{LT_2gamma}
R^2_{\mbox{L-T}} = (\frac{\tilde{G}_{Ep}}{\tilde{G}_{Mp}})^2 + 2 \Big(\tau + \frac{\tilde{G}_{Ep}}{\tilde{G}_{Mp}}\Big) Y_{2\gamma}~.
\end{equation}

From equations (\ref{pol_2gamma}) and (\ref{LT_2gamma}) we see that only the $Y_{2\gamma}$
term leads to a difference between the Rosenbluth separations and recoil polarization form factor ratio
and such difference is proportional to $Y_{2\gamma}$.

\subsection{Experiment E04-019}
Experiment E04-019 \cite{e04019} will measure the $\varepsilon$ dependence of the recoil polarization extractions of 
$\frac{G_{Ep}}{G_{Mp}}$. The recoil polarization will be measured using the big Focal Plane Polarimeter (FPP) in Hall C and it will be done 
at a fixed $Q^2$ value of 3.20 GeV$^2$. The value $Q^2 =$ 3.20 GeV$^2$ is chosen to overlap the high-precision cross section measurements 
of the Super-Rosenbluth E01-001 experiment. Measurements will be taken at 4 $\varepsilon$ points and the kinematics are chosen to be the 
same as those in the Super-Rosenbluth experiment so that the high-precision cross sections measured by the Super-Rosenbluth experiment can 
be used in the analysis. Based on equations (\ref{Vanderhaeghen_Pt}), (\ref{Vanderhaeghen_Pl}), and (\ref{Vanderhaeghen_sigma_R}) and by 
using the measured values of $\frac{P_{t}}{P_{l}}$, and $\sigma_R$ at a given $\varepsilon$ point, one can solve for $\tilde{G}_{Mp}$,
$Y_{2\gamma}$, and $R$. Note that because the E04-019 experiment will measure the $\varepsilon$ dependence and not the size
of $Y_{2\gamma}$, it will not by itself provide sufficient information to correct the recoil polarization for TPE effects.
Experiment E04-019 is scheduled to run along with the coming Hall C experiment $G^p_{E} -$III of JLAB in 2006.

\subsection{Experiment E05-017}
Experiment E05-017 \cite{e05017} is a high-precision Rosenbluth separation measurement similar to the Super-Rosenbluth E01-001 experiment
where protons will be detected with an improved precision. It aims to mapping the 
$\varepsilon$ and $Q^2$ dependence of TPE contributions to the elastic e-p cross section by providing high-precision Rosenbluth 
separations for 1$\leq Q^2 \leq$6 GeV$^2$ by taking more $\varepsilon$ and $Q^2$ points over a wide range. Such high precision 
measurements will allow for precise extraction of TPE effects from the difference between Rosenbluth and recoil polarization measurements. 
Moreover, it will make improved measurements of any nonlinearities in the $\varepsilon$ dependence of the cross sections than those 
reported in this work. By providing improved Rosenbluth separations measurements of $\frac{\mu_{p}G_{Ep}}{G_{Mp}}$, E05-017 will 
determine the $Q^2$ dependence of TPE. The overall goal is to increase the precision on $\frac{\mu_{p}G_{Ep}}{G_{Mp}}$ 
by a factor of 2-3 over the entire $Q^2$ range compared to a global Rosenbluth extractions which will allow for precise comparison 
with the recoil polarization results. Experiment E05-017 is approved to run in Hall C of JLAB. 

\subsection{Experiment E04-116}
Experiment E04-116 \cite{e04116} we will measure the $\varepsilon$ dependence of the charge asymmetry 
$R = \frac{\sigma(e^+)}{\sigma(e^-)}$, where $\sigma(e^+)$ and $\sigma(e^-)$ are the elastic cross 
sections of positron- and electron-proton scattering, respectively. 
Such measurements will allow for direct determination of TPE contribution to the elastic e-p scattering
cross section and will be done for 0.5$\leq Q^2 \leq$2.0 and 0.1$\leq \varepsilon \leq$0.9.

The crucial point here is that the radiative corrections can be classified into two categories: 
those whose their sign depends on the charge of the incident lepton, and those whose their sign does not 
depend on the charge of the incident lepton. The standard radiative corrections are all independent
of the sign of the lepton except for the interference term between the electron(positron) bremsstrahlung
and proton bremsstrahlung that is regularized by the infrared term from TPE. In other words, the standard
radiative corrections developed by Mo and Tsai \cite{mo69} include only the infrared divergent contributions
from TPE diagrams, which necessarily cancel the infrared divergent contributions from the interference 
term between electron and proton bremsstrahlung, and neglect contributions from multiple soft photons
exchange (Coulomb corrections). 

The relation of $R$ to TPE and hence to the elastic e-p scattering cross section is given below: 
The amplitude of the elastic e-p scattering up to $\alpha^2$ including only amplitudes that contribute to
the charge asymmetry can be written as:
\begin{equation} \label{eq:charge_asymmetry}
A_{ep \to ep} = e_{e}e_{p}A_{Born} + e^2_{e}e_{p}A_{e-bremsst} + e_{e}e^2_{p}A_{p-bremsst} + 
e^2_{e}e^2_{p}A_{2\gamma}~,
\end{equation}
where $e_{e}$($e_{p}$) is the charge of the electron(proton) and the amplitudes 
$A_{Born}$, $A_{e-bremsst}$, $A_{p-bremsst}$, and $A_{2\gamma}$ are the Born(single-photon exchange), 
electron-bremsstrahlung, proton-bremsstrahlung, and two-photon exchange amplitudes, respectively. By squaring
equation (\ref{eq:charge_asymmetry}) and keeping only corrections up to $\alpha$ that have odd powers of electron charge, we get:
\begin{equation} \label{eq:square_charge_asymmetry}
|A_{ep \to ep}|^2 = e^2_{e}e^2_{p}\Big(|A_{Born}|^2 + 2e_{e}e_{p}A_{Born}Re(A^\star_{2\gamma}) +
2e_{e}e_{p}Re(A_{e-bremsst}A^\star_{p-bremsst})\Big)~,
\end{equation}
where $Re$ stands for the real part of the amplitude. 

The term $2e_{e}e_{p}Re(A_{e-bremsst}A^\star_{p-bremsst}$) in equation (\ref{eq:square_charge_asymmetry})
describes the interference between the electron and proton bremsstrahlung and its infrared divergence
contribution cancels exactly the corresponding infrared divergence contribution from 
$2e_{e}e_{p}A_{Born}Re(A^\star_{2\gamma})$. Therefore, after correcting the elastic e-p scattering 
cross section for this interference term, the only radiative correction that leads to a charge asymmetry 
would have to come from TPE.
The effects of TPE and Coulomb corrections have the opposite sign for
electrons and positrons or $\sigma(e^{\pm}) = \sigma_{Born}(1 \mp \delta_{2\gamma})$ where 
$\delta_{2\gamma}$ is the TPE correction yielding a charge asymmetry 
$R = \frac{\sigma(e^+)}{\sigma(e^-)} \approx 1 - 2\delta_{2\gamma}$. By plotting $R$ as a function
of $\varepsilon$, any deviation of $R$ from 1.0 is a model-independent measure of TPE in 
elastic e-p scattering. Experiment E04-116 will run in Hall B of JLAB. It is approved for engineering run to check
backgrounds with secondary beam. 

A similar $e^{+}$-$e^{-}$ comparison is proposed to run at VEPP-3. The experiment will measure  
the ratio $R = \frac{\sigma(e^+)}{\sigma(e^-)}$ at low $\varepsilon$ and moderate $Q^2$, where no precision data
exist. For more details, the experiment is described in reference \cite{vepp_proposal}.

\chapter{Summary and Conclusion} \label{chap_summary_conclusion}
\pagestyle{plain}
High precision measurements of the elastic e-p scattering cross sections 
were made at the Hall A of the Thomas Jefferson National Accelerator Facility.
The Super-Rosenbluth experiment E01-001 ran in May 2002 where an unpolarized electron beam from 
the Continuous Electron Beam Accelerator Facility in the range of 1.912$ \le E \le$ 4.702 GeV 
was used and directed on a 4-cm-long unpolarized liquid-hydrogen target. 
Protons were detected, in contrast to previous measurements where the scattered electrons were detected, 
to dramatically decrease any $\varepsilon$ dependence systematic uncertainties and corrections. 
The left arm spectrometer was used to measure three $Q^2$ points of 2.64, 3.20, and 4.10 GeV$^2$. 
Simultaneously, measurements at $Q^2$ = 0.5 GeV$^2$ were carried out using the right arm spectrometer 
which served as a luminosity monitor to remove any uncertainties due to beam charge, current, and target 
density fluctuations. A total of 12 $\varepsilon$ points (5 $\varepsilon$ points for $Q^2$ = 2.64 GeV$^2$, 
4 $\varepsilon$ points for $Q^2$ = 3.20 GeV$^2$, and 3 $\varepsilon$ points for $Q^2$ = 4.10 GeV$^2$) 
were measured covering an angular range of 12.52$^o$$< \theta_{L} <$38.26$^o$ for the left arm, while 
the right arm was at $Q^2$ = 0.5 GeV$^2$, and used to simultaneously measure 5 $\varepsilon$ points 
covering an angular range of 58.29$^o$$< \theta_{R} <$64.98$^o$, allowing for high precision L-T separation
of the proton electric and magnetic form factors and hence their ratio $\frac{\mu_{p} G_{Ep}}{G_{Mp}}$.
The measured cross sections were corrected for internal and external radiative corrections using updated
procedure based on Mo and Tsai \cite{mo69,walker94,ent01}. The absolute uncertainty (scale uncertainty) in the measured cross 
sections is approximately 3\% for both arms and with relative uncertainties, random and slope, below 1\% for the left arm, 
and below 1\% random and 6\% slope for the right arm.
The extracted form factors, $\frac{G_{Ep}}{G_{D}}$ and $\frac{G_{Mp}}{\mu_{p} G_{D}}$, were determined for the left arm
at the three $Q^2$ points to 4\%-7\% and 1.5\%, respectively.
Moreover, the ratio $\frac{\mu_{p}G_{Ep}}{G_{Mp}}$ was determined at the three $Q^2$ points to 4\%-7\%, and found
to approximate form factor scaling or $\frac{\mu_{p} G_{Ep}}{G_{Mp}} \approx$ 1.0.

The results of this work are in agreement with the previous Rosenbluth data and inconsistent with the high-$Q^2$
recoil polarization results. The E01-001 experiment provided systematic uncertainties much smaller than the 
best previous Rosenbluth measurements \cite{andivahis94}, and comparable to those of the recoil polarization, 
clearly establishing the discrepancy between the Rosenbluth separations and recoil polarization results. 
Furthermore, the high precision of the results confirmed that the discrepancy is not an experimental error in the 
Rosenbluth measurements or technique, confirmed the reliability of the elastic e-p scattering cross sections extracted 
from previous Rosenbluth separations, provided a strong test of the conventional radiative corrections used, and 
constituted a precise measurement of the discrepancy.

A possible source for the discrepancy is the effect of a missing correction to the reduced cross sections due
to two-photon exchange (TPE) corrections. Therefore, it is absolutely crucial to 
understand the radiative corrections and in particular higher-order processes such as TPE theoretically and 
experimentally before we can be confident in our knowledge of the proton form factors. 
Consequently, and motivated by the results of the E01-001 experiment, the interest in the physics of TPE
and Coulomb distortion corrections and their impact on electron scattering observables has risen experimentally
and theoretically. In addition, the results of the E01-001 experiment have laid down the foundation for 
new experiments aimed at measuring the size of the TPE corrections.\\[0.08cm]

\pagestyle{myheadings}

\emph{\textbf{``I have noticed that nobody writes a book in his day, but is certain to declare on the morrow, 
that if this were to be altered it would be better, if this were to be added it would be more desirable, 
if this were to be moved forward it would be preferable and if this were to be omitted it would be more pleasing. 
This saying is one of the wisest maxims and it is evidence that the whole human race is seized by imperfection.''}}\\
$-$ Al-Imad Al-Asfahani (Al-Katib), 1125-1200 A.D.: A historian and one of the major writers. 
He accompanied Salah Al-Din Al-Ayyubi (Saladin) and recorded his history.
\clearpage
\pagestyle{myheadings}





\bibliography{thesis}
\vfill
\pagebreak

\end{document}